\documentclass[12pt,oneside]{book} 
\usepackage[T1]{fontenc}
\usepackage[utf8]{inputenc}
\usepackage{imakeidx}
\makeindex[title={\'{I}ndice~Remissivo},intoc,noautomatic=true]
\usepackage[style=philosophy-classic,shorthandintro=false,natbib=true,backend=biber,indexing=cite]{biblatex}
\DeclareNameAlias{sortname}{family-given}
\DeclareNameAlias{default}{family-given}
\usepackage[brazil]{babel}
\usepackage{csquotes,amsmath,amsfonts,amssymb,mathrsfs,fancyhdr,setspace}
\usepackage[svgnames]{xcolor}
\usepackage[scr]{rsfso}
\usepackage[colorlinks,allcolors=MidnightBlue]{hyperref}

\usepackage{fitch}
\usepackage{imakeidx}
\usepackage{nicefrac}
\usepackage{forest}
\usepackage{tikz}
\usetikzlibrary{patterns}
\usepackage{multirow}
\usepackage{booktabs}
\usepackage{newpxtext,newpxmath}

\title{\Huge\textbf{Consci\^encia e mec\^anica qu\^antica: uma abordagem filos\'ofica}}
\author{Raoni Arroyo}
\date{%
Dipartimento di Filosofia, Comunicazione e Spettacolo, Universit\`{a} degli Studi Roma Tre. Apoio: processo n$^{\text{o}}$ 2022/15992-8, Funda\c{c}\~{a}o de Amparo \`{a} Pesquisa do Estado de S\~{a}o Paulo (FAPESP). \\Centro de L\'{o}gica, Epistemologia e Hist\'{o}ria da Ciência (CLE), Universidade Estadual de Campinas (UNICAMP)
\vfill
\today }

\begin{document}
\frenchspacing\sloppy\raggedbottom

\maketitle
\tableofcontents\onehalfspacing

\pagestyle{fancy} 
\fancyhf{}
\setlength{\headheight}{15pt}
\fancyhead[C]{\leftmark} 
\fancyfoot[C]{\thepage}
\setlength{\headheight}{15.8pt}
\renewcommand{\headrulewidth}{0pt}

\chapter*{Introdu\c{c}\~{a}o: Um problema filos\'{o}fico na f\'{i}sica}
\addcontentsline{toc}{chapter}{Introdu\c{c}\~{a}o: Um problema filos\'{o}fico na f\'{i}sica}
\label{CapIntro}
Existem v\'{a}rias maneiras de enunciar o que \'{e} a mec\^{a}nica qu\^{a}ntica.\footnote{~A ilustra\c{c}\~{a}o contida na parte superior da p\'{a}gina 210 da obra \citetitle{zahn1702}, de \citeauthor{zahn1702} \citeyearpar{zahn1702} é, a meu ver, uma boa forma de visualizar a questão. Nela, v\'{a}rios observadores têm referenciais diferentes sobre um drag\~{a}o que voa no c\'{e}u. Assim como o drag\~{a}o na ilustra\c{c}\~{a}o, a mec\^{a}nica qu\^{a}ntica pode ser \textit{vista} de diversas formas.} Se este fosse um livro de hist\'{o}ria da f\'{i}sica, eu come\c{c}aria introduzindo o advento da mec\^{a}nica qu\^{a}ntica atrav\'{e}s da teoriza\c{c}\~{a}o de Max Planck, em 1900, sobre a radia\c{c}\~{a}o de corpo negro. No entanto, como \'{e} um livro de filosofia, escolhi situar o debate atrav\'{e}s do contraste entre as concep\c{c}\~{o}es filos\'{o}ficas das f\'{i}sicas cl\'{a}ssica e qu\^{a}ntica. Para tanto, aponto muito brevemente três teses principais, t\'{a}cita ou implicitamente assumidas por aquilo que se conhece como f\'{i}sica cl\'{a}ssica:

\begin{enumerate}
    \item\textit{Previsibilidade e determinismo causal}: todo evento \'{e} necessitado por eventos anteriores, ent\~{a}o podemos: (i) conhecer todas as condi\c{c}\~{o}es iniciais dos sistemas f\'{i}sicos, e, a partir delas, (ii) prever com certeza o seu comportamento futuro a partir de uma cadeia causal.

    \item \textit{Separabilidade}: objetos f\'{i}sicos existem separadamente e independentemente uns dos outros; deste modo, caso estejam espacialmente separados, o que quer que aconte\c{c}a com um n\~{a}o pode ter efeitos imediatos sobre o outro.
    
    \item \textit{Realismo objetivista}: objetos f\'{i}sicos na realidade externa existem com propriedades bem definidas em todos os momentos, e isso independe de qualquer intera\c{c}\~{a}o com seres humanos.
\end{enumerate}

Como procurarei expor ao longo deste livro, a mec\^{a}nica qu\^{a}ntica ---ou melhor, a maneira tradicional sobre como pensar no que a mec\^{a}nica qu\^{a}ntica nos diz sobre a realidade--- nos for\c{c}a a rejeitar tais teses. \'{E} importante deixar claro logo no in\'{i}cio que o termo \textit{mec\^{a}nica qu\^{a}ntica}, conforme empregado neste livro, refere-se \`{a} mec\^{a}nica qu\^{a}ntica \textit{usual} (ou \textit{padr\~{a}o}), conforme estudada nos cursos de f\'{i}sica ao redor do mundo \citep[ver][]{griffiths1995introduction, hughes1989structure, arroyo-olegario2021}. O conceito de \textit{medi\c{c}\~{a}o} ocupa um papel central na discuss\~{a}o acerca da interpreta\c{c}\~{a}o da mec\^{a}nica qu\^{a}ntica, estando presente desde os primeiros debates ontol\'{o}gicos da teoria conduzidos, mesmo que indiretamente, pelos f\'{i}sicos Niels Bohr e Werner Heisenberg. \'{E} um dos maiores problemas filos\'{o}ficos para a quest\~{a}o interpretativa da mec\^{a}nica qu\^{a}ntica, dando \`{a} mec\^{a}nica qu\^{a}ntica diversas interpreta\c{c}\~{o}es nas quais uma ontologia pr\'{o}pria parece estar relacionada a cada uma delas.

Na f\'{i}sica cl\'{a}ssica, a medi\c{c}\~{a}o \'{e} um aspecto que pode nos parecer intuitivamente simples e relativamente pouco problem\'{a}tico ---como o ato de medir o peso de um corpo maci\c{c}o tal como uma bola de bilhar. J\'{a} na mec\^{a}nica qu\^{a}ntica, a medi\c{c}\~{a}o n\~{a}o \'{e} um conceito consensual, havendo diversas posi\c{c}\~{o}es filos\'{o}ficas conflitantes sobre seu modo de opera\c{c}\~{a}o, de modo que quest\~{o}es como ``a medi\c{c}\~{a}o \textit{cria} ou \textit{revela} o valor observado?'' permeiam o debate filos\'{o}fico sobre conceito de medi\c{c}\~{a}o.

Argumentarei, ao longo deste livro, contr\'{a}rio \`{a} pr\'{a}tica usual, que \textit{todas} as caracter\'{i}sticas problem\'{a}ticas dos fundamentos da mec\^{a}nica qu\^{a}ntica se relacionam com a no\c{c}\~{a}o de medi\c{c}\~{a}o. A pr\'{a}tica usual \'{e} considerar a medi\c{c}\~{a}o como \textit{um} problema fundacional dentre muitos outros, tais como determinismo, localidade, ontologia, etc. No entanto, argumentarei em cada cap\'{i}tulo que esses problemas s\~{a}o subsidi\'{a}rios e dependentes da no\c{c}\~{a}o de \textit{medi\c{c}\~{a}o}. Assim, me alinho com \citet{gibbins1987particles}, para quem o problema da medi\c{c}\~{a}o \'{e} \textit{o} problema central da mec\^{a}nica qu\^{a}ntica. Dessa forma, considero que problemas fundamentais da microf\'{i}sica, tais como incerteza, complementaridade, localidade, contextualidade e infla\c{c}\~{a}o ontol\'{o}gica, s\~{a}o consequências da interpreta\c{c}\~{a}o do problema da medi\c{c}\~{a}o e, portanto, consequências da interpreta\c{c}\~{a}o da mec\^{a}nica qu\^{a}ntica. Tamb\'{e}m fica impl\'{i}cito que endosso a tese de \citet{friederich2014therapeutic}, segundo a qual uma interpreta\c{c}\~{a}o da mec\^{a}nica qu\^{a}ntica se caracteriza fundamentalmente pelo fornecimento de uma solu\c{c}\~{a}o ao problema da medi\c{c}\~{a}o ---ainda que esse assunto seja debatido somente de passagem ao final do cap\'{i}tulo \ref{CapvNeumann}.

Trago um debate essencialmente filos\'{o}fico, na medida em que trato do debate acerca da natureza das entidades e processos que regem uma das teorias f\'{i}sicas com maior sucesso emp\'{i}rico da hist\'{o}ria da ciência moderna. Neste livro, busco destacar alguns dos aspectos filos\'{o}ficos centrais no debate em torno do que se conhece como problema da medi\c{c}\~{a}o qu\^{a}ntica. Procuro, especificamente, discutir sobre a introdu\c{c}\~{a}o do conceito de consciência, dentro do debate da medi\c{c}\~{a}o, como um problema essencialmente ontol\'{o}gico. \'{E} importante esclarecer que, ao inv\'{e}s de defender uma ou outra posi\c{c}\~{a}o, procuro mostrar que existe um campo para a discuss\~{a}o filos\'{o}fica na interpreta\c{c}\~{a}o da mec\^{a}nica qu\^{a}ntica e, como a discuss\~{a}o filos\'{o}fica se d\'{a} por problemas, buscarei explicitar os aspectos problem\'{a}ticos em torno da interpreta\c{c}\~{a}o do conceito de medi\c{c}\~{a}o. A estrutura do livro \'{e} a seguinte.

No primeiro cap\'{i}tulo, inicio a discuss\~{a}o por diretrizes dadas pela hist\'{o}ria da filosofia da f\'{i}sica, isto \'{e}, pela gênese do problema que seu deu a partir das formula\c{c}\~{o}es de Bohr Heisenberg sobre o ato de medir. Nesse cap\'{i}tulo, sustentarei a seguinte tese: as no\c{c}\~{o}es de \textit{incerteza} e \textit{complementaridade}, fundamentais para aquilo que se conhece como a \textit{interpreta\c{c}\~{a}o ortodoxa} da mec\^{a}nica qu\^{a}ntica, s\~{a}o moldadas no final da d\'{e}cada de 1920 por quest\~{o}es embrion\'{a}rias ao \textit{problema da medi\c{c}\~{a}o}, cuja formaliza\c{c}\~{a}o aparecer\'{a} somente anos mais tarde.

No segundo cap\'{i}tulo, trato do famoso debate entre Bohr e Einstein, enfatizando a rela\c{c}\~{a}o do debate com concep\c{c}\~{o}es filos\'{o}ficas conflitantes acerca da realidade ---e do papel da medi\c{c}\~{a}o na mec\^{a}nica qu\^{a}ntica. Nesse cap\'{i}tulo, forne\c{c}o mais elementos para endossar a tese geral deste livro: os problemas nos fundamentos da mec\^{a}nica qu\^{a}ntica s\~{a}o problemas filos\'{o}ficos, quase a sua totalidade ---todos que aqui ser\~{a}o discutidos--- relacionados \`{a} no\c{c}\~{a}o de medi\c{c}\~{a}o. Uma parte desse cap\'{i}tulo foi publicada em formato de artigo \citep{arroyo2023felinos}.

O terceiro cap\'{i}tulo \'{e} central no livro. \'{E} nele que o problema da medi\c{c}\~{a}o aparece explicitamente, ainda que eu tenha deliberadamente filtrado quest\~{o}es relativas ao formalismo da mec\^{a}nica qu\^{a}ntica, e tratado diretamente com as quest\~{o}es conceituais que envolvem o problema. Trato especificamente de uma interpreta\c{c}\~{a}o da mec\^{a}nica qu\^{a}ntica, que \'{e} extremamente mal vista pela comunidade f\'{i}sica e filos\'{o}fica: a interpreta\c{c}\~{a}o que atribui \`{a} consciência humana poder causal na medi\c{c}\~{a}o. Por um lado, \'{e} compreens\'{i}vel que essa interpreta\c{c}\~{a}o seja t\~{a}o mal vista pela comunidade acadêmica: em nome dela, foram feitas muitas deturpa\c{c}\~{o}es de maneira intelectualmente pouco honesta; por outro lado, tento mostrar como \'{e} uma interpreta\c{c}\~{a}o perfeitamente consistente, e que n\~{a}o deve ser descartada do rol de interpreta\c{c}\~{o}es dispon\'{i}veis sem justificativas adicionais. Por fim, mostro como tal interpreta\c{c}\~{a}o \'{e} somente uma, em meio \`{a} vasta gama de op\c{c}\~{o}es de interpreta\c{c}\~{o}es da mec\^{a}nica qu\^{a}ntica. Assim, qualquer senten\c{c}a que comece com ``a mec\^{a}nica qu\^{a}ntica implica que\dots'' deve ser lida com bastante cautela, especialmente no tocante a aspectos filos\'{o}ficos. Parte deste cap\'{i}tulo foi publicado no formato de artigo \citep{arroyo-sversutti-2022},\footnote{~Agrade\c{c}o ao William Sversutti por ter autorizado que eu publicasse sa partes que escrevi do referido artigo neste livro.} e partes dos outros cap\'{i}tulos compuseram minha disserta\c{c}\~{a}o de mestrado \citep{arroyo2015masters}. Minha abordagem neste cap\'{i}tulo \'{e} majoritariamente hist\'{o}rica e conceitual, priorizando a gênese das interpreta\c{c}\~{o}es da consciência causal na mec\^{a}nica qu\^{a}ntica. Alternativas recentes, como as reunidas no volume editado por \citet{gao2022}, n\~{a}o foram abordadas neste livro. Deixei-as para outra oportunidade.

No quarto cap\'{i}tulo, exploro alguns horizontes poss\'{i}veis para uma fundamenta\c{c}\~{a}o mais rigorosa dos fundamentos da consciência na mec\^{a}nica qu\^{a}ntica, tendo em vista a interpreta\c{c}\~{a}o apresentada no cap\'{i}tulo anterior. Em espec\'{i}fico, considero a possibilidade de uma fundamenta\c{c}\~{a}o filos\'{o}fica, inspirada na ontologia de processos de Alfred Whitehead. Essa alternativa ontol\'{o}gica tem a vantagem de evitar os problemas da metaf\'{i}sica dualista ---\textit{viz.} o problema mente-corpo--- ao mesmo tempo que deixa aberta a possibilidade de causa\c{c}\~{a}o mental. Desnecess\'{a}rio dizer que a estrutura ontol\'{o}gica whiteheadiana tamb\'{e}m evita uma justifica\c{c}\~{a}o baseada em crit\'{e}rios religiosos para a causalidade da consciência, como \'{e} feito por alguns exemplos discutidos no cap\'{i}tulo \ref{CapvNeumann}.

Por fim, o cap\'{i}tulo \ref{Cap:formalismo} \'{e} uma esp\'{e}cie de apêndice, no qual s\~{a}o tratadas quest\~{o}es matem\'{a}ticas m\'{i}nimas relativas ao formalismo da mec\^{a}nica qu\^{a}ntica, de modo a tornar ainda mais preciso o problema da medi\c{c}\~{a}o. O formalismo apresentado ali n\~{a}o \'{e} necess\'{a}rio para o entendimento pleno das quest\~{o}es tratadas neste livro, mas serve a uma leitura mais aprofundada, embora ainda introdut\'{o}ria, da filosofia da f\'{i}sica. Esse cap\'{i}tulo tamb\'{e}m teve parte do seu conte\'{u}do publicado em formato de artigo \citep{arroyo2023felinos}.

\chapter{Quest\~{o}es de fundamento}
\label{CapCopenhague}

\noindent Uma caracter\'{i}stica not\'{a}vel da mec\^{a}nica qu\^{a}ntica n\~{a}o-relativista (doravante apenas \textit{mec\^{a}nica qu\^{a}ntica}) \'{e} sua quest\~{a}o interpretativa. \'{E} poss\'{i}vel interpretar a mec\^{a}nica qu\^{a}ntica de diversas maneiras. As diferen\c{c}as interpretativas, por sua vez, se mostram de diversos modos: podem ser estruturais, modificando, por exemplo, axiomas da teoria ou equa\c{c}\~{o}es de movimento; podem ser substanciais, na medida em que alteram o pr\'{o}prio objeto de estudo da f\'{i}sica; e tamb\'{e}m podem ser ontol\'{o}gicas, na medida em que diferen\c{c}as interpretativas podem significar diferen\c{c}as de concep\c{c}\~{o}es sobre como o mundo \'{e} e quais s\~{a}o as entidades que o comp\~{o}em.

As fronteiras entre a f\'{i}sica e a filosofia, e tamb\'{e}m entre teoria e interpreta\c{c}\~{a}o, se tornam borradas quando nos deparamos com os fundamentos da mec\^{a}nica qu\^{a}ntica. Seja como for, qualquer abordagem interpretativa tem um ponto de partida. De uma perspectiva da hist\'{o}ria da filosofia e da f\'{i}sica, o ponto de partida para as quest\~{o}es interpretativas tem um nome: a interpreta\c{c}\~{a}o de Copenhague. 

Por isso, acredito que seja um bom lugar para come\c{c}ar esta investiga\c{c}\~{a}o. Como veremos ao longo deste livro, todas as interpreta\c{c}\~{o}es analisadas aqui têm como ponto de partida, direta ou indiretamente, a interpreta\c{c}\~{a}o de Copenhague. Seja pelos experimentos mentais, ou pelas quest\~{o}es filos\'{o}ficas levantadas por dois dos fundadores da mec\^{a}nica qu\^{a}ntica: os f\'{i}sicos Niels Bohr\index{Bohr, Niels} e Werner Heisenberg\index{Heisenberg, Werner}.

Analiso separadamente as formula\c{c}\~{o}es de Heisenberg\index{Heisenberg, Werner} e Bohr\index{Bohr, Niels}, tentando delinear, da forma mais precisa quanto poss\'{i}vel, a defini\c{c}\~{a}o dos principais conceitos de tais autores, que abordam, respectivamente, o princ\'{i}pio da indetermina\c{c}\~{a}o\index{Heisenberg, Werner!princ\'{i}pio da indetermina\c{c}\~{a}o} e a complementaridade\index{Bohr, Niels!princ\'{i}pio da complementaridade}. Em seguida, discuto, tamb\'{e}m, sobre algumas das diferen\c{c}as filos\'{o}ficas fundamentais entre os dois autores que comp\~{o}em o cerne da interpreta\c{c}\~{a}o de Copenhague da mec\^{a}nica qu\^{a}ntica ---deixando de lado a discuss\~{a}o de outros autores, n\~{a}o menos importantes, como Born, Dirac, Hermann, Pauli, entre outros; para uma abordagem mais completa da hist\'{o}ria dos fundamentos da mec\^{a}nica qu\^{a}ntica, ver \citet{Jammer1974,becker2018}.

Em diversos manuais e livros did\'{a}ticos de f\'{i}sica, a mec\^{a}nica qu\^{a}ntica \'{e} exposta sob a \'{o}tica da interpreta\c{c}\~{a}o de Copenhague ---\textit{e.g.} em \citet*{schiff1949textbookQM,dicke1960QMtextbook, textbookQM1, textbookQM2}. Trata-se de uma interpreta\c{c}\~{a}o que, supostamente, adv\'{e}m diretamente das formula\c{c}\~{o}es de Bohr\index{Bohr, Niels} e Heisenberg\index{Heisenberg, Werner}, e \'{e} at\'{e} mesmo considerada a interpreta\c{c}\~{a}o ortodoxa da mec\^{a}nica qu\^{a}ntica. A no\c{c}\~{a}o de uma interpreta\c{c}\~{a}o unit\'{a}ria da mec\^{a}nica qu\^{a}ntica, chamada de ``interpreta\c{c}\~{a}o de Copenhague'', de acordo com \citet{howard2004invented}, fora introduzida por Heisenberg\index{Heisenberg, Werner}. At\'{e} os anos 1950, segundo \citet[p.~680]{howard2004invented}, existia apenas um chamado ``esp\'{i}rito de Copenhague'', que representaria ``\textelp{} um grupo de pensadores unidos pela determina\c{c}\~{a}o de defender a mec\^{a}nica qu\^{a}ntica como uma teoria completa e correta''; \citet{desp1999concep} considera a interpreta\c{c}\~{a}o de Copenhague uma ferramenta pr\'{a}tica para a solu\c{c}\~{a}o de problemas da f\'{i}sica qu\^{a}ntica. Para que possamos discutir com a literatura especializada, chamo de interpreta\c{c}\~{a}o de Copenhague a ado\c{c}\~{a}o dos pontos de vista do princ\'{i}pio da incerteza\index{Heisenberg, Werner!princ\'{i}pio da incerteza|seealso{princ\'{i}pio da indetermina\c{c}\~{a}o}} e da complementaridade\index{Bohr, Niels!princ\'{i}pio da complementaridade} ---conceitos que ser\~{a}o explicados adiante.

Jamais existiu consenso sobre uma interpreta\c{c}\~{a}o unit\'{a}ria da mec\^{a}nica qu\^{a}ntica e/ou suas implica\c{c}\~{o}es filos\'{o}ficas. Exemplo disso \'{e} o fato de que os pr\'{o}prios te\'{o}ricos fundadores da mec\^{a}nica qu\^{a}ntica, como Heisenberg\index{Heisenberg, Werner} e Bohr\index{Bohr, Niels}, frequentemente divergiam em quest\~{o}es filos\'{o}ficas, como procuro expor ao final deste cap\'{i}tulo. Ainda assim, conforme observa \citet{beller1996conceptual}, os dois f\'{i}sicos deliberadamente ocultariam suas diferen\c{c}as em nome de uma interpreta\c{c}\~{a}o unit\'{a}ria de Copenhague.

\'{E} preciso salientar que a mec\^{a}nica qu\^{a}ntica, estritamente falando, n\~{a}o oferece uma vis\~{a}o de mundo ou uma ontologia. A interpreta\c{c}\~{a}o de Copenhague considera que a mec\^{a}nica qu\^{a}ntica seja meramente um conjunto de regras para fazer predi\c{c}\~{o}es sobre tipos especiais de condi\c{c}\~{o}es experimentais. No entanto, considero que \'{e} poss\'{i}vel extrair uma ontologia associada \`{a} investiga\c{c}\~{a}o da mec\^{a}nica qu\^{a}ntica. Portanto, tratarei de ontologia mesmo que os proponentes da teoria n\~{a}o o tenham feito explicitamente.

\'{E} igualmente importante ressaltar que, por mais que a mec\^{a}nica qu\^{a}ntica apresente diversos problemas filos\'{o}ficos, sua capacidade de predi\c{c}\~{a}o \'{e} bastante grande, atingindo dezenas de casas decimais de precis\~{a}o \citep[ver][]{dacosta2019notasdeaula, friederich2014therapeutic}. Isto \'{e}, trata-se de uma teoria muito bem sucedida em termos da concord\^{a}ncia de suas predi\c{c}\~{o}es com resultados experimentais. Inegavelmente a mec\^{a}nica qu\^{a}ntica \textit{funciona para todos os prop\'{o}sitos pr\'{a}ticos}. Dito isso, passemos ao debate conceitual acerca da mec\^{a}nica qu\^{a}ntica.

\section{O princ\'{i}pio da indetermina\c{c}\~{a}o}

O famoso ``princ\'{i}pio da incerteza\index{Heisenberg, Werner!princ\'{i}pio da incerteza|seealso{princ\'{i}pio da indetermina\c{c}\~{a}o}}'' foi formulado por \citet{heisenberg1927uncert}\index{Heisenberg, Werner}. \'{E} um dos pontos centrais ---e mais famosos--- daquilo que se entende por interpreta\c{c}\~{a}o de Copenhague, sendo um dos aspectos que diferenciam radicalmente a f\'{i}sica cl\'{a}ssica da f\'{i}sica qu\^{a}ntica. Ademais, como j\'{a} disse anteriormente, veremos, ao longo deste livro, que todas as caracter\'{i}sticas que diferenciam radicalmente as f\'{i}sicas cl\'{a}ssica e qu\^{a}ntica se relacionam com o conceito de medi\c{c}\~{a}o.

De acordo com \citet[p.~65]{Jammer1974}, quando teve acesso ao manuscrito do (ainda n\~{a}o publicado) artigo de \citet{heisenberg1927uncert}\index{Heisenberg, Werner}, \citet{bohr1928quantumpostulate}\index{Bohr, Niels} teria apresentado uma s\'{e}rie de cr\'{i}ticas acerca da base conceitual sob a qual as rela\c{c}\~{o}es foram formuladas, ainda que a validade das rela\c{c}\~{o}es de Heisenberg\index{Heisenberg, Werner} ---ou seja, sua existência--- n\~{a}o fosse questionada. Nesta se\c{c}\~{a}o, \'{e} delineada, de acordo com a posi\c{c}\~{a}o de Heisenberg\index{Heisenberg, Werner}, uma defini\c{c}\~{a}o t\~{a}o precisa quanto poss\'{i}vel para o princ\'{i}pio da incerteza\index{Heisenberg, Werner!princ\'{i}pio da incerteza|seealso{princ\'{i}pio da indetermina\c{c}\~{a}o}}.

Grosso modo, o princ\'{i}pio da incerteza\index{Heisenberg, Werner!princ\'{i}pio da incerteza|seealso{princ\'{i}pio da indetermina\c{c}\~{a}o}} postula a impossibilidade de atribuir valores exatos para certas propriedades observ\'{a}veis dos objetos qu\^{a}nticos, tais como \textit{posi\c{c}\~{a}o} e \textit{momento}  (\textit{momentum}), simultaneamente, de modo que tal atribui\c{c}\~{a}o deva obedecer uma quantidade constante de \textit{incerteza}. Essa \'{e} a defini\c{c}\~{a}o paradigm\'{a}tica do princ\'{i}pio, encontrada frequentemente em manuais e livros did\'{a}ticos de mec\^{a}nica qu\^{a}ntica e representada sob a forma da seguinte desigualdade: \begin{equation}\Delta p \Delta q \ge \hbar/2\pi\end{equation} (em que ``$q$'' e ``$p$'' representam os desvios padr\~{a}o, isto \'{e}, as propriedades observ\'{a}veis e ``$\hbar$'' representa a constante reduzida de Planck). Essa apresenta\c{c}\~{a}o matem\'{a}tica do princ\'{i}pio \'{e} referente \`{a} formula\c{c}\~{a}o de \citet{kennard1927}. Muito embora a formula\c{c}\~{a}o conceitual tenha partido de \citet{heisenberg1927uncert}\index{Heisenberg, Werner}, a primeira formula\c{c}\~{a}o matem\'{a}tica do princ\'{i}pio deve-se a \citet{kennard1927} ---que menciona Heisenberg\index{Heisenberg, Werner} em seu artigo. Para mais detalhes hist\'{o}ricos, ver \citet[\S~2.3]{sep-qt-uncertainty}. As vari\'{a}veis ``tempo'' e ``energia'' podem igualmente expressar o argumento, sendo tamb\'{e}m observ\'{a}veis. No entanto, manterei o racioc\'{i}nio com os observ\'{a}veis ``posi\c{c}\~{a}o'' e ``momento'', frequentemente expressos sob a forma dos caracteres $q$ e $p$, respectivamente. O termo ``posi\c{c}\~{a}o'' \'{e} uma propriedade observ\'{a}vel que designa, como o nome intuitivamente sugere, a posi\c{c}\~{a}o de um objeto qu\^{a}ntico em movimento; o termo ``momento'' pode ser entendido como uma propriedade observ\'{a}vel que designa a dire\c{c}\~{a}o ou a velocidade do movimento de um objeto qu\^{a}ntico.

Duas quest\~{o}es surgem imediatamente: 

\begin{itemize}
    \item Quanto ao primeiro termo: o ``princ\'{i}pio da incerteza\index{Heisenberg, Werner!princ\'{i}pio da incerteza|seealso{princ\'{i}pio da indetermina\c{c}\~{a}o}}'' \'{e}, de fato, um \textit{princ\'{i}pio} da teoria qu\^{a}ntica?
    \item Quanto ao segundo termo: o ``princ\'{i}pio'' se refere a uma tese \textit{epistemol\'{o}gica} (de fato ``princ\'{i}pio da incerteza\index{Heisenberg, Werner!princ\'{i}pio da incerteza|seealso{princ\'{i}pio da indetermina\c{c}\~{a}o}}'') ou a uma tese \textit{ontol\'{o}gica} (como ``princ\'{i}pio da indetermina\c{c}\~{a}o\index{Heisenberg, Werner!princ\'{i}pio da indetermina\c{c}\~{a}o}'')?
\end{itemize}

Discuto, adiante, o que implica levar em considera\c{c}\~{a}o uma referência epistemol\'{o}gica ou ontol\'{o}gica. Para uma abordagem acerca da primeira quest\~{a}o, \'{e} necess\'{a}rio distinguir entre as ``rela\c{c}\~{o}es de incerteza'' e o ``princ\'{i}pio da incerteza\index{Heisenberg, Werner!princ\'{i}pio da incerteza|seealso{princ\'{i}pio da indetermina\c{c}\~{a}o}}''. Segundo \citet{pessoa2003conceitos}, cabe a seguinte distin\c{c}\~{a}o entre os dois termos:

\begin{quote}
O princ\'{i}pio [da incerteza], que se aplica a grandezas n\~{a}o compat\'{i}veis entre si \textelp{}, exprime o fato de que uma maior previsibilidade nos resultados da medi\c{c}\~{a}o de um dos observ\'{a}veis implica uma diminui\c{c}\~{a}o na previsibilidade do outro. Uma rela\c{c}\~{a}o de incerteza \'{e} qualquer rela\c{c}\~{a}o matem\'{a}tica que exprima quantitativamente o princ\'{i}pio. \citep[p.~77]{pessoa2003conceitos}.
\end{quote}

Na f\'{i}sica cl\'{a}ssica, todas as grandezas s\~{a}o compat\'{i}veis, o que n\~{a}o acontece na mec\^{a}nica qu\^{a}ntica. As rela\c{c}\~{o}es de incerteza s\~{a}o consequências do formalismo da mec\^{a}nica qu\^{a}ntica. De fato, essa \'{e} uma das cr\'{i}ticas tecidas por  \citet{popper1967qmobserver} em rela\c{c}\~{a}o ao princ\'{i}pio da incerteza\index{Heisenberg, Werner!princ\'{i}pio da incerteza|seealso{princ\'{i}pio da indetermina\c{c}\~{a}o}}: as rela\c{c}\~{o}es n\~{a}o poderiam alcan\c{c}ar o status de princ\'{i}pio da teoria qu\^{a}ntica por uma quest\~{a}o de prioridade l\'{o}gica. As rela\c{c}\~{o}es s\~{a}o derivadas da pr\'{o}pria teoria qu\^{a}ntica, de modo que seria imposs\'{i}vel fazer o caminho inverso e obter a teoria qu\^{a}ntica a partir das rela\c{c}\~{o}es de incerteza.

Para \citet[p.~13]{reichenbach144philfoundQM}, no entanto, o princ\'{i}pio \'{e} uma ``afirma\c{c}\~{a}o emp\'{i}rica''. Assim, a quest\~{a}o em torno da utiliza\c{c}\~{a}o ou n\~{a}o das rela\c{c}\~{o}es de incerteza sob o nome de ``princ\'{i}pio'' deveria se dar no sentido emp\'{i}rico do termo, na medida em que as rela\c{c}\~{o}es s\~{a}o apresentadas originalmente como um resultado experimental, ainda que formulada a partir de um experimento mental, como veremos a seguir. Da forma como interpretam \citet{sep-qt-uncertainty}, Heisenberg\index{Heisenberg, Werner} expressaria que rela\c{c}\~{o}es de incerteza seriam um princ\'{i}pio fundamental da natureza, isto \'{e}, imposto como uma lei emp\'{i}rica, ao inv\'{e}s de ser tomado como um resultado derivado do formalismo da teoria.

O princ\'{i}pio da incerteza\index{Heisenberg, Werner!princ\'{i}pio da incerteza|seealso{princ\'{i}pio da indetermina\c{c}\~{a}o}} \'{e} uma interpreta\c{c}\~{a}o agregada \`{a}s rela\c{c}\~{o}es (matem\'{a}ticas) de incerteza, frequentemente associada \`{a}quilo que se entende por interpreta\c{c}\~{a}o de Copenhague. De acordo com \citet{cassidy1998indeterminacy}, Heisenberg\index{Heisenberg, Werner} nunca teria endossado o ponto de vista de que suas rela\c{c}\~{o}es fossem de fato um princ\'{i}pio da mec\^{a}nica qu\^{a}ntica. Segundo o autor, para designar o argumento expresso atrav\'{e}s do suposto princ\'{i}pio da incerteza\index{Heisenberg, Werner!princ\'{i}pio da incerteza|seealso{princ\'{i}pio da indetermina\c{c}\~{a}o}}, como ficara popularmente conhecido, Heisenberg\index{Heisenberg, Werner} utilizava os termos ``rela\c{c}\~{o}es de imprecis\~{a}o'' (\textit{inaccuracy relations}, \textit{Ungenauigkeitsrelationen}) ou ``rela\c{c}\~{o}es de indetermina\c{c}\~{a}o'' (\textit{indeterminacy relations}, \textit{Unbestimmtheitsrelationen}).

Como n\~{a}o entrarei aqui na discuss\~{a}o relativa ao formalismo da teoria qu\^{a}ntica, a discuss\~{a}o que se seguir\'{a}, para o escopo desta obra, ser\'{a} relativa \`{a}quilo que se refere ao princ\'{i}pio. Adoto, por ora, a nomenclatura ``rela\c{c}\~{o}es de Heisenberg\index{Heisenberg, Werner}'' (ou somente as ``rela\c{c}\~{o}es'') para me referir ao que fora chamado at\'{e} aqui de princ\'{i}pio da incerteza\index{Heisenberg, Werner!princ\'{i}pio da incerteza|seealso{princ\'{i}pio da indetermina\c{c}\~{a}o}}. Desse modo, n\~{a}o me comprometerei ---ao menos de antem\~{a}o--- com alguma interpreta\c{c}\~{a}o, como as explicitadas acima.

A tentativa de responder \`{a} segunda quest\~{a}o esbarra na dificuldade de n\~{a}o haver uma \'{u}nica terminologia, na medida em que n\~{a}o existe um consenso para a interpreta\c{c}\~{a}o das rela\c{c}\~{o}es. Para uma melhor compreens\~{a}o do significado das rela\c{c}\~{o}es de Heisenberg\index{Heisenberg, Werner}, examinarei o racioc\'{i}nio do pr\'{o}prio autor. O t\'{i}tulo do artigo de 1927, no qual as rela\c{c}\~{o}es s\~{a}o formuladas, parcialmente traduzido para o português, seria: ``Sobre o conte\'{u}do `\textit{anschaulich}' da teoria qu\^{a}ntica cinem\'{a}tica e mec\^{a}nica''. De acordo com \citet{sep-qt-uncertainty}, o termo ``\textit{anschaulich}'' merece aten\c{c}\~{a}o especial. \'{E} uma palavra pr\'{o}pria da l\'{i}ngua alem\~{a}, cuja tradu\c{c}\~{a}o para outros idiomas \'{e} frequentemente amb\'{i}gua, de modo que a express\~{a}o ``conte\'{u}do \textit{anschaulich}'' tem diversas tradu\c{c}\~{o}es.

No volume organizado por Wheeler e Zurek, o t\'{i}tulo do artigo de \citet{heisenberg1927uncert}\index{Heisenberg, Werner} fora traduzido para o inglês como ``\textit{the physical content}'' (``o conte\'{u}do f\'{i}sico''); \citet{cassidy1992bioheisenb}, bi\'{o}grafo de Heisenberg\index{Heisenberg, Werner}, traduziu como ``\textit{the perceptible content}'' (``o conte\'{u}do percept\'{i}vel''). A tradu\c{c}\~{a}o literal mais aproximada seria ``conte\'{u}do visualiz\'{a}vel'', sendo a vis\~{a}o \'{e} frequentemente utilizada como uma met\'{a}fora para o entendimento da quest\~{a}o proposta. Esse \'{e} um motivo pelo qual parte da literatura entende a palavra ``\textit{anschaulich}'' como referindo-se \`{a} intuitividade \citep[ver][]{deronde2019probing}, muitas vezes ---e aqui voltamos \`{a} tradu\c{c}\~{a}o literal--- intercambi\'{a}vel com \textit{visualizabilidade} \citep[para uma abordagem relacionando visualizabilidade e inteligibilidade na mec\^{a}nica qu\^{a}ntica, ver][\S~7]{deregt2017}.  \citet{sep-qt-uncertainty} sugerem a tradu\c{c}\~{a}o ``conte\'{u}do intelig\'{i}vel''. Para \citet[p.~64]{heisenberg1927uncert}\index{Heisenberg, Werner}, o que garante \textit{anschaulich} a um conceito f\'{i}sico \'{e} sua correspondência biun\'{i}voca com uma opera\c{c}\~{a}o experimental especificamente designada para a aplica\c{c}\~{a}o de tal conceito. Assim, fica claro que a palavra ``\textit{anschaulich}'' n\~{a}o se refere a um conte\'{u}do puramente intelig\'{i}vel, que poderia ser entendido como um conte\'{u}do puramente conceitual, sem correspondente experimental. Desse modo, uma sugest\~{a}o \'{e} que a express\~{a}o tenha um significado mais pr\'{o}ximo ao ``conte\'{u}do manifesto'', da forma como enuncia atrav\'{e}s da seguinte passagem:

\begin{quote}
Quando algu\'{e}m quiser ter clareza sobre o que se deve entender pelas palavras ``posi\c{c}\~{a}o do objeto'', como, por exemplo, do el\'{e}tron (relativamente a um dado referencial), \'{e} preciso especificar experimentos definidos com o aux\'{i}lio dos quais se pretenda medir a ``posi\c{c}\~{a}o do el\'{e}tron''; caso contr\'{a}rio, a express\~{a}o n\~{a}o ter\'{a} significado. \citep[p.~64]{heisenberg1927uncert}\index{Heisenberg, Werner}.
\end{quote}

Em outras palavras, se trata de um postulado que declara que apenas as propriedades que forem \textit{em princ\'{i}pio observ\'{a}veis} devem se inserir na teoria. Para fins heur\'{i}sticos dessa exposi\c{c}\~{a}o, \'{e} com esse significado em mente que procederei daqui em diante. Tal atitude fora identificada como uma posi\c{c}\~{a}o operacionista dos conceitos f\'{i}sicos, frequentemente associada ao positivismo/empirismo l\'{o}gico.\footnote{~N\~{a}o \'{e} incomum encontrar proponentes do positivismo que sejam proponentes do operacionismo, motivo pelo qual as duas posturas foram consideradas como referindo-se \`{a} mesma e \'{u}nica postura, ou por vezes muito similares; ver \citet{sep-operationalism} para uma discuss\~{a}o panor\^{a}mica sobre o operacionismo como uma postura distinta do positivismo.} Ao mencionar o termo ``positivismo'', tem-se em mente, principalmente, a defesa dos aspectos empiricista e verificacionista da ciência, segundo os quais a experiência (ou a medi\c{c}\~{a}o) \'{e} condi\c{c}\~{a}o necess\'{a}ria para a formula\c{c}\~{a}o de enunciados cient\'{i}ficos. Tais termos ser\~{a}o discutidos no cap\'{i}tulo seguinte. Adoto, a partir daqui, a nomenclatura proposta por \citet[p.~74]{pessoa2003conceitos}, de ``postulado operacionista'' para me referir \`{a} passagem citada acima.

Para exemplificar esse postulado, \citet[p.~64]{heisenberg1927uncert}\index{Heisenberg, Werner} introduz um experimento de pensamento ---posteriormente conhecido como ``microsc\'{o}pio de Heisenberg\index{Heisenberg, Werner!microsc\'{o}pio de}''--- no qual se objetiva efetuar uma medi\c{c}\~{a}o de posi\c{c}\~{a}o sobre um el\'{e}tron a partir de um microsc\'{o}pio de raios $\gamma$ (gama). Os raios gama têm o menor comprimento de onda conhecido at\'{e} ent\~{a}o do espectro luminoso. A ideia de utiliz\'{a}-los para iluminar o el\'{e}tron vem de uma propriedade matem\'{a}tica do processo de tal medi\c{c}\~{a}o, segundo a qual se obt\'{e}m maior precis\~{a}o quanto menor for o comprimento de onda da luz que iluminar\'{a} o el\'{e}tron. Ent\~{a}o, para efetuar uma medi\c{c}\~{a}o, seria preciso iluminar o el\'{e}tron. No entanto, a tentativa de iluminar um el\'{e}tron, e assim medir sua posi\c{c}\~{a}o, deve envolver ao menos um f\'{o}ton, cuja intera\c{c}\~{a}o com o el\'{e}tron pode ser considerada uma colis\~{a}o de modo a implicar uma perturba\c{c}\~{a}o no momento do el\'{e}tron ---dist\'{u}rbio que \'{e} maior quando menor for o comprimento de onda da luz que colide com o el\'{e}tron--- e isso limitaria a precis\~{a}o do conhecimento sobre tal momento. Esse fenômeno \'{e} conhecido como ``efeito Compton''. Para um detalhamento f\'{i}sico-te\'{o}rico desse fenômeno, ver \citet[p.~8]{chibeni2005incerteza}. Com tal racioc\'{i}nio, Heisenberg\index{Heisenberg, Werner} \'{e} capaz de afirmar que:

\begin{quote}
No instante de tempo em que a posi\c{c}\~{a}o \'{e} determinada, isto \'{e}, no instante em que o f\'{o}ton \'{e} disperso pelo el\'{e}tron, o el\'{e}tron sofre uma mudan\c{c}a descont\'{i}nua no momento. Essa mudan\c{c}a \'{e} maior \textelp{} quanto mais exata for a determina\c{c}\~{a}o da posi\c{c}\~{a}o. No instante em que a posi\c{c}\~{a}o do el\'{e}tron \'{e} conhecida, seu momento poder\'{a} ser conhecido apenas por magnitudes que correspondam a essa mudan\c{c}a descont\'{i}nua; assim, quanto mais precisamente for determinada a posi\c{c}\~{a}o, menos precisamente o momento \'{e} conhecido, e vice-versa. \citep[p.~64]{heisenberg1927uncert}\index{Heisenberg, Werner}.
\end{quote}

Essa \'{e} a primeira formula\c{c}\~{a}o das rela\c{c}\~{o}es de Heisenberg\index{Heisenberg, Werner}, que implicam, \`{a} primeira vista, uma tese epistemol\'{o}gica, na medida em que se relaciona com uma limita\c{c}\~{a}o do conhecimento acerca dos valores observ\'{a}veis. Tal formula\c{c}\~{a}o induz a uma conclus\~{a}o preliminar acerca de uma dr\'{a}stica ruptura entre os conceitos ``cl\'{a}ssico'' e ``qu\^{a}ntico'': os conceitos (tais como posi\c{c}\~{a}o e momento) teriam, na teoria f\'{i}sica cl\'{a}ssica, defini\c{c}\~{o}es exatas (isto \'{e}, limitadas somente pela imprecis\~{a}o dos instrumentos de medida), o que n\~{a}o acontece na f\'{i}sica qu\^{a}ntica, visto que os conceitos agora obedecem a uma limita\c{c}\~{a}o imposta pela opera\c{c}\~{a}o experimental, impedindo, assim, que a ``defini\c{c}\~{a}o'' dos conceitos seja simultaneamente exata.

Uma tese sem\^{a}ntica est\'{a} impl\'{i}cita aqui. Como observam \citet{sep-qt-uncertainty}, o postulado operacionista especifica que um experimento garante significado a um conceito tal como ``posi\c{c}\~{a}o'', de modo que a atitude de, por exemplo, ``efetuar uma medi\c{c}\~{a}o de posi\c{c}\~{a}o sobre um el\'{e}tron'' acaba por atribuir significado \`{a} posi\c{c}\~{a}o do objeto qu\^{a}ntico em quest\~{a}o. A formula\c{c}\~{a}o das rela\c{c}\~{o}es de Heisenberg\index{Heisenberg, Werner} parece indicar, para al\'{e}m do que se pode conhecer acerca dos observ\'{a}veis, uma limita\c{c}\~{a}o acerca do que se pode dizer dos conceitos f\'{i}sicos em dada opera\c{c}\~{a}o experimental. Assim, os autores prop\~{o}em o uso da nomenclatura ``princ\'{i}pio de medi\c{c}\~{a}o=significado''.

No entanto, \citet[p.~73]{heisenberg1927uncert}\index{Heisenberg, Werner} exibe uma segunda formula\c{c}\~{a}o das rela\c{c}\~{o}es, de car\'{a}ter ontol\'{o}gico, quando afirma: ``acredito que se possa formular proveitosamente a origem da [no\c{c}\~{a}o de] `\'{o}rbita' cl\'{a}ssica da seguinte maneira: a `\'{o}rbita' passa a existir somente quando a observamos''. De acordo com tal formula\c{c}\~{a}o, a medi\c{c}\~{a}o n\~{a}o apenas garante significado para uma propriedade observ\'{a}vel de um objeto qu\^{a}ntico, mas, de fato, garante realidade f\'{i}sica para tal conceito. \citet{sep-qt-uncertainty} prop\~{o}em, para esse racioc\'{i}nio, o uso da nomenclatura ``princ\'{i}pio de \textit{medi\c{c}\~{a}o=cria\c{c}\~{a}o}'' ---que, como discutirei adiante, \citet{heisen1958physphil}\index{Heisenberg, Werner} afirma posteriormente que n\~{a}o se trataria de uma cria\c{c}\~{a}o, mas de uma atualiza\c{c}\~{a}o de potencialidades, remetendo aos conceitos de ``ato'' e ``potência'' dos anal\'{i}ticos posteriores de \citet[\S 99b28--29]{aristotelesorganon}. Para uma an\'{a}lise aprofundada do conceito de ``\textit{potentia}'' em Heisenberg\index{Heisenberg, Werner}, ver \citet{pangle2014heisenont,koznjak2020}.

De acordo com o quadro conceitual exposto acima, a medi\c{c}\~{a}o dos observ\'{a}veis (no caso, posi\c{c}\~{a}o e momento) parece proceder da seguinte maneira: quando a posi\c{c}\~{a}o \'{e} medida pelo princ\'{i}pio de medi\c{c}\~{a}o=significado, pode-se atribuir significado epistemol\'{o}gico ao conceito f\'{i}sico ``posi\c{c}\~{a}o do el\'{e}tron''; al\'{e}m disso, pelo princ\'{i}pio de \textit{medi\c{c}\~{a}o=cria\c{c}\~{a}o}, pode-se atribuir realidade f\'{i}sica \`{a} no\c{c}\~{a}o de posi\c{c}\~{a}o, tal que a rela\c{c}\~{a}o de incerteza impossibilitaria a medi\c{c}\~{a}o simult\^{a}nea do outro observ\'{a}vel (o momento) com uma precis\~{a}o arbitrariamente grande. Deve-se notar que a defini\c{c}\~{a}o de algumas das propriedades observ\'{a}veis (nesse exemplo, o momento) s\~{a}o imprecisas num sentido ontol\'{o}gico (de acordo com o princ\'{i}pio de \textit{medi\c{c}\~{a}o=cria\c{c}\~{a}o}), de modo que s\'{o} se pode atribuir \`{a} realidade do el\'{e}tron um momento impreciso. Aqui poderia caber uma obje\c{c}\~{a}o, \textit{viz.}, que a ideia da ``existência'' de posi\c{c}\~{a}o, momento, ou trajet\'{o}ria n\~{a}o \'{e} uma tese ontol\'{o}gica forte de ``existência no mundo'', mas uma ideia mais relativizada, \textit{i.e.}, ``existência na teoria''. Essa discuss\~{a}o ser\'{a} brevemente retomada no cap\'{i}tulo \ref{CapEPR}; no entanto j\'{a} posso adiantar que entendo ambos os sentidos como ontol\'{o}gicos. Para uma discuss\~{a}o mais aprofundada, ver \citet{arenhartarroyo2021manu,arroyodasilva2022}. Ainda assim, nosso alvo \'{e} a interpreta\c{c}\~{a}o epistêmica do princ\'{i}pio, conforme defendida por \textit{e.g.} \citet{chibeni2005incerteza}. Alinho-me com \citet{sep-qt-uncertainty} com a interpreta\c{c}\~{a}o \textit{ontol\'{o}gica} do princ\'{i}pio da incerteza\index{Heisenberg, Werner!princ\'{i}pio da incerteza|seealso{princ\'{i}pio da indetermina\c{c}\~{a}o}}:

\begin{quote}
    A quest\~{a}o \'{e}, ent\~{a}o, que status atribuiremos ao momento do el\'{e}tron pouco antes de sua medi\c{c}\~{a}o final. \'{E} real? De acordo com Heisenberg\index{Heisenberg, Werner}, n\~{a}o \'{e}. Antes da medi\c{c}\~{a}o final, o melhor que podemos atribuir ao el\'{e}tron \'{e} algum momento impreciso ou difuso. Esses termos s\~{a}o entendidos aqui em um sentido ontol\'{o}gico, caracterizando um atributo real do el\'{e}tron. \citep[\S~2.2]{sep-qt-uncertainty}.
\end{quote}

At\'{e} aqui, parece seguro definir as rela\c{c}\~{o}es de Heisenberg\index{Heisenberg, Werner} como a impossibilidade de medi\c{c}\~{a}o das propriedades observ\'{a}veis de um objeto qu\^{a}ntico com precis\~{a}o arbitrariamente grande. Anos mais tarde, Heisenberg\index{Heisenberg, Werner} exibe uma defini\c{c}\~{a}o de suas rela\c{c}\~{o}es de forma ainda mais precisa:

\begin{quote}
O princ\'{i}pio da incerteza\index{Heisenberg, Werner!princ\'{i}pio da incerteza|seealso{princ\'{i}pio da indetermina\c{c}\~{a}o}} se refere ao grau de indetermina\c{c}\~{a}o no poss\'{i}vel conhecimento presente de valores simult\^{a}neos de v\'{a}rias quantidades com as quais a teoria qu\^{a}ntica lida; ele n\~{a}o se restringe, por exemplo, \`{a} exatid\~{a}o de uma \'{u}nica medi\c{c}\~{a}o de posi\c{c}\~{a}o ou de velocidade. Assim, suponhamos que a velocidade de um el\'{e}tron livre \'{e} conhecida com precis\~{a}o, enquanto que sua posi\c{c}\~{a}o \'{e} completamente desconhecida. O princ\'{i}pio afirma que cada observa\c{c}\~{a}o subsequente da posi\c{c}\~{a}o ir\'{a} alterar o momento por um valor desconhecido e indetermin\'{a}vel tal que, ap\'{o}s a realiza\c{c}\~{a}o da experiência, nosso conhecimento do movimento do el\'{e}tron \'{e} restringido pela rela\c{c}\~{a}o de incerteza. Isso pode ser expresso em termos gerais e concisos ao dizer que cada experimento destr\'{o}i parte do conhecimento do sistema, que fora obtido por experimentos anteriores. Essa formula\c{c}\~{a}o torna claro que a rela\c{c}\~{a}o de incerteza n\~{a}o se refere ao passado; se a velocidade do el\'{e}tron \'{e} previamente conhecida e a posi\c{c}\~{a}o \'{e} medida com exatid\~{a}o, a posi\c{c}\~{a}o para os tempos anteriores a tal medi\c{c}\~{a}o pode ser calculada. Ent\~{a}o, para tais tempos \textelp{a rela\c{c}\~{a}o de incerteza} \'{e} menor do que o limite usual, mas esse conhecimento do passado \'{e} de car\'{a}ter puramente especulativo visto que nunca (devido \`{a} altera\c{c}\~{a}o desconhecida do momento causada pela medi\c{c}\~{a}o da posi\c{c}\~{a}o) pode ser usado como condi\c{c}\~{a}o inicial em qualquer c\'{a}lculo da progress\~{a}o futura do el\'{e}tron e, portanto, n\~{a}o pode ser objeto de verifica\c{c}\~{a}o experimental. \textit{\'{E} uma quest\~{a}o de cren\c{c}a pessoal se se pode ou n\~{a}o atribuir realidade f\'{i}sica ao c\'{a}lculo relativo \`{a} hist\'{o}ria passada do el\'{e}tron.} \citep[p.~20, ênfase adicionada]{heisen1930physical}\index{Heisenberg, Werner}.
\end{quote}

Nessa defini\c{c}\~{a}o, a ênfase \'{e} dada no fato de que os valores dos observ\'{a}veis podem ser conhecidos precisamente, o que parece contradizer a defini\c{c}\~{a}o cl\'{a}ssica das rela\c{c}\~{o}es de incerteza. No entanto, Heisenberg\index{Heisenberg, Werner} afirma que as rela\c{c}\~{o}es n\~{a}o se aplicariam para valores de medi\c{c}\~{o}es passadas, de modo que os valores passados n\~{a}o podem ser utilizados para os c\'{a}lculos futuros, pois cada nova medi\c{c}\~{a}o perturba descontinuamente o valor de um dos observ\'{a}veis de maneira, a princ\'{i}pio, incontrol\'{a}vel.

Como observa \citet[p.~68]{Jammer1974}, a limita\c{c}\~{a}o imposta pelas rela\c{c}\~{o}es de Heisenberg\index{Heisenberg, Werner} n\~{a}o imp\~{o}e uma restri\c{c}\~{a}o \`{a} defini\c{c}\~{a}o dos observ\'{a}veis visto que, se considerados isolados, podem ser medidos com precis\~{a}o arbitrariamente grande. As rela\c{c}\~{o}es se aplicam somente \`{a} tentativa de medi\c{c}\~{a}o simult\^{a}nea dos dois observ\'{a}veis.

Quanto ao estatuto ontol\'{o}gico relativo \`{a} ``hist\'{o}ria passada'' dos observ\'{a}veis (ou seja, dos valores ``precisos'' dos observ\'{a}veis em medi\c{c}\~{o}es passadas e isoladas), \citet[p.~20]{heisen1930physical}\index{Heisenberg, Werner} relega ao plano da ``cren\c{c}a pessoal'', visto n\~{a}o haver possibilidade de referir um aparato experimental pr\'{o}prio para verificar tal no\c{c}\~{a}o. Sua pr\'{o}pria ``cren\c{c}a pessoal'' \'{e} negar sua realidade f\'{i}sica se for levado em considera\c{c}\~{a}o o princ\'{i}pio de \textit{medi\c{c}\~{a}o=cria\c{c}\~{a}o}. O trecho destacado da cita\c{c}\~{a}o acima poderia ser interpretado como uma pista na dire\c{c}\~{a}o de que Heisenberg\index{Heisenberg, Werner} n\~{a}o queria se comprometer com a leitura ontol\'{o}gica das rela\c{c}\~{o}es de incerteza. No entanto ---e isso ficar\'{a} mais claro adiante---, o que est\'{a} em jogo \'{e} a \textit{existência} requerida da realidade f\'{i}sica do el\'{e}tron \textit{anteriormente} \`{a} medi\c{c}\~{a}o: lembre-se do princ\'{i}pio de medi\c{c}\~{a}o=cria\c{c}\~{a}o: antes da medi\c{c}\~{a}o, n\~{a}o h\'{a} raz\~{a}o de fato sobre certas propriedades do el\'{e}tron!

Ainda assim se mant\'{e}m a quest\~{a}o acerca do que as rela\c{c}\~{o}es de Heisenberg\index{Heisenberg, Werner} de fato expressam (ainda que as alternativas n\~{a}o sejam exclusivas): $(i)$ uma limita\c{c}\~{a}o experimental sobre o que se pode conhecer acerca dos objetos qu\^{a}nticos, uma incerteza; $(ii)$ uma restri\c{c}\~{a}o acerca do significado que se pode atribuir \`{a} defini\c{c}\~{a}o dos objetos qu\^{a}nticos, uma indefini\c{c}\~{a}o; $(iii)$ uma restri\c{c}\~{a}o ontol\'{o}gica quanto \`{a}s propriedades observ\'{a}veis dos objetos qu\^{a}nticos, uma indetermina\c{c}\~{a}o.

O extenso debate acerca da interpreta\c{c}\~{a}o das rela\c{c}\~{o}es de Heisenberg\index{Heisenberg, Werner} \'{e} refletido na pr\'{o}pria existência de diversas nomenclaturas para as rela\c{c}\~{o}es de Heisenberg\index{Heisenberg, Werner}. \citet[p.~61--62]{Jammer1974} identifica três termos distintos, utilizados por Heisenberg\index{Heisenberg, Werner} no artigo de 1927, para se referir ao argumento de suas rela\c{c}\~{o}es: (1) \textit{Ungenauigkeit}, que denota ``inexatid\~{a}o'' ou ``imprecis\~{a}o''; (2) \textit{Unbestimmtheit}, que denota ``indetermina\c{c}\~{a}o''; (3) \textit{Unsicherheit}, que denota ``incerteza''.

Da mesma forma, existem três usos distintos do argumento. Se a ênfase \'{e} dada na $(a)$ ausência de conhecimento subjetivo acerca das propriedades dos objetos qu\^{a}nticos, utiliza-se a acep\c{c}\~{a}o (1) ---h\'{a} uma incerteza de car\'{a}ter epistemol\'{o}gico. Se a ênfase \'{e} dada na $(b)$ ausência de conhecimento objetivo, independentemente de observador acerca das propriedades dos objetos qu\^{a}nticos, utiliza-se a acep\c{c}\~{a}o (2) ---h\'{a} uma indetermina\c{c}\~{a}o de car\'{a}ter ontol\'{o}gico. O termo (3) \'{e} utilizado de forma neutra, para quando esta ênfase n\~{a}o for dada.
De acordo com \citet{sep-qt-uncertainty}, Heisenberg\index{Heisenberg, Werner} transita livremente das implica\c{c}\~{o}es epistemol\'{o}gicas para as implica\c{c}\~{o}es ontol\'{o}gicas. Segundo \citet[p.~78]{pessoa2003conceitos}, o motivo pelo qual as rela\c{c}\~{o}es de Heisenberg\index{Heisenberg, Werner} transitam de uma tese epistemol\'{o}gica para uma tese ontol\'{o}gica \'{e} justamente a assun\c{c}\~{a}o do postulado operacionista.

De fato, tal postulado \'{e}, al\'{e}m do ponto de partida do argumento, a base conceitual das rela\c{c}\~{o}es de Heisenberg\index{Heisenberg, Werner}. Tanto as implica\c{c}\~{o}es epistemol\'{o}gicas quanto ontol\'{o}gicas das rela\c{c}\~{o}es se fundamentam no ato de medi\c{c}\~{a}o, entendida nesse contexto como uma opera\c{c}\~{a}o experimental. Se as rela\c{c}\~{o}es demonstram que n\~{a}o \'{e} poss\'{i}vel medir as propriedades observ\'{a}veis de um objeto qu\^{a}ntico de forma precisa e simult\^{a}nea, isto quer dizer que, em \'{u}ltima an\'{a}lise, tais propriedades nem sequer existem simultaneamente de forma determinada. Assim, se segue logicamente que, devido ao fato de n\~{a}o existirem de forma determinada, n\~{a}o podem ser conhecidas ou definidas de forma determinada. Desse modo, por mais que Heisenberg\index{Heisenberg, Werner} dê menos aten\c{c}\~{a}o \`{a}s implica\c{c}\~{o}es ontol\'{o}gicas desse argumento, elas parecem ocupar um lugar central no plano conceitual das rela\c{c}\~{o}es, tal que as implica\c{c}\~{o}es epistemol\'{o}gicas parecem derivar da implica\c{c}\~{a}o ontol\'{o}gica do princ\'{i}pio \textit{medi\c{c}\~{a}o=cria\c{c}\~{a}o}. Portanto, parece seguro caracterizar que, para Heisenberg\index{Heisenberg, Werner}, as rela\c{c}\~{o}es s\~{a}o entendidas como rela\c{c}\~{o}es de indetermina\c{c}\~{a}o. Isto \'{e}, se assumido o postulado operacionista, que parece ser o cerne do argumento de \citet{heisenberg1927uncert}\index{Heisenberg, Werner}, o sentido ontol\'{o}gico \'{e} condi\c{c}\~{a}o necess\'{a}ria para as implica\c{c}\~{o}es epistemol\'{o}gicas e sem\^{a}nticas.

No entanto, \citet[p.~76]{Jammer1974} considera ``estranha'' e at\'{e} mesmo ``inconsistente'' a atitude de classificar o racioc\'{i}nio de Heisenberg\index{Heisenberg, Werner} como positivista, conforme a ado\c{c}\~{a}o do postulado operacionista parece sugerir. A motiva\c{c}\~{a}o para o racioc\'{i}nio das rela\c{c}\~{o}es de indetermina\c{c}\~{a}o fora fortemente influenciada por uma conversa com Albert Einstein\index{Einstein, Albert}, como reconhece o pr\'{o}prio \citet[p.~95]{heisen1969partetodo}. Da forma como \citet[p.~78]{heisen1969partetodo} transcreve, o racioc\'{i}nio de Einstein\index{Einstein, Albert} seria o seguinte: ``em princ\'{i}pio \'{e} um grande erro tentar fundamentar uma teoria apenas nas grandezas observ\'{a}veis. Na realidade, d\'{a}-se exatamente o inverso. \'{E} a teoria que decide o que podemos observar''. Tal racioc\'{i}nio acerca do significado do termo ``observa\c{c}\~{a}o'' parece ser o oposto da proposta positivista para as ciências ---na qual as teorias cient\'{i}ficas deveriam ter como ponto de partida os dados observ\'{a}veis.

Em uma entrevista conduzida por Kuhn, Heisenberg\index{Heisenberg, Werner} esclarece esse ponto:
\begin{quote}
Ele [Einstein\index{Einstein, Albert}] explicou-me que o que se observa ou n\~{a}o \'{e} decidido pela teoria. Somente quando você tem a teoria completa, você pode dizer o que pode ser observado. A palavra observa\c{c}\~{a}o significa que você faz algo que \'{e} consistente com as leis f\'{i}sicas conhecidas. Ent\~{a}o se você n\~{a}o tem leis f\'{i}sicas, você n\~{a}o observa nada. Bem, você tem impress\~{o}es e você tem algo em sua chapa fotogr\'{a}fica, mas você n\~{a}o tem nenhuma maneira de ir da placa para os \'{a}tomos. Se você n\~{a}o tem nenhuma maneira de ir de placa para os \'{a}tomos, qual a utilidade da placa? \citep[sec.~XVIII]{heisenberg1963interviewkuhn}\index{Heisenberg, Werner}.
\end{quote}

A referida teoria (que deve preceder a observa\c{c}\~{a}o) seria, no entendimento de Heisenberg\index{Heisenberg, Werner}, a matem\'{a}tica.

\begin{quote}
Bem, n\'{o}s temos um esquema matem\'{a}tico consistente e esse esquema matem\'{a}tico consistente nos diz tudo o que pode ser observado. N\~{a}o existe algo na natureza que n\~{a}o possa ser descrito por esse esquema matem\'{a}tico. \textelp{} ondas e corp\'{u}sculos s\~{a}o, com certeza, um modo de express\~{a}o, e n\'{o}s chegamos a estes conceitos atrav\'{e}s da f\'{i}sica cl\'{a}ssica. A f\'{i}sica cl\'{a}ssica nos ensinou a falar acerca de part\'{i}culas e ondas, mas desde que a f\'{i}sica cl\'{a}ssica n\~{a}o \'{e} verdadeira l\'{a} [na f\'{i}sica qu\^{a}ntica], por que devemos nos ater tanto a estes conceitos? Por que n\~{a}o dizer simplesmente que n\~{a}o podemos usar esses conceitos com uma precis\~{a}o muito elevada? Da\'{i} as rela\c{c}\~{o}es de incerteza, e, por isso, n\'{o}s temos que abandonar estes conceitos at\'{e} certo ponto. Ent\~{a}o ficamos al\'{e}m desse limite da teoria cl\'{a}ssica, e devemos perceber que nossas palavras n\~{a}o s\~{a}o adequadas. Elas n\~{a}o têm de fato base na realidade f\'{i}sica e, portanto, um novo esquema matem\'{a}tico seria melhor que elas, porque o novo esquema matem\'{a}tico diz o que pode e o que n\~{a}o pode estar l\'{a}. A natureza de alguma forma segue tal esquema. \citep[sec.~XVIII]{heisenberg1963interviewkuhn}\index{Heisenberg, Werner}.
\end{quote}

O argumento original das rela\c{c}\~{o}es de Heisenberg\index{Heisenberg, Werner} (sob o exemplo do microsc\'{o}pio de raios gama), de acordo com \citet[p.~67]{redhead1987inc}, infere que ``uma part\'{i}cula descrita classicamente se `infecta' com as rela\c{c}\~{o}es de incerteza da mec\^{a}nica qu\^{a}ntica quando interage com um agente qu\^{a}ntico em uma medi\c{c}\~{a}o''. Isso parece indicar, no limite, a rejei\c{c}\~{a}o por parte de Heisenberg\index{Heisenberg, Werner} da descri\c{c}\~{a}o cl\'{a}ssica (tais como ondas e part\'{i}culas) para os objetos qu\^{a}nticos. Para \citet[p.~68]{Jammer1974}, isto \'{e} not\'{a}vel, visto que a formula\c{c}\~{a}o matem\'{a}tica da teoria, na concep\c{c}\~{a}o de Heisenberg\index{Heisenberg, Werner}, permitiria a predi\c{c}\~{a}o de todo e qualquer experimento, de modo que a utiliza\c{c}\~{a}o de termos cl\'{a}ssicos, tais como ``ondas'' ou ``part\'{i}culas'', seria obsoleta para a descri\c{c}\~{a}o do que ocorre em uma medi\c{c}\~{a}o qu\^{a}ntica ---ao menos diante de tal esquema matem\'{a}tico.

Pela defini\c{c}\~{a}o, ainda em linhas gerais, que busco apresentar para o princ\'{i}pio de \citet{heisenberg1927uncert}\index{Heisenberg, Werner}, chamarei de princ\'{i}pio da indetermina\c{c}\~{a}o\index{Heisenberg, Werner!princ\'{i}pio da indetermina\c{c}\~{a}o}, dada a ênfase nos pressupostos ontol\'{o}gicos subjacentes ao racioc\'{i}nio de sua formula\c{c}\~{a}o. Passemos \`{a} analise de alguns aspectos centrais da formula\c{c}\~{a}o da complementaridade\index{Bohr, Niels!princ\'{i}pio da complementaridade} de Bohr\index{Bohr, Niels} para definir com maior precis\~{a}o a no\c{c}\~{a}o de interpreta\c{c}\~{a}o de Copenhague.

\section{A complementaridade}

Juntamente com o princ\'{i}pio da indetermina\c{c}\~{a}o\index{Heisenberg, Werner!princ\'{i}pio da indetermina\c{c}\~{a}o} de Heisenberg\index{Heisenberg, Werner}, a no\c{c}\~{a}o de complementaridade\index{Bohr, Niels!princ\'{i}pio da complementaridade}, formulada por \citet{bohr1928quantumpostulate}\index{Bohr, Niels}, cont\'{e}m o cerne daquilo que se conhece por interpreta\c{c}\~{a}o de Copenhague, muitas vezes chamada de ``interpreta\c{c}\~{a}o da complementaridade\index{Bohr, Niels!princ\'{i}pio da complementaridade}'' ou ``interpreta\c{c}\~{a}o ortodoxa''. No entanto, o termo ``complementaridade\index{Bohr, Niels!princ\'{i}pio da complementaridade}'' tem, de acordo com \citet[p.~88--89]{Jammer1974}, usos muito distintos e fora aplicado a diversas outras \'{a}reas do conhecimento, tais como \'{e}tica, lingu\'{i}stica, psicologia e teologia. No contexto da f\'{i}sica ---sobre o qual me aterei exclusivamente--- o termo tem diversos usos filos\'{o}ficos distintos, com implica\c{c}\~{o}es epistemol\'{o}gicas (como o pr\'{o}prio Bohr\index{Bohr, Niels} parece sugerir), l\'{o}gicas e at\'{e} mesmo ontol\'{o}gicas. Buscarei evidenciar tais implica\c{c}\~{o}es ao longo deste cap\'{i}tulo.

Me aterei, a princ\'{i}pio, \`{a} formula\c{c}\~{a}o original de \citet{bohr1928quantumpostulate}\index{Bohr, Niels}, na tentativa de reconstruir uma defini\c{c}\~{a}o t\~{a}o precisa quanto poss\'{i}vel do termo complementaridade\index{Bohr, Niels!princ\'{i}pio da complementaridade}, entendendo que haver\'{a} uma s\'{e}rie de dificuldades, na medida em que, como apontam \citet[p.~95]{Jammer1974} e \citet[p.~142]{faye2012niels}, nem mesmo Bohr\index{Bohr, Niels} delineou uma defini\c{c}\~{a}o clara para aquilo que diz respeito ao conceito complementaridade\index{Bohr, Niels!princ\'{i}pio da complementaridade}.

O termo aparece pela primeira vez em uma palestra de \citet{bohr1928quantumpostulate}\index{Bohr, Niels} ministrada em 1927, na cidade italiana de Como, conhecida como ``\textit{Como lecture}'', e publicada no ano seguinte. A argumenta\c{c}\~{a}o conduzida por \citet{bohr1928quantumpostulate}\index{Bohr, Niels} se d\'{a} por duas premissas e uma conclus\~{a}o: 

\begin{itemize}
\item[] $(P_{1})$: Os conceitos cl\'{a}ssicos s\~{a}o indispens\'{a}veis para a descri\c{c}\~{a}o dos experimentos qu\^{a}nticos.

\item[] $(P_{2})$: A indivisibilidade dos fenômenos qu\^{a}nticos \'{e} um fato imposto pela natureza e deve ser aceito como tal. Isto \'{e}, como cada medi\c{c}\~{a}o envolve a troca de uma quantidade finita de energia (de ao menos um \textit{quantum}), nenhuma medi\c{c}\~{a}o seria rigorosamente idêntica \`{a} outra e, por isso, fala-se na indivisibilidade dos fenômenos qu\^{a}nticos.

\item[] $(C_{1})$: O uso dos conceitos cl\'{a}ssicos tem sua limita\c{c}\~{a}o na descri\c{c}\~{a}o dos fenômenos qu\^{a}nticos.
\end{itemize}

Iniciarei a an\'{a}lise desse argumento partindo da premissa $(P_{2})$. Uma das principais caracter\'{i}sticas que diferencia as teorias cl\'{a}ssica e qu\^{a}ntica seria a introdu\c{c}\~{a}o do postulado qu\^{a}ntico, contido na premissa de que ele:

\begin{quote}
\textelp{} atribui a qualquer processo atômico uma descontinuidade essencial, ou ainda uma individualidade, completamente estranha para as teorias cl\'{a}ssicas \textelp{}. \citep[p.~88]{bohr1928quantumpostulate}\index{Bohr, Niels}.
\end{quote}

\'{E} precisamente a essa descontinuidade inerente ao processo de medi\c{c}\~{a}o que Heisenberg\index{Heisenberg, Werner} se refere nas rela\c{c}\~{o}es de indetermina\c{c}\~{a}o. Tal postulado declara que toda e qualquer intera\c{c}\~{a}o entre (ao menos) dois sistemas \'{e} caracterizada pela troca de energia de (ao menos) um quantum, de modo que qualquer medi\c{c}\~{a}o envolve uma intera\c{c}\~{a}o entre o fenômeno qu\^{a}ntico e as agências de medi\c{c}\~{a}o.

O termo ``agência de medi\c{c}\~{a}o'' \'{e} utilizado com frequência nos escritos de Bohr\index{Bohr, Niels}, o que talvez indique uma posi\c{c}\~{a}o de neutralidade em rela\c{c}\~{a}o ao que, de fato, seria a causa da medi\c{c}\~{a}o, de modo a n\~{a}o se comprometer com as ambiguidades contidas em termos como ``observa\c{c}\~{a}o'' que poderiam remeter a um aspecto humano. Dado o postulado qu\^{a}ntico e suas consequências para o ato de medi\c{c}\~{a}o, Bohr\index{Bohr, Niels} \'{e} capaz de enunciar pela primeira vez o sentido do termo ``complementaridade\index{Bohr, Niels!princ\'{i}pio da complementaridade}'':

\begin{quote}
Por um lado, a defini\c{c}\~{a}o do estado de um sistema f\'{i}sico, como entendido comumente, alega a elimina\c{c}\~{a}o de todas as interferências externas. Mas, nesse caso, de acordo com o postulado qu\^{a}ntico, qualquer observa\c{c}\~{a}o ser\'{a} imposs\'{i}vel, e, acima de tudo, os conceitos de espa\c{c}o e tempo perdem imediatamente o seu significado. Por outro lado, se, para tornar a observa\c{c}\~{a}o poss\'{i}vel, temos que permitir certas intera\c{c}\~{o}es com agências apropriadas de medi\c{c}\~{a}o que n\~{a}o perten\c{c}am ao sistema, uma defini\c{c}\~{a}o n\~{a}o amb\'{i}gua do estado do sistema naturalmente n\~{a}o \'{e} mais poss\'{i}vel, e a causalidade, no sentido comum da palavra, est\'{a} fora de quest\~{a}o. A pr\'{o}pria natureza da teoria qu\^{a}ntica nos obriga, portanto, a considerar a coordena\c{c}\~{a}o espa\c{c}o-tempo e a alega\c{c}\~{a}o da causalidade, a uni\~{a}o que caracteriza as teorias cl\'{a}ssicas, como caracter\'{i}sticas complementares, mas exclusivas, da descri\c{c}\~{a}o, simbolizando a idealiza\c{c}\~{a}o da observa\c{c}\~{a}o e da defini\c{c}\~{a}o respectivamente. \citep[p.~89--90]{bohr1928quantumpostulate}\index{Bohr, Niels}.
\end{quote}

Diversas considera\c{c}\~{o}es podem ser extra\'{i}das do trecho acima, que \'{e} a primeira vez em que Bohr\index{Bohr, Niels} se refere ao termo ``complementaridade\index{Bohr, Niels!princ\'{i}pio da complementaridade}''. Chamo a aten\c{c}\~{a}o aos seguintes pontos da cita\c{c}\~{a}o acima: $(i)$ a ressignifica\c{c}\~{a}o do conceito cl\'{a}ssico de observa\c{c}\~{a}o; $(ii)$ o operacionismo e $(iii)$ as vari\'{a}veis complementares. O ponto $(i)$ deixa claro que, uma vez assumido o postulado qu\^{a}ntico, uma observa\c{c}\~{a}o passiva de um objeto isolado n\~{a}o seria poss\'{i}vel, uma vez que, na teoria qu\^{a}ntica, h\'{a} a troca de energia discreta (de ao menos um quantum) entre a agência de medi\c{c}\~{a}o e o objeto medido. Tal inter-rela\c{c}\~{a}o acaba por aparentemente desconstruir a linha, clara na teoria cl\'{a}ssica, que distingue sujeito e objeto.

O ponto $(ii)$, que chamo de operacionismo, parece ter as mesmas consequências do postulado operacionista proposto por \citet[p.~64]{heisenberg1927uncert}\index{Heisenberg, Werner} na formula\c{c}\~{a}o das rela\c{c}\~{o}es de indetermina\c{c}\~{a}o, na medida em que admite significado somente aos conceitos sobre os quais se possa indicar uma opera\c{c}\~{a}o experimental. Isto se torna not\'{a}vel em v\'{a}rias passagens da palestra de Como, quando, por exemplo, \citet[p.~91--92]{bohr1928quantumpostulate}\index{Bohr, Niels} admite que a ``\textelp{} radia\c{c}\~{a}o em espa\c{c}os livres assim como part\'{i}culas materiais isoladas s\~{a}o abstra\c{c}\~{o}es, suas propriedades na teoria qu\^{a}ntica s\~{a}o defin\'{i}veis e observ\'{a}veis apenas atrav\'{e}s de sua intera\c{c}\~{a}o com outros sistemas''.

Em um sentido ontol\'{o}gico mais forte, afirma que

\begin{quote}
\textelp{} uma realidade independente, no sentido f\'{i}sico usual [cl\'{a}ssico], n\~{a}o pode ser atribu\'{i}da nem ao fenômeno nem \`{a}s agências de observa\c{c}\~{a}o. \citep[p.~89]{bohr1928quantumpostulate}\index{Bohr, Niels}.
\end{quote}

Assim, o ponto $(ii)$ parece enfatizar, de acordo com \citet{sep-qt-uncertainty}, que o contexto experimental define aquilo que pode ser significativamente atribu\'{i}do \`{a} descri\c{c}\~{a}o de um objeto qu\^{a}ntico, ao inv\'{e}s de alterar propriedades pr\'{e}-existentes em tal objeto. De fato, a \'{u}ltima coloca\c{c}\~{a}o \'{e} uma interpreta\c{c}\~{a}o poss\'{i}vel da primeira formula\c{c}\~{a}o da complementaridade\index{Bohr, Niels!princ\'{i}pio da complementaridade} expressa por \citet{bohr1928quantumpostulate}\index{Bohr, Niels}. Entretanto, ao conflitar com o operacionismo do ponto $(ii)$ sublinhado acima, tal interpreta\c{c}\~{a}o fora veementemente combatida por Bohr\index{Bohr, Niels} na defesa da completude da mec\^{a}nica qu\^{a}ntica na segunda metade da d\'{e}cada de 30, assunto que tratarei em detalhe no cap\'{i}tulo seguinte.

Os dois pontos citados acima carregam not\'{a}veis consequências filos\'{o}ficas em rela\c{c}\~{a}o ao racioc\'{i}nio de Bohr\index{Bohr, Niels}. Por enquanto, deixarei de lado a discuss\~{a}o em torno de tais implica\c{c}\~{o}es, e enfatizarei o ponto $(iii)$ a fim de delinear uma defini\c{c}\~{a}o clara para o termo ``complementaridade\index{Bohr, Niels!princ\'{i}pio da complementaridade}''. O racioc\'{i}nio utilizado por Bohr\index{Bohr, Niels} nessa passagem \'{e} de que a complementaridade\index{Bohr, Niels!princ\'{i}pio da complementaridade} seria relativa a modos de descri\c{c}\~{a}o mutuamente exclusivos, que seriam: $(a)$ a descri\c{c}\~{a}o ou coordena\c{c}\~{a}o espa\c{c}o-temporal de um objeto qu\^{a}ntico e $(b)$ a descri\c{c}\~{a}o causal ou a alega\c{c}\~{a}o da causalidade de tal objeto.

Enquanto a no\c{c}\~{a}o $(a)$ \'{e}, de certa forma, mais clara, o item $(b)$ merece mais aten\c{c}\~{a}o. A op\c{c}\~{a}o de Bohr\index{Bohr, Niels} da defini\c{c}\~{a}o do item $(b)$, identificada como causalidade, se refere, segundo \citet[p.~95]{Jammer1974} ``aos teoremas de conserva\c{c}\~{a}o de energia e momento'', o que Patr\'{i}cia \citet[p.~171]{patricialeite2003causalidadeMQ} identifica como ``o determinismo causal do formalismo matem\'{a}tico''; de fato, assegura \citet[p.~170]{patricialeite2003causalidadeMQ}, o formalismo da teoria qu\^{a}ntica, sob a representa\c{c}\~{a}o matem\'{a}tica da evolu\c{c}\~{a}o temporal de uma fun\c{c}\~{a}o de onda, seria sempre determinista. A evolu\c{c}\~{a}o temporal dos sistemas qu\^{a}nticos ser\'{a} tratada em maiores detalhes no cap\'{i}tulo \ref{CapvNeumann} sob a nomenclatura de ``processo 2''. Em sua formula\c{c}\~{a}o original, as vari\'{a}veis complementares ---ou observ\'{a}veis ou vari\'{a}veis conjugadas--- $(a)$ e $(b)$ denotam a incompatibilidade de qualquer tentativa de, simultaneamente, se atribuir validade a uma descri\c{c}\~{a}o espa\c{c}o-temporal das leis matem\'{a}ticas.

Como aponta \citet[p.~102]{Jammer1974}, Bohr\index{Bohr, Niels} n\~{a}o utiliza os termos ``posi\c{c}\~{a}o'' e ``momento'', ou ``part\'{i}cula'' e ``onda'', na palestra de Como, ainda que pudesse tê-lo feito facilmente. De fato, como notam \citet{sep-qt-uncertainty}, as vari\'{a}veis de posi\c{c}\~{a}o e momento seriam os melhores exemplos para tratar da complementaridade\index{Bohr, Niels!princ\'{i}pio da complementaridade} de Bohr\index{Bohr, Niels}, num sentido de clareza ou praticidade, uma vez que s\~{a}o estas as vari\'{a}veis utilizadas nos debates em rela\c{c}\~{a}o \`{a} interpreta\c{c}\~{a}o de Bohr\index{Bohr, Niels}. Assim, unicamente porque os exemplos que se seguir\~{a}o pressup\~{o}em de alguma forma o uso das vari\'{a}veis posi\c{c}\~{a}o e momento, utilizarei por ora, por motivos de clareza, a ``vers\~{a}o de Pauli'' como sugere \citet[p.~102]{Jammer1974}, que intercambia a vari\'{a}vel $(a)$ por ``posi\c{c}\~{a}o'' e $(b)$ por ``momento''.

Uma das contribui\c{c}\~{o}es de Weizs\"{a}cker para a compreens\~{a}o do termo ``complementaridade\index{Bohr, Niels!princ\'{i}pio da complementaridade}'' de Bohr\index{Bohr, Niels} fora a distin\c{c}\~{a}o entre v\'{a}rias acep\c{c}\~{o}es do termo. A vers\~{a}o de Pauli seria chamada de ``complementaridade\index{Bohr, Niels!princ\'{i}pio da complementaridade} paralela'' visto que os conceitos de ``posi\c{c}\~{a}o'' e ``momento'' pertenceriam \`{a} mesma imagem intuitiva dos processos f\'{i}sicos, caso se queira definir completamente o estado de um sistema; a vers\~{a}o de Bohr\index{Bohr, Niels}, no entanto, seria chamada de complementaridade\index{Bohr, Niels!princ\'{i}pio da complementaridade} circular. Em simultaneidade, as vari\'{a}veis $(a)$ e $(b)$ constituem o significado cl\'{a}ssico do termo observa\c{c}\~{a}o.

Na teoria cl\'{a}ssica, dois modos de descri\c{c}\~{a}o $(a)$ e $(b)$ s\~{a}o combinados, uma vez que $(a)$ o estado de um sistema se desenvolve continuamente no espa\c{c}o e no tempo, e $(b)$ a mudan\c{c}a do estado de um sistema, causada pela intera\c{c}\~{a}o, \'{e} determinada pelos princ\'{i}pios de conserva\c{c}\~{a}o de momento e energia. Por isso, na mec\^{a}nica cl\'{a}ssica, um estado bem definido pode \textit{sempre} ser atribu\'{i}do a um sistema isolado, quer ele interaja ou n\~{a}o com outro sistema.

Na teoria qu\^{a}ntica, no entanto, em consequência do postulado qu\^{a}ntico, n\~{a}o seria poss\'{i}vel a medi\c{c}\~{a}o simult\^{a}nea das duas vari\'{a}veis, o que desproveria de sentido os conceitos $(a)$ e $(b)$, de acordo com o crit\'{e}rio operacionista assumido. Para tanto, Bohr\index{Bohr, Niels} prop\~{o}e que tais vari\'{a}veis componham uma descri\c{c}\~{a}o complementar, caso tomadas em situa\c{c}\~{o}es experimentais distintas, mutuamente exclusivas, mas, no entanto, necess\'{a}rias para uma descri\c{c}\~{a}o exaustiva dos fenômenos qu\^{a}nticos.

Da forma como descrito, o termo ``complementaridade\index{Bohr, Niels!princ\'{i}pio da complementaridade}'' de Bohr\index{Bohr, Niels} parece se referir a modos de descri\c{c}\~{a}o distintos, acompanhados de arranjos experimentais distintos, de modo que pode ser estendido \`{a}s vari\'{a}veis elas mesmas em termos de quais descri\c{c}\~{o}es complementares s\~{a}o formuladas, assim, por exemplo, uma coordenada de posi\c{c}\~{a}o e uma vari\'{a}vel de momento s\~{a}o chamadas complementares umas \`{a}s outras; neste sentido, o termo ``complementaridade\index{Bohr, Niels!princ\'{i}pio da complementaridade}'' \'{e} justificado somente se as vari\'{a}veis s\~{a}o utilizadas em descri\c{c}\~{o}es que correspondam a opera\c{c}\~{o}es experimentais complementares.

S\~{a}o precisamente tais modos complementares de descri\c{c}\~{a}o que devem ser realizados na terminologia da linguagem da teoria cl\'{a}ssica, de modo que podemos passar para a an\'{a}lise da primeira premissa $(P_{1})$. Isto se daria, a princ\'{i}pio, pela natureza da observa\c{c}\~{a}o que, segundo \citet[p.~89]{bohr1928quantumpostulate}\index{Bohr, Niels} ``em \'{u}ltima an\'{a}lise, toda observa\c{c}\~{a}o pode, de fato, ser reduzida \`{a}s nossas percep\c{c}\~{o}es sensoriais''. Uma observa\c{c}\~{a}o de um objeto qu\^{a}ntico parece representar a amplia\c{c}\~{a}o de um sinal microsc\'{o}pico (qu\^{a}ntico), por uma agência de medi\c{c}\~{a}o, para o n\'{i}vel macrosc\'{o}pico (cl\'{a}ssico), de tal forma que:

\begin{quote}
Ao tra\c{c}ar as observa\c{c}\~{o}es de volta \`{a}s nossas sensa\c{c}\~{o}es, novamente deve-se referir o postulado qu\^{a}ntico em conex\~{a}o com a percep\c{c}\~{a}o da agência de observa\c{c}\~{a}o [medi\c{c}\~{a}o], seja por meio de sua a\c{c}\~{a}o direta sobre o olho ou por meio de auxiliares adequados \textelp{}. \citep[p.~102]{bohr1928quantumpostulate}\index{Bohr, Niels}.
\end{quote}

Assim, raciocina \citet[p.~126]{bohr1928quantumpostulate}\index{Bohr, Niels}, na medida em que ``\textelp{} toda palavra na linguagem se refere a nossa percep\c{c}\~{a}o comum'', e que nossa percep\c{c}\~{a}o comum \'{e} relativa aos macroobjetos ---os objetos da teoria cl\'{a}ssica--- nossa linguagem deve ser cl\'{a}ssica.

Tentei, at\'{e} aqui, reconstruir a argumenta\c{c}\~{a}o de Bohr\index{Bohr, Niels} sobre o termo ``complementaridade\index{Bohr, Niels!princ\'{i}pio da complementaridade}''. Da forma como proposto por \citet{Jammer1974}, a reconstru\c{c}\~{a}o da premissa $P_2$ pode ser resumidamente enunciada passo a passo da seguinte maneira:

\begin{quote}
1. Indivisibilidade do quantum de a\c{c}\~{a}o (postulado qu\^{a}ntico).

2. Descontinuidade (ou indivisibilidade) dos processos qu\^{a}nticos.

3. Incontrolabilidade da intera\c{c}\~{a}o entre objeto e instrumento [de medi\c{c}\~{a}o].

4. Impossibilidade de uma (estrita) descri\c{c}\~{a}o espa\c{c}o-temporal, ao mesmo tempo, causal.

5. Ren\'{u}ncia ao modo cl\'{a}ssico de descri\c{c}\~{a}o. \citep[p.~101]{Jammer1974}.
\end{quote}

Passemos agora \`{a} an\'{a}lise cr\'{i}tica do conceito ``complementaridade\index{Bohr, Niels!princ\'{i}pio da complementaridade}''. O ponto 5 indicado na conclus\~{a}o $(C_{1})$ ---\textit{viz.}, que o uso linguagem cl\'{a}ssica \'{e} \textit{limitado} para a descri\c{c}\~{a}o dos fenômenos qu\^{a}nticos--- pode soar contradit\'{o}rio tendo em vista a necessidade, expressa por Bohr\index{Bohr, Niels}, do uso da linguagem cl\'{a}ssica para a explica\c{c}\~{a}o dos fenômenos qu\^{a}nticos. No entanto, para Bohr\index{Bohr, Niels}, o que caracteriza um modo cl\'{a}ssico de descri\c{c}\~{a}o \'{e} a existência de apenas uma descri\c{c}\~{a}o completa. No entendimento de Bohr\index{Bohr, Niels}, tal \'{u}nico modo se refere a uma \'{u}nica descri\c{c}\~{a}o, ao mesmo tempo causal e espa\c{c}o-temporal. Assim, se for levado em considera\c{c}\~{a}o que uma descri\c{c}\~{a}o cl\'{a}ssica jamais fornece uma descri\c{c}\~{a}o completa de um objeto qu\^{a}ntico no sentido da necessidade da exclusividade m\'{u}tua de (ao menos dois) modos cl\'{a}ssicos de descri\c{c}\~{a}o, a aparência de uma contradi\c{c}\~{a}o desaparece.

Ainda assim, outra dificuldade para a utiliza\c{c}\~{a}o da terminologia cl\'{a}ssica para a descri\c{c}\~{a}o dos fenômenos qu\^{a}nticos \'{e} exposta por \citet[p.~201--229]{howard1994conceptclassic}, na medida em que os conceitos cl\'{a}ssicos carregam pressupostos filos\'{o}ficos diferentes ou at\'{e} mesmo contradit\'{o}rios em rela\c{c}\~{a}o \`{a}queles assumidos pela mec\^{a}nica qu\^{a}ntica ---da forma como interpretada pela complementaridade\index{Bohr, Niels!princ\'{i}pio da complementaridade}. O comprometimento ontol\'{o}gico\index{ontologia!comprometimento ontol\'{o}gico} com a tese de que os entes possuem uma realidade objetiva independente \'{e} uma caracter\'{i}stica not\'{a}vel do referencial conceitual cl\'{a}ssico. Em outras palavras, os termos cl\'{a}ssicos trazem consigo a ideia de que os objetos que comp\~{o}em o mundo existem independentemente de qualquer intera\c{c}\~{a}o (medi\c{c}\~{a}o/observa\c{c}\~{a}o) ---o que parece claramente contradizer o postulado qu\^{a}ntico, assumido como ponto de partida para a interpreta\c{c}\~{a}o de Copenhague.

Tal comprometimento ontol\'{o}gico,\index{ontologia!comprometimento ontol\'{o}gico} presente na terminologia cl\'{a}ssica, fora chamado por \citet[p.~207]{howard1994conceptclassic} de ``princ\'{i}pio da separabilidade'', que seria uma nomenclatura abreviada de um princ\'{i}pio, atribu\'{i}do a Einstein\index{Einstein, Albert}, que prevê a ``existência mutuamente independente de coisas espacialmente distantes''. Dessa maneira, a assun\c{c}\~{a}o da separabilidade seria necess\'{a}ria para a no\c{c}\~{a}o de independência ontol\'{o}gica. Para \citet[p.~169]{einstein1971QMreality}, a separabilidade seria a condi\c{c}\~{a}o necess\'{a}ria para que conceitos f\'{i}sicos ou leis f\'{i}sicas fossem formuladas. \citet{sep-einstein-philscience} e \citet[p.~122]{krause2010indiscernibles} v\~{a}o al\'{e}m e consideram que o realismo einsteiniano \textit{\'{e}} o pr\'{o}prio princ\'{i}pio da separabilidade.

O princ\'{i}pio da separabilidade ser\'{a} tratado mais detalhadamente no cap\'{i}tulo \ref{CapEPR}. Por ora, limito-me a explicitar a forma como Bohr\index{Bohr, Niels} enuncia esse problema (bem como sua solu\c{c}\~{a}o):

\begin{quote}
A elucida\c{c}\~{a}o dos paradoxos da f\'{i}sica atômica tem divulgado o fato de que a intera\c{c}\~{a}o inevit\'{a}vel entre os objetos e os instrumentos de medi\c{c}\~{a}o define um limite absoluto \`{a} possibilidade de falar de um comportamento de objetos atômicos que seja independente dos meios de observa\c{c}\~{a}o. Estamos aqui diante de um problema epistemol\'{o}gico muito novo na filosofia natural, onde toda a descri\c{c}\~{a}o das experiências at\'{e} agora tem sido baseada na suposi\c{c}\~{a}o, j\'{a} inerente \`{a}s conven\c{c}\~{o}es comuns da linguagem, de que \'{e} poss\'{i}vel distinguir claramente entre o comportamento dos objetos e os meios de observa\c{c}\~{a}o. Essa suposi\c{c}\~{a}o n\~{a}o \'{e} apenas plenamente justificada por toda experiência cotidiana, mas constitui at\'{e} mesmo toda a base da f\'{i}sica cl\'{a}ssica. \textelp{} Como n\'{o}s estamos tratando, por\'{e}m, com fenômenos como processos atômicos individuais que, devido \`{a} sua pr\'{o}pria natureza, s\~{a}o essencialmente determinados pela intera\c{c}\~{a}o entre os objetos em quest\~{a}o e os instrumentos de medi\c{c}\~{a}o necess\'{a}rios para a defini\c{c}\~{a}o do arranjo experimental, somos, portanto, obrigados a examinar mais de perto a quest\~{a}o sobre o tipo de conhecimento que pode ser obtido em rela\c{c}\~{a}o aos objetos. A este respeito, devemos, por um lado, perceber que o escopo de cada experimento f\'{i}sico ---para adquirir conhecimento em condi\c{c}\~{o}es reprodut\'{i}veis e transmiss\'{i}veis--- n\~{a}o nos deixa escolha a n\~{a}o ser usar conceitos cotidianos, talvez refinados pela terminologia da f\'{i}sica cl\'{a}ssica, n\~{a}o s\'{o} em todos os relatos de constru\c{c}\~{a}o e de manipula\c{c}\~{a}o dos instrumentos de medi\c{c}\~{a}o, mas tamb\'{e}m na descri\c{c}\~{a}o dos resultados experimentais reais. Por outro lado, \'{e} igualmente importante entender que essa pr\'{o}pria circunst\^{a}ncia implica que nenhum resultado de um experimento relativo a um fenômeno, que, em princ\'{i}pio, est\'{a} fora do alcance da f\'{i}sica cl\'{a}ssica, pode ser interpretado como provedor de informa\c{c}\~{o}es sobre propriedades independentes dos objetos. \citep[p.~25--26]{bohr1938nature}\index{Bohr, Niels}.
\end{quote}

O argumento da passagem citada acima \'{e} o seguinte: $(i)$ a separabilidade deve ser abandonada em se tratando dos fenômenos qu\^{a}nticos; $(ii)$ a assun\c{c}\~{a}o da independência ---que pressup\~{o}e a separabilidade--- \'{e} inerente ao modo cl\'{a}ssico de descri\c{c}\~{a}o; $(iii)$ para comunicar os resultados dos experimentos qu\^{a}nticos, de modo a evitar ambiguidades, a linguagem cl\'{a}ssica deve ser utilizada; $(iv)$ a linguagem cl\'{a}ssica \'{e} fundada na assun\c{c}\~{a}o da independência que a teoria qu\^{a}ntica nega. Ao que parece, para Bohr\index{Bohr, Niels}, a utiliza\c{c}\~{a}o dos conceitos cl\'{a}ssicos \'{e} necess\'{a}ria para que haja uma comunica\c{c}\~{a}o dos experimentos qu\^{a}nticos livre de ambiguidades. Tal comunica\c{c}\~{a}o seria a base para aquilo que Bohr\index{Bohr, Niels} chama de objetividade: uma comunica\c{c}\~{a}o objetiva \'{e} uma comunica\c{c}\~{a}o livre de ambiguidades.

\begin{quote}
Nossa tarefa deve ser responder pela experiência de um modo independente do julgamento subjetivo, individual, e, por conseguinte, objetivo na medida em que pode ser inequivocamente comunicada na linguagem humana comum. \textelp{} \'{e} decisivo perceber que, por mais que os fenômenos ultrapassem o alcance da experiência comum, a descri\c{c}\~{a}o do arranjo experimental e o registro das observa\c{c}\~{o}es deve ser baseada na linguagem comum. \citep[p.~10--11]{bohr1958atomic}\index{Bohr, Niels}.
\end{quote}

De tal linha de racioc\'{i}nio, segue-se que, para que haja objetividade na descri\c{c}\~{a}o dos experimentos qu\^{a}nticos, \'{e} necess\'{a}ria a assun\c{c}\~{a}o da independência ontol\'{o}gica tanto do instrumento de medi\c{c}\~{a}o quanto do objeto qu\^{a}ntico ---e, por conseguinte, do princ\'{i}pio de separabilidade--- visto que a linguagem cl\'{a}ssica, necess\'{a}ria para a descri\c{c}\~{a}o objetiva dos fenômenos qu\^{a}nticos, \'{e} baseada em tais no\c{c}\~{o}es filos\'{o}ficas.

Essa problem\'{a}tica se desdobra, para \citet[p.~128--129]{faye2012niels} em dois pontos principais: $(i)$ se o aparelho \'{e} cl\'{a}ssico, o resultado deve ser cl\'{a}ssico e $(ii)$ a descri\c{c}\~{a}o \'{e} cl\'{a}ssica, pois a natureza da no\c{c}\~{a}o de observa\c{c}\~{a}o \'{e} cl\'{a}ssica. O ponto $(i)$ \'{e} caracterizado pelo seguinte argumento: o aparato escolhido para efetuar uma medi\c{c}\~{a}o \'{e} constitu\'{i}do de um objeto macrosc\'{o}pico, cujo funcionamento \'{e} baseado inteiramente em leis cl\'{a}ssicas, e os dados emp\'{i}ricos da medi\c{c}\~{a}o fornecidos por tal aparelho devem ser entendidos de acordo com seu funcionamento, de modo que tais dados emp\'{i}ricos s\'{o} podem ser descritos em termos dos conceitos cl\'{a}ssicos. A fragilidade do ponto $(i)$ \'{e} justamente sua contingência hist\'{o}rica, de modo que aparelhos mais avan\c{c}ados (menores) poderiam vir a descrever ``quanticamente'' um fenômeno qu\^{a}ntico. Esse racioc\'{i}nio tamb\'{e}m parece controverso, pois pressup\~{o}e que algum dia poder\'{i}amos perceber diretamente um aparelho qu\^{a}ntico de medi\c{c}\~{a}o ---o que parece esbarrar nas pr\'{o}prias limita\c{c}\~{o}es da percep\c{c}\~{a}o humana.

O ponto $(ii)$, no entanto, parece ser mais fundamental. Para \citet[p.~127--129]{faye2012niels}, a f\'{i}sica cl\'{a}ssica desenvolveu m\'{e}todos para ordenar a experiência humana de uma forma objetiva. No mundo macrosc\'{o}pico \'{e} aparentemente poss\'{i}vel conectar descri\c{c}\~{o}es causais com descri\c{c}\~{o}es espa\c{c}o-temporais, da mesma forma que, aparentemente, \'{e} poss\'{i}vel distinguir entre um sistema utilizado como instrumento para observa\c{c}\~{a}o e um sistema a ser observado. Assim, ao que parece, a natureza da observa\c{c}\~{a}o que ordena e estrutura nossa experiência humana cotidiana assim procede, sendo a \'{u}nica garantia de que tal experiência possa vir a ser considerada objetiva. \'{E} precisamente porque os conceitos cl\'{a}ssicos se referem \`{a}s formas de percep\c{c}\~{a}o, sobre as quais n\'{o}s ---enquanto sujeitos humanos--- apreendemos o mundo exterior, que eles s\~{a}o indispens\'{a}veis para que a descri\c{c}\~{a}o de um fenômeno possa ser estruturada e comunicada de forma intelig\'{i}vel.

Da forma como \citet{faye2012niels} prop\~{o}e, a distin\c{c}\~{a}o entre sujeito e objeto seria uma pr\'{e}-condi\c{c}\~{a}o para o conhecimento objetivo, isto \'{e}, um conhecimento que n\~{a}o seja dependente da vis\~{a}o do sujeito sobre um determinado objeto ---o que seria poss\'{i}vel somente em termos de uma descri\c{c}\~{a}o espa\c{c}o-temporal e causal, de acordo com nossa percep\c{c}\~{a}o. Isso \'{e} not\'{a}vel se levarmos em considera\c{c}\~{a}o a redu\c{c}\~{a}o de \citet[p.~89]{bohr1928quantumpostulate}\index{Bohr, Niels} do ato de medi\c{c}\~{a}o \`{a}s nossas percep\c{c}\~{o}es cotidianas. Ou ainda, da forma como \citet[p.~80]{favrholdt1994realismbohr} ilustra a situa\c{c}\~{a}o, \'{e} ``\textelp{} porque somos seres macrosc\'{o}picos, nossa linguagem \'{e} necessariamente adaptada ao mundo macrosc\'{o}pico''. Bohr\index{Bohr, Niels} explicita a situa\c{c}\~{a}o da seguinte maneira:

\begin{quote}
A exigência de que seja poss\'{i}vel comunicar os resultados experimentais, de uma forma inequ\'{i}voca, implica que o arranjo experimental e os resultados da observa\c{c}\~{a}o devem ser expressos na linguagem comum adaptada para nossa orienta\c{c}\~{a}o no ambiente. Assim, a descri\c{c}\~{a}o de fenômenos qu\^{a}nticos exige uma distin\c{c}\~{a}o, em princ\'{i}pio, entre os objetos sob investiga\c{c}\~{a}o e o aparelho de medi\c{c}\~{a}o, por meio do qual as condi\c{c}\~{o}es experimentais s\~{a}o definidas. \citep[p.~78]{bohr1958atomic}\index{Bohr, Niels}.
\end{quote}

A linguagem cl\'{a}ssica seria ent\~{a}o utilizada pela assun\c{c}\~{a}o da separabilidade que sua terminologia carrega, e justificada pela necessidade da comunica\c{c}\~{a}o objetiva dos experimentos qu\^{a}nticos. De acordo com \citet[p.~209]{howard1994conceptclassic}, n\~{a}o se trataria de uma contingência hist\'{o}rica, pass\'{i}vel de ser superada por algum aprimoramento lingu\'{i}stico, mas justamente de uma necessidade metodol\'{o}gica. O racioc\'{i}nio segue da seguinte maneira: a separabilidade ---cl\'{a}ssica, do instrumento macrosc\'{o}pico--- \'{e} condi\c{c}\~{a}o necess\'{a}ria para que possamos dizer que um objeto qu\^{a}ntico tem \textit{tais e tais} propriedades bem definidas; isso n\~{a}o seria poss\'{i}vel caso objeto e instrumento fossem insepar\'{a}veis ou ontologicamente interdependentes. Sem a separabilidade, n\~{a}o ter\'{i}amos raz\~{o}es suficientes para justificar que consideramos os resultados das medi\c{c}\~{o}es como relatos de propriedades intr\'{i}nsecas do objeto. Ao que parece, Bohr\index{Bohr, Niels} enfatiza a necessidade de que a agência de medi\c{c}\~{a}o seja considerada cl\'{a}ssica ---isto \'{e}, fora do alcance do postulado qu\^{a}ntico (o referido ``quantum de a\c{c}\~{a}o'') e, portanto, separado ou independente--- no que tange \`{a} comunicabilidade dos seus resultados:

\begin{quote}
O novo recurso essencial na an\'{a}lise dos fenômenos qu\^{a}nticos \'{e}, no entanto, a introdu\c{c}\~{a}o de uma distin\c{c}\~{a}o fundamental entre o aparelho de medi\c{c}\~{a}o e os objetos sob investiga\c{c}\~{a}o. Essa \'{e} uma consequência direta da necessidade de considerar as fun\c{c}\~{o}es dos instrumentos de medi\c{c}\~{a}o em termos puramente cl\'{a}ssicos, excluindo, em princ\'{i}pio, qualquer rela\c{c}\~{a}o com o quantum de a\c{c}\~{a}o. \citep[p.~3--4]{bohr1958bcausality}\index{Bohr, Niels}.
\end{quote}

Isso n\~{a}o significa, no entanto, que a ontologia da f\'{i}sica cl\'{a}ssica deva ser estendida \`{a} mec\^{a}nica qu\^{a}ntica como um todo. O postulado qu\^{a}ntico mant\'{e}m a implica\c{c}\~{a}o de que as vari\'{a}veis complementares, ainda que descritas \`{a} maneira cl\'{a}ssica, s\'{o} podem ser aplicadas significativamente em rela\c{c}\~{a}o a uma opera\c{c}\~{a}o experimental e n\~{a}o ---como pressup\~{o}e a ontologia cl\'{a}ssica--- a despeito de qualquer opera\c{c}\~{a}o experimental. Isto significa que a complementaridade\index{Bohr, Niels!princ\'{i}pio da complementaridade} recusa qualquer descri\c{c}\~{a}o utilizada para indicar propriedades por tr\'{a}s dos fenômenos, existentes em si mesmos, inerentes e portadores de uma independência ontol\'{o}gica de qualquer opera\c{c}\~{a}o experimental. Assim, a utiliza\c{c}\~{a}o da no\c{c}\~{a}o filos\'{o}fica da separabilidade, impl\'{i}cita nos conceitos cl\'{a}ssicos para a descri\c{c}\~{a}o dos fenômenos qu\^{a}nticos \'{e} limitada, de modo que n\~{a}o estende a ontologia cl\'{a}ssica para os objetos qu\^{a}nticos. Da forma como diz Faye, a teoria qu\^{a}ntica e a teoria cl\'{a}ssica devem ser ``comensur\'{a}veis'', num sentido kuhniano, no que diz respeito ao seu significado emp\'{i}rico.:

\begin{quote}
As duas teorias podem ser baseadas em suposi\c{c}\~{o}es amplamente divergentes a respeito de determinados aspectos da realidade f\'{i}sica e, portanto, as teorias podem envolver diferentes compromissos ontol\'{o}gicos,\index{ontologia!comprometimento ontol\'{o}gico} mas o conte\'{u}do emp\'{i}rico da linguagem na qual estes pressupostos s\~{a}o expressos \'{e} o mesmo ou \'{e} similar. \citep[p.~118]{faye2012niels}.
\end{quote}

Ao que parece, h\'{a} aqui em jogo uma no\c{c}\~{a}o sem\^{a}ntica na qual o uso dos conceitos da f\'{i}sica cl\'{a}ssica \'{e} necess\'{a}rio para uma descri\c{c}\~{a}o exaustiva (ou seja, completa) da realidade f\'{i}sica que, de acordo com \citet{faye2012niels}, implicaria restri\c{c}\~{a}o do dom\'{i}nio de aplicabilidade dos conceitos cl\'{a}ssicos e n\~{a}o no seu abandono, uma vez que, para que os conceitos cl\'{a}ssicos possam ser aplicados \`{a} descri\c{c}\~{a}o qu\^{a}ntica, o significado de tais conceitos cl\'{a}ssicos deve ser compat\'{i}vel com a teoria qu\^{a}ntica. Essa passagem parece sugerir que Bohr\index{Bohr, Niels} contrastaria com a posi\c{c}\~{a}o historicista da ciência que que a teoria qu\^{a}ntica seria uma supera\c{c}\~{a}o da mec\^{a}nica cl\'{a}ssica, de modo que as duas teorias seriam incomensur\'{a}veis, isto \'{e}, totalmente incompat\'{i}veis. Bohr\index{Bohr, Niels} chama esse princ\'{i}pio metodol\'{o}gico de princ\'{i}pio da correspondência\index{Bohr, Niels!princ\'{i}pio da correspondência}, cuja formula\c{c}\~{a}o \'{e} enunciada da seguinte maneira:

\begin{quote}
A necessidade de fazer um uso extensivo \textelp{} dos conceitos cl\'{a}ssicos, sobre a qual a interpreta\c{c}\~{a}o de toda a experiência em \'{u}ltima an\'{a}lise depende, deu origem \`{a} formula\c{c}\~{a}o do chamado princ\'{i}pio da correspondência\index{Bohr, Niels!princ\'{i}pio da correspondência}, que expressa nossos esfor\c{c}os de utilizar todos os conceitos cl\'{a}ssicos ao atribuir-lhes uma re-interpreta\c{c}\~{a}o te\'{o}rico-qu\^{a}ntica adequada. \citep[p.~8]{bohr1962atomic}\index{Bohr, Niels}.
\end{quote}

A vis\~{a}o comum sobre a interpreta\c{c}\~{a}o de Copenhague seria a de relegar \`{a}s agências de medi\c{c}\~{a}o um comportamento inteiramente cl\'{a}ssico, isto \'{e}, considerar que as agências de medi\c{c}\~{a}o (frequentemente um aparelho) s\~{a}o um objeto macrosc\'{o}pico e, portanto, para todos os efeitos, cl\'{a}ssico. Isso fica expl\'{i}cito na seguinte passagem de Bohr\index{Bohr, Niels}:

\begin{quote}
Em arranjos experimentais reais, o cumprimento de tais exigências [de uma descri\c{c}\~{a}o inequ\'{i}voca do aparelho e dos resultados da medi\c{c}\~{a}o] \'{e} assegurada pelo uso, como aparelho medidor, de corpos r\'{i}gidos suficientemente pesados que permitam uma descri\c{c}\~{a}o totalmente cl\'{a}ssica das relativas posi\c{c}\~{o}es e velocidades. \citep[p.~3]{bohr1958bcausality}\index{Bohr, Niels}.
\end{quote}

Tal interpreta\c{c}\~{a}o comum, que concebe o aparelho de medi\c{c}\~{a}o como inteiramente cl\'{a}ssico, \'{e} chamada por \citet[p.~210]{howard1994conceptclassic} de ``interpreta\c{c}\~{a}o coincidente'' e afirma que a divis\~{a}o cl\'{a}ssica/qu\^{a}ntica coincide com a divis\~{a}o aparelho medidor/objeto medido. Nela, o crit\'{e}rio para delinear os limites do mundo cl\'{a}ssico para o mundo qu\^{a}ntico seria o ``tamanho'' do aparelho medidor que, por se tratar de um objeto macrosc\'{o}pico, deveria pertencer ao mundo cl\'{a}ssico.

De fato, o argumento do ``tamanho'' do objeto de medi\c{c}\~{a}o \'{e} apenas uma das caracter\'{i}sticas da interpreta\c{c}\~{a}o coincidente. Outra caracter\'{i}stica, igualmente importante, seria a irreversibilidade dos efeitos ampliados pelos instrumentos medidores. Uma das caracter\'{i}sticas dos objetos qu\^{a}nticos \'{e} sua reversibilidade no tempo ---uma propriedade que n\~{a}o \'{e} observada nos macrocorpos. Nos \'{u}ltimos, a caracter\'{i}stica observada \'{e} sua irreversibilidade, ou seja, a dura\c{c}\~{a}o ou permanência dos efeitos nos objetos. No entanto, optarei por apresentar o argumento de \citet{howard1994conceptclassic} frente \`{a} chamada interpreta\c{c}\~{a}o coincidente da complementaridade\index{Bohr, Niels!princ\'{i}pio da complementaridade} de Bohr\index{Bohr, Niels} apenas com o primeiro aspecto, do ``tamanho'' do aparelho medidor pelas consequências filos\'{o}ficas que tal argumento desencadear\'{a} nos cap\'{i}tulos seguintes no que tange ao problema do macrorrealismo ou macroobjetivismo \citep[ver][235--237]{desp1999concep}.

A interpreta\c{c}\~{a}o coincidente desencadearia, no entanto, uma s\'{e}rie de problemas filos\'{o}ficos como, por exemplo, a introdu\c{c}\~{a}o de um dualismo na ontologia do processo de medi\c{c}\~{a}o, uma vez que os objetos contidos na ontologia cl\'{a}ssica (no caso, os aparelhos medidores) devem interagir fisicamente com os objetos contidos na ontologia qu\^{a}ntica (no caso, os objetos qu\^{a}nticos) ao passo que perten\c{c}am a teorias f\'{i}sicas fundamentalmente diferentes. Uma s\'{e}ria inconsistência, relacionada indiretamente \`{a} problem\'{a}tica da interpreta\c{c}\~{a}o coincidente, seria a descontinuidade introduzida na teoria pelo postulado qu\^{a}ntico da forma como Bohr\index{Bohr, Niels} enuncia na seguinte passagem:

\begin{quote}
De acordo com a teoria qu\^{a}ntica, a impossibilidade de ignorar a intera\c{c}\~{a}o com o mecanismo de medi\c{c}\~{a}o significa que cada observa\c{c}\~{a}o introduz um novo elemento incontrol\'{a}vel. Na verdade, isto decorre das considera\c{c}\~{o}es expostas de que a medi\c{c}\~{a}o das coordenadas de posi\c{c}\~{a}o de uma part\'{i}cula \'{e} acompanhada n\~{a}o s\'{o} por uma mudan\c{c}a finita nas vari\'{a}veis din\^{a}micas, mas tamb\'{e}m a fixa\c{c}\~{a}o de sua posi\c{c}\~{a}o significa uma ruptura completa na descri\c{c}\~{a}o causal de seu comportamento din\^{a}mico, enquanto que a determina\c{c}\~{a}o de seu momento implica sempre uma lacuna no conhecimento de sua propaga\c{c}\~{a}o espacial. Essa situa\c{c}\~{a}o real\c{c}a de forma not\'{a}vel o car\'{a}ter complementar da descri\c{c}\~{a}o dos fenômenos atômicos, que surge como uma consequência inevit\'{a}vel da oposi\c{c}\~{a}o entre o postulado qu\^{a}ntico e a distin\c{c}\~{a}o entre o objeto e a agência de medi\c{c}\~{a}o, inerente \`{a} nossa pr\'{o}pria ideia de observa\c{c}\~{a}o. \citep[p.~103]{bohr1928quantumpostulate}\index{Bohr, Niels}.
\end{quote}

Essa ``ruptura'' ou ``lacuna'' parece ser uma dentre as mais s\'{e}rias dificuldades filos\'{o}ficas da posi\c{c}\~{a}o de Bohr\index{Bohr, Niels}. Tal dificuldade \'{e} agravada da forma como \citet[p.~11]{bohr1962atomic}\index{Bohr, Niels} enuncia em outro momento: ``a magnitude do dist\'{u}rbio causado pela medi\c{c}\~{a}o \'{e} sempre desconhecida''. Da forma como enunciada, a descontinuidade impl\'{i}cita no processo de medi\c{c}\~{a}o, de acordo com \citet[p.~99]{Jammer1974} ``n\~{a}o seria considerada como o resultado da troca de uma descri\c{c}\~{a}o para seu modo complementar, mas como o resultado de uma propriedade f\'{i}sica operacional''. A situa\c{c}\~{a}o se torna ainda mais problem\'{a}tica, caso levarmos em considera\c{c}\~{a}o a afirma\c{c}\~{a}o, de cunho essencialmente ontol\'{o}gico, de Bohr\index{Bohr, Niels}, de que n\~{a}o se deve atribuir uma \textit{realidade independente} aos objetos qu\^{a}nticos fora do seu contexto operacional. Essa dificuldade d\'{a} margem ao famoso problema da medi\c{c}\~{a}o qu\^{a}ntica.

O problema da medi\c{c}\~{a}o ser\'{a} analisado em detalhe nos cap\'{i}tulos seguintes, e \'{e} a inconsistência mais s\'{e}ria daquilo que se entende por interpreta\c{c}\~{a}o de Copenhague. Deixarei a an\'{a}lise e discuss\~{a}o dessa problem\'{a}tica para os cap\'{i}tulos seguintes. Por ora, me aterei ao delineamento dos termos que ser\~{a}o utilizados para a discuss\~{a}o subsequente acerca de tal problema. Pelo que foi considerado aqui, parece seguro delinear uma defini\c{c}\~{a}o para o termo complementaridade\index{Bohr, Niels!princ\'{i}pio da complementaridade} de acordo com a seguinte nota\c{c}\~{a}o de Jammer:

\begin{quote}
Uma determinada teoria $T$ admite uma interpreta\c{c}\~{a}o de complementaridade\index{Bohr, Niels!princ\'{i}pio da complementaridade} se as seguintes condi\c{c}\~{o}es forem satisfeitas: (1) $T$ cont\'{e}m (ao menos) duas descri\c{c}\~{o}es $D_1$ e $D_2$, de seu conte\'{u}do; (2) $D_1$ e $D_2$, referem-se ao mesmo universo de discurso U (no caso de Bohr\index{Bohr, Niels}, a microf\'{i}sica); (3) nem $D_1$ nem $D_2$, se tomados individualmente, respondem exaustivamente todos os fenômenos de U; (4) $D_1$ e $D_2$ s\~{a}o mutuamente exclusivos, no sentido de que a sua combina\c{c}\~{a}o numa \'{u}nica descri\c{c}\~{a}o engendraria contradi\c{c}\~{o}es l\'{o}gicas. \citep[p.~104]{Jammer1974}.
\end{quote}

Os pontos (1) a (3) s\~{a}o equivalentes a uma descri\c{c}\~{a}o sucinta daquilo que foi exposto at\'{e} aqui. O ponto (4), no entanto, merece aten\c{c}\~{a}o, uma vez que dele emerge um problema de ordem l\'{o}gica. O termo complementaridade\index{Bohr, Niels!princ\'{i}pio da complementaridade} se refere tamb\'{e}m \`{a} incompatibilidade dos modos cl\'{a}ssicos de descri\c{c}\~{a}o quando h\'{a} a tentativa de que sua combina\c{c}\~{a}o leve a um \'{u}nico modo de descri\c{c}\~{a}o para os fenômenos qu\^{a}nticos. No entanto, em l\'{o}gica cl\'{a}ssica, a conjun\c{c}\~{a}o de duas f\'{o}rmulas verdadeiras \'{e} tamb\'{e}m uma f\'{o}rmula v\'{a}lida, de modo que $D_1$ e $D_2$ (no caso da complementaridade\index{Bohr, Niels!princ\'{i}pio da complementaridade} aplicada \`{a} teoria qu\^{a}ntica, correspondendo respectivamente \`{a}s descri\c{c}\~{o}es ondulat\'{o}rias e corpusculares dos objetos qu\^{a}nticos) s\~{a}o formas v\'{a}lidas. Sendo assim, sua combina\c{c}\~{a}o tamb\'{e}m deveria ser v\'{a}lida. Portanto, como apontam da Costa e Krause:

\begin{quote}
\textelp{} se $\alpha$ e $\beta$ s\~{a}o as duas teses ou teoremas de uma teoria (fundada na l\'{o}gica cl\'{a}ssica), ent\~{a}o $\alpha\wedge\beta$ tamb\'{e}m \'{e} uma tese (ou um teorema) dessa teoria. Isto \'{e} o que entendemos intuitivamente quando dizemos que, com base na l\'{o}gica cl\'{a}ssica, uma proposi\c{c}\~{a}o ``verdadeira'' n\~{a}o pode ``excluir'' outra proposi\c{c}\~{a}o ``verdadeira''. \textelp{} Isso corresponde ao fato de que, em l\'{o}gica cl\'{a}ssica, se $\alpha$ \'{e} consequência de um conjunto de afirma\c{c}\~{o}es $\Delta$ e $\beta$ \'{e} tamb\'{e}m uma consequência de $\Delta$, ent\~{a}o $\alpha\wedge\beta$ ($\alpha$ e $\beta$) \'{e} tamb\'{e}m uma consequência do $\Delta$. Se $\beta$ \'{e} a nega\c{c}\~{a}o de $\alpha$ (ou vice-versa), ent\~{a}o essa regra implica que a partir do conjunto de f\'{o}rmulas $\Delta$ deduzimos uma contradi\c{c}\~{a}o $\alpha\wedge\neg\alpha$ (ou $\neg\beta\wedge\beta$). Al\'{e}m disso, quando $\alpha$ e $\beta$ s\~{a}o incompat\'{i}veis em algum sentido, $\alpha\wedge\beta$ constitui uma impossibilidade. \citep[p.~107]{NewtDec2006logicofcomplem}.
\end{quote}

Isso indica que a no\c{c}\~{a}o de complementaridade\index{Bohr, Niels!princ\'{i}pio da complementaridade} formulada por Bohr\index{Bohr, Niels} poderia encontrar dificuldades, caso a l\'{o}gica cl\'{a}ssica seja utilizada como a linguagem subjacente da teoria, visto que, da forma como enunciado, o conceito de ``complementaridade\index{Bohr, Niels!princ\'{i}pio da complementaridade}'' levaria a uma contradi\c{c}\~{a}o ---o que tornaria o conceito inconsistente. Para \citet[p.~112]{NewtDec2006logicofcomplem}, talvez a \'{u}nica solu\c{c}\~{a}o para tal problema seria a modifica\c{c}\~{a}o da l\'{o}gica subjacente na linguagem da complementaridade\index{Bohr, Niels!princ\'{i}pio da complementaridade} para um sistema no qual uma contradi\c{c}\~{a}o estrita (tal como $\alpha\wedge\neg\alpha$) n\~{a}o seria deduzida dos pares complementares, ou seja, da f\'{o}rmula $\alpha\wedge\beta$ (sob as condi\c{c}\~{o}es expostas acima, respectivamente correspondentes \`{a}s vari\'{a}veis $D_1$ e $D_2$). Para uma breve formula\c{c}\~{a}o de uma l\'{o}gica desse tipo, ver \citet[p.~112--116]{NewtDec2006logicofcomplem}. N\~{a}o me comprometerei aqui com um sistema l\'{o}gico em particular, mas me limitarei \`{a} exposi\c{c}\~{a}o dos problemas que surgem ao utilizar o racioc\'{i}nio cl\'{a}ssico (l\'{o}gico e f\'{i}sico) para a mec\^{a}nica qu\^{a}ntica. A discuss\~{a}o em torno desse ponto se estender\'{a} nos cap\'{i}tulos seguintes.

A despeito de todas as dificuldades que, como vimos, a interpreta\c{c}\~{a}o de Copenhague apresenta, procurei at\'{e} aqui precisar uma defini\c{c}\~{a}o desse conceito para que possamos discutir adiante sobre o problema da medi\c{c}\~{a}o. No entanto, tal defini\c{c}\~{a}o n\~{a}o ter\'{a} precis\~{a}o arbitrariamente grande na medida em que (1), como j\'{a} disse anteriormente, o pr\'{o}prio Bohr\index{Bohr, Niels} n\~{a}o delineou uma defini\c{c}\~{a}o precisa e nem mesmo os comentadores apresentam consenso sobre a complementaridade\index{Bohr, Niels!princ\'{i}pio da complementaridade} de Bohr\index{Bohr, Niels}; assim, \'{e} poss\'{i}vel interpret\'{a}-la desde uma concep\c{c}\~{a}o antirrealista (sendo essa a maneira tradicional) at\'{e} uma concep\c{c}\~{a}o \textit{realista} acerca da mec\^{a}nica qu\^{a}ntica. Como exemplo de uma leitura que endossa o antirrealismo de Bohr\index{Bohr, Niels}, pode-se referir a obra de \citet{faye2012niels}. J\'{a} a obra de  \citet{folse1985philosophybohr} oferece, em contraponto, uma leitura realista dos escritos de Bohr\index{Bohr, Niels}. O debate entre Faye e Folse acerca da postura de Bohr\index{Bohr, Niels} quanto ao \index{realismo cient\'{i}fico}realismo cient\'{i}fico pode ser encontrado em \citet{faye1994antirealistbohr} e \citet{folse1994realistbohr}. A discuss\~{a}o acerca do \'{u}ltimo ponto ser\'{a} realizada no cap\'{i}tulo \ref{CapEPR} sob a \'{o}tica do posicionamento de Bohr\index{Bohr, Niels} sobre as cr\'{i}ticas de incompletude de sua interpreta\c{c}\~{a}o. Sobre o primeiro ponto, talvez o mais pr\'{o}ximo de uma defini\c{c}\~{a}o que \citet[p.~10]{bohr1962atomic}\index{Bohr, Niels} chega \'{e} que o postulado qu\^{a}ntico nos obriga a adotar um novo modo de descri\c{c}\~{a}o descrita como complementar.

Assim, para que eu possa prosseguir com a discuss\~{a}o, adotarei, por ora, para fins pr\'{a}ticos, essa defini\c{c}\~{a}o (ainda que incompleta) que Bohr\index{Bohr, Niels} oferece sobre a complementaridade\index{Bohr, Niels!princ\'{i}pio da complementaridade}: tenha em mente essa defini\c{c}\~{a}o em todas as ocorrências de tal termo neste livro.

\section{Uma interpreta\c{c}\~{a}o fragmentada}

Com o arcabou\c{c}o conceitual exposto at\'{e} ent\~{a}o, \'{e} oportuno discutir sobre as diferen\c{c}as filos\'{o}ficas dos considerados principais autores daquilo que se entende por interpreta\c{c}\~{a}o de Copenhague. Ainda que uma an\'{a}lise exaustiva acerca do debate filos\'{o}fico entre os dois autores esteja fora do escopo deste livro, apontarei algumas considera\c{c}\~{o}es not\'{a}veis sobre determinados aspectos de suas divergências.

Um dos pontos essenciais dentre as (diversas) diferen\c{c}as filos\'{o}ficas entre Heisenberg\index{Heisenberg, Werner} e Bohr\index{Bohr, Niels} seria, para \citet[p.~521]{camilleri2007heisenbohr}, o fato de que, por um lado, Heisenberg\index{Heisenberg, Werner} enfatiza a necessidade do entendimento do significado do formalismo da teoria qu\^{a}ntica enquanto, por outro lado, Bohr\index{Bohr, Niels} enfatiza a necessidade de uma descri\c{c}\~{a}o completa dos fenômenos qu\^{a}nticos. Assim, como uma forma preliminar, podemos discutir a diferen\c{c}a entre Heisenberg\index{Heisenberg, Werner} e Bohr\index{Bohr, Niels} acerca da delinea\c{c}\~{a}o dos limites da teoria e da interpreta\c{c}\~{a}o da mec\^{a}nica qu\^{a}ntica.

Para Heisenberg\index{Heisenberg, Werner}, o formalismo matem\'{a}tico da teoria deveria ser suficientemente elaborado para que pudesse ser feita uma descri\c{c}\~{a}o exaustiva dos fenômenos, pois sua concep\c{c}\~{a}o era a de que n\~{a}o existiria algo que n\~{a}o pudesse ser expresso de acordo com uma formula\c{c}\~{a}o matem\'{a}tica ---o que, como aponta Heisenberg\index{Heisenberg, Werner}, n\~{a}o seria o caso para Bohr\index{Bohr, Niels}:

\begin{quote}
\textelp{} a clareza matem\'{a}tica n\~{a}o tinha em si qualquer virtude para Bohr\index{Bohr, Niels}. Ele temia que a estrutura matem\'{a}tica formal fosse obscurecer o n\'{u}cleo f\'{i}sico do problema, e, em qualquer caso, ele estava convencido de que uma explica\c{c}\~{a}o f\'{i}sica completa deve absolutamente preceder a formula\c{c}\~{a}o matem\'{a}tica. \citep[p.~98]{heisen1967interpretQM}.
\end{quote}

Tal controv\'{e}rsia se daria somente no plano da ordena\c{c}\~{a}o ou ``precedência'' dos conceitos; a discuss\~{a}o acerca da import\^{a}ncia e do alcance, tanto do formalismo quanto da interpreta\c{c}\~{a}o da teoria qu\^{a}ntica, n\~{a}o seria, de acordo com \citet[p.~67]{Jammer1974}, o aspecto central do debate entre Bohr\index{Bohr, Niels} e Heisenberg\index{Heisenberg, Werner} em rela\c{c}\~{a}o \`{a} interpreta\c{c}\~{a}o das rela\c{c}\~{o}es de indetermina\c{c}\~{a}o. A chave de leitura para a compreens\~{a}o desse debate seria, portanto, a diferen\c{c}a no ponto de partida escolhido por cada autor: ao passo que Heisenberg\index{Heisenberg, Werner} partiria do formalismo, o ponto de partida da interpreta\c{c}\~{a}o de Bohr\index{Bohr, Niels} acerca das rela\c{c}\~{o}es seria, de acordo com \citet[p.~66--69]{Jammer1974}, a dualidade onda-part\'{i}cula\index{dualidade onda-part\'{i}cula} ---isto \'{e}, a impossibilidade de reduzir a descri\c{c}\~{a}o dos objetos qu\^{a}nticos aos aspectos exclusivamente corpusculares ou ondulat\'{o}rios, visto que ambas as formas s\~{a}o encontradas nos experimentos qu\^{a}nticos. 

Bohr\index{Bohr, Niels} haveria encontrado indica\c{c}\~{o}es de que o argumento de Heisenberg\index{Heisenberg, Werner} conectaria descri\c{c}\~{o}es de part\'{i}culas com descri\c{c}\~{o}es de ondas que, assim, ``pressup\~{o}em implicitamente a \index{dualidade onda-part\'{i}cula}dualidade onda-part\'{i}cula''.\footnote{~Ainda que a dualidade onda-part\'{i}cula seja um aspecto central dos fundamentos da mec\^{a}nica qu\^{a}ntica, optei por n\~{a}o abord\'{a}-lo neste livro, visto que essa problem\'{a}tica recai na quest\~{a}o sobre a linguagem a ser utilizada para uma descri\c{c}\~{a}o dos fenômenos qu\^{a}nticos. Para uma abordagem centrada na problem\'{a}tica da assim chamada \textit{dualidade onda-part\'{i}cula}\index{dualidade onda-part\'{i}cula}, consulte \citet{pessoa2003conceitos,pessoa2006conceitos2}.} De fato, como enfatiza \citet[p.~15]{chibeni2005incerteza}, o experimento mental do microsc\'{o}pio de raios gama pressup\~{o}e uma ontologia de part\'{i}culas enquanto utiliza, ao mesmo tempo, conceitos ondulat\'{o}rios (como uma fun\c{c}\~{a}o de onda) para a representa\c{c}\~{a}o matem\'{a}tica dos objetos qu\^{a}nticos.

Outro argumento apresentado em \citet[p.~69]{Jammer1974}, seria o de que, originalmente, quaisquer deriva\c{c}\~{o}es das rela\c{c}\~{o}es de Heisenberg\index{Heisenberg, Werner} a partir dos experimentos mentais (como o do microsc\'{o}pio de raios gama) precisariam utilizar as equa\c{c}\~{o}es de Einstein\index{Einstein, Albert}--de Broglie, que conectam descri\c{c}\~{o}es da f\'{i}sica de part\'{i}culas com a f\'{i}sica ondulat\'{o}ria. No entanto, considero que os argumentos anteriores, sem a necessidade de adentrar numa discuss\~{a}o acerca do formalismo da teoria qu\^{a}ntica, s\~{a}o suficientes para expor o ponto de vista de Bohr\index{Bohr, Niels}.

Heisenberg\index{Heisenberg, Werner} e Bohr\index{Bohr, Niels} concordavam com o fato de que a interpreta\c{c}\~{a}o da teoria qu\^{a}ntica deveria utilizar a terminologia da f\'{i}sica cl\'{a}ssica. No entanto, ao passo que Heisenberg\index{Heisenberg, Werner} afirmava a insuficiência dos termos da f\'{i}sica de ondas ou da f\'{i}sica de part\'{i}culas para uma explica\c{c}\~{a}o completa dos fenômenos qu\^{a}nticos ---insuficiência essa expressa nas pr\'{o}prias rela\c{c}\~{o}es de indetermina\c{c}\~{a}o---, Bohr\index{Bohr, Niels} afirmava a necessidade do uso de ambas as teorias. Para Bohr\index{Bohr, Niels}, no entanto, o significado do termo `explica\c{c}\~{a}o' deveria ser revisado.

Em seu sentido cl\'{a}ssico, uma explica\c{c}\~{a}o seria um modo \'{u}nico, suficiente, para o esgotamento da descri\c{c}\~{a}o de um objeto. Segundo \citet[p.~15--16]{bohr1962atomic}\index{Bohr, Niels}, essa acep\c{c}\~{a}o do termo seria empregada por Heisenberg\index{Heisenberg, Werner} ao afirmar que um esquema matem\'{a}tico seria mais adequado para a explica\c{c}\~{a}o dos fenômenos qu\^{a}nticos do que uma ressignifica\c{c}\~{a}o dos conceitos cl\'{a}ssicos (quer sejam da f\'{i}sica de part\'{i}culas ou da f\'{i}sica ondulat\'{o}ria) j\'{a} utilizados para a descri\c{c}\~{a}o dos objetos qu\^{a}nticos. Contrariamente, \citet[p.~96]{bohr1962atomic}\index{Bohr, Niels} define uma nova acep\c{c}\~{a}o do termo explica\c{c}\~{a}o afirmando que ``devemos, em geral, estar preparados para aceitar o fato de que uma elucida\c{c}\~{a}o completa do mesmo e \'{u}nico objeto pode requerer diversos pontos de vista que desafiam uma descri\c{c}\~{a}o \'{u}nica'', em que os ``diversos pontos de vista'' seriam os aspectos complementares da descri\c{c}\~{a}o qu\^{a}ntica.

A quest\~{a}o do dist\'{u}rbio descont\'{i}nuo do ato da medi\c{c}\~{a}o seria uma indica\c{c}\~{a}o da impossibilidade de defini\c{c}\~{a}o simult\^{a}nea das propriedades observ\'{a}veis de um objeto qu\^{a}ntico, ou seja, de um modo \'{u}nico de explica\c{c}\~{a}o para os fenômenos qu\^{a}nticos. Dito de outra forma, o indeterminismo expresso pelas rela\c{c}\~{o}es de Heisenberg\index{Heisenberg, Werner}, para Bohr\index{Bohr, Niels}, seria um exemplo matem\'{a}tico da ruptura ou descontinuidade pr\'{o}pria do ato de medi\c{c}\~{a}o, o que obrigaria a formula\c{c}\~{a}o de pontos de vista diversos, complementares, para uma descri\c{c}\~{a}o exaustiva do objeto qu\^{a}ntico ---a linguagem de tal descri\c{c}\~{a}o deve permanecer, de acordo com a opera\c{c}\~{a}o experimental (complementar) em quest\~{a}o, na terminologia cl\'{a}ssica---, sendo a indetermina\c{c}\~{a}o expressa pelas rela\c{c}\~{o}es de Heisenberg\index{Heisenberg, Werner}, o pre\c{c}o a se pagar, caso haja a tentativa de aplica\c{c}\~{a}o simult\^{a}nea dos termos cl\'{a}ssicos mutuamente exclusivos.

Ao que parece, a descontinuidade impl\'{i}cita nos processos de medi\c{c}\~{a}o \'{e} um fator chave para que eu possa delinear algumas das divergências filos\'{o}ficas fundamentais entre Heisenberg\index{Heisenberg, Werner} e Bohr\index{Bohr, Niels}.

Enquanto para Heisenberg\index{Heisenberg, Werner} tal descontinuidade seria expressa atrav\'{e}s de uma formula\c{c}\~{a}o matem\'{a}tica, sob a nomenclatura de ``redu\c{c}\~{a}o do pacote de onda'',\footnote{~Que futuramente ficou conhecida como o ``colapso''. Tratarei desse assunto nos pr\'{o}ximos cap\'{i}tulos.} para Bohr\index{Bohr, Niels}, a situa\c{c}\~{a}o seria totalmente diferente. Na medida em que Bohr\index{Bohr, Niels} n\~{a}o considera que o formalismo matem\'{a}tico da teoria qu\^{a}ntica tenha um significado por si ---ou seja, considera que o formalismo precisa ser interpretado--- ou mesmo que represente algo real no sentido f\'{i}sico do termo, o problema implicado pela chamada redu\c{c}\~{a}o do pacote de onda n\~{a}o seria um problema, caso fosse uma no\c{c}\~{a}o limitada ao formalismo em si mesmo.

Talvez esse seja o motivo pelo qual \citet{folse1994realistbohr} considere Bohr\index{Bohr, Niels} um antirrealista quando diz respeito \`{a}s teorias, isto \'{e}, ao formalismo, e um realista no que tange \`{a}s entidades emp\'{i}ricas, na medida em que considera um objeto qu\^{a}ntico uma entidade real (quando observada). Assim, o problema da medi\c{c}\~{a}o (cuja contrapartida no formalismo seria a pr\'{o}pria no\c{c}\~{a}o de redu\c{c}\~{a}o do pacote de ondas, na terminologia de Heisenberg\index{Heisenberg, Werner}) parece ainda se aplicar na interpreta\c{c}\~{a}o de Bohr\index{Bohr, Niels}, visto que a ruptura impl\'{i}cita no ato de medi\c{c}\~{a}o \'{e} algo que se mant\'{e}m.

De acordo com \citet[p.~522]{camilleri2007heisenbohr}, essa diferen\c{c}a da precedência do formalismo matem\'{a}tico implica maneiras diferentes de visualizar o pr\'{o}prio problema da descontinuidade referido acima (o que ele chama de ``o paradoxo impl\'{i}cito da mec\^{a}nica qu\^{a}ntica''). Pois, se Heisenberg\index{Heisenberg, Werner} define um sistema qu\^{a}ntico nos termos de uma f\'{o}rmula matem\'{a}tica, como uma fun\c{c}\~{a}o de onda, essa defini\c{c}\~{a}o independe da experimenta\c{c}\~{a}o. Ainda que n\~{a}o se possa atribuir realidade f\'{i}sica \`{a} fun\c{c}\~{a}o de onda (pelo princ\'{i}pio de \textit{medi\c{c}\~{a}o=cria\c{c}\~{a}o}), essa representa\c{c}\~{a}o seria aplic\'{a}vel para a descri\c{c}\~{a}o de um objeto qu\^{a}ntico em termos de propens\~{o}es ou possibilidades. \citet[p.~53]{heisen1958physphil}\index{Heisenberg, Werner} enfatiza que essa realidade se daria num plano potencial ---em contraste ao plano atual dos fenômenos emp\'{i}ricos---, remontando ao pensamento aristot\'{e}lico de potência e ato\footnote{~Ver \citet[p.~257--258]{desp1999concep} e \citet{heisen1958physphil}\index{Heisenberg, Werner}.}

Por outro lado, a defini\c{c}\~{a}o de um sistema qu\^{a}ntico, independentemente de sua rela\c{c}\~{a}o com um contexto operacional, n\~{a}o teria significado na sem\^{a}ntica de Bohr\index{Bohr, Niels}, que busca na pr\'{o}pria experimenta\c{c}\~{a}o as condi\c{c}\~{o}es de possibilidade de defini\c{c}\~{a}o dos objetos qu\^{a}nticos. Assim, ao passo em que para Heisenberg\index{Heisenberg, Werner} a descontinuidade \'{e} fruto de um dist\'{u}rbio interacional entre a agência de medi\c{c}\~{a}o e o objeto qu\^{a}ntico medido, Bohr\index{Bohr, Niels} enfatiza que tal descontinuidade seria uma limita\c{c}\~{a}o na definibilidade, e n\~{a}o um dist\'{u}rbio f\'{i}sico.

Ainda assim, a tese de que ocorre um dist\'{u}rbio f\'{i}sico aparece dentre as teses principais da interpreta\c{c}\~{a}o de Copenhague. \citet[p.~87--98]{pessoa2003conceitos} elenca, em dez t\'{o}picos, as principais teses atribu\'{i}das \`{a}quilo que se chama de ``interpreta\c{c}\~{a}o ortodoxa'', dos quais sublinharei apenas um: o dist\'{u}rbio interacional, que afirma que h\'{a} uma intera\c{c}\~{a}o f\'{i}sica entre o objeto observado e a agência de medi\c{c}\~{a}o que observa tal objeto.

Esse ponto \'{e} uma das vias para se chegar ao problema da medi\c{c}\~{a}o, motivo pelo qual a interpreta\c{c}\~{a}o de Copenhague foi duramente criticada nos anos 1930, sob a acusa\c{c}\~{a}o de incompletude. No cap\'{i}tulo \ref{CapEPR}, analisarei os debates sobre a completude da mec\^{a}nica qu\^{a}ntica, enfatizando o comprometimento ontol\'{o}gico\index{ontologia!comprometimento ontol\'{o}gico} dos pontos de vista de Einstein\index{Einstein, Albert} e Bohr\index{Bohr, Niels} em rela\c{c}\~{a}o ao dist\'{u}rbio interacional e ao problema da medi\c{c}\~{a}o.

Procurei, neste cap\'{i}tulo, esbo\c{c}ar alguns pontos centrais da interpreta\c{c}\~{a}o de Copenhague, bem como seus aspectos filosoficamente problem\'{a}ticos. Devo enfatizar que de modo algum busco aqui uma descri\c{c}\~{a}o exaustiva dos conceitos de indetermina\c{c}\~{a}o e complementaridade\index{Bohr, Niels!princ\'{i}pio da complementaridade}, mas meramente uma defini\c{c}\~{a}o para possibilitar a discuss\~{a}o feita nos cap\'{i}tulos seguintes. Na realidade, uma descri\c{c}\~{a}o completa de tais conceitos ---especialmente a no\c{c}\~{a}o de complementaridade\index{Bohr, Niels!princ\'{i}pio da complementaridade}--- n\~{a}o \'{e} uma tarefa f\'{a}cil: conforme aponta \citet[p.~88]{Jammer1974} nem mesmo os interlocutores contempor\^{a}neos a Bohr\index{Bohr, Niels} foram capazes de compreender completamente sua interpreta\c{c}\~{a}o da teoria qu\^{a}ntica. Como procurei evidenciar ao longo deste cap\'{i}tulo, grande parte de tal deficiência se d\'{a} pelo fato de que Bohr\index{Bohr, Niels} jamais teria oferecido uma descri\c{c}\~{a}o formal para a no\c{c}\~{a}o de medi\c{c}\~{a}o, apesar de ser uma no\c{c}\~{a}o central em suas ideias.

Com o que foi exposto at\'{e} aqui, poderemos entender melhor alguns aspectos filos\'{o}ficos nos fundamentos da mec\^{a}nica qu\^{a}ntica, especificamente do conceito de medi\c{c}\~{a}o. Destaco como a interpreta\c{c}\~{a}o de Copenhague oferece uma vis\~{a}o de mundo bastante contraintuitiva em rela\c{c}\~{a}o \`{a} nossa percep\c{c}\~{a}o ordin\'{a}ria da realidade \`{a} nossa volta, principalmente no que diz respeito \`{a} suposi\c{c}\~{a}o ---ou at\'{e} mesmo \`{a} certeza--- ontol\'{o}gica da existência independente dos objetos que comp\~{o}em o mundo \`{a} nossa volta e do determinismo causal impl\'{i}cito na linearidade dos eventos que experienciamos cotidianamente. No pr\'{o}ximo cap\'{i}tulo, analiso em detalhes o debate entre Einstein\index{Einstein, Albert} e Bohr\index{Bohr, Niels}, que suscitou diversas quest\~{o}es filos\'{o}ficas acerca da problem\'{a}tica da medi\c{c}\~{a}o.

\chapter{Vis\~{o}es de mundo em conflito}
\label{CapEPR}

Neste cap\'{i}tulo, analisarei um dos debates filos\'{o}ficos centrais no que se refere \`{a}s quest\~{o}es de princ\'{i}pios ou fundamentos da mec\^{a}nica qu\^{a}ntica, especificamente em rela\c{c}\~{a}o ao debate entre Albert Einstein\index{Einstein, Albert} e Niels Bohr\index{Bohr, Niels}. Saliento que as pressuposi\c{c}\~{o}es ontol\'{o}gicas de ambos os autores, que se mostrar\~{a}o claras ao longo do debate aqui proposto, s\~{a}o fundamentais para a compreens\~{a}o de tal debate; da mesma forma, s\~{a}o fundamentais para compreender o momento em que se insere  o problema da medi\c{c}\~{a}o qu\^{a}ntica, que ser\'{a} discutido detalhadamente no cap\'{i}tulo \ref{CapvNeumann}.

Para tanto, caracterizo os termos utilizados, procurando, inicialmente, destacar de que modo uma quest\~{a}o concernente \`{a} interpreta\c{c}\~{a}o da mec\^{a}nica qu\^{a}ntica se insere na problem\'{a}tica filos\'{o}fica. Em seguida, busco uma defini\c{c}\~{a}o para o termo \textit{ontologia} que utilizarei ao longo do texto, o que me permitir\'{a} adentrar nos aspectos ontol\'{o}gicos do debate entre Bohr\index{Bohr, Niels} e Einstein\index{Einstein, Albert}, a fim de especificar os pressupostos ontol\'{o}gicos por detr\'{a}s da argumenta\c{c}\~{a}o de cada autor. Assim, ser\'{a} poss\'{i}vel delinear a quest\~{a}o da medi\c{c}\~{a}o qu\^{a}ntica como um debate essencialmente filos\'{o}fico. Com o advento da mec\^{a}nica qu\^{a}ntica, principalmente no final dos anos 1920, a preocupa\c{c}\~{a}o acerca dos fundamentos da realidade ---fundamentos conceitualmente formados no modelo da f\'{i}sica cl\'{a}ssica--- emergiu da pr\'{o}pria comunidade f\'{i}sica, instigando debates acerca das implica\c{c}\~{o}es ontol\'{o}gicas da mec\^{a}nica qu\^{a}ntica.

A no\c{c}\~{a}o de ``crise'' apresentada na obra de \citet[p.~119--120]{kuhn1998estrutura} parece refletir a problem\'{a}tica que surge com o advento da teoria qu\^{a}ntica no s\'{e}culo XX. A revis\~{a}o paradigm\'{a}tica que a mec\^{a}nica qu\^{a}ntica prop\~{o}e no terreno da f\'{i}sica pode ser abordada por diversos aspectos. Limito-me, aqui, a discutir aquilo que, na teoria kuhniana, constitui as diferen\c{c}as ``substanciais'', ou seja, as diferen\c{c}as ontol\'{o}gicas num sentido de diferentes ``mobili\'{a}rios do mundo''. Segundo \citet[p.~56]{preston2008kuhn}, paradigmas sucessivos ``\textelp{} envolvem diferentes ontologias como por exemplo, diferentes listas dos tipos de objetos que o mundo cont\'{e}m''. Nesse sentido, analiso a problem\'{a}tica de visualizar a concep\c{c}\~{a}o da mec\^{a}nica qu\^{a}ntica sob a \'{o}tica da f\'{i}sica cl\'{a}ssica como ontol\'{o}gica.

Para compreender um pouco melhor o recorte aqui proposto, assumirei uma distin\c{c}\~{a}o utilizada por \citet[p.~9]{cushing1994historicalQM}, ainda que grosseira, entre \textit{formalismo} e \textit{interpreta\c{c}\~{a}o}, segundo a qual o formalismo \'{e} o c\'{a}lculo simb\'{o}lico utilizado para fazer predi\c{c}\~{o}es te\'{o}ricas e experimentais, enquanto a interpreta\c{c}\~{a}o seria ``\textelp{} a hist\'{o}ria correspondente ao mobili\'{a}rio do mundo ---uma ontologia)''. Como a rela\c{c}\~{a}o entre ontologia e interpreta\c{c}\~{a}o n\~{a}o \'{e} trivial, tomarei um breve desvio por assunto antes de prosseguir. As quest\~{o}es sobre os limites entre teoria e interpreta\c{c}\~{a}o certamente s\~{a}o muito mais complexas do que a que indiquei aqui \citep[ver][]{cirkovic2005physics, Maudlin1995measurementproblem, arroyo2020phd}. Na verdade, h\'{a} quem argumente que essa fronteira \'{e} equivocada: \citet[p.~244]{muller1997-2}, por exemplo, diz que a ``distin\c{c}\~{a}o formalismo/interpreta\c{c}\~{a}o'' pode ``descansar em paz'' \citep[ver tamb\'{e}m][]{muller1997-1,muller1999-a}.

Seja como for, um formalismo puramente matem\'{a}tico n\~{a}o deve conseguir descrever os fen\^{o}menos f\'{i}sicos da mec\^{a}nica qu\^{a}ntica, motivo pelo qual aquilo que entende-se por mec\^{a}nica qu\^{a}ntica deve ser mais do que um formalismo puro \citep[ver][]{muller2015, arroyo-olegario2021}.\footnote{~Mas compare com os problemas apontados em \citet{arenhart-arroyo2023instante}.} Deve haver, portanto, um componente ontol\'{o}gico da f\'{i}sica para que a mec\^{a}nica qu\^{a}ntica possa ser dotada de ---no m\'{i}nimo--- adequa\c{c}\~{a}o emp\'{i}rica \citep[ver][]{vanfraassen1980scientific}. Por exemplo, o compromisso com estados qu\^{a}nticos, e outros objetos f\'{i}sicos \textit{e.g.} el\'{e}trons. Isso parece ir al\'{e}m do que um formalismo matem\'{a}tico \textit{e.g.}, a \'{a}lgebra linear, indica e/ou se compromete. Assim, quando uma teoria cient\'{i}fica come\c{c}a a se questionar sobre ``como \'{e} o mundo'' ---ou como o mundo \textit{poderia ser}--- caso tal teoria fosse verdadeira, ela come\c{c}a a se preocupar com aquilo que \citet[p.~242]{vanfrass1991quantum} chamou de ``quest\~{a}o de interpreta\c{c}\~{a}o'' \citep[ver tamb\'{e}m][]{ruetsche2015shaky}. Quest\~{o}es sobre o status ontol\'{o}gico da descri\c{c}\~{a}o qu\^{a}ntica ---indissoci\'{a}veis, como vimos, da no\c{c}\~{a}o de medi\c{c}\~{a}o conforme a concep\c{c}\~{a}o tradicional--- s\~{a}o quest\~{o}es interpretativas, e vice-versa. Reiterando, tal distin\c{c}\~{a}o \'{e} um terreno pantanoso: por vezes tomada como garantida \citep[\textit{e.g.,}][p.~78, nota.~3]{french2013}; por outras vezes, totalmente negligeciada  \citep[p.~294]{Ruetsche2018GetRealQM}. Ainda assim, a distin\c{c}\~{a}o oferecida por \citet[p.~9]{cushing1994historicalQM} deve bastar para uma aproxima\c{c}\~{a}o inicial ao tema.

Desse modo, assumo que o debate em rela\c{c}\~{a}o \`{a} interpreta\c{c}\~{a}o do formalismo da teoria qu\^{a}ntica se trata de um debate filos\'{o}fico, especificamente ontol\'{o}gico, na medida em que lida com as entidades que comp\~{o}em o mundo ---entidades essas dadas pela teoria, isto \'{e}, pelo debate te\'{o}rico (cient\'{i}fico). Portanto, os dois momentos do debate acerca da mec\^{a}nica qu\^{a}ntica (filos\'{o}fico e cient\'{i}fico) configuram inst\^{a}ncias diversas, por mais que estejam intrinsecamente conectados entre si.

Ainda assim, enfatizo que minha discuss\~{a}o se limitar\'{a}, neste livro, aos aspectos filos\'{o}ficos, especificamente ontol\'{o}gicos do debate. Para me referir ao debate ontol\'{o}gico de uma teoria f\'{i}sica, \'{e} preciso antes categorizar o termo ``ontologia''. Procurarei delinear brevemente uma defini\c{c}\~{a}o para esse termo, que usarei ao longo deste livro.

\section{As ontologias da ci\^{e}ncia e a ontologia do mundo}

\citet{sep-logic-ontology} elencou, dentre os principais usos na hist\'{o}ria da filosofia, quatro principais sentidos ou acep\c{c}\~{o}es do termo ``ontologia'', dos quais seleciono, para o prop\'{o}sito da discuss\~{a}o, apenas dois. S\~{a}o eles: o estudo acerca do que h\'{a}, que chamarei de \index{ontologia!tradicional}$\mathscr{O}_T$, e o estudo do comprometimento ontol\'{o}gico\index{ontologia!comprometimento ontol\'{o}gico}, que chamarei de \index{ontologia!naturalizada}$\mathscr{O}_N$.

O sentido \index{ontologia!tradicional}$\mathscr{O}_T$ \'{e} comumente chamado sentido \textit{tradicional} do termo ``ontologia'', o que remete \`{a}s discuss\~{o}es, desde Arist\'{o}teles, acerca de uma ``filosofia primeira'' cuja parte central seria a ontologia. Assim, o sentido \index{ontologia!tradicional}$\mathscr{O}_T$, ou tradicional, carrega a pressuposi\c{c}\~{a}o de ser \textit{a \'{u}nica} ontologia, isto \'{e}, a descri\c{c}\~{a}o mais geral do ser enquanto ser \textit{assim como ele \'{e}}.

Diferentemente, ao mencionar o sentido \index{ontologia!naturalizada}$\mathscr{O}_N$, ou \textit{naturalizado}, tem-se impl\'{i}cito, principalmente, o pensamento de \citet{quine1966paradox}, no qual me apoio para me referir \`{a} exist\^{e}ncia de entidades, atrav\'{e}s da linguagem utilizada para descrever as teorias cient\'{i}ficas, o que se torna expl\'{i}cito quando as senten\c{c}as s\~{a}o trazidas para uma linguagem formal.

Conforme argumentado por \citet{russell1905ondenoting}, algumas express\~{o}es lingu\'{i}sticas envolvem quantifica\c{c}\~{a}o existencial. Por exemplo, a frase ``um objeto qu\^{a}ntico'' carrega, implicitamente, o sentido: ``existe algo tal que esse algo \'{e} um objeto qu\^{a}ntico''. Como observou \citet{davidson1980sentences}, at\'{e} mesmo senten\c{c}as de a\c{c}\~{a}o pressup\~{o}em o quantificador existencial ($\exists$); assim, se o termo ``ontologia'' for entendido no sentido \index{ontologia!naturalizada}$\mathscr{O}_N$, pode-se dizer que uma senten\c{c}a como ``uma medi\c{c}\~{a}o efetuada sobre um el\'{e}tron'' compromete-se com a exist\^{e}ncia de uma entidade subat\^{o}mica.\index{ontologia!comprometimento ontol\'{o}gico}

Se a linguagem de uma teoria traz consigo um comprometimento racional com a exist\^{e}ncia de uma entidade, pode-se dizer que a teoria se compromete ontologicamente\index{ontologia!comprometimento ontol\'{o}gico} com essa entidade. \'{E} importante notar que tal afirma\c{c}\~{a}o n\~{a}o diz qual ontologia \'{e} correta, mas diz como o comprometimento ontol\'{o}gico\index{ontologia!comprometimento ontol\'{o}gico} com certas entidades ocorre ---e, portanto, com uma ontologia num sentido \index{ontologia!tradicional}$\mathscr{O}_T$ que as suporte. \'{E} nesse sentido que \citet[p.~66]{quine1966paradox} expressa sua m\'{a}xima: ``\textelp{} ser \'{e} ser o valor de uma vari\'{a}vel''. 

\'{E} interessante fazermos uma pausa aqui e chamar a aten\c{c}\~{a}o para uma quest\~{a}o delicada. A m\'{a}xima Quineana poderia ser interpretada de modo a considerar que as vari\'{a}veis em quest\~{a}o seriam vari\'{a}veis dentro da linguagem da l\'{o}gica cl\'{a}ssica, exclusivamente. No entanto, conforme procurei expor no cap\'{i}tulo anterior, podem existir dificuldades no caso de considerar a l\'{o}gica cl\'{a}ssica como a \'{u}nica l\'{o}gica adequada para o pleno entendimento da totalidade dos fen\^{o}menos e problemas da mec\^{a}nica qu\^{a}ntica ---tese com a qual n\~{a}o compartilho.

Diante essa problem\'{a}tica, diversos apontamentos acerca de quais desses princ\'{i}pios da l\'{o}gica cl\'{a}ssica podem ser revisados para a mec\^{a}nica qu\^{a}ntica foram formulados: (i) o princ\'{i}pio de n\~{a}o contradi\c{c}\~{a}o, da forma como sugerem \citet*[p.~127--226]{dallachiara2009quantumlogic}; (ii) o princ\'{i}pio do terceiro exclu\'{i}do,\footnote{~Na referida tradu\c{c}\~{a}o, traduzido como ``princ\'{i}pio do ter\c{c}o exclu\'{i}do''.} conforme sugere \citet[p.~131]{heisen1958physphil}\index{Heisenberg, Werner}; (iii) a lei de distributividade, da forma como sugerem \citet{birkhoff1936logic}. N\~{a}o discutirei aqui qual dos sistemas l\'{o}gicos n\~{a}o cl\'{a}ssicos seria o mais adequado ao contexto da mec\^{a}nica qu\^{a}ntica (nem mesmo compartilho da ideia de que a l\'{o}gica cl\'{a}ssica seja \textit{inadequada} para a mec\^{a}nica qu\^{a}ntica), isto \'{e}, n\~{a}o me comprometo com algum sistema n\~{a}o cl\'{a}ssico em particular.

Ao inv\'{e}s disso, me aterei \`{a} posi\c{c}\~{a}o de \citet*[757]{DecNewtBue2007paraconsistency}, para os quais outras l\'{o}gicas \textit{podem ajudar} na compreens\~{a}o de certos aspectos da realidade qu\^{a}ntica que n\~{a}o s\~{a}o facilmente explic\'{a}veis quando tratadas \`{a} maneira da l\'{o}gica cl\'{a}ssica, diferentemente das posi\c{c}\~{o}es normativas de que a l\'{o}gica da mec\^{a}nica qu\^{a}ntica n\~{a}o deve ser a l\'{o}gica cl\'{a}ssica. Embora seja de dif\'{i}cil caracteriza\c{c}\~{a}o, \'{e} poss\'{i}vel esbo\c{c}ar uma descri\c{c}\~{a}o do paradigma l\'{o}gico-cl\'{a}ssico. Quando utilizo o termo ``l\'{o}gica n\~{a}o cl\'{a}ssica'', tenho em mente precisamente uma l\'{o}gica pautada pelos princ\'{i}pios de \textit{identidade}, \textit{terceiro exclu\'{i}do} e \textit{n\~{a}o contradi\c{c}\~{a}o} ---o que equivaleria \`{a}quilo que \citet[p.~8]{newton1980ensaios} chama de ``grande l\'{o}gica''. Ainda assim, pode haver l\'{o}gicas n\~{a}o cl\'{a}ssicas que conservem os princ\'{i}pios supracitados. Uma discuss\~{a}o aprofundada sobre esse assunto pode ser encontrada em \citet{newton1993logicaindutiva}.

Da forma como procurei enfatizar no cap\'{i}tulo anterior, a complementaridade\index{Bohr, Niels!princ\'{i}pio da complementaridade} de Bohr\index{Bohr, Niels} seria um dos casos em que uma l\'{o}gica n\~{a}o cl\'{a}ssica ajudaria significativamente na compreens\~{a}o dos conceitos envolvidos. Assim, visto que considero a possibilidade da utiliza\c{c}\~{a}o de sistemas l\'{o}gicos n\~{a}o cl\'{a}ssicos para a interpreta\c{c}\~{a}o da mec\^{a}nica qu\^{a}ntica, adoto aqui a relativiza\c{c}\~{a}o do princ\'{i}pio de Quine, proposta por \citet[p.~284]{dacosta2002principia}: ``penso que ser \'{e} ser o valor de uma vari\'{a}vel em uma dada linguagem com uma determinada l\'{o}gica''.

Feitas tais considera\c{c}\~{o}es acerca da l\'{o}gica subjacente, retornei \`{a} quest\~{a}o dos dois sentidos para a ontologia. \`{A} primeira vista, os sentidos \index{ontologia!tradicional}$\mathscr{O}_T$ e \index{ontologia!naturalizada}$\mathscr{O}_N$ do termo ``ontologia'' s\~{a}o excludentes. No entanto, tomarei a posi\c{c}\~{a}o de \citet{arekrause2012indist}, que compatibilizam as duas acep\c{c}\~{o}es do termo, no preciso sentido em que \index{ontologia!naturalizada}$\mathscr{O}_N$ n\~{a}o implica aquilo que de fato existe ou n\~{a}o, mas somente as entidades com as quais as teorias cient\'{i}ficas se comprometem. Desse modo, pode-se dizer que, se o sentido \index{ontologia!naturalizada}$\mathscr{O}_N$ est\'{a} diretamente associado a uma ou outra teoria cient\'{i}fica, ent\~{a}o depende de aspectos da investiga\c{c}\~{a}o emp\'{i}rica. Assim, se de \index{ontologia!naturalizada}$\mathscr{O}_N$ resulta que certos pressupostos implicam no comprometimento ontol\'{o}gico\index{ontologia!comprometimento ontol\'{o}gico} com certo tipo de entidade, deve-se ou aceitar uma resposta para uma quest\~{a}o do tipo \index{ontologia!tradicional}$\mathscr{O}_T$ acerca de tal entidade ou revisar tais pressupostos filos\'{o}ficos.

Dito de outro modo, o estudo da ontologia associada a uma teoria cient\'{i}fica, num sentido \index{ontologia!naturalizada}$\mathscr{O}_N$, isto \'{e}, a an\'{a}lise sobre os objetos que comp\~{o}em o mundo adotados por essa teoria, n\~{a}o exclui a possibilidade da formula\c{c}\~{a}o de uma ontologia num sentido \index{ontologia!tradicional}$\mathscr{O}_T$ baseado no mobili\'{a}rio ontol\'{o}gico que a teoria fornece. Assim, por mais que os dois sentidos mencionados n\~{a}o sejam excludentes, no que tange aos prop\'{o}sitos da presente an\'{a}lise, basta dizer que assumo, da mesma forma que \citet[p.~48]{arekrause2012indist}, que ``\'{e} leg\'{i}timo investigar a ontologia de uma teoria (ou associada a uma teoria)'' ---num sentido localizado e descritivo, conforme explicitado anteriormente no sentido \index{ontologia!naturalizada}$\mathscr{O}_N$, de modo que n\~{a}o tratarei aqui uma ontologia num sentido \index{ontologia!tradicional}$\mathscr{O}_T$ ---ainda que o cap\'{i}tulo \ref{CapWhitehead} traga algumas considera\c{c}\~{o}es a respeito de \index{ontologia!tradicional}$\mathscr{O}_T$.

Por fim, \'{e} oportuno enfatizar que n\~{a}o utilizo o termo \index{ontologia!e metaf\'{i}sica}\textit{metaf\'{i}sica}. Me alinho com uma tend\^{e}ncia recente na \index{ontologia!e metaf\'{i}sica}metaf\'{i}sica anal\'{i}tica, seguindo autores tais como \citet{arenhart2012ontological}, \citet{Hofweber2016MetOnt}, \citet{tahko2015introduction}, \citet{thomsonjones2017existencenature}, \citet{RaoJonas2019dualismQM} e \citet{arenhartarroyo2021manu}, para quem a \index{ontologia!e metaf\'{i}sica}ontologia trata de quest\~{o}es relativas \`{a} \textit{exist\^{e}ncia} de certas entidades, enquanto a \index{ontologia!e metaf\'{i}sica}\textit{metaf\'{i}sica} ou perfil metaf\'{i}sico trata sobre quest\~{o}es relativas \`{a} \textit{natureza} de tais entidades. Este cap\'{i}tulo trata exclusivamente da ontologia da mec\^{a}nica qu\^{a}ntica, portanto, de \index{ontologia!naturalizada}$\mathscr{O}_N$.

Em suma, pode-se classificar a terminologia apresentada aqui da seguinte maneira: \index{ontologia!tradicional}$\mathscr{O}_T$ diz o que h\'{a}, de fato, no mundo em que voc\^{e} e eu vivemos; \index{ontologia!naturalizada}$\mathscr{O}_N$ diz o que h\'{a} \textit{modulo} uma teoria cient\'{i}fica em quest\~{a}o; e a tese do \index{realismo cient\'{i}fico}realismo cient\'{i}fico \'{e} a correspond\^{e}ncia de \index{ontologia!naturalizada}$\mathscr{O}_N$ em \index{ontologia!tradicional}$\mathscr{O}_T$.

Neste cap\'{i}tulo, argumentarei que o cerne do debate entre interpreta\c{c}\~{o}es da teoria qu\^{a}ntica estaria em uma concep\c{c}\~{a}o de realidade, do tipo \index{ontologia!tradicional}$\mathscr{O}_T$, que seria um tipo de escolha filos\'{o}fica feita por cada um de seus proponentes. Mais ainda, argumentarei que essa escolha tem implica\c{c}\~{o}es do tipo \index{ontologia!naturalizada}$\mathscr{O}_N$. Analiso, neste cap\'{i}tulo, o debate entre Einstein\index{Einstein, Albert} e Bohr\index{Bohr, Niels} para visualizar essa quest\~{a}o.

\section{A realidade da mec\^{a}nica qu\^{a}ntica}

As teses associadas \`{a} interpreta\c{c}\~{a}o de Copenhague, analisadas no cap\'{i}tulo \ref{CapCopenhague}, foram por muito tempo consideradas uma atitude dominante entre os f\'{i}sicos. No entanto, Einstein\index{Einstein, Albert} nunca teria condescendido \`{a} atitude dessa interpreta\c{c}\~{a}o frente aos pressupostos ontol\'{o}gicos \index{ontologia!naturalizada}$\mathscr{O}_N$ que ela carregava. Pode-se destacar suas retic\^{e}ncias em rela\c{c}\~{a}o ao indeterminismo implicado pelo princ\'{i}pio da indetermina\c{c}\~{a}o\index{Heisenberg, Werner!princ\'{i}pio da indetermina\c{c}\~{a}o} de Heisenberg\index{Heisenberg, Werner} e \`{a}s considera\c{c}\~{o}es acerca da causalidade propostas pela complementaridade\index{Bohr, Niels!princ\'{i}pio da complementaridade} de Bohr\index{Bohr, Niels}, mas Einstein\index{Einstein, Albert} se opunha, sobretudo, \`{a} tese do \textit{dist\'{u}rbio interacional}. Isso pois Einstein\index{Einstein, Albert} teria prefer\^{e}ncias ontol\'{o}gicas \index{ontologia!tradicional}$\mathscr{O}_T$ nas quais os estados n\~{a}o observados devem possuir propriedades bem definidas.

Vale recapitular que o argumento do dist\'{u}rbio interacional afirma que h\'{a}, em um processo de medi\c{c}\~{a}o, uma intera\c{c}\~{a}o f\'{i}sica entre o objeto observado e a ag\^{e}ncia de medi\c{c}\~{a}o que observa tal objeto. Tal argumento \'{e} considerado um argumento calcado na concep\c{c}\~{a}o \textit{cl\'{a}ssica}, na medida em que pressup\~{o}e princ\'{i}pio da \textit{separabilidade} como \index{ontologia!tradicional}$\mathscr{O}_T$. Isto \'{e}, o argumento pressup\~{o}e que todos os objetos especialmente distintos existem em distintos estados f\'{i}sicos. Dito de outro modo, um aparelho de medi\c{c}\~{a}o s\'{o} poderia perturbar um objeto que j\'{a} esteja l\'{a} para ser perturbado. Essa afirmativa, como vimos no cap\'{i}tulo \ref{CapCopenhague}, parece indicar um compromisso com uma ideia essencialmente \textit{cl\'{a}ssica} de medi\c{c}\~{a}o. No entanto, a interpreta\c{c}\~{a}o de Copenhague afirma que, a princ\'{i}pio, o conhecimento emp\'{i}rico de tais estados \'{e} impossibilitado pelo postulado qu\^{a}ntico. Assim, a afirma\c{c}\~{a}o do dist\'{u}rbio interacional \'{e} confusa e abriu espa\c{c}o para muitas cr\'{i}ticas na d\'{e}cada de 30. Dentre elas, e talvez a principal, viria por parte de Einstein\index{Einstein, Albert}.

Para alguns historiadores da f\'{i}sica, como \citet[p.~120]{Jammer1974}, o debate entre Bohr\index{Bohr, Niels} e Eintein seria ``um dos grandes debates na hist\'{o}ria da f\'{i}sica''. Ademais, para \citet[p.~126]{folse1994realistbohr}, o pensamento de Bohr\index{Bohr, Niels} s\'{o} poderia ser considerado totalmente maduro ap\'{o}s discuss\~{o}es estabelecidas com Einstein\index{Einstein, Albert}, principalmente no que diz respeito ao conceito de medi\c{c}\~{a}o. Isto \'{e}, se antes de tal debate Bohr\index{Bohr, Niels} haveria endossado a tese do dist\'{u}rbio interacional, depois dele, certamente, isso j\'{a} n\~{a}o mais seria o caso. O debate entre Einstein\index{Einstein, Albert} e Bohr\index{Bohr, Niels} em rela\c{c}\~{a}o \`{a} completude da mec\^{a}nica qu\^{a}ntica \'{e} um \'{o}timo exemplo de como as diferen\c{c}as numa ontologia \index{ontologia!naturalizada}$\mathscr{O}_N$ direcionam ou ao menos influenciam a concep\c{c}\~{a}o da interpreta\c{c}\~{a}o da teoria qu\^{a}ntica de cada autor.

Para visualizar essa tese, iniciarei com a an\'{a}lise do famoso artigo de \citet*[, doravante abreviado como ``EPR'']{epr1963epr}\index{Einstein, Albert!argumento EPR}. O artigo, redigido por Podolsky, questiona a atitude da interpreta\c{c}\~{a}o de Copenhague frente \`{a} no\c{c}\~{a}o de medi\c{c}\~{a}o, como busco analisar adiante.

De acordo com a interpreta\c{c}\~{a}o de Copenhague, as propriedades dos objetos qu\^{a}nticos n\~{a}o teriam valores definidos simultaneamente, devido \`{a} impossibilidade da medi\c{c}\~{a}o de tais quantidades. Ou seja, o estado de um objeto individual em qualquer tempo determinado n\~{a}o teria valores definidos para todas as suas quantidades f\'{i}sicas. \citet*{epr1963epr}\index{Einstein, Albert!argumento EPR} prop\~{o}em um contraexemplo, mediante um experimento de pensamento (\textit{Gedankenexperiment}), em que medi\c{c}\~{o}es precisas e simult\^{a}neas \textit{pudessem} de fato ser efetuadas sobre as propriedades observ\'{a}veis.

Tal racioc\'{i}nio \'{e} frequentemente referido sob a nomenclatura de paradoxo EPR. No entanto, seguirei a proposta de \citet[p.~187--188]{Jammer1974} de optar pelo termo \textit{argumento EPR} visto que os pr\'{o}prios autores jamais consideraram sua tese como um paradoxo, nem no sentido medieval, de insolubilidade, nem no sentido moderno de uma antinomia sint\'{a}tica ou sem\^{a}ntica. O primeiro autor a considerar o \index{Einstein, Albert!argumento EPR}argumento EPR como paradoxal foi \citet[p.~556]{schrodinger1935cat}\index{Schr\"{o}dinger, Erwin} no sentido etimol\'{o}gico do termo \textit{paradoxo}, isto \'{e}, no sentido de ser contr\'{a}rio \`{a} opini\~{a}o corrente na medida em que o \index{Einstein, Albert!argumento EPR}argumento EPR implicaria a ren\'{u}ncia do princ\'{i}pio de localidade, um princ\'{i}pio deveras intuitivo na \'{e}poca (e at\'{e} mesmo nos dias de hoje), ou seja, favor\'{a}vel \`{a} opini\~{a}o corrente. 

Tratarei aqui do argumento conforme exposto por EPR, deixando de lado, portanto, formula\c{c}\~{o}es posteriores tal como a de \citet[p.~611--623]{bohm1951quantumtheory}. Outro aviso antes de prosseguirmos: n\~{a}o tratarei com profundidade da no\c{c}\~{a}o de ``localidade''; antes, guio minha apresenta\c{c}\~{a}o para um princ\'{i}pio ontol\'{o}gico ainda mais forte ---do qual a no\c{c}\~{a}o de localidade \'{e} subsidi\'{a}ria--- \textit{viz.}, o princ\'{i}pio da separabilidade (ou princ\'{i}pio da exist\^{e}ncia independente).

O \index{Einstein, Albert!argumento EPR}argumento EPR se baseia, de acordo com \citet[p.~184]{Jammer1974}, em quatro premissas principais, em que as duas primeiras s\~{a}o formuladas, e as duas \'{u}ltimas s\~{a}o assumidas. Seguirei a reconstru\c{c}\~{a}o de \citet[p.~184]{Jammer1974}, embora n\~{a}o seja a ordena\c{c}\~{a}o do artigo original de \citet*[138]{epr1963epr}\index{Einstein, Albert!argumento EPR}. S\~{a}o elas:

\begin{enumerate}
\item \textit{Crit\'{e}rio de realidade:} os elementos de realidade f\'{i}sica n\~{a}o podem ser determinados por considera\c{c}\~{o}es filos\'{o}ficas \textit{a priori}, mas t\^{e}m de ser encontrados por meio de resultados experimentais e medi\c{c}\~{o}es. ``\textelp{} Se, sem perturbar de forma alguma um sistema, podemos prever com seguran\c{c}a (isto \'{e}, com uma probabilidade igual \`{a} unidade) o valor de uma quantidade f\'{i}sica, ent\~{a}o existe um elemento da realidade f\'{i}sica correspondente a essa quantidade f\'{i}sica'' \citep*[138]{epr1963epr}\index{Einstein, Albert!argumento EPR}.\footnote{~Comparar com a tradu\c{c}\~{a}o de Pessoa Jr. em \citet[p.~213]{lehner2011trad}: ``[s]e, sem de modo algum perturbar um sistema, pudermos prever com certeza (ou seja, com probabilidade igual \`{a} unidade) o valor de uma quantidade f\'{i}sica, ent\~{a}o existe um elemento de realidade f\'{i}sica correspondente a essa quantidade f\'{i}sica''.}

\item \textit{Crit\'{e}rio de completude:} uma teoria \'{e} completa se e somente se ``\textelp{} cada elemento da realidade f\'{i}sica tem uma contrapartida na teoria f\'{i}sica'' \citep*[138]{epr1963epr}\index{Einstein, Albert!argumento EPR}.

\item \textit{Assun\c{c}\~{a}o da localidade:} se ``no momento da medi\c{c}\~{a}o de \textelp{} dois sistemas que j\'{a} n\~{a}o mais interagem, nenhuma mudan\c{c}a real pode ocorrer no segundo sistema em consequ\^{e}ncia de qualquer coisa que possa ser feito com o primeiro sistema'' \citep*[140]{epr1963epr}\index{Einstein, Albert!argumento EPR}.\label{EPR:localidade}

\item \textit{Assun\c{c}\~{a}o da validade:} ``\textelp{} as previs\~{o}es estat\'{i}sticas da mec\^{a}nica qu\^{a}ntica ---na medida em que sejam relevantes para o argumento em si--- s\~{a}o confirmadas pela experi\^{e}ncia'' \citep[p.~184]{Jammer1974}.
\end{enumerate}

\'{E} not\'{a}vel que a formula\c{c}\~{a}o do crit\'{e}rio de realidade carrega pressuposi\c{c}\~{o}es do tipo \index{ontologia!tradicional}$\mathscr{O}_T$, na medida em que considera a realidade f\'{i}sica \textit{algo} cuja exist\^{e}ncia espa\c{c}o-temporal seja objetiva e independente. Esse tipo de pressuposi\c{c}\~{a}o \'{e} frequentemente associada aos conceitos de \textit{realidade f\'{i}sica} da f\'{i}sica cl\'{a}ssica. De acordo com \citet[p.~184]{Jammer1974}, a estrutura do argumento seria tal que, sob a base fornecida por 1), as assun\c{c}\~{o}es 3) e 4) implicariam que a mec\^{a}nica qu\^{a}ntica n\~{a}o satisfaria o crit\'{e}rio 2), que \'{e} o crit\'{e}rio de completude. Como um corol\'{a}rio, a descri\c{c}\~{a}o fornecida por tal teoria seria, ent\~{a}o, incompleta.

Enunciados os crit\'{e}rios, passarei \`{a} an\'{a}lise do experimento de pensamento. Dois objetos qu\^{a}nticos individuais, $A$ e $B$, separados espacialmente depois de interagirem um com o outro, seriam medidos. Devo enfatizar que estou tratando aqui do experimento mental cl\'{a}ssico EPR, e n\~{a}o de suas reformula\c{c}\~{o}es mais recentes ---tal como a de \citet{bohm1951quantumtheory}.

De acordo com o entendimento de EPR, a mec\^{a}nica qu\^{a}ntica, conforme a interpreta\c{c}\~{a}o de Copenhague, prev\^{e} que o sistema $I$ perturba o sistema $II$ de forma descont\'{i}nua. Antes da medi\c{c}\~{a}o, os observ\'{a}veis $A$ e $B$ n\~{a}o possuiriam propriedades bem definidas e, ap\'{o}s a medi\c{c}\~{a}o em algum deles, uma quantidade f\'{i}sica poderia ser determinada sobre o outro. E justamente essa seria a forma como operaria o princ\'{i}pio da indetermina\c{c}\~{a}o\index{Heisenberg, Werner!princ\'{i}pio da indetermina\c{c}\~{a}o}, segundo o qual o conhecimento pleno e simult\^{a}neo dos observ\'{a}veis $A$ e $B$ n\~{a}o seria poss\'{i}vel, visto que, da forma como tal rela\c{c}\~{a}o fora interpretada por EPR, a medi\c{c}\~{a}o de uma quantidade f\'{i}sica de algum dos pares implica perturba\c{c}\~{a}o ou dist\'{u}rbio do outro. Nesse sentido, $A$ e $B$ seriam observ\'{a}veis com quantidades f\'{i}sicas incompat\'{i}veis.

Tendo em vista esses pontos, pode-se passar ao \index{Einstein, Albert!argumento EPR}argumento EPR. Se as ``quantidades f\'{i}sicas incompat\'{i}veis'' ---$A$ e $B$--- t\^{e}m realidade simult\^{a}nea e se a descri\c{c}\~{a}o qu\^{a}ntica da realidade \'{e} completa, ent\~{a}o a mec\^{a}nica qu\^{a}ntica deveria fornecer valores precisos e simult\^{a}neos para os observ\'{a}veis incompat\'{i}veis $A$ e $B$. No entanto, de acordo com o princ\'{i}pio da indetermina\c{c}\~{a}o\index{Heisenberg, Werner!princ\'{i}pio da indetermina\c{c}\~{a}o}, a mec\^{a}nica qu\^{a}ntica n\~{a}o fornece tais valores precisos simult\^{a}neos para os valores das propriedades de, por exemplo, posi\c{c}\~{a}o e momento de um objeto qu\^{a}ntico e, por isso, tais propriedades s\~{a}o referidas como quantidades incompat\'{i}veis. Uma das maneiras de entender o porqu\^{e} da incompatibilidade de tais quantidades \'{e} pela tese do dist\'{u}rbio: um sistema perturba instantaneamente as propriedades do outro sistema, mesmo que espacialmente distantes.

Se isso fosse o caso, a assun\c{c}\~{a}o da \textit{localidade} (mencionada acima) tamb\'{e}m seria violada. Assim, ou a descri\c{c}\~{a}o qu\^{a}ntica da realidade n\~{a}o \'{e} completa, ou as quantidades f\'{i}sicas incompat\'{i}veis n\~{a}o podem ter realidade simult\^{a}nea. Abaixo, o \index{Einstein, Albert!argumento EPR}argumento EPR \'{e} reproduzido sob a forma de uma redu\c{c}\~{a}o ao absurdo. A disjun\c{c}\~{a}o ``ou'' do argumento \'{e} originalmente introduzida sob a forma de uma contradi\c{c}\~{a}o:

\begin{itemize}
    \item $C$: A descri\c{c}\~{a}o qu\^{a}ntica da realidade \'{e} completa;
    \item $RS$: Quantidades f\'{i}sicas ``incompat\'{i}veis'' podem ter realidade simult\^{a}nea;
    \item $\psi_{AB}$: A mec\^{a}nica qu\^{a}ntica fornece valores precisos e simult\^{a}neos para as quantidades `incompat\'{i}veis' $A$ e $B$.
\end{itemize}

\begin{equation*}
\begin{fitch}
\fa (RS \lor C) & P \\
\fj \neg \psi_{AB} & P \\
\fa \neg(RS\land C) & 1--2 \\
\fa \neg C \land \neg RS & 3\\
\fa \fh C\rightarrow RS & H: EPR \\
\fa \fa C\rightarrow\neg RS & 4 \\
\fa \fa C\rightarrow(RS\land\neg RS) & 5--6 \\
\fa \neg C & 7: RAA
\end{fitch}
\end{equation*}

Brevemente: a primeira premissa diz respeito \`{a} defini\c{c}\~{a}o de completude; a segunda premissa descreve a mec\^{a}nica qu\^{a}ntica. No terceiro passo, tem-se um \textit{modus tollens} a partir de 1 e 2; no quarto passo tem-se a aplica\c{c}\~{a}o da lei de de Morgan a partir de 3. O quinto passo \'{e} a hip\'{o}tese referente ao crit\'{e}rio de realidade, conforme exposto no \index{Einstein, Albert!argumento EPR}argumento EPR. No sexto passo, tem-se uma aplica\c{c}\~{a}o do silogismo disjuntivo a partir do passo 4; o passo 7 apresenta uma contradi\c{c}\~{a}o a partir de 5 e 6; o oitavo e \'{u}ltimo passo apresenta uma conclus\~{a}o por redu\c{c}\~{a}o ao absurdo. O uso do termo \textit{contradi\c{c}\~{a}o}, conforme empregado no racioc\'{i}nio, precisamente ap\'{o}s o condicional da etapa 7 da reconstru\c{c}\~{a}o acima, deve ser entendido \`{a} maneira da l\'{o}gica cl\'{a}ssica. \'{E} preciso qualificar tal afirma\c{c}\~{a}o, pois considero, anteriormente, a leg\'{i}tima possibilidade da utiliza\c{c}\~{a}o de l\'{o}gicas n\~{a}o cl\'{a}ssicas na interpreta\c{c}\~{a}o da mec\^{a}nica qu\^{a}ntica.

Tal situa\c{c}\~{a}o ocorre na medida em que a discuss\~{a}o acerca de uma interpreta\c{c}\~{a}o da mec\^{a}nica qu\^{a}ntica acontece no plano metalingu\'{i}stico, que corresponde a uma por\c{c}\~{a}o restrita da linguagem natural. \index{metalinguagem e metametalinguagem}Em tal metalinguagem, as regras sem\^{a}nticas s\~{a}o pressupostas e, portanto, n\~{a}o formalizadas; assim, a discuss\~{a}o metalingu\'{i}stica acontece em linguagem natural e, por conseguinte, obedece \`{a}s regras desse discurso que tem a l\'{o}gica cl\'{a}ssica como linguagem subjacente. Para uma discuss\~{a}o mais aprofundada sobre isso, ver \citet[p.~50--55]{church1956introduction} e \citet{krauseare2016logical}. Apresento o \index{Einstein, Albert!argumento EPR}argumento EPR de modo formalizado por quest\~{o}es de clareza; a discuss\~{a}o que apresento em torno da sem\^{a}ntica do argumento, no entanto, continua obedecendo \`{a}s regras metalingu\'{i}sticas da linguagem natural: a l\'{o}gica cl\'{a}ssica.

Ademais, como aponta \citet[p.~306]{murdoch1994bohreinstein}, o argumento original, conforme formalizado acima, tem uma estrutura inv\'{a}lida. Como o crit\'{e}rio de realidade adotado por EPR implica realidade simult\^{a}nea das quantidades f\'{i}sicas incompat\'{i}veis, deve-se negar a completude da descri\c{c}\~{a}o qu\^{a}ntica da realidade. Como EPR se comprometem com a tese da realidade independente como \index{ontologia!tradicional}$\mathscr{O}_T$, fica claro que todos os objetos qu\^{a}nticos possuem realidade independente ---logo, simult\^{a}nea. Isso ocorre, pois a no\c{c}\~{a}o de ``realidade simult\^{a}nea'' depende da no\c{c}\~{a}o de ``realidade objetiva''--- ou seja, dois objetos devem, primeiro, existir objetivamente para que possam ter realidade simult\^{a}nea. Assim, \citet*[141]{epr1963epr}\index{Einstein, Albert!argumento EPR} s\~{a}o ``for\c{c}ados a concluir'' que a descri\c{c}\~{a}o dos objetos, conforme a mec\^{a}nica qu\^{a}ntica (\textit{modulo} interpreta\c{c}\~{a}o de Copenhague) n\~{a}o \'{e} completa.

No mesmo ano, em resposta a EPR, \citet[p.~145--146]{bohr1935epr}\footnote{~Cuidado: devido a um erro de diagrama\c{c}\~{a}o, a reimpress\~{a}o na famosa edi\c{c}\~{a}o de 1983 possui as p\'{a}ginas 148--149 invertidas! Hoje em dia \'{e} mais seguro consultar o artigo digital, conforme dispon\'{i}vel na \textit{Physical Review} (\textsc{doi:} \href{https://doi.org/10.1103/PhysRev.47.777}{10.1103/PhysRev.47.777}). Isso n\~{a}o ocorre na tradu\c{c}\~{a}o brasileira de 1981, conforme publicada no volume 2 do peri\'{o}dico \textit{Cadernos de Hist\'{o}ria e Filosofia da Ci\^{e}ncia}.}\index{Einstein, Albert!argumento EPR!Resposta de Bohr} escreve um artigo argumentando em defesa do princ\'{i}pio da indetermina\c{c}\~{a}o\index{Heisenberg, Werner!princ\'{i}pio da indetermina\c{c}\~{a}o}. Nele, afirma que:

\begin{quote}
A aparente contradi\c{c}\~{a}o [apontada no artigo de EPR] s\'{o} evidencia uma inadequa\c{c}\~{a}o essencial da perspectiva filos\'{o}fica usual [cl\'{a}ssica] de fornecer uma descri\c{c}\~{a}o racional dos fen\^{o}menos f\'{i}sicos do tipo que estamos interessados na mec\^{a}nica qu\^{a}ntica. De fato, a intera\c{c}\~{a}o finita entre objeto e as ag\^{e}ncias de medi\c{c}\~{a}o, condicionadas pela pr\'{o}pria exist\^{e}ncia do quantum de a\c{c}\~{a}o, implica ---devido \`{a} impossibilidade de controlar a rea\c{c}\~{a}o provocada pelo objeto nos instrumentos de medi\c{c}\~{a}o, se estes devem servir a seus prop\'{o}sitos--- a necessidade de uma ren\'{u}ncia final ao ideal cl\'{a}ssico de causalidade e uma revis\~{a}o radical de nossa atitude perante o problema da realidade f\'{i}sica. \citep[p.~145--146]{bohr1935epr}\index{Einstein, Albert!argumento EPR!Resposta de Bohr}.\footnote{~Comparar com a tradu\c{c}\~{a}o de \citet[p.~30]{chibeni1997aspectos}: ``[a] aparente contradi\c{c}\~{a}o [apontada por EPR] na verdade revela apenas uma inadequa\c{c}\~{a}o essencial do ponto de vista usual da filosofia natural para um tratamento racional dos fen\^{o}menos f\'{i}sicos do tipo dos que nos ocupamos na mec\^{a}nica qu\^{a}ntica. De fato, a intera\c{c}\~{a}o finita entre objeto e agentes de mensura\c{c}\~{a}o, condicionada pela pr\'{o}pria exist\^{e}ncia do quantum de a\c{c}\~{a}o, acarreta \textelp{} a necessidade de uma ren\'{u}ncia final do ideal cl\'{a}ssico de causalidade e uma revis\~{a}o radical de nossa atitude com rela\c{c}\~{a}o ao problema da realidade f\'{i}sica''.}
\end{quote}
Pode-se observar que \'{e} precisamente em rela\c{c}\~{a}o ao crit\'{e}rio de realidade, assumido por EPR, frequentemente chamado de ``cl\'{a}ssico'', que \citet{bohr1935epr}\index{Einstein, Albert!argumento EPR!Resposta de Bohr}\index{Bohr, Niels} se posiciona contrariamente na passagem acima. Ao rejeitar a tese \index{ontologia!naturalizada}$\mathscr{O}_N$ de Einstein\index{Einstein, Albert}, Bohr\index{Bohr, Niels} acaba por elaborar ainda mais sua pr\'{o}pria \index{ontologia!naturalizada}$\mathscr{O}_N$; no entanto, tal rejei\c{c}\~{a}o \'{e} comumente vista como a necessidade de uma revis\~{a}o ontol\'{o}gica para as teorias f\'{i}sicas, ou ainda uma revis\~{a}o na sem\^{a}ntica, isto \'{e}, uma revis\~{a}o nos limites de aplica\c{c}\~{a}o e na defini\c{c}\~{a}o dos conceitos envolvidos, tal como o conceito de ``realidade f\'{i}sica''. Nesse mesmo artigo, diz Bohr\index{Bohr, Niels}:

\begin{quote}
A possibilidade de atribuir significado inequ\'{i}voco a express\~{o}es tais como ``realidade f\'{i}sica'' n\~{a}o pode, evidentemente, ser deduzida a partir de concep\c{c}\~{o}es filos\'{o}ficas a priori, mas ---como os autores do artigo citado [EPR] enfatizam--- deve ser fundamentada no recurso direto a experi\^{e}ncias e medi\c{c}\~{o}es. \citep[p.~145]{bohr1935epr}\index{Einstein, Albert!argumento EPR!Resposta de Bohr}\index{Bohr, Niels}.
\end{quote}
Segundo esse racioc\'{i}nio, se toda medi\c{c}\~{a}o \'{e} limitada \`{a} informa\c{c}\~{a}o que  se obt\'{e}m por meio dela, essa limita\c{c}\~{a}o se estende ao significado que se pode atribuir por meio dela ---o que \'{e} uma consequ\^{e}ncia direta da atitude operacionista, tamb\'{e}m uma \index{ontologia!naturalizada}$\mathscr{O}_N$, assumida por \citet[p.~89--90]{bohr1928quantumpostulate} nos fundamentos da interpreta\c{c}\~{a}o de Copenhague. Assim, a pr\'{o}pria ideia de uma \index{ontologia!tradicional}$\mathscr{O}_T$ n\~{a}o seria significativa, isto \'{e}, uma realidade \textit{em si}, com o estabelecimento das suas propriedades intr\'{i}nsecas, fora do contexto do aparato medidor utilizado. Para visualizar melhor esse aspecto do argumento de Bohr\index{Bohr, Niels}, utilizo a reconstru\c{c}\~{a}o do contra-argumento feita por \citet[p.~304]{murdoch1994bohreinstein}:

\begin{itemize}
\item Observ\'{a}veis complementares (como posi\c{c}\~{a}o e momento) n\~{a}o podem ser medidas simultaneamente; s\~{a}o necess\'{a}rias opera\c{c}\~{o}es experimentais mutuamente exclusivas para a sua medi\c{c}\~{a}o;
\item Uma medi\c{c}\~{a}o envolve uma intera\c{c}\~{a}o inelimin\'{a}vel entre o objeto e as ag\^{e}ncias de medi\c{c}\~{a}o;
\item A intera\c{c}\~{a}o com a medi\c{c}\~{a}o \'{e} indetermin\'{a}vel. Qualquer tentativa de medi-la necessitaria de mudan\c{c}as no arranjo experimental e ao menos mais uma intera\c{c}\~{a}o, o que impossibilitaria a medi\c{c}\~{a}o original;
\item Portanto, os resultados das medi\c{c}\~{o}es sucessivas de observ\'{a}veis complementares n\~{a}o podem ser atribu\'{i}dos por extrapola\c{c}\~{a}o ao mesmo instante.
\end{itemize}

De acordo com essa linha de racioc\'{i}nio, o tipo de experimento que EPR propuseram n\~{a}o seria poss\'{i}vel, pois os termos como \textit{posi\c{c}\~{a}o} ou  \textit{momento} s\'{o} teriam significado quando associados a uma opera\c{c}\~{a}o experimental e, uma vez que s\'{o} podem ser designados experimentos mutuamente exclusivos para verificar o valor de verdade de tais termos, n\~{a}o se poderia atribuir significado a uma senten\c{c}a como \textit{valores definidos simultaneamente de posi\c{c}\~{a}o e momento}.

Tal atitude indica, no limite, que as opera\c{c}\~{o}es experimentais deveriam ser condi\c{c}\~{o}es necess\'{a}rias para a defini\c{c}\~{a}o de senten\c{c}as tais como \textit{a posi\c{c}\~{a}o (ou momento) exata}. Na medida em que as opera\c{c}\~{o}es experimentais necess\'{a}rias para a defini\c{c}\~{a}o das propriedades observ\'{a}veis dos objetos qu\^{a}nticos s\~{a}o mutuamente exclusivas, as condi\c{c}\~{o}es para suas defini\c{c}\~{o}es tamb\'{e}m o seriam. Dito de outro modo, a tese \index{ontologia!naturalizada}$\mathscr{O}_N$ impl\'{i}cita por tr\'{a}s desse racioc\'{i}nio \'{e} que o contexto experimental deveria determinar e limitar a express\~{a}o ``realidade f\'{i}sica''.

De fato, \'{e} intuitiva a concep\c{c}\~{a}o de que o mundo que nos circunda possui um estatuto ontol\'{o}gico de exist\^{e}ncia independente. Isto \'{e}, que os objetos que o comp\~{o}em (\'{a}tomos, part\'{i}culas, pr\'{e}dios e montanhas) se limitariam a \textit{estar l\'{a}} de forma objetiva, a despeito da observa\c{c}\~{a}o de qualquer sujeito. Se as coisas fossem assim, ent\~{a}o as propriedades desses objetos existiriam e teriam propriedades bem definidas antes ou ap\'{o}s uma medi\c{c}\~{a}o, ou seja, a despeito de qualquer poss\'{i}vel medi\c{c}\~{a}o ou observa\c{c}\~{a}o. \'{E} justamente essa a defini\c{c}\~{a}o da no\c{c}\~{a}o de ``realidade objetiva'' utilizada no \index{Einstein, Albert!argumento EPR}argumento EPR.

Essa no\c{c}\~{a}o \'{e} compat\'{i}vel com a acep\c{c}\~{a}o \index{ontologia!tradicional}$\mathscr{O}_T$ do termo ontologia, pois pretende-se uma descri\c{c}\~{a}o \textit{da realidade}, e n\~{a}o somente um construto da ci\^{e}ncia. Isto \'{e}, se trata de uma tese que p\~{o}e-se \`{a} frente da investiga\c{c}\~{a}o te\'{o}rica e molda aquilo que pode (ou n\~{a}o) ser teorizado pela ci\^{e}ncia. Pode-se ver aqui a conflu\^{e}ncia entre duas posi\c{c}\~{o}es filos\'{o}ficas: (i) o \index{realismo metaf\'{i}sico}realismo metaf\'{i}sico e (ii) o \index{realismo cient\'{i}fico}realismo cient\'{i}fico. Grosso modo, tais acep\c{c}\~{o}es do termo realismo se comprometem com as seguintes teses: (i) h\'{a} uma (\'{u}nica) realidade f\'{i}sica que existe objetivamente, independente de qualquer teoria, vontade, consci\^{e}ncia ou observador e (ii) \'{e} tarefa da ci\^{e}ncia descrever corretamente essa realidade por meio das melhores teorias. A mec\^{a}nica qu\^{a}ntica, no entanto, tem sido, at\'{e} hoje, um \'{o}timo campo de debate para essas duas acep\c{c}\~{o}es do termo \textit{realismo}, na medida em que admite interpreta\c{c}\~{o}es contr\'{a}rias e favor\'{a}veis. A seguir, analisarei em linhas gerais o debate entre \index{realismo metaf\'{i}sico}realismo e antirrealismo no debate entre Einstein\index{Einstein, Albert} e Bohr\index{Bohr, Niels}, e como as fugazes fronteiras entre os realismos metaf\'{i}sico e cient\'{i}fico nesse debate.

\section{Realidade, separabilidade e indetermina\c{c}\~{a}o}
\subsection{Realidade}

Em uma carta endere\c{c}ada a Schr\"{o}dinger, datada de 19 de Junho de 1935, Einstein\index{Einstein, Albert} afirma que:

\begin{quote}
Por raz\~{o}es de linguagem, esse [artigo EPR]\index{Einstein, Albert!argumento EPR} foi escrito por Podolsky depois de muita discuss\~{a}o. Ainda assim, o artigo n\~{a}o saiu da forma como eu originalmente gostaria; \textit{ao contr\'{a}rio, o ponto essencial foi, por assim dizer, obscurecido pelo formalismo.} \citep[p.~35, nota 9, \^{e}nfase adicionada]{einstein-to-schrodinger1935}.\footnote{~Comparar o trecho enfatizado com a tradu\c{c}\~{a}o de \citet[35]{chibeni1997aspectos}: ``[Podolsky] sepultou o ponto central pela erudi\c{c}\~{a}o''.}
\end{quote}
Conforme entendido por Einstein, a tarefa da f\'{i}sica \'{e} em si um empreendimento filos\'{o}fico, na medida em que visa (ou deveria visar) a \textit{descri\c{c}\~{a}o da realidade} ---sem a qual ``\textelp{} a f\'{i}sica ent\~{a}o s\'{o} poderia reivindicar o interesse do com\'{e}rcio e da engenharia; todo o empreendimento seria um desastre lament\'{a}vel'' \citep[p.~39]{einstein190letter-schrod}. A maior \^{e}nfase do artigo EPR\index{Einstein, Albert!argumento EPR} foi dada \`{a} discuss\~{a}o sobre a atribui\c{c}\~{a}o de valores bem definidos simultaneamente para os pares observ\'{a}veis (como posi\c{c}\~{a}o e momento), discuss\~{a}o essa sobre a qual, na mesma carta, \citet{einstein-to-schrodinger1935}\index{Einstein, Albert} expressa seu descontentamento atrav\'{e}s da express\~{a}o ``\textit{ist mir Wurst}'' ---traduzida por \citet[p.~38]{fine1986shaky} como ``\textit{I couldn't care less}'' e por \citet[p.~56]{chibeni1997aspectos} como ``n\~{a}o ligo a m\'{i}nima''. 

O referido ``ponto essencial'',\footnote{~Ver \citet[p.~35, nota~9]{fine1986shaky}.} omitido no artigo EPR, \'{e} retomado por \citet{einstein-to-schrodinger1935} em seguida de maneira bastante simples. Considere a situa\c{c}\~{a}o na qual temos uma bola que pode ser distribu\'{i}da em duas caixas.

\begin{quote}
Agora descrevo um estado de coisas da seguinte forma: a probabilidade \'{e} $\nicefrac{1}{2}$ de que a bola esteja na primeira caixa. Essa \'{e} uma descri\c{c}\~{a}o completa?

\textit{N\~{a}o:} Uma descri\c{c}\~{a}o completa \'{e}: a bola est\'{a} (ou n\~{a}o est\'{a}) na primeira caixa. \'{E} assim que a caracteriza\c{c}\~{a}o do estado de coisas deve aparecer em uma descri\c{c}\~{a}o completa.

\textit{Sim:} Antes de abri-las, a bola n\~{a}o est\'{a} de forma alguma em uma das duas caixas. \textit{Estar em uma caixa definitiva s\'{o} acontece quando levanto as tampas.} \citep[p.~69, \^{e}nfase adicionada]{einstein-to-schrodinger1935}.\index{Schr\"{o}dinger, Erwin}\index{Einstein, Albert}
\end{quote}
Vale destacar que o c\'{e}lebre ``paradoxo do gato de Schr\"{o}dinger''\index{Schr\"{o}dinger, Erwin!paradoxo do gato} nasce dessas correspond\^{e}ncias entre Einstein\index{Einstein, Albert} e Schr\"{o}dinger ---especificamente do exemplo acima das caixas \citep[ver][cap\'{i}tulo~5]{fine1986shaky}.\index{Einstein, Albert} O que est\'{a} em jogo \'{e} a pr\'{o}pria no\c{c}\~{a}o de realidade, na qual est\'{a} condicionada \`{a} no\c{c}\~{a}o de ``medi\c{c}\~{a}o'' (\textit{viz.}, ``abrir a caixa''). Se a mec\^{a}nica qu\^{a}ntica oferece uma descri\c{c}\~{a}o completa da realidade, uma das consequ\^{e}ncias \'{e} que os objetos dos quais ela trata s\'{o} possuem realidade f\'{i}sica quando est\~{a}o sendo ``medidos''. Lembre-se que essa \'{e} uma consequ\^{e}ncia do princ\'{i}pio ``medi\c{c}\~{a}o=cria\c{c}\~{a}o'' que vimos no cap\'{i}tulo \ref{CapCopenhague}. Isso, para \citet{einstein1949autobiographical},\index{Einstein, Albert} violaria n\~{a}o somente nosso senso comum, mas o prop\'{o}sito da f\'{i}sica:

\begin{quote}
A f\'{i}sica \'{e} uma tentativa conceitual de compreender a realidade conforme ela \'{e} pensada independentemente de ser observada. Nesse sentido, fala-se de ``realidade f\'{i}sica''. Na f\'{i}sica pr\'{e}-qu\^{a}ntica, n\~{a}o havia d\'{u}vida sobre como isso deveria ser compreendido. \citep[81]{einstein1949autobiographical}.\index{Einstein, Albert}
\end{quote}
No entanto, \'{e} precisamente essa ``realidade f\'{i}sica'' (ou ``realidade independente'', como veremos a seguir) que o modo ---at\'{e} ent\~{a}o--- usual de interpretar a mec\^{a}nica qu\^{a}ntica questionaria. Sendo uma descri\c{c}\~{a}o, ao mesmo tempo \textit{completa} e \textit{estat\'{i}stica}, a conclus\~{a}o de que objetos n\~{a}o existem independentemente de suas medi\c{c}\~{o}es com todas as suas propriedades bem-definidas parece inescap\'{a}vel. E \'{e} precisamente \textit{esse} o ponto essencial das cr\'{i}ticas de Einstein\index{Einstein, Albert} \`{a} mec\^{a}nica qu\^{a}ntica (ou melhor: da maneira usual de interpretar a mec\^{a}nica qu\^{a}ntica).

\begin{quote}
O que n\~{a}o me satisfaz nessa teoria [qu\^{a}ntica], do ponto de vista dos princ\'{i}pios, \'{e} sua atitude em rela\c{c}\~{a}o ao que me parece ser o objetivo program\'{a}tico de toda a f\'{i}sica: a descri\c{c}\~{a}o completa de qualquer situa\c{c}\~{a}o real (individual), conforme ela supostamente existe independentemente de qualquer ato de observa\c{c}\~{a}o \textelp{}. \citep[667]{einstein1949remarks}.\index{Einstein, Albert}
\end{quote}
Assim, devo enfatizar novamente que o ``ponto essencial'' de Einstein\index{Einstein, Albert} n\~{a}o era tanto sobre a mec\^{a}nica qu\^{a}ntica asserir ou n\~{a}o valores simult\^{a}neos para quantidades incompat\'{i}veis. O ponto essencial era a realidade f\'{i}sica de algo \textit{depender} da medi\c{c}\~{a}o. Ou seja, o ponto essencial era contra a ideia de \textit{medi\c{c}\~{a}o=cria\c{c}\~{a}o}. A partir desse ponto, temos dois caminhos a seguir. O primeiro diz respeito \`{a} no\c{c}\~{a}o de ``separabilidade'', e o segundo diz respeito \`{a} indetermina\c{c}\~{a}o. Comecemos pelo primeiro.

\subsection{Separabilidade}
Para aprofundar a discuss\~{a}o, seguirei aqui a reconstru\c{c}\~{a}o dos argumentos de \citet{einstein1950later} proposta por \citet[p.~309]{murdoch1994bohreinstein}, segundo a qual o argumento EPR pode ser estruturado da seguinte maneira:

\begin{enumerate}
\item O estado f\'{i}sico de um objeto qu\^{a}ntico pode ser descrito tanto pelo vetor $|\psi\rangle$ ou $|\varphi\rangle$, e tal descri\c{c}\~{a}o depende do tipo de medi\c{c}\~{a}o realizada em outro objeto, distante, $A$;

\item O estado f\'{i}sico de um objeto n\~{a}o depende do tipo de medi\c{c}\~{a}o realizada no outro objeto ou sobre o estado f\'{i}sico do outro objeto (princ\'{i}pio da separabilidade);

\item O objeto $B$ est\'{a} no mesmo estado f\'{i}sico, quer seja descrito por $|\psi\rangle$ ou $|\varphi\rangle$;

\item Um vetor de estado fornece uma descri\c{c}\~{a}o completa do estado f\'{i}sico de um objeto apenas se descrever exclusivamente esse estado, isto \'{e}, exclusivamente $|\psi\rangle$ ou $|\varphi\rangle$ pode descrever completamente o estado de um dado objeto (a condi\c{c}\~{a}o de completude);

\item Na situa\c{c}\~{a}o EPR, o estado f\'{i}sico do objeto $B$ pode ser descrito quer por $|\psi\rangle$ ou $|\varphi\rangle$;

\item Nem $|\psi\rangle$ nem $|\varphi\rangle$ fornecem uma descri\c{c}\~{a}o completa do estado f\'{i}sico de $B$;

\item Portanto, a mec\^{a}nica qu\^{a}ntica n\~{a}o fornece uma descri\c{c}\~{a}o completa do estado f\'{i}sico de um objeto qu\^{a}ntico.
\end{enumerate}

Uma an\'{a}lise exaustiva do argumento de Einstein\index{Einstein, Albert} n\~{a}o \'{e} prop\'{o}sito deste livro, motivo pelo qual assumirei que a reconstru\c{c}\~{a}o feita por \citet[p.~309]{murdoch1994bohreinstein} \'{e} suficiente. No entanto, \'{e} relevante para minha an\'{a}lise a discuss\~{a}o sobre algumas implica\c{c}\~{o}es filos\'{o}ficas, especialmente nos pontos 2 e 4 da reconstru\c{c}\~{a}o acima.

Em outros textos, \citet[p.~681--682]{einstein1949remarks} argumenta que o referido princ\'{i}pio da separabilidade, contido na premissa 2, se divide em dois outros aspectos principais: o \textit{princ\'{i}pio da localidade} e o \textit{princ\'{i}pio da exist\^{e}ncia independente}. De acordo com o primeiro, o que acontece em uma determinada localiza\c{c}\~{a}o no espa\c{c}o independe do que acontece em outra determinada localiza\c{c}\~{a}o no espa\c{c}o, ou seja, n\~{a}o h\'{a} uma a\c{c}\~{a}o \`{a} dist\^{a}ncia imediata ou instant\^{a}nea entre objetos que ocupam diferentes lugares no espa\c{c}o. De acordo com o \'{u}ltimo aspecto, o que existe em uma determinada localiza\c{c}\~{a}o do espa\c{c}o independe daquilo que existe em outra determinada localiza\c{c}\~{a}o no espa\c{c}o, isto \'{e}, o princ\'{i}pio da exist\^{e}ncia independente afirma que n\~{a}o h\'{a} uma conex\~{a}o  ontol\'{o}gica imediata ou instant\^{a}nea entre objetos que ocupam diferentes lugares no espa\c{c}o. Para \citet[p.~310]{murdoch1994bohreinstein}, esse seria o ponto crucial omitido no artigo EPR, sugerindo ainda, que sua omiss\~{a}o seria o principal motivo pelo qual o argumento fora t\~{a}o suscet\'{i}vel a cr\'{i}ticas.

J\'{a} no princ\'{i}pio de completude, contido no ponto 4 da reconstru\c{c}\~{a}o de \citet[p.~309]{murdoch1994bohreinstein}, Einstein\index{Einstein, Albert} assume a exist\^{e}ncia de somente uma descri\c{c}\~{a}o completa de um sistema f\'{i}sico. Os argumentos sobre completude s\~{a}o encontrados em detalhe nas notas autobiogr\'{a}ficas de \citet[p.~83--87]{einstein1949autobiographical}, nas quais h\'{a} a afirma\c{c}\~{a}o de que se uma fun\c{c}\~{a}o de onda fornece uma descri\c{c}\~{a}o completa da realidade ---segundo os termos da sua pr\'{o}pria no\c{c}\~{a}o de completude explicitada acima---, ent\~{a}o existiriam casos em que a medi\c{c}\~{a}o deveria ser considerada como um ato de cria\c{c}\~{a}o, ao inv\'{e}s de um ato de revela\c{c}\~{a}o do valor de um objeto medido.

Dito de outro modo, uma descri\c{c}\~{a}o completa de um aspecto f\'{i}sico da realidade seria uma descri\c{c}\~{a}o do estado real de um objeto real. Assim, se uma descri\c{c}\~{a}o completa n\~{a}o fornece um valor definido para uma propriedade observ\'{a}vel do objeto em quest\~{a}o, significa que tal objeto n\~{a}o tem um valor definido para a propriedade observ\'{a}vel. No entanto, uma medi\c{c}\~{a}o subsequente mostraria um valor definido para tal propriedade, precisamente daquela que n\~{a}o tinha um valor definido. Como consequ\^{e}ncia, se assumido o princ\'{i}pio de completude, a medi\c{c}\~{a}o cria a quantidade definida de uma propriedade observ\'{a}vel ---e, por conseguinte, num sentido mais forte, a sua realidade f\'{i}sica--- ao inv\'{e}s de revelar uma propriedade (ou a realidade f\'{i}sica de tal propriedade) pr\'{e}-existente. Esse aspecto da medi\c{c}\~{a}o se refere ao princ\'{i}pio da \textit{medi\c{c}\~{a}o=cria\c{c}\~{a}o}.

Essa conclus\~{a}o seria, no entanto, conflitante com a vis\~{a}o einsteiniana de mundo, de acordo com a qual, a exist\^{e}ncia da realidade f\'{i}sica independe ontologicamente de uma medi\c{c}\~{a}o. Para \citet[p.~667]{einstein1949remarks}, a meta de uma teoria f\'{i}sica deveria ser a de fornecer ``\textelp{} a descri\c{c}\~{a}o completa de qualquer situa\c{c}\~{a}o real (e individual, que supostamente existe independentemente de qualquer ato de observa\c{c}\~{a}o ou comprova\c{c}\~{a}o)''. Assim, seguindo a linha de racioc\'{i}nio aqui proposta, o princ\'{i}pio da separabilidade e o princ\'{i}pio da completude seriam princ\'{i}pios mutuamente exclusivos. \citet[p.~682]{einstein1949remarks} teria optado por manter apenas o princ\'{i}pio da separabilidade e, da forma como interpreta a posi\c{c}\~{a}o de Bohr\index{Bohr, Niels}, a interpreta\c{c}\~{a}o de Copenhague optaria por manter apenas o princ\'{i}pio da completude.

Em suma, Einstein\index{Einstein, Albert} teria ao menos tr\^{e}s raz\~{o}es principais para discordar de Bohr\index{Bohr, Niels}: em primeiro lugar, seria a rejei\c{c}\~{a}o da tese verificacionista assumida por Bohr\index{Bohr, Niels}; em segundo lugar, estaria a rejei\c{c}\~{a}o da tese da \textit{medi\c{c}\~{a}o=cria\c{c}\~{a}o}; em terceiro lugar estaria a rejei\c{c}\~{a}o do princ\'{i}pio da completude como um todo, na medida em que \'{e} mutuamente exclusivo em rela\c{c}\~{a}o ao princ\'{i}pio da separabilidade, princ\'{i}pio esse muito caro para a vis\~{a}o einsteiniana, por negar uma a\c{c}\~{a}o \`{a} dist\^{a}ncia ou uma conex\~{a}o ontol\'{o}gica simult\^{a}nea entre as propriedades de dois objetos espacialmente separados. Volto a enfatizar que essa seria a leitura de Einstein\index{Einstein, Albert} sobre a interpreta\c{c}\~{a}o de Copenhague, e, principalmente, do pensamento de Bohr\index{Bohr, Niels} ---o que, como apresentarei adiante, n\~{a}o corresponde necessariamente \`{a} tese do pr\'{o}prio Bohr\index{Bohr, Niels}.

Vale relembrar que a proposta no artigo EPR seria a an\'{a}lise de uma situa\c{c}\~{a}o em que seria poss\'{i}vel atribuir valores bem definidos para as propriedades observ\'{a}veis de dois objetos $A$ e $B$. Na vis\~{a}o de Bohr\index{Bohr, Niels}, a tentativa para essa atribui\c{c}\~{a}o de valores seria, a princ\'{i}pio, equivocada, na medida em que qualquer afirma\c{c}\~{a}o sobre os valores bem definidos de tais propriedades s\'{o} seria dotada de significado em condi\c{c}\~{o}es experimentais mutuamente exclusivas. Assim, para \citet[p.~311--312]{murdoch1994bohreinstein}, no caso EPR, as condi\c{c}\~{o}es experimentais que permitiriam uma afirma\c{c}\~{a}o dotada de significado sobre a propriedade $x$ de um objeto $A$ excluiriam as condi\c{c}\~{o}es experimentais que permitiriam uma afirma\c{c}\~{a}o dotada de significado sobre o valor bem definido da propriedade $y$ desse mesmo objeto.

Da mesma forma, as condi\c{c}\~{o}es experimentais escolhidas para determinar o estado de $A$ constituiriam as condi\c{c}\~{o}es para que se pudesse fazer qualquer tipo de infer\^{e}ncia significativa sobre o objeto $B$, uma vez que a premissa do princ\'{i}pio da separabilidade \'{e} rejeitada. Logicamente, \'{e} rejeitada tamb\'{e}m a (sub)conclus\~{a}o 3 de sua reconstru\c{c}\~{a}o do argumento de Einstein\index{Einstein, Albert}, isto \'{e}, a rejei\c{c}\~{a}o de que os valores das propriedades observ\'{a}veis de $B$, quer seja $x$ ou $y$, independe dos valores das propriedades observ\'{a}veis de $A$. Assim,

\begin{quote}
\textelp{} nenhuma utiliza\c{c}\~{a}o bem definida do conceito de ``estado'' pode ser  feita, como referindo-se ao objeto separado do corpo com o qual tenha estado em contato, at\'{e} que as condi\c{c}\~{o}es externas envolvidas na defini\c{c}\~{a}o desse conceito sejam inequivocamente fixadas por um controle mais adequado do corpo auxiliar. \citep[p.~21]{bohr1958atomic}.
\end{quote}
A situa\c{c}\~{a}o proposta sugere que \'{e} correta a interpreta\c{c}\~{a}o de \citet[p.~682]{einstein1949remarks} de que Bohr\index{Bohr, Niels} rejeitaria o princ\'{i}pio de localidade. A argumenta\c{c}\~{a}o de Bohr\index{Bohr, Niels} n\~{a}o parece implicar exist\^{e}ncia de uma interdepend\^{e}ncia causal ou mec\^{a}nica entre os objetos $A$ e $B$ no que se refere ao ato da medi\c{c}\~{a}o, mas, ao inv\'{e}s disso, que a medi\c{c}\~{a}o efetuada em $A$ determina as condi\c{c}\~{o}es sobre aquilo que pode ser dito significativamente sobre $B$. Assim, n\~{a}o se trataria de uma rejei\c{c}\~{a}o do princ\'{i}pio de localidade como um princ\'{i}pio causal, mas da rejei\c{c}\~{a}o do princ\'{i}pio de localidade como um princ\'{i}pio sem\^{a}ntico.

Ou seja, seria o caso de afirmar que h\'{a} uma interdepend\^{e}ncia sem\^{a}ntica ---mas n\~{a}o causal--- por meio de uma opera\c{c}\~{a}o experimental ou medi\c{c}\~{a}o entre os objetos $A$ e $B$. A rejei\c{c}\~{a}o por parte de Bohr\index{Bohr, Niels} do princ\'{i}pio de localidade \'{e} amplamente conhecida e difundida nos livros did\'{a}ticos sobre mec\^{a}nica qu\^{a}ntica, ainda que por muitas vezes a \^{e}nfase n\~{a}o seja dada no aspecto sem\^{a}ntico de tal princ\'{i}pio.

No entanto, \citet{howard1985einsteinseparability}. a localidade seria \textit{apenas um} dos dois aspectos que comp\~{o}em um princ\'{i}pio maior: o princ\'{i}pio da \textit{separabilidade}. O outro aspecto do princ\'{i}pio da separabilidade seria o princ\'{i}pio da exist\^{e}ncia independente, em rela\c{c}\~{a}o ao qual a posi\c{c}\~{a}o de Bohr\index{Bohr, Niels} \'{e} menos clara. Como foi exposto anteriormente, o princ\'{i}pio da separabilidade (cujo princ\'{i}pio da exist\^{e}ncia independente seria um de seus aspectos) \'{e} mutuamente exclusivo em rela\c{c}\~{a}o ao princ\'{i}pio da completude que, por sua vez, implicaria a tese da \textit{medi\c{c}\~{a}o=cria\c{c}\~{a}o}, tese que Bohr\index{Bohr, Niels} parece rejeitar:

\begin{quote}
\textelp{} a discuss\~{a}o dos problemas epistemol\'{o}gicos na f\'{i}sica at\^{o}mica atraiu tanta aten\c{c}\~{a}o como nunca e, ao comentar sobre as vis\~{o}es de Einstein\index{Einstein, Albert} no que diz respeito \`{a} incompletude de modo de descri\c{c}\~{a}o da mec\^{a}nica qu\^{a}ntica, entrei mais diretamente em quest\~{o}es de terminologia. Nesse contexto, eu adverti especialmente contra frases, muitas vezes encontradas na literatura f\'{i}sica, como ``perturba\c{c}\~{a}o de fen\^{o}menos atrav\'{e}s da observa\c{c}\~{a}o'' ou ``cria\c{c}\~{a}o de atributos f\'{i}sicos para objetos at\^{o}micos atrav\'{e}s de medi\c{c}\~{o}es.'' Essas frases, que podem servir para lembrar dos aparentes paradoxos na teoria qu\^{a}ntica, s\~{a}o ao mesmo tempo capazes de causar confus\~{a}o, uma vez que palavras como ``fen\^{o}menos'' e ``observa\c{c}\~{o}es'', assim como ``atributos'' e ``medi\c{c}\~{o}es'', s\~{a}o utilizados de forma pouco compat\'{i}vel com a linguagem comum e defini\c{c}\~{a}o pr\'{a}tica. \citep[p.~63--64]{bohr1958atomic}.
\end{quote}
Essa rejei\c{c}\~{a}o seria logicamente acompanhada pela defesa de que o ato da medi\c{c}\~{a}o seria um ato de revela\c{c}\~{a}o de valores pr\'{e}-existentes do objeto medido sem que, no entanto, como observa \citet[p.~312]{murdoch1994bohreinstein}, ``\textelp{} tal valor pr\'{e}-existente revelado seja um valor absoluto, mas uma propriedade relativa ao arranjo experimental escolhido''.

Por esse motivo, \citet[p.~312]{murdoch1994bohreinstein} classifica a atitude de Bohr\index{Bohr, Niels} em um terreno m\'{e}dio, entre a posi\c{c}\~{a}o de \citet[p.~667]{einstein1949remarks}, segundo a qual uma  medi\c{c}\~{a}o revela de forma passiva valores pr\'{e}-existentes de uma realidade f\'{i}sica que existe de forma totalmente independente da medi\c{c}\~{a}o, e a posi\c{c}\~{a}o de \citet[p.~73]{heisenberg1927uncert}, segundo a qual uma medi\c{c}\~{a}o cria de forma ativa os valores de uma realidade f\'{i}sica que passa a existir com o ato da medi\c{c}\~{a}o. Dito de outra forma, segundo o racioc\'{i}nio de \citet[p.~312]{murdoch1994bohreinstein}, a posi\c{c}\~{a}o de Bohr\index{Bohr, Niels} poderia ser considerada como uma tese sem\^{a}ntica, que estaria entre uma tese epistemol\'{o}gica (expressa por aquilo que chamarei de \textit{medi\c{c}\~{a}o=revela\c{c}\~{a}o}) e uma tese ontol\'{o}gica (expressa pela \textit{medi\c{c}\~{a}o=cria\c{c}\~{a}o}).

Da forma como a problem\'{a}tica foi delineada, a posi\c{c}\~{a}o de Bohr\index{Bohr, Niels} estaria diretamente relacionada com os limites da definibilidade dos conceitos f\'{i}sicos, isto \'{e}, com o significado de tais conceitos. Na medida em que os limites ou significados seriam dados mediante a experi\^{e}ncia emp\'{i}rica, \citet[p.~313]{murdoch1994bohreinstein} aproxima esta posi\c{c}\~{a}o a uma atitude operacionista.

Uma concep\c{c}\~{a}o operacionista de significado estabelece que os termos que denotam um conceito f\'{i}sico ou quantidade te\'{o}rica t\^{e}m significado nas opera\c{c}\~{o}es experimentais utilizadas para medir tal conceito ou quantidade. Uma concep\c{c}\~{a}o operacionista de significado estabelece que os termos utilizados para denotar um conceito f\'{i}sico ou quantidade te\'{o}rica tem valor de verdade ou valor cognitivo, isto \'{e}, podem dizer que algo \'{e} verdadeiro ou falso, se e somente se tal valor de verdade pode ser confirmado por uma opera\c{c}\~{a}o experimental. Ainda assim, a leitura operacionista seria confirmada por Bohr\index{Bohr, Niels} na ocasi\~{a}o de uma resposta a Phillip Frank\footnote{~Ver \citet{beller1996conceptual} e \citet[p.~20]{fine1986shaky}.} que, em uma carta, questiona se a interpreta\c{c}\~{a}o de Bohr\index{Bohr, Niels} poderia ser aproximada \`{a} atitude operacionista.

\citet[p.~314]{murdoch1994bohreinstein} vai al\'{e}m e categoriza a concep\c{c}\~{a}o de significado de Bohr\index{Bohr, Niels} como verificacionista, na medida em que a atribui\c{c}\~{a}o do significado dos termos se d\'{a} mediante condi\c{c}\~{o}es de verifica\c{c}\~{a}o (em oposi\c{c}\~{a}o \`{a}s concep\c{c}\~{o}es segundo as quais as condi\c{c}\~{o}es para significado ou valor de verdade seriam independentes da verifica\c{c}\~{a}o experimental). De fato, s\~{a}o posi\c{c}\~{o}es muito pr\'{o}ximas. Segundo o racioc\'{i}nio de \citet[p.~314]{murdoch1994bohreinstein}, o operacionismo seria um subconjunto do verificacionismo, diferindo no fato de que o \'{u}ltimo, em um sentido mais amplo, iguala a no\c{c}\~{a}o de significado com a no\c{c}\~{a}o de uso, de modo que o significado de um termo deve ser suportado por condi\c{c}\~{o}es de verdade cuja verificabilidade e comunicabilidade s\~{a}o poss\'{i}veis. Por outro lado, a atitude operacionista afirma que um termo cujo valor de verdade \'{e} imposs\'{i}vel de ser determinado n\~{a}o \'{e} um termo que pode ser utilizado. Dessa forma, \citet[p.~314]{murdoch1994bohreinstein} identifica, na base verificacionista da posi\c{c}\~{a}o de Bohr\index{Bohr, Niels}, uma atitude mais pr\'{o}xima ao pragmatismo ao inv\'{e}s de um empirismo radical, como o operacionismo.

Sob tal perspectiva, Bohr\index{Bohr, Niels} consideraria que a no\c{c}\~{a}o cl\'{a}ssica de \textit{valores simultaneamente bem definidos} (\textit{e.g.}, para posi\c{c}\~{a}o e momento) seria uma idealiza\c{c}\~{a}o, cujo significado pressup\~{o}e uma a\c{c}\~{a}o virtualmente nula do postulado qu\^{a}ntico; da mesma forma, a no\c{c}\~{a}o de \textit{simultaneidade} (aplicada a fen\^{o}menos espacialmente separados) seria uma idealiza\c{c}\~{a}o cujo significado pressup\~{o}e uma velocidade infinita. A vis\~{a}o verificacionista e pragm\'{a}tica de significado assumida por Bohr\index{Bohr, Niels} estaria implicada por tr\'{a}s dessa vis\~{a}o na medida em que os conceitos n\~{a}o s\~{a}o revisados ---da forma como \citet[p.~699]{einstein1949remarks} propusera em rela\c{c}\~{a}o \`{a} formula\c{c}\~{a}o de novos conceitos---, mas, antes, ressignificados, isto \'{e}, restringidos a um escopo de aplica\c{c}\~{a}o (ainda) mais limitado.

A contrapartida metodol\'{o}gica para essa atitude seria o princ\'{i}pio da correspond\^{e}ncia\index{Bohr, Niels!princ\'{i}pio da correspond\^{e}ncia}, segundo o qual a f\'{i}sica qu\^{a}ntica seria uma generaliza\c{c}\~{a}o da f\'{i}sica cl\'{a}ssica. Ainda que o significado preciso de tal princ\'{i}pio ---crucial para a interpreta\c{c}\~{a}o de Copenhague--- seja motivo de debate, vale ressaltar que estou utilizando a compreens\~{a}o do pr\'{o}prio fundador do princ\'{i}pio, Bohr\index{Bohr, Niels}, para o racioc\'{i}nio que se segue; vale ressaltar, tamb\'{e}m, que a literatura posterior a Bohr\index{Bohr, Niels} n\~{a}o entende necessariamente o princ\'{i}pio da mesma forma como ele entendeu \citep[para mais discuss\~{o}es sobre o princ\'{i}pio da correspond\^{e}ncia, ver]{sep-bohr-correspondence}.\index{Bohr, Niels!princ\'{i}pio da correspond\^{e}ncia}

Assim, a rejei\c{c}\~{a}o de Bohr\index{Bohr, Niels} em rela\c{c}\~{a}o ao referido princ\'{i}pio da exist\^{e}ncia independente parece ser parcial. Ao passo que n\~{a}o se pode designar uma opera\c{c}\~{a}o experimental para determinar se de fato o estado f\'{i}sico de um objeto $B$ independe do estado f\'{i}sico de um objeto $A$ distante, a leitura verificacionista de Bohr\index{Bohr, Niels} parece indicar que tal princ\'{i}pio parece ser desprovido de significado. No entanto, a afirma\c{c}\~{a}o da tese da \textit{medi\c{c}\~{a}o=revela\c{c}\~{a}o} parece sugerir que o princ\'{i}pio da exist\^{e}ncia independente n\~{a}o \'{e} totalmente negado.

Se essa leitura for correta, uma not\'{a}vel implica\c{c}\~{a}o ontol\'{o}gica do pensamento de Bohr\index{Bohr, Niels} no que se refere ao comprometimento ontol\'{o}gico\index{ontologia!comprometimento ontol\'{o}gico} com uma realidade independente parece emergir, isto \'{e}, uma leitura realista do pensamento desse autor seria possibilitada. Para \citet[p.~198]{faye2012niels}, as diversas defini\c{c}\~{o}es e discuss\~{o}es acerca de uma defini\c{c}\~{a}o para a concep\c{c}\~{a}o filos\'{o}fica do \index{realismo metaf\'{i}sico}realismo t\^{e}m em comum dois pontos essenciais: ``(1) o mundo existe independentemente de nossas mentes; e (2) a verdade \'{e} uma no\c{c}\~{a}o n\~{a}o epist\^{e}mica; isto \'{e}, uma proposi\c{c}\~{a}o n\~{a}o \'{e} verdadeira porque \'{e} prov\'{a}vel ou cognosc\'{i}vel''. Segundo \citet[p.~128]{folse1994realistbohr}, \citet[p.~98]{faye1994antirealistbohr} defenderia uma interpreta\c{c}\~{a}o de Bohr\index{Bohr, Niels} classificada como um antirrealismo objetivo, na medida em que Bohr\index{Bohr, Niels} aceitaria (1) e rejeitaria (2).

O antirrealismo da leitura de Faye emergiria da nega\c{c}\~{a}o da transcend\^{e}ncia das condi\c{c}\~{o}es de verdade, isto \'{e}, da nega\c{c}\~{a}o do significado de todas as afirma\c{c}\~{o}es indecid\'{i}veis (as afirma\c{c}\~{o}es sobre as quais \'{e} poss\'{i}vel verificar o valor de verdade mediante uma opera\c{c}\~{a}o experimental) cujo alcance epist\^{e}mico est\'{a} fora de qualquer poss\'{i}vel sujeito cognoscente; em outras palavras, da nega\c{c}\~{a}o de que o significado seja intr\'{i}nseco ao objeto em si mesmo:

\begin{quote}
\textelp{Senten\c{c}as} decid\'{i}veis s\~{a}o aquelas que s\~{a}o ou determinadamente verdadeiras ou determinadamente falsas devido \`{a} nossa posse de meios cognitivos em princ\'{i}pio adequados ou evid\^{e}ncias perceptuais pelas quais podemos verificar ou falsific\'{a}-las. Em outras palavras, tais senten\c{c}as t\^{e}m condi\c{c}\~{o}es de verdade cuja verifica\c{c}\~{a}o \'{e} acess\'{i}vel. A classe complementar de declara\c{c}\~{o}es \'{e} aquela cujos membros s\~{a}o indecid\'{i}veis, portanto, n\~{a}o t\^{e}m valores de verdade determinados, devido ao fato de que tais senten\c{c}as t\^{e}m condi\c{c}\~{o}es de verdade cuja verifica\c{c}\~{a}o \'{e} transcendente. No entanto, em oposi\c{c}\~{a}o ao antirrealista, o realista diria que at\'{e} mesmo essas senten\c{c}as indecid\'{i}veis t\^{e}m um valor de verdade determinado; o que acontece \'{e} que somos incapazes de descobrir qual. Assim, tanto o realista quanto o antirrealista objetivo operam com uma no\c{c}\~{a}o de objetividade. \citep[p.~199]{faye2012niels}.
\end{quote}
Por outro lado, o termo \textit{objetivo} da nomenclatura ``antirrealismo objetivo'' de \citet{faye2012niels} emerge como uma implica\c{c}\~{a}o de (1), na medida em que as afirma\c{c}\~{o}es decid\'{i}veis (as afirma\c{c}\~{o}es sobre as quais se possam verificar o valor de verdade mediante uma opera\c{c}\~{a}o experimental) tenham suas condi\c{c}\~{o}es de verdade garantidas pela realidade independente, por mais que o sentido de tal afirma\c{c}\~{a}o (como o estado de um objeto) seja desconhecido por qualquer poss\'{i}vel sujeito cognoscente. 

Da forma como \citet[p.~128--130]{folse1994realistbohr} interpreta tal leitura, Faye n\~{a}o excluiria a possibilidade de que, para Bohr\index{Bohr, Niels}, um objeto n\~{a}o observado possua de fato valores bem definidos para suas propriedades f\'{i}sicas como, por exemplo, posi\c{c}\~{a}o ou momento. No entanto, uma afirma\c{c}\~{a}o acerca dos valores simultaneamente bem definidos de tais propriedades n\~{a}o seria uma afirma\c{c}\~{a}o bem formulada na sem\^{a}ntica da complementaridade\index{Bohr, Niels!princ\'{i}pio da complementaridade} e, portanto, seria sem sentido.

Contudo, deve ficar claro que, como observa \citet[p.~208]{faye2012niels} em rela\c{c}\~{a}o a (1), n\~{a}o h\'{a} evid\^{e}ncia textual que sustente a tese de que Bohr\index{Bohr, Niels} atribuiria valores intr\'{i}nsecos \`{a}s propriedades n\~{a}o observadas dos objetos qu\^{a}nticos. Quando \citet[p.~200]{faye2012niels} menciona (1), parece faz\^{e}-lo enfatizando a objetividade dos conceitos, em um campo sem\^{a}ntico, qui\c{c}\'{a} epistemol\'{o}gico, mas, certamente, n\~{a}o ontol\'{o}gico.

O antirrealista objetivo, em rela\c{c}\~{a}o \`{a}s declara\c{c}\~{o}es sobre a realidade f\'{i}sica, toma como ponto de partida as circunst\^{a}ncias publicamente acess\'{i}veis ao especificar sua no\c{c}\~{a}o de verdade \textelp{}. O antirrealismo objetivo \'{e}, ent\~{a}o, a posi\c{c}\~{a}o que sustenta que a verdade \'{e} um conceito que se relaciona com circunst\^{a}ncias cuja ocorr\^{e}ncia ou n\~{a}o-ocorr\^{e}ncia \'{e}, a princ\'{i}pio, empiricamente acess\'{i}vel \`{a}s nossas capacidades cognitivas.

A vis\~{a}o sobre (1), em rela\c{c}\~{a}o ao pensamento de Bohr\index{Bohr, Niels}, \'{e} compartilhada por \citet[p.~128]{folse1994realistbohr}. Por mais que \citet[p.~204]{faye2012niels} e \citet[p.~128]{folse1994realistbohr} concordem com a vis\~{a}o de que Bohr\index{Bohr, Niels} ocuparia um terreno m\'{e}dio entre os dois extremos do idealismo e do \index{realismo metaf\'{i}sico}realismo ---o que tamb\'{e}m coaduna com a leitura de \citet[p.~312]{murdoch1994bohreinstein}---, Folse defende uma leitura realista do pensamento de Bohr\index{Bohr, Niels}. \citet[p.~128--131]{folse1994realistbohr} argumenta que o ponto (2) n\~{a}o seria t\~{a}o decisivo quanto o ponto (1) na medida em que o comprometimento ontol\'{o}gico\index{ontologia!comprometimento ontol\'{o}gico} com uma realidade independente seria mais fundamental do que uma tese epistemol\'{o}gica, relativa ao dom\'{i}nio do significado dos conceitos utilizados mediante nosso conhecimento. Em outras palavras, \citet{folse1994realistbohr} considera que a aceita\c{c}\~{a}o de (1) seria suficiente para uma interpreta\c{c}\~{a}o realista do pensamento de Bohr\index{Bohr, Niels}, tendo em vista o comprometimento ontol\'{o}gico\index{ontologia!comprometimento ontol\'{o}gico} com a exist\^{e}ncia de uma realidade independente.

No entanto, \citet[p.~207--211]{faye2012niels} exp\~{o}e s\'{e}rias restri\c{c}\~{o}es \`{a} interpreta\c{c}\~{a}o realista de Folse, das quais sublinharei apenas uma. Quando \citet[p.~257]{folse1985philosophybohr} afirma que a intera\c{c}\~{a}o de um objeto com os instrumentos de medi\c{c}\~{a}o produz ou causa o fen\^{o}meno, acaba por admitir a ocorr\^{e}ncia da tese da \textit{medi\c{c}\~{a}o=cria\c{c}\~{a}o} ---uma implica\c{c}\~{a}o que, como vimos, \'{e} rejeitada por Bohr\index{Bohr, Niels}. Al\'{e}m disso, tal ocorr\^{e}ncia parece ser incompat\'{i}vel com o comprometimento ontol\'{o}gico\index{ontologia!comprometimento ontol\'{o}gico} com uma realidade independente. Isto \'{e}, a atribui\c{c}\~{a}o de um poder \textit{criador} ao ato da medi\c{c}\~{a}o parece ser irreconcili\'{a}vel com a afirma\c{c}\~{a}o de que tais propriedades, criadas, \textit{j\'{a} estavam l\'{a}} mesmo antes do ato criador. Por fim, se a tese de Folse for correta, ent\~{a}o deve haver alguma evid\^{e}ncia textual ---o que n\~{a}o h\'{a}--- em que Bohr\index{Bohr, Niels} assume que objetos at\^{o}micos possuam intrinsecamente propriedades bem definidas, mas que, no entanto, n\~{a}o podem ser verificadas empiricamente, dado que uma opera\c{c}\~{a}o experimental n\~{a}o \'{e} capaz de revelar aquilo que est\'{a} por tr\'{a}s do fen\^{o}meno.

O fato de que Bohr\index{Bohr, Niels} acreditava que os objetos qu\^{a}nticos seriam reais \'{e} consensual, mas, segundo \citet{sep-qm-copenhagen} ainda h\'{a} muito debate na literatura das \'{u}ltimas d\'{e}cadas a respeito do tipo de realidade que eles t\^{e}m, isto \'{e}, se s\~{a}o ou n\~{a}o algo diferente e para al\'{e}m da observa\c{c}\~{a}o, de modo que tal quest\~{a}o permanece aberta.

Bohr\index{Bohr, Niels} parece deliberadamente evitar o comprometimento com as teses realistas e com as teses idealistas atrav\'{e}s do princ\'{i}pio da correspond\^{e}ncia\index{Bohr, Niels!princ\'{i}pio da correspond\^{e}ncia}, isto \'{e}, pela afirma\c{c}\~{a}o de que um objeto (tal como o aparelho medidor) \'{e} considerado um objeto cl\'{a}ssico em um determinado conjunto de circunst\^{a}ncias, a saber, para os prop\'{o}sitos da medi\c{c}\~{a}o. No entanto, esta afirma\c{c}\~{a}o acaba por esbarrar em outro problema, talvez ainda mais s\'{e}rio.

A separabilidade assumida para o ato da medi\c{c}\~{a}o seria parcialmente arbitr\'{a}ria. Para que se possa dizer que ocorreu uma medi\c{c}\~{a}o, o objeto medido n\~{a}o pode ser parte da ag\^{e}ncia de medi\c{c}\~{a}o, ou seja, \'{e} necess\'{a}ria uma distin\c{c}\~{a}o entre duas entidades, de modo que, para fins pr\'{a}ticos, um instrumento de medi\c{c}\~{a}o \'{e} um instrumento de medi\c{c}\~{a}o, e um objeto \'{e} um objeto. Como observa \citet[p.~139]{faye2012niels}, se a separa\c{c}\~{a}o \'{e} assumida, sua intera\c{c}\~{a}o \'{e}, do ponto de vista do ato da medi\c{c}\~{a}o, indeterminada, pois ``\textelp{} a intera\c{c}\~{a}o s\'{o} pode ser determinada se o aparelho de medi\c{c}\~{a}o for considerado simultaneamente como um aparelho e como um objeto, o que \'{e} logicamente imposs\'{i}vel''.

O que daria o tom de arbitrariedade na distin\c{c}\~{a}o proposta seria o ponto de demarca\c{c}\~{a}o da separabilidade, que j\'{a} seria conhecida por Bohr\index{Bohr, Niels} desde o primeiro artigo em que exp\~{o}e a complementaridade\index{Bohr, Niels!princ\'{i}pio da complementaridade}, no qual afirma que:

\begin{quote}
\textelp{} o conceito de observa\c{c}\~{a}o \'{e} arbitr\'{a}rio pois depende de quais objetos s\~{a}o inclu\'{i}dos no sistema para ser observado. \textelp{} em qual ponto o conceito de observa\c{c}\~{a}o ---envolvendo o postulado qu\^{a}ntico, com a sua \textit{irracionalidade} inerente--- deve ser utilizado \'{e} uma quest\~{a}o de conveni\^{e}ncia. \citep[p.~89]{bohr1928quantumpostulate}.
\end{quote}
A ``quest\~{a}o de conveni\^{e}ncia'' do crit\'{e}rio de demarca\c{c}\~{a}o para a separabilidade do processo de medi\c{c}\~{a}o foi tida como a resposta de Bohr\index{Bohr, Niels} frente ao problema da medi\c{c}\~{a}o, sobre o qual discutirei no cap\'{i}tulo \ref{CapvNeumann} ---solu\c{c}\~{a}o esta criticada por diversos pensadores da \'{e}poca.

\citet[p.~410--414]{heisenberg1979deterministic}\footnote{~Material publicado postumamente.} argumentou que, como a linha de demarca\c{c}\~{a}o entre o objeto qu\^{a}ntico a ser investigado, representado matematicamente por uma fun\c{c}\~{a}o de onda, e o instrumento de medi\c{c}\~{a}o, descrito por meio de conceitos cl\'{a}ssicos, seria arbitr\'{a}ria, ent\~{a}o todos os sistemas (incluindo o instrumento de medi\c{c}\~{a}o) deveriam ser considerados sistemas qu\^{a}nticos, isto \'{e}, as leis qu\^{a}nticas deveriam se aplicar de forma irrestrita.

Sob a mesma linha de racioc\'{i}nio, \citet{vNeum1955mathematical} elaborou uma concep\c{c}\~{a}o de medi\c{c}\~{a}o qu\^{a}ntica a partir do formalismo da teoria, segundo a qual todos os observ\'{a}veis t\^{e}m um tratamento qu\^{a}ntico. Diferentemente de Bohr\index{Bohr, Niels} e Einstein\index{Einstein, Albert}, von Neumann formulou uma teoria formal da medi\c{c}\~{a}o, na qual o problema da medi\c{c}\~{a}o aparece de forma mais clara e distinta, como analisarei no cap\'{i}tulo \ref{CapvNeumann}. Para nos aprofundarmos na teoria da medi\c{c}\~{a}o de \citet{vNeum1955mathematical}, fa\c{c}o algumas considera\c{c}\~{o}es gerais sobre a teoria medi\c{c}\~{a}o em mec\^{a}nica qu\^{a}ntica ---que tamb\'{e}m ser\'{a} o assunto do cap\'{i}tulo \ref{CapvNeumann}.

Antes de passarmos a esse assunto, conv\'{e}m explorarmos, por fim, a indetermina\c{c}\~{a}o.

\subsection{Indetermina\c{c}\~{a}o}
A tese da medi\c{c}\~{a}o=cria\c{c}\~{a}o \'{e}, de fato, central na mec\^{a}nica qu\^{a}ntica padr\~{a}o. Antes da medi\c{c}\~{a}o, diz-se que os valores ---ou que as \textit{propriedades}--- dos objetos qu\^{a}nticos s\~{a}o \textit{indeterminados}. Vejamos o que isso quer dizer.

Segundo a ``vis\~{a}o natural'' de \citet{armstrong1961} todo objeto f\'{i}sico \'{e} determinado. Para entender o que isso significa, \'{e} preciso ter em mente a distin\c{c}\~{a}o entre propriedades \textit{determinadas} e \textit{determin\'{a}veis}:
\begin{description}
\item[Propriedades determin\'{a}veis:] Especifica um conjunto de propriedades. \textit{E.g.} forma, cor, tamanho.
\item[Propriedade determinada:] Especifica um elemento do conjunto determin\'{a}vel. \textit{E.g.} quadrado, azul, 1m.
\end{description}
Por exemplo, gatos e canecas de caf\'{e} t\^{e}m a propriedade de \textit{e.g.} cor determin\'{a}vel; Ulisses, meu gato laranja, tem a propriedade determin\'{a}vel de cor determinada: ele \'{e} \textit{laranja}. A vis\~{a}o natural pode ser enunciada da seguinte maneira:

\begin{quote}
    Um objeto f\'{i}sico \'{e} determinado em todos os aspectos, tem uma cor, temperatura, tamanho etc perfeitamente precisos. N\~{a}o faz sentido dizer que um objeto f\'{i}sico \'{e} azul claro, mas n\~{a}o tem um tom definido de azul claro. \citep[p.~59]{armstrong1961}.
\end{quote}
A \textit{indetermina\c{c}\~{a}o} sobre a qual a medi\c{c}\~{a}o=cria\c{c}\~{a}o se refere \'{e} justamente a falha da vis\~{a}o natural:

\begin{quote}
Indetermina\c{c}\~{a}o \'{e} a situa\c{c}\~{a}o em que um objeto possui uma propriedade determin\'{a}vel, mas nenhum valor determinado para essa determin\'{a}vel. \citep[p.~76]{lewis2016quaont}.
\end{quote}
Existem diversos exemplos de indetermina\c{c}\~{a}o desse tipo na literatura filos\'{o}fica. Por exemplo, inst\^{a}ncias do famoso \textit{paradoxo de sorites}\footnote{~O nome vem da palavra ``\textit{soros}'' em grego, que significa ``monte'', e \'{e} usualmente atribu\'{i}do a Eubulides, do S\'{e}culo IV a.C.; ver \citet{sep-sorites-paradox}.} incluem especificar qual o gr\~{a}o de areia que delimita um monte. Voc\^{e} pode tentar fazer esse experimento em praticamente qualquer laborat\'{o}rio de metaf\'{i}sica: pegue um gr\~{a}o de areia, e coloque-o sobre outro gr\~{a}o de areia; repita esse procedimento algumas vezes. A quest\~{a}o \'{e}: em qual ponto um monte de areia foi formado? A fronteira \'{e} \textit{indeterminada}, segundo o paradoxo de sorites. Outras inst\^{a}ncias disso \'{e} a indetermina\c{c}\~{a}o da fronteira exata de florestas, cidades, e at\'{e} mesmo montanhas ou nuvens. Esse n\~{a}o \'{e}, contudo, o tipo de indetermina\c{c}\~{a}o em jogo na mec\^{a}nica qu\^{a}ntica. \citet{torza2022} oferece um bom guia sobre casos de indetermina\c{c}\~{a}o na literatura filos\'{o}fica:
\begin{quote}
\textelp{} (i) os objetos `nebulosos' do mundo macrosc\'{o}pico, como nuvens, montanhas e pessoas; (ii) conting\^{e}ncias futuras e o futuro em aberto; e (iii) indetermina\c{c}\~{a}o qu\^{a}ntica. Exemplos putativos de iii incluem (iii.a) a falha da determina\c{c}\~{a}o de valores de observ\'{a}veis qu\^{a}nticos; (iii.b) a individualidade vaga de objetos qu\^{a}nticos; e (iii.c) a indetermina\c{c}\~{a}o de contagem decorrente na teoria qu\^{a}ntica de campos. \citep[p.~338]{torza2022}.
\end{quote}
O caso exemplificado pela tese medi\c{c}\~{a}o=cria\c{c}\~{a}o s\~{a}o casos do tipo (iii.a). Vejamos mais a fundo como isso ocorre usando o famoso ``teorema de Bell\index{Bell, John Stewart!teorema de Bell}'' ---pano de fundo, diga-se de passagem, para o Pr\^{e}mio Nobel de F\'{i}sica concedido a \citet*{nobelprize2022} pelas realiza\c{c}\~{o}es experimentais do referido teorema. Seguirei aqui a exposi\c{c}\~{a}o de \citet{chen2022}, por tratar-se da exposi\c{c}\~{a}o mais intuitiva e amig\'{a}vel a profissionais da filosofia sem treinamento em f\'{i}sica (que \'{e} justamente o p\'{u}blico alvo deste livro).\footnote{~De fato, n\~{a}o estou s\'{o} com essa avalia\c{c}\~{a}o da exposi\c{c}\~{a}o de \citet{chen2022}; ver \citet{andreoletti-vervoort2022} para mais pessoas endossando essa impress\~{a}o.}

O teorema de Bell\index{Bell, John Stewart!teorema de Bell} \'{e} uma expans\~{a}o do \index{Einstein, Albert!argumento EPR}argumento EPR (de fato, um experimento mental), que explicita de maneira particularmente como a tese medi\c{c}\~{a}o=cria\c{c}\~{a}o opera ---e como \'{e}, de fato, \textit{inevit\'{a}vel}--- na mec\^{a}nica qu\^{a}ntica padr\~{a}o. Vamos l\'{a}.

O famoso teorema de Bell\index{Bell, John Stewart!teorema de Bell}, conforme apresentado em \citet{bell1964EPR}, \'{e} uma generaliza\c{c}\~{a}o do argumento EPR\index{Einstein, Albert!argumento EPR}, ent\~{a}o. Como vimos na p\'{a}gina \pageref{EPR:localidade}, uma das assun\c{c}\~{o}es centrais do argumento EPR\index{Einstein, Albert!argumento EPR} \'{e} a \textit{localidade}. Relembremo-nos como ela \'{e} enunciada. Dados dois sistemas,
\begin{quote}
    \textelp{} nenhuma mudan\c{c}a real pode ocorrer no segundo sistema em consequ\^{e}ncia de qualquer coisa que possa ser feito com o primeiro sistema'' \citep*[140]{epr1963epr}\index{Einstein, Albert!argumento EPR}.
\end{quote}
Essa premissa parece bastante razo\'{a}vel. Repare, afinal, que praticamente todas as intera\c{c}\~{o}es que temos no mundo \`{a} nossa volta s\~{a}o intera\c{c}\~{o}es locais desse tipo. Digamos que voc\^{e} est\'{a} preparando uma pipoca. Uma pipoca estourada tem um efeito na panela e nas imedia\c{c}\~{o}es do fog\~{a}o. Algum tempo depois, tem efeito na cozinha (especialmente se a panela estiver destampada!), mas \textit{nada acontece} ---concomitantemente ao estouro da pipoca--- do outro lado da rua \textit{em decorr\^{e}ncia} do estouro da pipoca.

Claro, algu\'{e}m pode saber que voc\^{e} est\'{a} preparando pipoca porque ouviu o barulho. Mas repare que essa ainda \'{e} uma intera\c{c}\~{a}o \textit{local}: as ondas sonoras precisaram passar pelo espa\c{c}o entre a pipoca e a pessoa que ouviu o barulho, e isso leva algum tempo (indicado pela dist\^{a}ncia entre as duas coisas e a velocidade do som). \'{E}, portanto, uma intera\c{c}\~{a}o local. Aposto que voc\^{e} consegue pensar em diversos outros casos desse tipo. Chamemos essa premissa de {\bf\textsc{loc}}, definida da seguinte maneira:
\begin{description}
    \item[\textsc{loc}~$\overset{\underset{\mathrm{def}}{}}{=}$] Eventos espacialmente distantes n\~{a}o podem ter influ\^{e}ncia imediata.
\end{description}

As outras assun\c{c}\~{o}es do teorema s\~{a}o as predi\c{c}\~{o}es da mec\^{a}nica qu\^{a}ntica, dentre as quais \citet{chen2022} destaca tr\^{e}s. Chamemos elas de \textsc{pred$_{\text{\bf\textsc{mq}}}$1}, \textsc{pred$_{\text{\bf\textsc{mq}}}$2}, e \textsc{pred$_{\text{\bf\textsc{mq}}}$3}, respectivamente. Todas elas s\~{a}o relativos a objetos qu\^{a}nticos $1$ e $2$ (\textit{e.g.} f\'{o}tons) disparados a partir de uma fonte em dire\c{c}\~{o}es diametralmente opostas, e que encontrar\~{a}o dois filtros (\textit{e.g.}, polarizadores) que podem ou n\~{a}o impedir sua passagem (\textit{i.e.}, absorv\^{e}-los). S\~{a}o elas:

\begin{description}
\item[\textsc{pred$_{\text{\bf\textsc{mq}}}$1}~$\overset{\underset{\mathrm{def}}{}}{=}$] Dois f\'{o}tons ($1$ e $2$) na mesma orienta\c{c}\~{a}o e os polarizadores no mesmo \^{a}ngulo \textit{sempre concordam com os resultados} ($100\%$). Isto \'{e}, se $1$ passa pelo polarizador, sabemos que $2$ tamb\'{e}m passar\'{a} ---e vice-versa: se $1$ \'{e} absorvido, sabemos que $2$ tamb\'{e}m ser\'{a}.
\end{description}

\begin{figure}[ht!]
    \centering

 
\tikzset{
pattern size/.store in=\mcSize, 
pattern size = 5pt,
pattern thickness/.store in=\mcThickness, 
pattern thickness = 0.3pt,
pattern radius/.store in=\mcRadius, 
pattern radius = 1pt}
\makeatletter
\pgfutil@ifundefined{pgf@pattern@name@_bpic98rad}{
\pgfdeclarepatternformonly[\mcThickness,\mcSize]{_bpic98rad}
{\pgfqpoint{0pt}{-\mcThickness}}
{\pgfpoint{\mcSize}{\mcSize}}
{\pgfpoint{\mcSize}{\mcSize}}
{
\pgfsetcolor{\tikz@pattern@color}
\pgfsetlinewidth{\mcThickness}
\pgfpathmoveto{\pgfqpoint{0pt}{\mcSize}}
\pgfpathlineto{\pgfpoint{\mcSize+\mcThickness}{-\mcThickness}}
\pgfusepath{stroke}
}}
\makeatother

 
\tikzset{
pattern size/.store in=\mcSize, 
pattern size = 5pt,
pattern thickness/.store in=\mcThickness, 
pattern thickness = 0.3pt,
pattern radius/.store in=\mcRadius, 
pattern radius = 1pt}
\makeatletter
\pgfutil@ifundefined{pgf@pattern@name@_tj6gttkvg}{
\pgfdeclarepatternformonly[\mcThickness,\mcSize]{_tj6gttkvg}
{\pgfqpoint{0pt}{-\mcThickness}}
{\pgfpoint{\mcSize}{\mcSize}}
{\pgfpoint{\mcSize}{\mcSize}}
{
\pgfsetcolor{\tikz@pattern@color}
\pgfsetlinewidth{\mcThickness}
\pgfpathmoveto{\pgfqpoint{0pt}{\mcSize}}
\pgfpathlineto{\pgfpoint{\mcSize+\mcThickness}{-\mcThickness}}
\pgfusepath{stroke}
}}
\makeatother
\tikzset{every picture/.style={line width=0.75pt}} 

\begin{tikzpicture}[x=0.75pt,y=0.75pt,yscale=-.8,xscale=.8]

\draw   (225,84) -- (179.5,84) .. controls (172.73,84) and (167.25,70.57) .. (167.25,54) .. controls (167.25,37.43) and (172.73,24) .. (179.5,24) -- (225,24)(237.25,54) .. controls (237.25,70.57) and (231.77,84) .. (225,84) .. controls (218.23,84) and (212.75,70.57) .. (212.75,54) .. controls (212.75,37.43) and (218.23,24) .. (225,24) .. controls (231.77,24) and (237.25,37.43) .. (237.25,54) ;
\draw    (167.25,54) .. controls (165.58,55.67) and (163.92,55.67) .. (162.25,54) .. controls (160.58,52.33) and (158.92,52.33) .. (157.25,54) .. controls (155.58,55.67) and (153.92,55.67) .. (152.25,54) .. controls (150.58,52.33) and (148.92,52.33) .. (147.25,54) .. controls (145.58,55.67) and (143.92,55.67) .. (142.25,54) .. controls (140.58,52.33) and (138.92,52.33) .. (137.25,54) .. controls (135.58,55.67) and (133.92,55.67) .. (132.25,54) .. controls (130.58,52.33) and (128.92,52.33) .. (127.25,54) .. controls (125.58,55.67) and (123.92,55.67) .. (122.25,54) .. controls (120.58,52.33) and (118.92,52.33) .. (117.25,54) .. controls (115.58,55.67) and (113.92,55.67) .. (112.25,54) .. controls (110.58,52.33) and (108.92,52.33) .. (107.25,54) .. controls (105.58,55.67) and (103.92,55.67) .. (102.25,54) -- (100.75,54) -- (92.75,54) ;
\draw [shift={(89.75,54)}, rotate = 360] [fill={rgb, 255:red, 0; green, 0; blue, 0 }  ][line width=0.08]  [draw opacity=0] (10.72,-5.15) -- (0,0) -- (10.72,5.15) -- (7.12,0) -- cycle    ;
\draw    (225,54) .. controls (226.67,52.33) and (228.33,52.33) .. (230,54) .. controls (231.67,55.67) and (233.33,55.67) .. (235,54) .. controls (236.67,52.33) and (238.33,52.33) .. (240,54) .. controls (241.67,55.67) and (243.33,55.67) .. (245,54) .. controls (246.67,52.33) and (248.33,52.33) .. (250,54) .. controls (251.67,55.67) and (253.33,55.67) .. (255,54) .. controls (256.67,52.33) and (258.33,52.33) .. (260,54) .. controls (261.67,55.67) and (263.33,55.67) .. (265,54) .. controls (266.67,52.33) and (268.33,52.33) .. (270,54) .. controls (271.67,55.67) and (273.33,55.67) .. (275,54) .. controls (276.67,52.33) and (278.33,52.33) .. (280,54) .. controls (281.67,55.67) and (283.33,55.67) .. (285,54) .. controls (286.67,52.33) and (288.33,52.33) .. (290,54) -- (291.5,54) -- (299.5,54) ;
\draw [shift={(302.5,54)}, rotate = 180] [fill={rgb, 255:red, 0; green, 0; blue, 0 }  ][line width=0.08]  [draw opacity=0] (10.72,-5.15) -- (0,0) -- (10.72,5.15) -- (7.12,0) -- cycle    ;
\draw  [pattern=_bpic98rad,pattern size=9pt,pattern thickness=0.75pt,pattern radius=0pt, pattern color={rgb, 255:red, 0; green, 0; blue, 0}] (74.75,54) .. controls (74.75,29.7) and (81.47,10) .. (89.75,10) .. controls (98.03,10) and (104.75,29.7) .. (104.75,54) .. controls (104.75,78.3) and (98.03,98) .. (89.75,98) .. controls (81.47,98) and (74.75,78.3) .. (74.75,54) -- cycle ;
\draw  [pattern=_tj6gttkvg,pattern size=9pt,pattern thickness=0.75pt,pattern radius=0pt, pattern color={rgb, 255:red, 0; green, 0; blue, 0}] (287.5,54) .. controls (287.92,29.7) and (294.98,10) .. (303.27,10) .. controls (311.55,10) and (317.92,29.7) .. (317.5,54) .. controls (317.08,78.3) and (310.02,98) .. (301.73,98) .. controls (293.45,98) and (287.08,78.3) .. (287.5,54) -- cycle ;

\draw (182.25,88) node [anchor=north west][inner sep=0.75pt]  [xscale=0.8,yscale=0.8] [align=left] {Fonte};
\draw (123,70.5) node [anchor=north west][inner sep=0.75pt]  [xscale=0.8,yscale=0.8] [align=left] {1};
\draw (258.25,70.5) node [anchor=north west][inner sep=0.75pt]  [xscale=0.8,yscale=0.8] [align=left] {2};
\draw (80.75,102.4) node [anchor=north west][inner sep=0.75pt]  [xscale=0.8,yscale=0.8]  {$0^{o}$};
\draw (293.5,102.4) node [anchor=north west][inner sep=0.75pt]  [xscale=0.8,yscale=0.8]  {$0^{o}$};

\end{tikzpicture}
    \caption{{\bf\textsc{pred$_{\text{\bf\textsc{mq}}}$1}} com f\'{o}tons absorvidos}
    \label{fig:bell-0-abs}
\end{figure}

\begin{figure}[ht!]
    \centering

 
\tikzset{
pattern size/.store in=\mcSize, 
pattern size = 5pt,
pattern thickness/.store in=\mcThickness, 
pattern thickness = 0.3pt,
pattern radius/.store in=\mcRadius, 
pattern radius = 1pt}
\makeatletter
\pgfutil@ifundefined{pgf@pattern@name@_h12iepk5o}{
\pgfdeclarepatternformonly[\mcThickness,\mcSize]{_h12iepk5o}
{\pgfqpoint{0pt}{-\mcThickness}}
{\pgfpoint{\mcSize}{\mcSize}}
{\pgfpoint{\mcSize}{\mcSize}}
{
\pgfsetcolor{\tikz@pattern@color}
\pgfsetlinewidth{\mcThickness}
\pgfpathmoveto{\pgfqpoint{0pt}{\mcSize}}
\pgfpathlineto{\pgfpoint{\mcSize+\mcThickness}{-\mcThickness}}
\pgfusepath{stroke}
}}
\makeatother

 
\tikzset{
pattern size/.store in=\mcSize, 
pattern size = 5pt,
pattern thickness/.store in=\mcThickness, 
pattern thickness = 0.3pt,
pattern radius/.store in=\mcRadius, 
pattern radius = 1pt}
\makeatletter
\pgfutil@ifundefined{pgf@pattern@name@_5mnlv7bot}{
\pgfdeclarepatternformonly[\mcThickness,\mcSize]{_5mnlv7bot}
{\pgfqpoint{0pt}{-\mcThickness}}
{\pgfpoint{\mcSize}{\mcSize}}
{\pgfpoint{\mcSize}{\mcSize}}
{
\pgfsetcolor{\tikz@pattern@color}
\pgfsetlinewidth{\mcThickness}
\pgfpathmoveto{\pgfqpoint{0pt}{\mcSize}}
\pgfpathlineto{\pgfpoint{\mcSize+\mcThickness}{-\mcThickness}}
\pgfusepath{stroke}
}}
\makeatother
\tikzset{every picture/.style={line width=0.75pt}} 

\begin{tikzpicture}[x=0.75pt,y=0.75pt,yscale=-.8,xscale=.8]

\draw   (280.25,84) -- (234.75,84) .. controls (227.98,84) and (222.5,70.57) .. (222.5,54) .. controls (222.5,37.43) and (227.98,24) .. (234.75,24) -- (280.25,24)(292.5,54) .. controls (292.5,70.57) and (287.02,84) .. (280.25,84) .. controls (273.48,84) and (268,70.57) .. (268,54) .. controls (268,37.43) and (273.48,24) .. (280.25,24) .. controls (287.02,24) and (292.5,37.43) .. (292.5,54) ;
\draw    (222.5,54) .. controls (220.83,55.67) and (219.17,55.67) .. (217.5,54) .. controls (215.83,52.33) and (214.17,52.33) .. (212.5,54) .. controls (210.83,55.67) and (209.17,55.67) .. (207.5,54) .. controls (205.83,52.33) and (204.17,52.33) .. (202.5,54) .. controls (200.83,55.67) and (199.17,55.67) .. (197.5,54) .. controls (195.83,52.33) and (194.17,52.33) .. (192.5,54) .. controls (190.83,55.67) and (189.17,55.67) .. (187.5,54) .. controls (185.83,52.33) and (184.17,52.33) .. (182.5,54) .. controls (180.83,55.67) and (179.17,55.67) .. (177.5,54) .. controls (175.83,52.33) and (174.17,52.33) .. (172.5,54) .. controls (170.83,55.67) and (169.17,55.67) .. (167.5,54) .. controls (165.83,52.33) and (164.17,52.33) .. (162.5,54) .. controls (160.83,55.67) and (159.17,55.67) .. (157.5,54) .. controls (155.83,52.33) and (154.17,52.33) .. (152.5,54) .. controls (150.83,55.67) and (149.17,55.67) .. (147.5,54) .. controls (145.83,52.33) and (144.17,52.33) .. (142.5,54) .. controls (140.83,55.67) and (139.17,55.67) .. (137.5,54) .. controls (135.83,52.33) and (134.17,52.33) .. (132.5,54) .. controls (130.83,55.67) and (129.17,55.67) .. (127.5,54) .. controls (125.83,52.33) and (124.17,52.33) .. (122.5,54) .. controls (120.83,55.67) and (119.17,55.67) .. (117.5,54) .. controls (115.83,52.33) and (114.17,52.33) .. (112.5,54) .. controls (110.83,55.67) and (109.17,55.67) .. (107.5,54) .. controls (105.83,52.33) and (104.17,52.33) .. (102.5,54) .. controls (100.83,55.67) and (99.17,55.67) .. (97.5,54) .. controls (95.83,52.33) and (94.17,52.33) .. (92.5,54) .. controls (90.83,55.67) and (89.17,55.67) .. (87.5,54) .. controls (85.83,52.33) and (84.17,52.33) .. (82.5,54) -- (78.5,54) -- (70.5,54) ;
\draw [shift={(67.5,54)}, rotate = 360] [fill={rgb, 255:red, 0; green, 0; blue, 0 }  ][line width=0.08]  [draw opacity=0] (10.72,-5.15) -- (0,0) -- (10.72,5.15) -- (7.12,0) -- cycle    ;
\draw    (280.25,54) .. controls (281.92,52.33) and (283.58,52.33) .. (285.25,54) .. controls (286.92,55.67) and (288.58,55.67) .. (290.25,54) .. controls (291.92,52.33) and (293.58,52.33) .. (295.25,54) .. controls (296.92,55.67) and (298.58,55.67) .. (300.25,54) .. controls (301.92,52.33) and (303.58,52.33) .. (305.25,54) .. controls (306.92,55.67) and (308.58,55.67) .. (310.25,54) .. controls (311.92,52.33) and (313.58,52.33) .. (315.25,54) .. controls (316.92,55.67) and (318.58,55.67) .. (320.25,54) .. controls (321.92,52.33) and (323.58,52.33) .. (325.25,54) .. controls (326.92,55.67) and (328.58,55.67) .. (330.25,54) .. controls (331.92,52.33) and (333.58,52.33) .. (335.25,54) .. controls (336.92,55.67) and (338.58,55.67) .. (340.25,54) .. controls (341.92,52.33) and (343.58,52.33) .. (345.25,54) .. controls (346.92,55.67) and (348.58,55.67) .. (350.25,54) .. controls (351.92,52.33) and (353.58,52.33) .. (355.25,54) .. controls (356.92,55.67) and (358.58,55.67) .. (360.25,54) .. controls (361.92,52.33) and (363.58,52.33) .. (365.25,54) .. controls (366.92,55.67) and (368.58,55.67) .. (370.25,54) .. controls (371.92,52.33) and (373.58,52.33) .. (375.25,54) .. controls (376.92,55.67) and (378.58,55.67) .. (380.25,54) .. controls (381.92,52.33) and (383.58,52.33) .. (385.25,54) .. controls (386.92,55.67) and (388.58,55.67) .. (390.25,54) .. controls (391.92,52.33) and (393.58,52.33) .. (395.25,54) .. controls (396.92,55.67) and (398.58,55.67) .. (400.25,54) .. controls (401.92,52.33) and (403.58,52.33) .. (405.25,54) .. controls (406.92,55.67) and (408.58,55.67) .. (410.25,54) .. controls (411.92,52.33) and (413.58,52.33) .. (415.25,54) .. controls (416.92,55.67) and (418.58,55.67) .. (420.25,54) -- (424.25,54) -- (432.25,54) ;
\draw [shift={(435.25,54)}, rotate = 180] [fill={rgb, 255:red, 0; green, 0; blue, 0 }  ][line width=0.08]  [draw opacity=0] (10.72,-5.15) -- (0,0) -- (10.72,5.15) -- (7.12,0) -- cycle    ;
\draw  [pattern=_h12iepk5o,pattern size=9pt,pattern thickness=0.75pt,pattern radius=0pt, pattern color={rgb, 255:red, 0; green, 0; blue, 0}] (130,54) .. controls (130,29.7) and (136.72,10) .. (145,10) .. controls (153.28,10) and (160,29.7) .. (160,54) .. controls (160,78.3) and (153.28,98) .. (145,98) .. controls (136.72,98) and (130,78.3) .. (130,54) -- cycle ;
\draw  [pattern=_5mnlv7bot,pattern size=9pt,pattern thickness=0.75pt,pattern radius=0pt, pattern color={rgb, 255:red, 0; green, 0; blue, 0}] (342.75,54) .. controls (343.17,29.7) and (350.23,10) .. (358.52,10) .. controls (366.8,10) and (373.17,29.7) .. (372.75,54) .. controls (372.33,78.3) and (365.27,98) .. (356.98,98) .. controls (348.7,98) and (342.33,78.3) .. (342.75,54) -- cycle ;

\draw (237.5,88) node [anchor=north west][inner sep=0.75pt]  [xscale=0.8,yscale=0.8] [align=left] {Fonte};
\draw (100.75,70.5) node [anchor=north west][inner sep=0.75pt]  [xscale=0.8,yscale=0.8] [align=left] {1};
\draw (391,70.5) node [anchor=north west][inner sep=0.75pt]  [xscale=0.8,yscale=0.8] [align=left] {2};
\draw (136,102.4) node [anchor=north west][inner sep=0.75pt]  [xscale=0.8,yscale=0.8]  {$0^{o}$};
\draw (348.75,102.4) node [anchor=north west][inner sep=0.75pt]  [xscale=0.8,yscale=0.8]  {$0^{o}$};

\end{tikzpicture}
 \caption{{\bf\textsc{pred$_{\text{\bf\textsc{mq}}}$1}} com f\'{o}tons passando}
    \label{fig:bell-0-pass}
\end{figure}

\begin{description}
\item[\textsc{pred$_{\text{\bf\textsc{mq}}}$2}~$\overset{\underset{\mathrm{def}}{}}{=}$] Dois f\'{o}tons ($1$ e $2$) na mesma orienta\c{c}\~{a}o e com os polarizadores diferindo num \^{a}ngulo de $30^o$ \textit{discordam em $25\%$ dos resultados}. Isto \'{e}, se $1$ passa pelo polarizador, sabemos que $2$ tem $25\%$ de chances de n\~{a}o passar. 
\end{description}

\begin{figure}[ht!]
    \centering

 
\tikzset{
pattern size/.store in=\mcSize, 
pattern size = 5pt,
pattern thickness/.store in=\mcThickness, 
pattern thickness = 0.3pt,
pattern radius/.store in=\mcRadius, 
pattern radius = 1pt}
\makeatletter
\pgfutil@ifundefined{pgf@pattern@name@_rxnyth7xk}{
\pgfdeclarepatternformonly[\mcThickness,\mcSize]{_rxnyth7xk}
{\pgfqpoint{0pt}{-\mcThickness}}
{\pgfpoint{\mcSize}{\mcSize}}
{\pgfpoint{\mcSize}{\mcSize}}
{
\pgfsetcolor{\tikz@pattern@color}
\pgfsetlinewidth{\mcThickness}
\pgfpathmoveto{\pgfqpoint{0pt}{\mcSize}}
\pgfpathlineto{\pgfpoint{\mcSize+\mcThickness}{-\mcThickness}}
\pgfusepath{stroke}
}}
\makeatother

 
\tikzset{
pattern size/.store in=\mcSize, 
pattern size = 5pt,
pattern thickness/.store in=\mcThickness, 
pattern thickness = 0.3pt,
pattern radius/.store in=\mcRadius, 
pattern radius = 1pt}
\makeatletter
\pgfutil@ifundefined{pgf@pattern@name@_rlbdehkrg}{
\pgfdeclarepatternformonly[\mcThickness,\mcSize]{_rlbdehkrg}
{\pgfqpoint{0pt}{-\mcThickness}}
{\pgfpoint{\mcSize}{\mcSize}}
{\pgfpoint{\mcSize}{\mcSize}}
{
\pgfsetcolor{\tikz@pattern@color}
\pgfsetlinewidth{\mcThickness}
\pgfpathmoveto{\pgfqpoint{0pt}{\mcSize}}
\pgfpathlineto{\pgfpoint{\mcSize+\mcThickness}{-\mcThickness}}
\pgfusepath{stroke}
}}
\makeatother
\tikzset{every picture/.style={line width=0.75pt}} 

\begin{tikzpicture}[x=0.75pt,y=0.75pt,yscale=-.8,xscale=.8]

\draw   (280.25,84) -- (234.75,84) .. controls (227.98,84) and (222.5,70.57) .. (222.5,54) .. controls (222.5,37.43) and (227.98,24) .. (234.75,24) -- (280.25,24)(292.5,54) .. controls (292.5,70.57) and (287.02,84) .. (280.25,84) .. controls (273.48,84) and (268,70.57) .. (268,54) .. controls (268,37.43) and (273.48,24) .. (280.25,24) .. controls (287.02,24) and (292.5,37.43) .. (292.5,54) ;
\draw    (222.5,54) .. controls (220.83,55.67) and (219.17,55.67) .. (217.5,54) .. controls (215.83,52.33) and (214.17,52.33) .. (212.5,54) .. controls (210.83,55.67) and (209.17,55.67) .. (207.5,54) .. controls (205.83,52.33) and (204.17,52.33) .. (202.5,54) .. controls (200.83,55.67) and (199.17,55.67) .. (197.5,54) .. controls (195.83,52.33) and (194.17,52.33) .. (192.5,54) .. controls (190.83,55.67) and (189.17,55.67) .. (187.5,54) .. controls (185.83,52.33) and (184.17,52.33) .. (182.5,54) .. controls (180.83,55.67) and (179.17,55.67) .. (177.5,54) .. controls (175.83,52.33) and (174.17,52.33) .. (172.5,54) .. controls (170.83,55.67) and (169.17,55.67) .. (167.5,54) .. controls (165.83,52.33) and (164.17,52.33) .. (162.5,54) .. controls (160.83,55.67) and (159.17,55.67) .. (157.5,54) .. controls (155.83,52.33) and (154.17,52.33) .. (152.5,54) .. controls (150.83,55.67) and (149.17,55.67) .. (147.5,54) .. controls (145.83,52.33) and (144.17,52.33) .. (142.5,54) .. controls (140.83,55.67) and (139.17,55.67) .. (137.5,54) .. controls (135.83,52.33) and (134.17,52.33) .. (132.5,54) .. controls (130.83,55.67) and (129.17,55.67) .. (127.5,54) .. controls (125.83,52.33) and (124.17,52.33) .. (122.5,54) .. controls (120.83,55.67) and (119.17,55.67) .. (117.5,54) .. controls (115.83,52.33) and (114.17,52.33) .. (112.5,54) .. controls (110.83,55.67) and (109.17,55.67) .. (107.5,54) .. controls (105.83,52.33) and (104.17,52.33) .. (102.5,54) .. controls (100.83,55.67) and (99.17,55.67) .. (97.5,54) .. controls (95.83,52.33) and (94.17,52.33) .. (92.5,54) .. controls (90.83,55.67) and (89.17,55.67) .. (87.5,54) .. controls (85.83,52.33) and (84.17,52.33) .. (82.5,54) -- (78.5,54) -- (70.5,54) ;
\draw [shift={(67.5,54)}, rotate = 360] [fill={rgb, 255:red, 0; green, 0; blue, 0 }  ][line width=0.08]  [draw opacity=0] (10.72,-5.15) -- (0,0) -- (10.72,5.15) -- (7.12,0) -- cycle    ;
\draw    (280.25,54) .. controls (281.92,52.33) and (283.58,52.33) .. (285.25,54) .. controls (286.92,55.67) and (288.58,55.67) .. (290.25,54) .. controls (291.92,52.33) and (293.58,52.33) .. (295.25,54) .. controls (296.92,55.67) and (298.58,55.67) .. (300.25,54) .. controls (301.92,52.33) and (303.58,52.33) .. (305.25,54) .. controls (306.92,55.67) and (308.58,55.67) .. (310.25,54) .. controls (311.92,52.33) and (313.58,52.33) .. (315.25,54) .. controls (316.92,55.67) and (318.58,55.67) .. (320.25,54) .. controls (321.92,52.33) and (323.58,52.33) .. (325.25,54) .. controls (326.92,55.67) and (328.58,55.67) .. (330.25,54) .. controls (331.92,52.33) and (333.58,52.33) .. (335.25,54) .. controls (336.92,55.67) and (338.58,55.67) .. (340.25,54) .. controls (341.92,52.33) and (343.58,52.33) .. (345.25,54) -- (346.75,54) -- (354.75,54) ;
\draw [shift={(357.75,54)}, rotate = 180] [fill={rgb, 255:red, 0; green, 0; blue, 0 }  ][line width=0.08]  [draw opacity=0] (10.72,-5.15) -- (0,0) -- (10.72,5.15) -- (7.12,0) -- cycle    ;
\draw  [pattern=_rxnyth7xk,pattern size=9pt,pattern thickness=0.75pt,pattern radius=0pt, pattern color={rgb, 255:red, 0; green, 0; blue, 0}] (130,54) .. controls (130,29.7) and (136.72,10) .. (145,10) .. controls (153.28,10) and (160,29.7) .. (160,54) .. controls (160,78.3) and (153.28,98) .. (145,98) .. controls (136.72,98) and (130,78.3) .. (130,54) -- cycle ;
\draw  [pattern=_rlbdehkrg,pattern size=9pt,pattern thickness=0.75pt,pattern radius=0pt, pattern color={rgb, 255:red, 0; green, 0; blue, 0}] (344.76,46.5) .. controls (357.28,25.67) and (373.24,12.14) .. (380.42,16.28) .. controls (387.59,20.42) and (383.26,40.67) .. (370.74,61.5) .. controls (358.22,82.33) and (342.26,95.86) .. (335.08,91.72) .. controls (327.91,87.58) and (332.24,67.33) .. (344.76,46.5) -- cycle ;

\draw (237.5,88) node [anchor=north west][inner sep=0.75pt]  [xscale=0.8,yscale=0.8] [align=left] {Fonte};
\draw (100.75,70.5) node [anchor=north west][inner sep=0.75pt]  [xscale=0.8,yscale=0.8] [align=left] {1};
\draw (391,70.5) node [anchor=north west][inner sep=0.75pt]  [xscale=0.8,yscale=0.8] [align=left] {2};
\draw (136,102.4) node [anchor=north west][inner sep=0.75pt]  [xscale=0.8,yscale=0.8]  {$0^{o}$};
\draw (344.75,102.4) node [anchor=north west][inner sep=0.75pt]  [xscale=0.8,yscale=0.8]  {$30^{o}$};

\end{tikzpicture} \caption{{\bf\textsc{pred$_{\text{\bf\textsc{mq}}}$2}} com resultado em desacordo ($25\%$)}
    \label{fig:bell-30}
\end{figure}

\begin{description}
\item[\textsc{pred$_{\text{\bf\textsc{mq}}}$3}~$\overset{\underset{\mathrm{def}}{}}{=}$] Dois f\'{o}tons ($1$ e $2$) na mesma orienta\c{c}\~{a}o e com os polarizadores diferindo num \^{a}ngulo de $60^o$ \textit{discordam em $75\%$ dos resultados}. Isto \'{e}, se $1$ passa pelo polarizador, sabemos que $2$ tem $75\%$ de chances de n\~{a}o passar.
\end{description}

\begin{figure}[ht!]
    \centering

 
\tikzset{
pattern size/.store in=\mcSize, 
pattern size = 5pt,
pattern thickness/.store in=\mcThickness, 
pattern thickness = 0.3pt,
pattern radius/.store in=\mcRadius, 
pattern radius = 1pt}
\makeatletter
\pgfutil@ifundefined{pgf@pattern@name@_zzep3i0tx}{
\pgfdeclarepatternformonly[\mcThickness,\mcSize]{_zzep3i0tx}
{\pgfqpoint{0pt}{-\mcThickness}}
{\pgfpoint{\mcSize}{\mcSize}}
{\pgfpoint{\mcSize}{\mcSize}}
{
\pgfsetcolor{\tikz@pattern@color}
\pgfsetlinewidth{\mcThickness}
\pgfpathmoveto{\pgfqpoint{0pt}{\mcSize}}
\pgfpathlineto{\pgfpoint{\mcSize+\mcThickness}{-\mcThickness}}
\pgfusepath{stroke}
}}
\makeatother

 
\tikzset{
pattern size/.store in=\mcSize, 
pattern size = 5pt,
pattern thickness/.store in=\mcThickness, 
pattern thickness = 0.3pt,
pattern radius/.store in=\mcRadius, 
pattern radius = 1pt}
\makeatletter
\pgfutil@ifundefined{pgf@pattern@name@_2752x0yab}{
\pgfdeclarepatternformonly[\mcThickness,\mcSize]{_2752x0yab}
{\pgfqpoint{0pt}{-\mcThickness}}
{\pgfpoint{\mcSize}{\mcSize}}
{\pgfpoint{\mcSize}{\mcSize}}
{
\pgfsetcolor{\tikz@pattern@color}
\pgfsetlinewidth{\mcThickness}
\pgfpathmoveto{\pgfqpoint{0pt}{\mcSize}}
\pgfpathlineto{\pgfpoint{\mcSize+\mcThickness}{-\mcThickness}}
\pgfusepath{stroke}
}}
\makeatother
\tikzset{every picture/.style={line width=0.75pt}} 

\begin{tikzpicture}[x=0.75pt,y=0.75pt,yscale=-.8,xscale=.8]

\draw   (280.25,84) -- (234.75,84) .. controls (227.98,84) and (222.5,70.57) .. (222.5,54) .. controls (222.5,37.43) and (227.98,24) .. (234.75,24) -- (280.25,24)(292.5,54) .. controls (292.5,70.57) and (287.02,84) .. (280.25,84) .. controls (273.48,84) and (268,70.57) .. (268,54) .. controls (268,37.43) and (273.48,24) .. (280.25,24) .. controls (287.02,24) and (292.5,37.43) .. (292.5,54) ;
\draw    (222.5,54) .. controls (220.83,55.67) and (219.17,55.67) .. (217.5,54) .. controls (215.83,52.33) and (214.17,52.33) .. (212.5,54) .. controls (210.83,55.67) and (209.17,55.67) .. (207.5,54) .. controls (205.83,52.33) and (204.17,52.33) .. (202.5,54) .. controls (200.83,55.67) and (199.17,55.67) .. (197.5,54) .. controls (195.83,52.33) and (194.17,52.33) .. (192.5,54) .. controls (190.83,55.67) and (189.17,55.67) .. (187.5,54) .. controls (185.83,52.33) and (184.17,52.33) .. (182.5,54) .. controls (180.83,55.67) and (179.17,55.67) .. (177.5,54) .. controls (175.83,52.33) and (174.17,52.33) .. (172.5,54) .. controls (170.83,55.67) and (169.17,55.67) .. (167.5,54) .. controls (165.83,52.33) and (164.17,52.33) .. (162.5,54) .. controls (160.83,55.67) and (159.17,55.67) .. (157.5,54) .. controls (155.83,52.33) and (154.17,52.33) .. (152.5,54) .. controls (150.83,55.67) and (149.17,55.67) .. (147.5,54) .. controls (145.83,52.33) and (144.17,52.33) .. (142.5,54) .. controls (140.83,55.67) and (139.17,55.67) .. (137.5,54) .. controls (135.83,52.33) and (134.17,52.33) .. (132.5,54) .. controls (130.83,55.67) and (129.17,55.67) .. (127.5,54) .. controls (125.83,52.33) and (124.17,52.33) .. (122.5,54) .. controls (120.83,55.67) and (119.17,55.67) .. (117.5,54) .. controls (115.83,52.33) and (114.17,52.33) .. (112.5,54) .. controls (110.83,55.67) and (109.17,55.67) .. (107.5,54) .. controls (105.83,52.33) and (104.17,52.33) .. (102.5,54) .. controls (100.83,55.67) and (99.17,55.67) .. (97.5,54) .. controls (95.83,52.33) and (94.17,52.33) .. (92.5,54) .. controls (90.83,55.67) and (89.17,55.67) .. (87.5,54) .. controls (85.83,52.33) and (84.17,52.33) .. (82.5,54) -- (78.5,54) -- (70.5,54) ;
\draw [shift={(67.5,54)}, rotate = 360] [fill={rgb, 255:red, 0; green, 0; blue, 0 }  ][line width=0.08]  [draw opacity=0] (10.72,-5.15) -- (0,0) -- (10.72,5.15) -- (7.12,0) -- cycle    ;
\draw    (280.25,54) .. controls (281.92,52.33) and (283.58,52.33) .. (285.25,54) .. controls (286.92,55.67) and (288.58,55.67) .. (290.25,54) .. controls (291.92,52.33) and (293.58,52.33) .. (295.25,54) .. controls (296.92,55.67) and (298.58,55.67) .. (300.25,54) .. controls (301.92,52.33) and (303.58,52.33) .. (305.25,54) .. controls (306.92,55.67) and (308.58,55.67) .. (310.25,54) .. controls (311.92,52.33) and (313.58,52.33) .. (315.25,54) .. controls (316.92,55.67) and (318.58,55.67) .. (320.25,54) .. controls (321.92,52.33) and (323.58,52.33) .. (325.25,54) .. controls (326.92,55.67) and (328.58,55.67) .. (330.25,54) .. controls (331.92,52.33) and (333.58,52.33) .. (335.25,54) .. controls (336.92,55.67) and (338.58,55.67) .. (340.25,54) .. controls (341.92,52.33) and (343.58,52.33) .. (345.25,54) -- (346.75,54) -- (354.75,54) ;
\draw [shift={(357.75,54)}, rotate = 180] [fill={rgb, 255:red, 0; green, 0; blue, 0 }  ][line width=0.08]  [draw opacity=0] (10.72,-5.15) -- (0,0) -- (10.72,5.15) -- (7.12,0) -- cycle    ;
\draw  [pattern=_zzep3i0tx,pattern size=9pt,pattern thickness=0.75pt,pattern radius=0pt, pattern color={rgb, 255:red, 0; green, 0; blue, 0}] (130,54) .. controls (130,29.7) and (136.72,10) .. (145,10) .. controls (153.28,10) and (160,29.7) .. (160,54) .. controls (160,78.3) and (153.28,98) .. (145,98) .. controls (136.72,98) and (130,78.3) .. (130,54) -- cycle ;
\draw  [pattern=_2752x0yab,pattern size=9pt,pattern thickness=0.75pt,pattern radius=0pt, pattern color={rgb, 255:red, 0; green, 0; blue, 0}] (350.25,41.01) .. controls (371.51,29.23) and (392.1,25.49) .. (396.24,32.67) .. controls (400.38,39.84) and (386.51,55.21) .. (365.25,66.99) .. controls (343.99,78.77) and (323.4,82.51) .. (319.26,75.33) .. controls (315.12,68.16) and (328.99,52.79) .. (350.25,41.01) -- cycle ;

\draw (237.5,88) node [anchor=north west][inner sep=0.75pt]  [xscale=0.8,yscale=0.8] [align=left] {Fonte};
\draw (100.75,70.5) node [anchor=north west][inner sep=0.75pt]  [xscale=0.8,yscale=0.8] [align=left] {1};
\draw (391,70.5) node [anchor=north west][inner sep=0.75pt]  [xscale=0.8,yscale=0.8] [align=left] {2};
\draw (136,102.4) node [anchor=north west][inner sep=0.75pt]  [xscale=0.8,yscale=0.8]  {$0^{o}$};
\draw (344.75,102.4) node [anchor=north west][inner sep=0.75pt]  [xscale=0.8,yscale=0.8]  {$60^{o}$};

\end{tikzpicture} \caption{{\bf\textsc{pred$_{\text{\bf\textsc{mq}}}$3}} com resultado em desacordo ($75\%$)}
    \label{fig:bell-60}
\end{figure}

O argumento de Bell, conforme reconstru\'{i}do por \citet{chen2022}, tem duas partes. A primeira \'{e} o argumento EPR\index{Einstein, Albert!argumento EPR} em si, agora refraseado na terminologia que acabamos de introduzir:
\begin{equation}
{\bf\text{Parte I}}=\big\langle\text{\bf\textsc{predi\c{c}\~{a}o}}_{\text{\bf\textsc{mq}}}\text{1}\land\text{\bf\textsc{loc}}\big\rangle\to\text{\bf\textsc{det}}
\end{equation}

Suponha que o f\'{o}ton $1$ passa pelo polarizador. De acordo com a {\bf\textsc{pred$_{\text{\bf\textsc{mq}}}$1}}, n\~{a}o precisamos medir $2$ para sabermos que ele tamb\'{e}m passou. Contudo, de acordo com a {\bf\textsc{loc}}, o estado de $1$ n\~{a}o pode afetar o estado de $2$. Assim a conjun\c{c}\~{a}o $\big\langle\text{\bf\textsc{predi\c{c}\~{a}o}}_{\text{\bf\textsc{mq}}}\text{1}\land\text{\bf\textsc{loc}}\big\rangle$ implica que, se podemos inferir com precis\~{a}o o valor de $2$ pelo valor da medida de $1$, o valor de 2 deveria ser pr\'{e}-determinado, pr\'{e}-existente, ou  {\bf\textsc{det}} ---o que seria an\'{a}logo \`{a} tese da ``vis\~{a}o natural'' de \citet{armstrong1961} conforme vimos anteriormente.

No entanto, a mec\^{a}nica qu\^{a}ntica n\~{a}o oferece uma descri\c{c}\~{a}o de \textit{qual} estado os f\'{o}tons assumem. Apenas probabilidades. Isto \'{e}, n\~{a}o diz se os f\'{o}tons ir\~{a}o passar ou n\~{a}o: diz que \textit{caso} o primeiro passe, o segundo \textit{tamb\'{e}m} passa com probabilidade $100\%$ ---e o mesmo vale para a absor\c{c}\~{a}o. Sem uma descri\c{c}\~{a}o de {\bf\textsc{det}}, portanto, o argumento EPR\index{Einstein, Albert!argumento EPR} conclui que a mec\^{a}nica qu\^{a}ntica \'{e} incompleta \citep*{epr1963epr}. A ``Parte I'' do teorema de Bell\index{Bell, John Stewart!teorema de Bell} \citep[conforme a reconstru\c{c}\~{a}o de][a qual estou seguindo aqui]{chen2022} \textit{\'{e}}, repito, o argumento EPR\index{Einstein, Albert!argumento EPR}. 

Notavelmente, {\bf\textsc{det}} ---que \'{e} uma conclus\~{a}o derivada da Parte I--- nos permite montar uma \textit{tabela} com os resultados das tr\^{e}s predi\c{c}\~{o}es da mec\^{a}nica qu\^{a}ntica. Para cada par, e para cada f\'{o}ton, existe uma determinada propriedade/disposi\c{c}\~{a}o para reagir de uma determinada maneira a uma determinada orienta\c{c}\~{a}o do polarizador. Por exemplo, para o polarizador orientado em $30$ graus, um f\'{o}ton deve ter uma das duas propriedades: ou ele passa ou ele \'{e} absorvido. Ele tem a propriedade antes mesmo do experimento ser conduzido.

Ademais, {\bf\textsc{det}} nos permite montar a tabela \ref{tabela:bell} com apenas $8$ possibilidades. O motivo disso \'{e} que, em cada par, o f\'{o}ton esquerdo e o f\'{o}ton direito devem concordar em como reagir a qualquer orienta\c{c}\~{a}o espec\'{i}fica do polarizador, de acordo com as predi\c{c}\~{o}es da mec\^{a}nica qu\^{a}ntica. Vamos adotar a seguinte nota\c{c}\~{a}o, seguindo \citet{chen2022}: $P$ para a propriedade de passar e $A$ para a propriedade de ser absorvido; e os \'{i}ndices indicam a angula\c{c}\~{a}o do polarizador. Assim, por exemplo, $P_{30}$ indica que o f\'{o}ton passou pelo polarizador orientado a $30$ graus, e $A_{0}$ que o f\'{o}ton foi absorvido pelo polarizador orientado a $0$ graus. Ent\~{a}o eis a tabela.

    \begin{table}[ht!]

\centering
\begin{tabular}{cccc} 
\toprule
  & Esquerda                                                    & Direita           & \%                         \\ 
\hline
1 & \begin{tabular}[c]{@{}l@{}}$P_0, P_{30}, P_{60}$\\\end{tabular} & $P_0, P_{30}, P_{60}$ & \multirow{2}{*}{$\alpha$}  \\
2 & $A_0, A_{30}, A_{60}$                                           & $A_0, A_{30}, A_{60}$ &                            \\
3 & $A_0, P_{30}, P_{60}$                                           & $A_0, P_{30}, P_{60}$ & \multirow{2}{*}{$\beta$}   \\
4 & $P_0, A_{30}, A_{60}$                                           & $P_0, A_{30}, A_{60}$ &                            \\
5 & $P_0, A_{30}, P_{60}$                                           & $P_0, A_{30}, P_{60}$ & \multirow{2}{*}{$\gamma$}  \\
6 & $A_0, P_{30}, A_{60}$                                           & $A_0, P_{30}, A_{60}$ &                            \\
7 & $P_0, P_{30}, A_{60}$                                           & $P_0, P_{30}, A_{60}$ & \multirow{2}{*}{$\delta$}  \\
8 & $A_0, A_{30}, P_{60}$                                           & $A_0, A_{30}, P_{60}$ &                            \\
\bottomrule
\end{tabular}
\caption{Resultados poss\'{i}veis}
\label{tabela:bell}
\end{table}

Em particular, temos que $\alpha+\beta+\gamma+\delta=100\%$, e isso nos permite formular a seguinte desigualdade (chamada ``desigualdade de Bell''):
\begin{equation}\label{eq:ineq1}
    \alpha,\beta,\gamma,\delta\geq 0 \to (\beta+\gamma)+(\gamma+\delta)\geq\beta+\delta 
\end{equation}

Para verificar essa desigualdade, basta checar as possibilidades:
\begin{enumerate}
    \item Suponha que $\gamma=0$, $\beta=0$, e $\delta=0$. Ent\~{a}o $0+0=0$.
    \item Suponha que $\gamma=1$, $\beta=0$, e $\delta=0$. Ent\~{a}o $1+1>0$.
    \item Suponha que $\gamma=0$, $\beta=1$, e $\delta=0$. Ent\~{a}o $1+0=1$.
    \item Suponha que $\gamma=0$, $\beta=0$, e $\delta=1$. Ent\~{a}o $0+1=1$.
    \item Suponha que $\gamma=1$, $\beta=1$, e $\delta=0$. Ent\~{a}o $2+1>1$.
    \item Suponha que $\gamma=0$, $\beta=1$, e $\delta=1$. Ent\~{a}o $1+1=2$.
    \item Suponha que $\gamma=1$, $\beta=0$, e $\delta=1$. Ent\~{a}o $1+2>1$.
    \item Suponha que $\gamma=1$, $\beta=1$, e $\delta=1$. Ent\~{a}o $2+2>2$.
\end{enumerate}

Agora podemos preencher essa desigualdade com as predi\c{c}\~{o}es da mec\^{a}nica qu\^{a}ntica. A \textsc{pred$_{\text{\bf\textsc{mq}}}$2} implica que $\beta+\gamma=\text{25\% de desacordo}$ e que, da mesma maneira, $\gamma+\delta=\text{25\% de desacordo}$. J\'{a} a  \textsc{pred$_{\text{\bf\textsc{mq}}}$3} implica que $\beta+\delta=\text{75\% de desacordo}$. Isso tudo pode ser verificado chegando a tabela \ref{tabela:bell}. Com isso, chegamos \`{a} segunda parte do argumento de Bell:
\begin{equation}
    {\bf\text{Parte II}}=\big\langle\text{\bf\textsc{det}}\land(\text{\bf\textsc{pred}}_{\text{\bf\textsc{mq}}}\text{2}\land\text{\bf\textsc{pred}}_{\text{\bf\textsc{mq}}}\text{3})\big\rangle\to\text{\bf\textsc{inconsist\^{e}ncia}}
\end{equation}

Para ver a inconsist\^{e}ncia, basta checar a tabela, e formaremos a seguinte desigualdade
\begin{equation}
\underbrace{\beta+\gamma=\text{25\%}}_{\text{\bf\textsc{pred}}_{\text{\bf\textsc{mq}}}\text{\bf2}}
+
\underbrace{\gamma+\delta=\text{25\%}}_{\text{\bf\textsc{pred}}_{\text{\bf\textsc{mq}}}\text{\bf2}}
<
\underbrace{\beta+\delta=\text{75\%}}_{\text{\bf\textsc{pred}}_{\text{\bf\textsc{mq}}}\text{\bf3}}
\end{equation}
o que se traduz para
\begin{equation}
    50<75\to\perp 
\end{equation}
o que \'{e} uma inconsist\^{e}ncia. Logo, a conjun\c{c}\~{a}o
\begin{equation}
\overbrace{\Big[(\beta+\gamma)+(\gamma+\delta)\geq\beta+\delta\Big]}^{\text{\bf\textsc{det}}}\land \overbrace{\Big[(\beta+\gamma)+(\gamma+\delta)<\beta+\delta\Big]}^{\text{\bf\textsc{pred}}_{\text{\bf\textsc{mq}}}}\to\perp 
\end{equation}
\'{e} inconsistente. Isso quer dizer que alguma das premissas do teorema devem ser abandonadas. Nas palavras de \citet{lewis2019}:

\begin{quote}
    Tomado como verdadeiro, o teorema de Bell\index{Bell, John Stewart!teorema de Bell} parece mostrar que a mec\^{a}nica qu\^{a}ntica \'{e} imposs\'{i}vel---que nenhum modelo f\'{i}sico poderia, em princ\'{i}pio, produzir a distribui\c{c}\~{a}o dos resultados de medi\c{c}\~{a}o previstos pelo algoritmo matem\'{a}tico no cora\c{c}\~{a}o da mec\^{a}nica qu\^{a}ntica. Mas a mec\^{a}nica qu\^{a}ntica est\'{a} bem confirmada; essa distribui\c{c}\~{a}o de resultados de medi\c{c}\~{a}o \'{e} \textit{realmente observada}, e o que \'{e} real n\~{a}o pode ser imposs\'{i}vel! Assim, a maneira de ler o teorema de Bell\index{Bell, John Stewart!teorema de Bell} \'{e} como uma redu\c{c}\~{a}o ao absurdo: como a prova de Bell leva a uma conclus\~{a}o absurda, uma de suas suposi\c{c}\~{o}es deve ser falsa. \citep[p.~36]{lewis2019}.
\end{quote}
Como vimos, uma das suposi\c{c}\~{o}es feitas pelo teorema foi a localidade (que, como vimos, implica na determina\c{c}\~{a}o e na separabilidade). A mec\^{a}nica qu\^{a}ntica padr\~{a}o, com a medi\c{c}\~{a}o=cria\c{c}\~{a}o, mostra-se incompat\'{i}vel com tais teses conforme procurei demonstrar com o teorema de Bell\index{Bell, John Stewart!teorema de Bell}.

Aqui podemos entender, finalmente, com precis\~{a}o a ideia que Einstein trouxe em sua carta a Schr\"{o}dinger, a saber, da disposi\c{c}\~{a}o da bola em uma de duas caixas. Se a descri\c{c}\~{a}o qu\^{a}ntica for completa, n\~{a}o h\'{a} raz\~{a}o de fato sobre o estado da localiza\c{c}\~{a}o da bola \textit{at\'{e} que} uma medi\c{c}\~{a}o seja feita; relembremo-nos de suas palavras: ``[a bola] estar em uma caixa definitiva s\'{o} acontece quando levanto as tampas'' \citep[p.~69]{einstein-to-schrodinger1935}. Caso contr\'{a}rio, se a bola \textit{j\'{a} estivesse em alguma das caixas} antes da medi\c{c}\~{a}o, a descri\c{c}\~{a}o qu\^{a}ntica seria incompleta (porque n\~{a}o diz em \textit{qual} caixa ela est\'{a}, somente diz as probabilidades de estar em cada uma das caixas. Portanto, a completude da mec\^{a}nica qu\^{a}ntica vem com o pre\c{c}o de negar n\~{a}o somente a localidade, mas a determina\c{c}\~{a}o, que \'{e} \textit{consequ\^{e}ncia}, como vimos, da conjun\c{c}\~{a}o entre a localidade e as predi\c{c}\~{o}es da mec\^{a}nica qu\^{a}ntica. 

A ideia de medi\c{c}\~{a}o=cria\c{c}\~{a}o (ao menos na maneira usual de se formular e interpretar a mec\^{a}nica qu\^{a}ntica) veio para ficar.

No pr\'{o}ximo cap\'{i}tulo expandirei a forma como a no\c{c}\~{a}o de ``medi\c{c}\~{a}o'', t\~{a}o cara para a mec\^{a}nica qu\^{a}ntica padr\~{a}o, pode vir a operar. Veremos, tamb\'{e}m, como ela gera o problema da medi\c{c}\~{a}o, que conectar\'{a} todas as problem\'{a}ticas expostas at\'{e} ent\~{a}o.

\chapter{A consciência colapsa}
\label{CapvNeumann}

\noindent Como vimos no cap\'{i}tulo \ref{CapCopenhague}, o problema da medi\c{c}\~{a}o na mec\^{a}nica qu\^{a}ntica tem sua gênese j\'{a} nas primeiras discuss\~{o}es em torno da interpreta\c{c}\~{a}o de Copenhague, na medida em que a posi\c{c}\~{a}o geral de Bohr seria que as propriedades f\'{i}sicas dos objetos qu\^{a}nticos dependeriam fundamentalmente das condi\c{c}\~{o}es experimentais, isto \'{e}, de medi\c{c}\~{a}o, efetuadas sobre tais objetos ---posicionamento que aparece explicitamente no debate suscitado por EPR.

De acordo com \citet[p.~473]{Jammer1974}, a concep\c{c}\~{a}o ortodoxa de medi\c{c}\~{a}o envolve os objetos a serem medidos e os instrumentos macrosc\'{o}picos de medi\c{c}\~{a}o que, embora necess\'{a}rios para que uma medi\c{c}\~{a}o seja realizada, ``\textelp{} n\~{a}o s\~{a}o explicados pela teoria qu\^{a}ntica em si mesma, mas considerados como logicamente anteriores \`{a} teoria''. Assim, na vis\~{a}o de Bohr\index{Bohr, Niels}, n\~{a}o existiria a necessidade de uma teoria da medi\c{c}\~{a}o qu\^{a}ntica, na medida em que a assun\c{c}\~{a}o do princ\'{i}pio da correspondência\index{Bohr, Niels!princ\'{i}pio da correspondência} supostamente permitiria uma interpreta\c{c}\~{a}o da mec\^{a}nica qu\^{a}ntica que deliberadamente se afastaria do problema da medi\c{c}\~{a}o.

Ainda que o princ\'{i}pio da correspondência\index{Bohr, Niels!princ\'{i}pio da correspondência} de Bohr n\~{a}o possa ser substitu\'{i}do por uma teoria formalizada da medi\c{c}\~{a}o, o tratamento duplo em rela\c{c}\~{a}o ao processo de medi\c{c}\~{a}o seria, como salienta \citet[p.~472]{Jammer1974}, uma das caracter\'{i}sticas mais obscuras da interpreta\c{c}\~{a}o de Copenhague, especificamente no que se refere \`{a} arbitrariedade da classifica\c{c}\~{a}o dos dom\'{i}nios cl\'{a}ssico e qu\^{a}ntico. Ademais, identifico, ao longo deste livro, alguns aspectos do problema da medi\c{c}\~{a}o na interpreta\c{c}\~{a}o de Bohr.

Como enfatizei at\'{e} aqui, o conceito de medi\c{c}\~{a}o se relaciona com todos os aspectos filos\'{o}ficos problem\'{a}ticos da mec\^{a}nica qu\^{a}ntica expostos neste livro. Juntamente com \citet[p.~104]{gibbins1987particles}, considero que a medi\c{c}\~{a}o \'{e} um aspecto ligado \`{a} maioria dos paradoxos da mec\^{a}nica qu\^{a}ntica ---ao menos aqueles investigados at\'{e} aqui. No cap\'{i}tulo \ref{CapCopenhague}, apresentei a discuss\~{a}o filos\'{o}fica suscitada pela medi\c{c}\~{a}o das propriedades observ\'{a}veis ---posi\c{c}\~{a}o e momento--- de um objeto qu\^{a}ntico. Da mesma forma, no cap\'{i}tulo \ref{CapEPR}, apresentei o debate filos\'{o}fico que emerge dos efeitos da medi\c{c}\~{a}o de um objeto $A$ em um objeto espacialmente distante $B$. Assim, conforme procurei elucidar, parece razoavelmente justificada a posi\c{c}\~{a}o de \citet[p.~104]{gibbins1987particles} de que ``\textelp{} o problema da medi\c{c}\~{a}o \'{e} \textit{o} problema central da filosofia da mec\^{a}nica qu\^{a}ntica''.

Neste cap\'{i}tulo, analisarei detalhadamente a no\c{c}\~{a}o de medi\c{c}\~{a}o em mec\^{a}nica qu\^{a}ntica, bem como o problema da medi\c{c}\~{a}o qu\^{a}ntica. Para tanto, iniciarei a discuss\~{a}o pontuando as diferen\c{c}as entre a f\'{i}sica cl\'{a}ssica e a f\'{i}sica qu\^{a}ntica em rela\c{c}\~{a}o ao conceito de medi\c{c}\~{a}o. Em seguida, analisarei a formula\c{c}\~{a}o da teoria da medi\c{c}\~{a}o de \citet{vNeum1955mathematical}\index{Neumann, John von} e suas extens\~{o}es ontol\'{o}gicas. Ao final do cap\'{i}tulo, pontuarei algumas atitudes alternativas \`{a}s formula\c{c}\~{o}es apresentadas ao longo deste livro.

\section{Medi\c{c}\~{a}o: cl\'{a}ssica e qu\^{a}ntica}

Muito embora a f\'{i}sica tenha sido considerada a ciência da medi\c{c}\~{a}o por \citet{campbell1928measurement}, \citet[p.~471]{Jammer1974} afirma que haveria pouco interesse, por parte dos f\'{i}sicos, anteriormente ao advento da mec\^{a}nica qu\^{a}ntica, em explorar mais profundamente o conceito de medi\c{c}\~{a}o. Para \citet[p.~102]{gibbins1987particles}, isso ocorre, pois a descri\c{c}\~{a}o do processo de medi\c{c}\~{a}o \'{e} um procedimento pouco problem\'{a}tico na f\'{i}sica cl\'{a}ssica.

A no\c{c}\~{a}o cl\'{a}ssica de medi\c{c}\~{a}o (bem como sua representa\c{c}\~{a}o matem\'{a}tica) envolveria, de acordo com \citet[p.~471]{Jammer1974}, dois processos, sendo um f\'{i}sico e um psicof\'{i}sico: o processo f\'{i}sico denota uma intera\c{c}\~{a}o que chamarei $\mathcal{I}_1$ entre um objeto que denominarei $\mathcal{X}$ a ser observado (tal como um corpo maci\c{c}o ou uma corrente el\'{e}trica) e um instrumento de medi\c{c}\~{a}o que denominarei $\mathcal{M}$ (tal como uma balan\c{c}a ou um amper\'{i}metro), de modo que $(\mathcal{IX}\leftrightarrow\mathcal{M})$; o processo psicof\'{i}sico denota uma intera\c{c}\~{a}o que chamarei $\mathcal{I}_2$ entre $\mathcal{M}$ e um observador $\mathcal{O}$ (seus \'{o}rg\~{a}os dos sentidos e, em \'{u}ltima an\'{a}lise, sua consciência).

\`{A} primeira vista, tal afirma\c{c}\~{a}o parece estranha na medida em que, da forma como \citet[p.~471]{Jammer1974} generaliza a no\c{c}\~{a}o de f\'{i}sica cl\'{a}ssica, a realidade f\'{i}sica cl\'{a}ssica seria composta por entidades desprovidas de qualidades sensoriais, isto \'{e}, de corpos extensos e seu movimento no espa\c{c}o, ou seja, n\~{a}o haveria espa\c{c}o para a introdu\c{c}\~{a}o da consciência humana como uma parte fundamental na teoria; no entanto, na medida em que a teoria cl\'{a}ssica adquire validade atrav\'{e}s da testabilidade de suas predi\c{c}\~{o}es, a introdu\c{c}\~{a}o desse conceito parece ser mais plaus\'{i}vel, visto que uma opera\c{c}\~{a}o tal como um teste deve envolver, em \'{u}ltima an\'{a}lise, a consciência humana.

Se aceitarmos a defini\c{c}\~{a}o do processo f\'{i}sico como $(\mathcal{IX}\leftrightarrow\mathcal{M})$, deve-se aceitar, por consequência l\'{o}gica, uma a\c{c}\~{a}o do objeto sobre o instrumento de medi\c{c}\~{a}o de forma $(\mathcal{IX}\rightarrow\mathcal{M})$ e, ao mesmo tempo, uma a\c{c}\~{a}o do aparelho medidor sobre o objeto de forma $(\mathcal{IM}\rightarrow\mathcal{X})$. No entanto, a ordem de magnitude da a\c{c}\~{a}o $(\mathcal{IM}\rightarrow\mathcal{X})$ seria t\~{a}o menor do que a a\c{c}\~{a}o de $(\mathcal{IX}\rightarrow\mathcal{M})$, a ponto de ser considerada como elimin\'{a}vel na intera\c{c}\~{a}o $\mathcal{I}_1$. O aspecto psicof\'{i}sico da medi\c{c}\~{a}o cl\'{a}ssica tamb\'{e}m seria abandonado sob a alega\c{c}\~{a}o de que a rela\c{c}\~{a}o entre $\mathcal{M}$ e O estaria fora dos dom\'{i}nios de uma teoria f\'{i}sica.

A a\c{c}\~{a}o do objeto no instrumento de medi\c{c}\~{a}o, no entanto, n\~{a}o poderia ser negligenciada, na medida em que o resultado $\mathcal{M}$, tal como a ponteiro de uma balan\c{c}a indicando um valor $y$, deve depender de $\mathcal{X}$, de modo que a medi\c{c}\~{a}o cl\'{a}ssica seria, de acordo com \citet[p.~471--472]{Jammer1974}, reduzida \`{a} a\c{c}\~{a}o $(\mathcal{IX}\rightarrow\mathcal{M})$. Dito de outro modo, como sugere \citet[p.~102]{gibbins1987particles}, a intera\c{c}\~{a}o $(\mathcal{M}\rightarrow\mathcal{X})$ pode ser arbitrariamente pequena, o que sugere que a medi\c{c}\~{a}o cl\'{a}ssica pode ser descrita com uma precis\~{a}o arbitrariamente grande. Esta atitude permitiria \`{a} f\'{i}sica cl\'{a}ssica o fornecimento de uma abordagem inteiramente objetiva no tratamento dos processos f\'{i}sicos, isto \'{e}, consider\'{a}-los de forma independente da medi\c{c}\~{a}o e, consequentemente, eliminar da teoria o papel da consciência do observador impl\'{i}cito em $\mathcal{I}_2$.

Com o advento da mec\^{a}nica qu\^{a}ntica, mais precisamente com o postulado qu\^{a}ntico, que prevê a necessidade da intera\c{c}\~{a}o finita (isto \'{e}, de ao menos um quantum) entre $\mathcal{M}$ e $\mathcal{X}$, a magnitude da a\c{c}\~{a}o $(\mathcal{IM}\rightarrow\mathcal{X})$ seria igualmente relevante a a\c{c}\~{a}o $(\mathcal{IX}\rightarrow\mathcal{M})$. Como consequência, de acordo com \citet[p.~472]{Jammer1974}, a condi\c{c}\~{a}o para a consistência da concep\c{c}\~{a}o cl\'{a}ssica de medi\c{c}\~{a}o n\~{a}o seria mais aplic\'{a}vel, uma vez que o projeto cl\'{a}ssico de uma abordagem independente da medi\c{c}\~{a}o \'{e} invi\'{a}vel na mec\^{a}nica qu\^{a}ntica, isto \'{e}, n\~{a}o se pode atribuir \`{a} intera\c{c}\~{a}o $(\mathcal{M}\rightarrow\mathcal{X})$ uma grandeza arbitrariamente pequena ---o que \'{e}, como vimos no cap\'{i}tulo \ref{CapCopenhague}, uma das vias para se chegar ao princ\'{i}pio da indetermina\c{c}\~{a}o\index{Heisenberg, Werner!princ\'{i}pio da indetermina\c{c}\~{a}o}.

Um dos aspectos problem\'{a}ticos da medi\c{c}\~{a}o qu\^{a}ntica \'{e} a produ\c{c}\~{a}o de um resultado macrosc\'{o}pico, determinado, fruto da intera\c{c}\~{a}o $\mathcal{I}_1$. Esse \'{e} o problema tornado claro pela assim chamada ``decoerência'' ---que, ao contr\'{a}rio do que muita gente pensa n\~{a}o soluciona o problema da medi\c{c}\~{a}o; \citet[\S~2]{bacciagaluppi2020} \'{e} categ\'{o}rico ao afirmar que ``\textelp{} na presen\c{c}a de fenômenos de decoerência, o problema da medi\c{c}\~{a}o continua existindo ou, na verdade, piora''.

Esse aspecto n\~{a}o nos interessa aqui, pois \'{e} ontologicamente neutro em rela\c{c}\~{a}o \`{a}s teses da \textit{medi\c{c}\~{a}o=revela\c{c}\~{a}o} e \textit{medi\c{c}\~{a}o=cria\c{c}\~{a}o}. O aspecto problem\'{a}tico que desejo enfatizar aqui tem seu recorte nas interpreta\c{c}\~{o}es que adotam a tese da \textit{medi\c{c}\~{a}o=cria\c{c}\~{a}o}: enquanto n\~{a}o houver a intera\c{c}\~{a}o $\mathcal{I}_1$, nenhum evento pode ser considerado atual, mas t\~{a}o somente potencial. Explicitados esses pontos, passarei \`{a} an\'{a}lise da teoria da medi\c{c}\~{a}o qu\^{a}ntica de von Neumann\index{Neumann, John von}.

\section{O problema da medi\c{c}\~{a}o}

De acordo com \citet[p.~474]{Jammer1974}, a teoria da medi\c{c}\~{a}o de \citet{vNeum1955mathematical}\index{Neumann, John von} se assemelha \`{a} interpreta\c{c}\~{a}o de Copenhague, na medida em que tamb\'{e}m atribui um papel fundamental \`{a} descontinuidade presente no ato da medi\c{c}\~{a}o, mas, de forma contr\'{a}ria a Bohr, considera o instrumento de medi\c{c}\~{a}o $\mathcal{M}$ um sistema qu\^{a}ntico-mec\^{a}nico. O racioc\'{i}nio de \citet{vNeum1955mathematical}\index{Neumann, John von} fornece, para \citet[p.~109]{gibbins1987particles}, as condi\c{c}\~{o}es necess\'{a}rias para a formula\c{c}\~{a}o de uma teoria da medi\c{c}\~{a}o em mec\^{a}nica qu\^{a}ntica, sendo a base conceitual para diversas outras teorias da medi\c{c}\~{a}o.

O ponto de partida de \citet[p.~349--351]{vNeum1955mathematical}\index{Neumann, John von} seria a suposi\c{c}\~{a}o de que existem dois tipos de processos ou mudan\c{c}as dos estados qu\^{a}nticos: o processo 1, chamado de ``mudan\c{c}as arbitr\'{a}rias por medi\c{c}\~{a}o'', e o processo 2, chamado de ``mudan\c{c}as autom\'{a}ticas''. O processo 1 \'{e} enunciado como ``o ato descont\'{i}nuo, n\~{a}o causal e instant\^{a}neo de experimentos ou medi\c{c}\~{o}es''; o processo 2 \'{e} enunciado como a ``mudan\c{c}a causal e cont\'{i}nua no curso do tempo''. Ao passo que o processo 2 \'{e} descrito pelas leis de movimento da mec\^{a}nica qu\^{a}ntica,\footnote{~Frequentemente descrita pela ``Equa\c{c}\~{a}o de Schr\"{o}dinger\index{Schr\"{o}dinger, Erwin!Equa\c{c}\~{a}o de Schr\"{o}dinger}\index{Schr\"{o}dinger, Erwin}'', como aponto no cap\'{i}tulo \ref{Cap:formalismo}.} o processo 1 n\~{a}o \'{e}. O processo 1 \'{e} irredut\'{i}vel e, portanto, n\~{a}o pode ser reduzido ao processo 2.

Enquanto o processo 2 envolve uma evolu\c{c}\~{a}o cont\'{i}nua e determinista, o processo 1, ao contr\'{a}rio, envolve uma descontinuidade indeterminista e irrevers\'{i}vel. O processo 1 descreve a transforma\c{c}\~{a}o do estado de um sistema f\'{i}sico ap\'{o}s o ato da medi\c{c}\~{a}o, isto \'{e}, transforma o estado inicial de tal sistema (descrito pelo processo 2) em um estado inteiramente novo, n\~{a}o previs\'{i}vel pelas leis din\^{a}micas de movimento especificadas pelo processo 2. Isto \'{e} not\'{a}vel, pois ao passo que o processo 2 afirma que o estado final do sistema qu\^{a}ntico em quest\~{a}o seja indeterminado em rela\c{c}\~{a}o \`{a}s suas propriedades calcul\'{a}veis pela equa\c{c}\~{a}o de movimento, o processo 1 afirma um valor determinado para tal estado final, registrado pelo ato da medi\c{c}\~{a}o.

O problema da medi\c{c}\~{a}o foi ent\~{a}o delineado pela primeira vez de modo claro: \'{e} o problema da conjun\c{c}\~{a}o entre os dois processos que seriam, para \citet[p.~417]{vNeum1955mathematical}\index{Neumann, John von}, uma ``peculiar natureza dual do procedimento da mec\^{a}nica qu\^{a}ntica''. Mais adiante, afirma:

\begin{quote}
\textelp{} a mec\^{a}nica qu\^{a}ntica descreve os eventos que ocorrem nas partes observadas do mundo ---contanto que elas n\~{a}o interajam com a parte observante--- com o aux\'{i}lio do processo 2; mas assim que uma intera\c{c}\~{a}o ocorre, isto \'{e}, uma medi\c{c}\~{a}o, \'{e} requerido a aplica\c{c}\~{a}o do processo 1. \citep[p.~420]{vNeum1955mathematical}\index{Neumann, John von}.
\end{quote}

Frequentemente, o experimento da dupla fenda \'{e} trazido para ilustrar tal afirma\c{c}\~{a}o.\footnote{~Ver \citet{pessoa2003conceitos} para maiores detalhes.} Considere um aparato no qual el\'{e}trons s\~{a}o disparados da fonte em dire\c{c}\~{a}o \`{a} tela detectora. Isso \'{e} feito com uma intensidade t\~{a}o baixa, que somente um el\'{e}tron por vez \'{e} disparado, de modo que sempre haja no m\'{a}ximo um el\'{e}tron em todo o aparato. Para chegar nela, os el\'{e}trons precisam passar por uma das fendas, $F_1$ ou $F_2$, conforme representado na figura \ref{fig:DF}.

\begin{figure}[ht!]
    \centering

\tikzset{every picture/.style={line width=0.75pt}} 

\begin{tikzpicture}[x=0.75pt,y=0.75pt,yscale=-1,xscale=1]

\draw [line width=1.5]    (117.95,164.19) -- (168.05,164.19) ;
\draw [line width=1.5]    (201.45,164.19) -- (301.65,164.19) ;
\draw [line width=1.5]    (335.05,164.19) -- (385.16,164.19) ;
\draw   (248.93,84.5) .. controls (248.93,82.95) and (250.1,81.7) .. (251.55,81.7) .. controls (253,81.7) and (254.17,82.95) .. (254.17,84.5) .. controls (254.17,86.04) and (253,87.3) .. (251.55,87.3) .. controls (250.1,87.3) and (248.93,86.04) .. (248.93,84.5) -- cycle ;
\draw  [draw opacity=0][dash pattern={on 0.84pt off 2.51pt}] (267.28,91.5) .. controls (267.28,101.25) and (260.24,109.15) .. (251.55,109.15) .. controls (242.86,109.15) and (235.82,101.25) .. (235.82,91.5) -- (251.55,91.5) -- cycle ; \draw [dash pattern={on 0.84pt off 2.51pt}] [dash pattern={on 0.84pt off 2.51pt}]  (267.28,91.5) .. controls (267.28,101.25) and (260.24,109.15) .. (251.55,109.15) .. controls (242.86,109.15) and (235.82,101.25) .. (235.82,91.5) ;  
\draw  [draw opacity=0][dash pattern={on 0.84pt off 2.51pt}] (284.95,91.99) .. controls (284.95,91.99) and (284.95,91.99) .. (284.95,91.99) .. controls (284.95,111.64) and (270,127.57) .. (251.55,127.57) .. controls (233.1,127.57) and (218.15,111.64) .. (218.15,91.99) -- (251.55,91.99) -- cycle ; \draw [dash pattern={on 0.84pt off 2.51pt}] [dash pattern={on 0.84pt off 2.51pt}]  (284.95,91.99) .. controls (284.95,111.64) and (270,127.57) .. (251.55,127.57) .. controls (233.1,127.57) and (218.15,111.64) .. (218.15,91.99) ;  
\draw  [draw opacity=0][dash pattern={on 0.84pt off 2.51pt}] (301.01,91.69) .. controls (301.01,91.69) and (301.01,91.69) .. (301.01,91.69) .. controls (301.01,120.41) and (278.86,143.7) .. (251.55,143.7) .. controls (224.24,143.7) and (202.1,120.41) .. (202.1,91.69) -- (251.55,91.69) -- cycle ; \draw [dash pattern={on 0.84pt off 2.51pt}] [dash pattern={on 0.84pt off 2.51pt}]  (301.01,91.69) .. controls (301.01,91.69) and (301.01,91.69) .. (301.01,91.69) .. controls (301.01,120.41) and (278.86,143.7) .. (251.55,143.7) .. controls (224.24,143.7) and (202.1,120.41) .. (202.1,91.69) ;  
\draw  [draw opacity=0][dash pattern={on 0.84pt off 2.51pt}] (318.35,91.36) .. controls (318.35,91.36) and (318.35,91.36) .. (318.35,91.36) .. controls (318.35,91.36) and (318.35,91.36) .. (318.35,91.36) .. controls (318.35,130.77) and (288.44,162.72) .. (251.55,162.72) .. controls (214.66,162.72) and (184.75,130.77) .. (184.75,91.36) -- (251.55,91.36) -- cycle ; \draw [dash pattern={on 0.84pt off 2.51pt}] [dash pattern={on 0.84pt off 2.51pt}]  (318.35,91.36) .. controls (318.35,91.36) and (318.35,91.36) .. (318.35,91.36) .. controls (318.35,91.36) and (318.35,91.36) .. (318.35,91.36) .. controls (318.35,130.77) and (288.44,162.72) .. (251.55,162.72) .. controls (214.66,162.72) and (184.75,130.77) .. (184.75,91.36) ;  
\draw [line width=1.5]    (117.95,276.84) -- (385.16,276.84)(117.95,279.84) -- (385.16,279.84) ;
\draw  [draw opacity=0][dash pattern={on 0.84pt off 2.51pt}] (200.15,164.81) .. controls (200.15,164.81) and (200.15,164.81) .. (200.15,164.81) .. controls (200.15,164.81) and (200.15,164.81) .. (200.15,164.81) .. controls (200.15,174.56) and (193.11,182.46) .. (184.42,182.46) .. controls (175.73,182.46) and (168.69,174.56) .. (168.69,164.81) -- (184.42,164.81) -- cycle ; \draw [dash pattern={on 0.84pt off 2.51pt}] [dash pattern={on 0.84pt off 2.51pt}]  (200.15,164.81) .. controls (200.15,164.81) and (200.15,164.81) .. (200.15,164.81) .. controls (200.15,174.56) and (193.11,182.46) .. (184.42,182.46) .. controls (175.73,182.46) and (168.69,174.56) .. (168.69,164.81) ;  
\draw  [draw opacity=0][dash pattern={on 0.84pt off 2.51pt}] (217.82,165.3) .. controls (217.82,165.3) and (217.82,165.3) .. (217.82,165.3) .. controls (217.82,165.3) and (217.82,165.3) .. (217.82,165.3) .. controls (217.82,184.95) and (202.87,200.88) .. (184.42,200.88) .. controls (165.97,200.88) and (151.02,184.95) .. (151.02,165.3) -- (184.42,165.3) -- cycle ; \draw [dash pattern={on 0.84pt off 2.51pt}] [dash pattern={on 0.84pt off 2.51pt}]  (217.82,165.3) .. controls (217.82,165.3) and (217.82,165.3) .. (217.82,165.3) .. controls (217.82,184.95) and (202.87,200.88) .. (184.42,200.88) .. controls (165.97,200.88) and (151.02,184.95) .. (151.02,165.3) ;  
\draw  [draw opacity=0][dash pattern={on 0.84pt off 2.51pt}] (233.87,165) .. controls (233.87,193.72) and (211.73,217.01) .. (184.42,217.01) .. controls (157.11,217.01) and (134.97,193.72) .. (134.97,165) -- (184.42,165) -- cycle ; \draw [dash pattern={on 0.84pt off 2.51pt}] [dash pattern={on 0.84pt off 2.51pt}]  (233.87,165) .. controls (233.87,193.72) and (211.73,217.01) .. (184.42,217.01) .. controls (157.11,217.01) and (134.97,193.72) .. (134.97,165) ;  
\draw  [draw opacity=0][dash pattern={on 0.84pt off 2.51pt}] (251.22,164.67) .. controls (251.22,164.67) and (251.22,164.67) .. (251.22,164.67) .. controls (251.22,164.67) and (251.22,164.67) .. (251.22,164.67) .. controls (251.22,204.08) and (221.31,236.03) .. (184.42,236.03) .. controls (147.53,236.03) and (117.62,204.08) .. (117.62,164.67) -- (184.42,164.67) -- cycle ; \draw [dash pattern={on 0.84pt off 2.51pt}] [dash pattern={on 0.84pt off 2.51pt}]  (251.22,164.67) .. controls (251.22,164.67) and (251.22,164.67) .. (251.22,164.67) .. controls (251.22,164.67) and (251.22,164.67) .. (251.22,164.67) .. controls (251.22,204.08) and (221.31,236.03) .. (184.42,236.03) .. controls (147.53,236.03) and (117.62,204.08) .. (117.62,164.67) ;  
\draw  [draw opacity=0][dash pattern={on 0.84pt off 2.51pt}] (267.96,164.75) .. controls (267.96,164.75) and (267.96,164.75) .. (267.96,164.75) .. controls (267.96,164.75) and (267.96,164.75) .. (267.96,164.75) .. controls (267.96,213.93) and (230.56,253.8) .. (184.42,253.8) .. controls (156.25,253.8) and (131.33,238.94) .. (116.2,216.16) -- (184.42,164.75) -- cycle ; \draw [dash pattern={on 0.84pt off 2.51pt}] [dash pattern={on 0.84pt off 2.51pt}]  (267.96,164.75) .. controls (267.96,164.75) and (267.96,164.75) .. (267.96,164.75) .. controls (267.96,213.93) and (230.56,253.8) .. (184.42,253.8) .. controls (156.25,253.8) and (131.33,238.94) .. (116.2,216.16) ;  
\draw  [draw opacity=0][dash pattern={on 0.84pt off 2.51pt}] (288.84,164.67) .. controls (288.84,226.27) and (242.09,276.21) .. (184.42,276.21) .. controls (157.64,276.21) and (133.21,265.44) .. (114.72,247.73) -- (184.42,164.67) -- cycle ; \draw [dash pattern={on 0.84pt off 2.51pt}] [dash pattern={on 0.84pt off 2.51pt}]  (288.84,164.67) .. controls (288.84,226.27) and (242.09,276.21) .. (184.42,276.21) .. controls (157.64,276.21) and (133.21,265.44) .. (114.72,247.73) ;  

\draw  [draw opacity=0][dash pattern={on 0.84pt off 2.51pt}] (334.42,164.81) .. controls (334.42,164.81) and (334.42,164.81) .. (334.42,164.81) .. controls (334.42,174.56) and (327.37,182.46) .. (318.68,182.46) .. controls (309.99,182.46) and (302.95,174.56) .. (302.95,164.81) -- (318.68,164.81) -- cycle ; \draw [dash pattern={on 0.84pt off 2.51pt}] [dash pattern={on 0.84pt off 2.51pt}]  (334.42,164.81) .. controls (334.42,174.56) and (327.37,182.46) .. (318.68,182.46) .. controls (309.99,182.46) and (302.95,174.56) .. (302.95,164.81) ;  
\draw  [draw opacity=0][dash pattern={on 0.84pt off 2.51pt}] (352.08,165.3) .. controls (352.08,184.95) and (337.13,200.88) .. (318.68,200.88) .. controls (300.23,200.88) and (285.28,184.95) .. (285.28,165.3) -- (318.68,165.3) -- cycle ; \draw [dash pattern={on 0.84pt off 2.51pt}] [dash pattern={on 0.84pt off 2.51pt}]  (352.08,165.3) .. controls (352.08,184.95) and (337.13,200.88) .. (318.68,200.88) .. controls (300.23,200.88) and (285.28,184.95) .. (285.28,165.3) ;  
\draw  [draw opacity=0][dash pattern={on 0.84pt off 2.51pt}] (368.14,165) .. controls (368.14,193.72) and (345.99,217.01) .. (318.68,217.01) .. controls (291.37,217.01) and (269.23,193.72) .. (269.23,165) -- (318.68,165) -- cycle ; \draw [dash pattern={on 0.84pt off 2.51pt}] [dash pattern={on 0.84pt off 2.51pt}]  (368.14,165) .. controls (368.14,193.72) and (345.99,217.01) .. (318.68,217.01) .. controls (291.37,217.01) and (269.23,193.72) .. (269.23,165) ;  
\draw  [draw opacity=0][dash pattern={on 0.84pt off 2.51pt}] (385.48,164.67) .. controls (385.48,164.67) and (385.48,164.67) .. (385.48,164.67) .. controls (385.48,164.67) and (385.48,164.67) .. (385.48,164.67) .. controls (385.48,204.08) and (355.58,236.03) .. (318.68,236.03) .. controls (281.79,236.03) and (251.88,204.08) .. (251.88,164.67) -- (318.68,164.67) -- cycle ; \draw [dash pattern={on 0.84pt off 2.51pt}] [dash pattern={on 0.84pt off 2.51pt}]  (385.48,164.67) .. controls (385.48,164.67) and (385.48,164.67) .. (385.48,164.67) .. controls (385.48,204.08) and (355.58,236.03) .. (318.68,236.03) .. controls (281.79,236.03) and (251.88,204.08) .. (251.88,164.67) ;  
\draw  [draw opacity=0][dash pattern={on 0.84pt off 2.51pt}] (386.9,216.52) .. controls (371.77,239.3) and (346.85,254.17) .. (318.68,254.17) .. controls (272.55,254.17) and (235.15,214.3) .. (235.15,165.12) -- (318.68,165.12) -- cycle ; \draw [dash pattern={on 0.84pt off 2.51pt}] [dash pattern={on 0.84pt off 2.51pt}]  (386.9,216.52) .. controls (371.77,239.3) and (346.85,254.17) .. (318.68,254.17) .. controls (272.55,254.17) and (235.15,214.3) .. (235.15,165.12) ;  
\draw  [draw opacity=0][dash pattern={on 0.84pt off 2.51pt}] (388.38,247.73) .. controls (369.89,265.44) and (345.47,276.21) .. (318.68,276.21) .. controls (261.01,276.21) and (214.26,226.27) .. (214.26,164.67) -- (318.68,164.67) -- cycle ; \draw [dash pattern={on 0.84pt off 2.51pt}] [dash pattern={on 0.84pt off 2.51pt}]  (388.38,247.73) .. controls (369.89,265.44) and (345.47,276.21) .. (318.68,276.21) .. controls (261.01,276.21) and (214.26,226.27) .. (214.26,164.67) ;  

\draw [line width=1.5]    (117.95,360.79) -- (385.16,360.79)(117.95,363.79) -- (385.16,363.79) ;
\draw   (132.5,361.17) .. controls (141.99,361.17) and (150.17,355.71) .. (158.72,349.97) .. controls (167.27,344.23) and (175.45,338.76) .. (184.94,338.76) .. controls (194.43,338.76) and (202.62,344.23) .. (211.17,349.97) .. controls (219.48,355.55) and (227.45,360.87) .. (236.6,361.16) ;
\draw   (266.76,361.17) .. controls (276.25,361.17) and (284.43,355.71) .. (292.98,349.97) .. controls (301.54,344.23) and (309.72,338.76) .. (319.21,338.76) .. controls (328.7,338.76) and (336.88,344.23) .. (345.43,349.97) .. controls (353.74,355.55) and (361.71,360.87) .. (370.86,361.16) ;
\draw [line width=1.5]    (117.62,443.63) -- (384.83,443.63)(117.62,446.63) -- (384.83,446.63) ;
\draw   (133.55,443.44) .. controls (135.92,443.44) and (137.97,434.02) .. (140.1,424.12) .. controls (142.24,414.21) and (144.29,404.79) .. (146.66,404.79) .. controls (149.03,404.79) and (151.08,414.21) .. (153.22,424.12) .. controls (155.35,434.02) and (157.4,443.44) .. (159.77,443.44) .. controls (162.14,443.44) and (164.19,434.02) .. (166.33,424.12) .. controls (168.46,414.21) and (170.51,404.79) .. (172.88,404.79) .. controls (175.25,404.79) and (177.3,414.21) .. (179.44,424.12) .. controls (181.58,434.02) and (183.62,443.44) .. (185.99,443.44) .. controls (188.37,443.44) and (190.41,434.02) .. (192.55,424.12) .. controls (194.69,414.21) and (196.73,404.79) .. (199.11,404.79) .. controls (201.48,404.79) and (203.52,414.21) .. (205.66,424.12) .. controls (207.8,434.02) and (209.84,443.44) .. (212.22,443.44) .. controls (214.59,443.44) and (216.63,434.02) .. (218.77,424.12) .. controls (220.91,414.21) and (222.96,404.79) .. (225.33,404.79) .. controls (227.7,404.79) and (229.75,414.21) .. (231.88,424.12) .. controls (234.02,434.02) and (236.07,443.44) .. (238.44,443.44) .. controls (240.81,443.44) and (242.86,434.02) .. (245,424.12) .. controls (247.13,414.21) and (249.18,404.79) .. (251.55,404.79) .. controls (253.92,404.79) and (255.97,414.21) .. (258.11,424.12) .. controls (260.24,434.02) and (262.29,443.44) .. (264.66,443.44) .. controls (267.03,443.44) and (269.08,434.02) .. (271.22,424.12) .. controls (273.36,414.21) and (275.4,404.79) .. (277.77,404.79) .. controls (280.15,404.79) and (282.19,414.21) .. (284.33,424.12) .. controls (286.47,434.02) and (288.51,443.44) .. (290.89,443.44) .. controls (293.26,443.44) and (295.3,434.02) .. (297.44,424.12) .. controls (299.58,414.21) and (301.62,404.79) .. (304,404.79) .. controls (306.37,404.79) and (308.41,414.21) .. (310.55,424.12) .. controls (312.69,434.02) and (314.74,443.44) .. (317.11,443.44) .. controls (319.48,443.44) and (321.53,434.02) .. (323.66,424.12) .. controls (325.8,414.21) and (327.85,404.79) .. (330.22,404.79) .. controls (332.59,404.79) and (334.64,414.21) .. (336.78,424.12) .. controls (338.91,434.02) and (340.96,443.44) .. (343.33,443.44) .. controls (345.7,443.44) and (347.75,434.02) .. (349.89,424.12) .. controls (352.02,414.21) and (354.07,404.79) .. (356.44,404.79) .. controls (358.81,404.79) and (360.86,414.21) .. (363,424.12) .. controls (365.14,434.02) and (367.18,443.44) .. (369.55,443.44) .. controls (369.73,443.44) and (369.9,443.39) .. (370.08,443.29) ;

\draw (174.92,156.4) node [anchor=north west][inner sep=0.75pt]  [font=\small]  {$F_{1}$};
\draw (230.72,64.9) node [anchor=north west][inner sep=0.75pt]  [font=\small]  {$\text{Fonte}$};
\draw (235.72,282.74) node [anchor=north west][inner sep=0.75pt]  [font=\small]  {$\text{Tela}$};
\draw (309.18,156.4) node [anchor=north west][inner sep=0.75pt]  [font=\small]  {$F_{2}$};
\draw (188.22,366.69) node [anchor=north west][inner sep=0.75pt]  [font=\small]  {$\text{Padr\~{a}o estat\'{i}stico 1}$};
\draw (188.22,449.53) node [anchor=north west][inner sep=0.75pt]  [font=\small]  {$\text{Padr\~{a}o estat\'{i}stico 2}$};

\end{tikzpicture}
\caption{Esquema gr\'{a}fico do experimento da fenda dupla.}
    \label{fig:DF}
\end{figure}

De maneira bastante simplificada, o padr\~{a}o estat\'{i}stico 2 emerge quando a trajet\'{o}ria dos el\'{e}trons est\'{a} sendo descrita pelo processo 2 (\textit{e.g.}, pela Equa\c{c}\~{a}o de Schr\"{o}dinger\index{Schr\"{o}dinger, Erwin!Equa\c{c}\~{a}o de Schr\"{o}dinger}\index{Schr\"{o}dinger, Erwin}). \'{E} um padr\~{a}o de interferência que resulta da superposi\c{c}\~{a}o entre o el\'{e}tron ter passado por $F_1$ e o el\'{e}tron ter passado por $F_2$. Sempre que uma medi\c{c}\~{a}o \'{e} efetuada, no entanto, (\textit{i.e.}, sempre que o processo 1 passa a descrever o experimento) o padr\~{a}o estat\'{i}stico 2 \'{e} modificado para o padr\~{a}o estat\'{i}stico 1. Ele representa que metade dos el\'{e}trons s\~{a}o detectados passando por $F_1$ e metade por $F_2$. Explicar tal mudan\c{c}a \'{e} parte do que est\'{a} em jogo com o problema da medi\c{c}\~{a}o. Para \citet{pessoajr2022}, o problema da medi\c{c}\~{a}o \textit{\'{e}} o problema do colapso. Vejamos como podemos entender essa afirma\c{c}\~{a}o em um cen\'{a}rio mais amplo.

Em uma taxonomia amplamente difundida, \citet{Maudlin1995measurementproblem} define o problema da medi\c{c}\~{a}o como a conjun\c{c}\~{a}o problem\'{a}tica entre as seguintes suposi\c{c}\~{o}es sobre a descri\c{c}\~{a}o que a mec\^{a}nica qu\^{a}ntica d\'{a} aos sistemas f\'{i}sicos:

\begin{itemize}
    \item[$\alpha$)]\label{trilema:A} \'{E} uma descri\c{c}\~{a}o \textit{completa}. Isto \'{e}, a suposi\c{c}\~{a}o de que a mec\^{a}nica qu\^{a}ntica descreve todos os aspectos f\'{i}sicos do sistema f\'{i}sico em quest\~{a}o.
    \item[$\beta$)]\label{trilema:B} \'{E} uma descri\c{c}\~{a}o \textit{linear}. Essa suposi\c{c}\~{a}o afirma que a descri\c{c}\~{a}o qu\^{a}ntica dos sistemas f\'{i}sicos deve ocorrer exclusivamente por processos lineares.
    \item[$\gamma$)]\label{trilema:C} \'{E} uma descri\c{c}\~{a}o que fornece resultados \textit{\'{u}nicos}.
\end{itemize}

Sem entrar em detalhes acerca de quest\~{o}es da matem\'{a}tica subjacente \`{a} discuss\~{a}o das interpreta\c{c}\~{o}es da mec\^{a}nica qu\^{a}ntica, pode-se entender a raz\~{a}o pela qual a conjun\c{c}\~{a}o entre $\alpha$, $\beta$ e $\gamma$ \'{e} problem\'{a}tica com o seguinte racioc\'{i}nio. Suponha que $|\psi\rangle$ \'{e} uma descri\c{c}\~{a}o qu\^{a}ntica do sistema qu\^{a}ntico $S$, que pode ter os valores $S_0$ ou $S_1$. Se assumirmos a premissa $\alpha$, ent\~{a}o a descri\c{c}\~{a}o de $S$ por $|\psi\rangle$ \'{e} completa, isto \'{e}, n\~{a}o h\'{a} nada a se dizer de $S$, em termos f\'{i}sicos, al\'{e}m daquilo que \'{e} dito por $|\psi\rangle$. Como uma caracter\'{i}stica da descri\c{c}\~{a}o linear \'{e} a admiss\~{a}o de uma soma de resultados como um resultado, ao assumirmos $\beta$ tem-se que $S_0+S_1$ \'{e} uma descri\c{c}\~{a}o poss\'{i}vel de $S$ em termos de $|\psi\rangle$. No entanto $\gamma$ pede que tenha-se, exclusivamente, $S_0$ \textit{ou} $S_1$ como resultado de $S$ pela descri\c{c}\~{a}o $|\psi\rangle$.

Assim, ao menos uma das três suposi\c{c}\~{o}es acima deve ser negada. As interpreta\c{c}\~{o}es da mec\^{a}nica qu\^{a}ntica dividem-se, em \textit{qual} dessas suposi\c{c}\~{o}es \'{e} negada. As interpreta\c{c}\~{o}es do primeiro grupo s\~{a}o as que negam $\alpha$. S\~{a}o as interpreta\c{c}\~{o}es que introduzem \textit{vari\'{a}veis ocultas} no formalismo da medi\c{c}\~{a}o. Num segundo grupo, estariam as interpreta\c{c}\~{o}es que negam $\beta$ e introduzem outras leis din\^{a}micas para a mec\^{a}nica qu\^{a}ntica, como o colapso. Por fim, no terceiro grupo est\~{a}o as que negam $\gamma$, e introduzem o conceito de ``ramifica\c{c}\~{a}o''. Essa \'{e}, de modo bastante geral, uma breve taxonomia das interpreta\c{c}\~{o}es da mec\^{a}nica qu\^{a}ntica.\footnote{~Isso n\~{a}o quer dizer que tal taxonomia n\~{a}o apresente problemas. \citet{muller2023}, por exemplo, oferece um exame cr\'{i}tico de tal taxonomia (por \textit{trilema}), bem como uma taxonomia mais detalhada (por \textit{polilema}) dos problemas da medi\c{c}\~{a}o. Aqui atenho-me ao chamado ``problema da realidade dos resultados'', sendo o primeiro dos seis problemas da medi\c{c}\~{a}o elencados por \citet{muller2023}.}

A interpreta\c{c}\~{a}o de von Neumann\index{Neumann, John von} est\'{a} dentre as interpreta\c{c}\~{o}es do segundo grupo, que negam $\beta$. Para adequar a discuss\~{a}o que se segue a essa taxonomia, farei a seguinte escolha terminol\'{o}gica. Aquilo que von Neumann\index{Neumann, John von} chamou de ``processo 2'' ser\'{a} chamado daqui pra frente de ``evolu\c{c}\~{a}o linear'', e aquilo que ele chamou de ``processo 1'' ser\'{a} chamado, daqui adiante, de ``colapso'' (tamb\'{e}m referido, em algumas cita\c{c}\~{o}es, como ``redu\c{c}\~{a}o''). O colapso \'{e} uma lei din\^{a}mica n\~{a}o-linear, associada \`{a} evolu\c{c}\~{a}o linear em processos de medi\c{c}\~{a}o. O colapso, assim, \'{e} uma solu\c{c}\~{a}o ao problema da medi\c{c}\~{a}o ---solu\c{c}\~{a}o essa que, como veremos, introduz \textit{diversos} outros problemas. As interpreta\c{c}\~{o}es do primeiro e terceiro grupo ser\~{a}o consideradas brevemente no cap\'{i}tulo \ref{CapPaisagem}.

\section{A interpreta\c{c}\~{a}o da consciência causal\index{interpreta\c{c}\~{a}o da consciência!causal}}

Antes de adentrar nas especificidades dessa particular interpreta\c{c}\~{a}o da mec\^{a}nica qu\^{a}ntica, devo tecer alguns breves coment\'{a}rios de natureza sociol\'{o}gica. \'{E} not\'{a}vel que têm sido feitas muitas apropria\c{c}\~{o}es indevidas, que deturpam os assuntos que envolvem a mec\^{a}nica qu\^{a}ntica. Isso foi tratado com maestria nos trabalhos de \citet{osvaldo2011misticismoquantico}, \citet{fred2011coachquantico} e \citet{sandrofred2016misticismo}. No entanto, como mostram \citet{BarOas2016Consc}, a interpreta\c{c}\~{a}o da consciência causal\index{interpreta\c{c}\~{a}o da consciência!causal} n\~{a}o foi at\'{e} o presente falseada experimentalmente; e, mais ainda, conforme argumentam \citet{RaoJonas2019dualismQM}, n\~{a}o existem boas raz\~{o}es filos\'{o}ficas para que tal interpreta\c{c}\~{a}o seja descartada do rol de interpreta\c{c}\~{o}es poss\'{i}veis para a mec\^{a}nica qu\^{a}ntica. Tratarei dessa interpreta\c{c}\~{a}o especificamente para esclarecer quais s\~{a}o os usos leg\'{i}timos da consciência na mec\^{a}nica qu\^{a}ntica, e dimension\'{a}-la como \textit{mais uma} interpreta\c{c}\~{a}o ---e n\~{a}o \textit{``A''} interpreta\c{c}\~{a}o da mec\^{a}nica qu\^{a}ntica, como encontra-se em literaturas menos respons\'{a}veis sobre o assunto.\footnote{~Ver, por exemplo, \citet{goswami1993selfaware}.}

A mec\^{a}nica qu\^{a}ntica considera a uni\~{a}o $\langle \text{objeto} + \text{aparato}\rangle$ um \'{u}nico sistema, chamado sistema composto. No racioc\'{i}nio de \citet{vNeum1955mathematical}\index{Neumann, John von}, o sistema composto obtido por $\mathcal{I}_1$ n\~{a}o seria suficiente para completar uma medi\c{c}\~{a}o. Se todos os objetos materiais (microsc\'{o}picos ou macrosc\'{o}picos) s\~{a}o constitu\'{i}dos por objetos qu\^{a}nticos, ent\~{a}o a intera\c{c}\~{a}o entre um objeto qu\^{a}ntico (a ser medido) e um aparelho de amplifica\c{c}\~{a}o (a supostamente medir) n\~{a}o completaria uma medi\c{c}\~{a}o, mas ficaria atrelada \`{a} evolu\c{c}\~{a}o linear.

Poder-se-ia sugerir que ao aparato $\mathcal{M}$ fosse acoplado um segundo aparato de medi\c{c}\~{a}o $\mathcal{M}^{\prime}$, na inten\c{c}\~{a}o de completar uma medi\c{c}\~{a}o no sistema composto. Essa proposta, no entanto, levaria a uma regress\~{a}o infinita de aparatos medidores na medida em que $\mathcal{M}^{\prime}$ se relacionaria com $\mathcal{M}$ da mesma maneira que $\mathcal{M}$ se relaciona com $\mathcal{X}$ no caso do sistema composto $\langle \text{objeto} + \text{aparato}\rangle$, isto \'{e}, n\~{a}o conseguiria completar uma medi\c{c}\~{a}o.

Esse aspecto problem\'{a}tico foi nomeado por \citet[p.~167]{desp1999concep} de ``cadeia de von Neumann\index{Neumann, John von}''. \'{E} preciso salientar que tal regress\~{a}o infinita \'{e} uma dificuldade filos\'{o}fica bastante s\'{e}ria para uma teoria, sendo um dos c\'{e}lebres paradoxos cl\'{a}ssicos, conhecido atrav\'{e}s do termo em latim ``\textit{reductio ad infinitum}''. Assim, o ato da medi\c{c}\~{a}o deve ser uma opera\c{c}\~{a}o finita, o que seria poss\'{i}vel, ao que parece, somente por um ato de medi\c{c}\~{a}o, em $\mathcal{M}$, em ``\textelp{} um ato descont\'{i}nuo, n\~{a}o causal e instant\^{a}neo'', isto \'{e}, correspondente ao colapso. A quest\~{a}o ontol\'{o}gica (\index{ontologia!naturalizada}$\mathscr{O}_N$) dessa discuss\~{a}o reside justamente nas respostas para a quest\~{a}o sobre onde e como o referido ``ato'' do colapso acontece: \citet[p.~418--420]{vNeum1955mathematical}\index{Neumann, John von} afirma, em um longo par\'{a}grafo (que reproduzirei integralmente), que o ato da medi\c{c}\~{a}o seria causado pela percep\c{c}\~{a}o do observador:

\begin{quote}
Primeiro, \'{e} inerentemente e totalmente correto que a medi\c{c}\~{a}o ou o processo relacionado \`{a} percep\c{c}\~{a}o subjetiva \'{e} uma nova entidade em rela\c{c}\~{a}o ao ambiente f\'{i}sico e n\~{a}o \'{e} redut\'{i}vel a ele ---de fato, a percep\c{c}\~{a}o subjetiva nos leva para a vida intelectual interior do indiv\'{i}duo, que \'{e} extra observ\'{a}vel por sua pr\'{o}pria natureza (j\'{a} que deve ser assumida por qualquer observa\c{c}\~{a}o ou experimento conceb\'{i}vel). (Veja a discuss\~{a}o nos par\'{a}grafos acima). No entanto, \'{e} uma exigência fundamental do ponto de vista cient\'{i}fico ---o chamado ``princ\'{i}pio do paralelismo psico-f\'{i}sico''--- que deva ser poss\'{i}vel descrever o processo extra f\'{i}sico da percep\c{c}\~{a}o subjetiva como se ele fosse pertencente, na realidade, ao mundo f\'{i}sico ---isto \'{e}, atribuir \`{a}s suas partes processos f\'{i}sicos equivalentes no ambiente objetivo, no espa\c{c}o comum. (\'{E} claro que nesse processo relacionando surge a frequente necessidade de localizar alguns desses processos em pontos situados dentro da por\c{c}\~{a}o do espa\c{c}o ocupada pelos nossos pr\'{o}prios corpos. Mas isso n\~{a}o altera o fato de que eles perten\c{c}am ao ``mundo sobre n\'{o}s'', o ambiente objetivo referido anteriormente.) Num exemplo simples, estes conceitos podem ser aplicados do seguinte modo: desejamos medir uma temperatura. Se quisermos, podemos prosseguir com esse processo numericamente at\'{e} que tenhamos a temperatura do ambiente do recipiente de merc\'{u}rio atrav\'{e}s do termômetro, e ent\~{a}o dizer: essa temperatura foi medida pelo termômetro. Mas podemos levar o c\'{a}lculo adiante e, a partir das propriedades do merc\'{u}rio, que podem ser explicadas em termos cin\'{e}ticos e moleculares, podemos calcular seu aquecimento, expans\~{a}o, e o comprimento resultante da coluna de merc\'{u}rio, e em seguida dizer: esse \'{e} o comprimento visto pelo observador. Indo ainda mais longe, e levando a fonte de luz em considera\c{c}\~{a}o, n\'{o}s poder\'{i}amos encontrar o reflexo do quanta de luz sobre a coluna opaca de merc\'{u}rio, e o caminho do quanta de luz remanescente at\'{e} o olho do observador, sua refrac\c{c}\~{a}o na lente do olho, e a forma\c{c}\~{a}o uma imagem sobre a retina, e em seguida n\'{o}s dir\'{i}amos: essa imagem \'{e} registada pela retina do observador. E se o nosso conhecimento fisiol\'{o}gico fosse mais preciso do que \'{e} hoje, poder\'{i}amos ir ainda mais longe, tra\c{c}ando as rea\c{c}\~{o}es qu\'{i}micas que produzem a impress\~{a}o dessa imagem na retina, no nervo \'{o}tico e no c\'{e}rebro, e ent\~{a}o, no final, dizer: essas mudan\c{c}as qu\'{i}micas de suas c\'{e}lulas cerebrais s\~{a}o percebidas pelo observador. Mas em qualquer caso, n\~{a}o importa o qu\~{a}o longe calcularmos ---do recipiente de merc\'{u}rio, com a escala do termômetro, para a retina, ou no c\'{e}rebro--- em algum momento devemos dizer: ``e isso \'{e} percebido pelo observador''. Ou seja, devemos sempre dividir o mundo em duas partes, uma sendo o sistema observado e a outra sendo o observador. No primeiro caso, podemos acompanhar todos os processos f\'{i}sicos (pelo menos a princ\'{i}pio) com uma precis\~{a}o arbitrariamente grande. No \'{u}ltimo caso, isso \'{e} insignificante. A fronteira entre os dois \'{e} bastante arbitr\'{a}ria. Em particular, vimos nas quatro possibilidades diferentes do exemplo acima que o observador, nesse sentido, n\~{a}o deve ser identificado com o corpo do observador real: num dos casos do exemplo acima, inclu\'{i}mos at\'{e} mesmo o termômetro, enquanto em outro exemplo, at\'{e} mesmo os olhos e as vias do nervo \'{o}ptico n\~{a}o foram inclu\'{i}dos. Levar esse limite profundamente de forma arbitr\'{a}ria para o interior do corpo do observador \'{e} o teor real do princ\'{i}pio do paralelismo psico-f\'{i}sico ---mas isso n\~{a}o altera o fato de que em cada m\'{e}todo da descri\c{c}\~{a}o a fronteira deva ser posta em algum lugar, se n\~{a}o for para o m\'{e}todo continuar vagamente, isto \'{e}, se uma compara\c{c}\~{a}o com a experiência deve ser poss\'{i}vel. De fato a experiência s\'{o} faz declara\c{c}\~{o}es deste tipo: um observador realizou certa observa\c{c}\~{a}o (subjetiva); e nunca alguma como esta: uma grandeza f\'{i}sica tem um determinado valor. \citep[p.~418--420]{vNeum1955mathematical}\index{Neumann, John von}.
\end{quote}

Embora von Neumann\index{Neumann, John von} n\~{a}o tenha mencionado a palavra consciência, parece ser un\^{a}nime, dentre as diversas leituras dessa famosa passagem, que \citet[p.~420]{vNeum1955mathematical}\index{Neumann, John von} se refere \`{a} ``consciência do observador'' quando enuncia o poder causal da ``percep\c{c}\~{a}o subjetiva do observador''. Em outra passagem, \citet[p.~421]{vNeum1955mathematical}\index{Neumann, John von} enuncia o observador como um ``\textit{ego} abstrato'', isto \'{e}, um ``eu'', uma subjetividade abstrata. Assim, para \citet[p.~418--421]{vNeum1955mathematical}\index{Neumann, John von}, somente algo fora do sistema composto por $\mathcal{X} \wedge \mathcal{M}$ ---tal como a consciência do observador $\mathcal{O}$--- poderia dar cabo \`{a} tal cadeia infinita, reintroduzindo a intera\c{c}\~{a}o psicof\'{i}sica $\mathcal{I}_2$ na teoria da medi\c{c}\~{a}o.

A principal motiva\c{c}\~{a}o hist\'{o}rica para essa interpreta\c{c}\~{a}o, de acordo com \citet[p.~480]{Jammer1974}, seria uma s\'{e}rie de longas conversas que \citet[nota~218]{vNeum1955mathematical}\index{Neumann, John von} mantinha com Le\'{o} Szil\'{a}rd, que teria publicado um estudo influente sobre a interven\c{c}\~{a}o de um ser inteligente em um sistema termodin\^{a}mico. O estudo de \citet{szilard1929entropy}, para Jammer,

\begin{quote}
\textelp{} marcou o in\'{i}cio de especula\c{c}\~{o}es instigantes sobre o efeito de uma interven\c{c}\~{a}o f\'{i}sica da mente sobre a mat\'{e}ria e, assim, abriu o caminho para a afirma\c{c}\~{a}o de longo alcance de von Neumann\index{Neumann, John von}, sobre a impossibilidade de formular uma teoria completa e consistente de medi\c{c}\~{a}o mec\^{a}nica qu\^{a}ntica sem referência \`{a} consciência humana. \citep[p.~480]{Jammer1974}.
\end{quote}

A fim de discutir tal situa\c{c}\~{a}o, \citet[p.~421]{vNeum1955mathematical}\index{Neumann, John von} divide o universo de discurso em 3 partes correspondentes \`{a} nota\c{c}\~{a}o I, II e III, de modo que ``I'' corresponde ao objeto (ou sistema) a ser observado, ``II'' corresponde ao instrumento de medi\c{c}\~{a}o e ``III'' ao observador, isto \'{e}, seu ego abstrato. Em todos os casos, o resultado da medi\c{c}\~{a}o em I efetuada por II+III \'{e} o mesmo do que a medi\c{c}\~{a}o em I+II efetuada por III. No primeiro caso, a evolu\c{c}\~{a}o linear se aplica a I e, no segundo caso, a I+II. Em todos os casos, a evolu\c{c}\~{a}o linear n\~{a}o se aplica a III, isto \'{e}, III \'{e} a \'{u}nica parte para qual o colapso se aplica em todos os casos.\footnote{~Ver tamb\'{e}m \citet[p.~78]{breuer2001vneumannincomplet}.}

Utilizarei o famoso experimento mental do gato\index{Schr\"{o}dinger, Erwin!paradoxo do gato} de \citet[p.~157]{schrodinger1935cat}\index{Schr\"{o}dinger, Erwin} para ilustrar tal problem\'{a}tica, uma vez que se trata de uma situa\c{c}\~{a}o idealizada poucos anos mais tarde da publica\c{c}\~{a}o de \citet{vNeum1955mathematical}\index{Neumann, John von}, para explicitar a dificuldade do ``problema da medi\c{c}\~{a}o'' na mec\^{a}nica qu\^{a}ntica. O experimento mental do gato\index{Schr\"{o}dinger, Erwin!paradoxo do gato} de \citet{schrodinger1935cat}\index{Schr\"{o}dinger, Erwin} seria uma extrapola\c{c}\~{a}o da descri\c{c}\~{a}o qu\^{a}ntica da realidade. Antes de introduzir o ---agora famoso--- experimento de pensamento, \citet[p.~157]{schrodinger1935cat}\index{Schr\"{o}dinger, Erwin!paradoxo do gato} precede do seguinte aviso: ``pode-se at\'{e} mesmo imaginar casos rid\'{i}culos'' (\textit{sic}). Porque o experimento \'{e} demasiadamente gr\'{a}fico com felinos, precederei de uma analogia mais ``humanizada'', como recomendada por \citet{putnam2012}\footnote{~\citet[183]{putnam2012} afirma que essa \'{e} uma ideia retirada de John Bell (``\textit{Sally's cat}''), mas confesso que falhei em encontrar a referência para tal.} do experimento de pensamento do gato de Schr\"{o}dinger. Agora, o experimento envolve a possibilidade do felino em jejum ser alimentado ou mantido em jejum ---a depender de uma cadeia de eventos que inicia-se com o decaimento de um objeto qu\^{a}ntico. Nas palavras (modificadas) dele:

\begin{quote}
Um gato\index{Schr\"{o}dinger, Erwin!paradoxo do gato} [em jejum] \'{e} posicionado em uma c\^{a}mara de a\c{c}o, juntamente com o seguinte dispositivo diab\'{o}lico (que deve ser resguardado contra a interferência direta do gato\index{Schr\"{o}dinger, Erwin!paradoxo do gato}): um contador Geiger [detector de radia\c{c}\~{a}o] com um pouco de subst\^{a}ncia radioativa, \textit{t\~{a}o} pouco que, \textit{talvez} no curso de uma hora, um dos \'{a}tomos decai ---mas tamb\'{e}m, com igual probabilidade, talvez nenhuma; se isso acontece, o detector \'{e} acionado e, atrav\'{e}s de um dispositivo el\'{e}trico, [alimenta o gato]. Se o sistema for deixado a si mesmo por uma hora, poder-se-ia dizer que o gato\index{Schr\"{o}dinger, Erwin!paradoxo do gato} [permanece em jejum] \textit{se} enquanto isso nenhum \'{a}tomo decaiu. O primeiro decaimento atômico o teria [alimentado]. A fun\c{c}\~{a}o de onda de todo o sistema poderia expressar isso por ter nela o gato\index{Schr\"{o}dinger, Erwin!paradoxo do gato} [em jejum] e o gato\index{Schr\"{o}dinger, Erwin!paradoxo do gato} [alimentado] (desculpe a express\~{a}o) misturado ou espalhado em partes iguais. \citep[p.~157, ênfase original]{schrodinger1935cat}\index{Schr\"{o}dinger, Erwin}.
\end{quote}

O n\'{u}cleo do argumento est\'{a} contido na ideia de que, at\'{e} que uma observa\c{c}\~{a}o direta seja efetuada sobre o sistema em quest\~{a}o (o que corresponde, nessa interpreta\c{c}\~{a}o, ao colapso), a descri\c{c}\~{a}o do formalismo qu\^{a}ntico n\~{a}o forneceria nada al\'{e}m de \textit{possibilidades}, com igual probabilidade, de dois estados atuais que s\~{a}o \textit{contr\'{a}rios}. Na literatura tradicional, esse racioc\'{i}nio \'{e} frequentemente expresso por meio da senten\c{c}a ``estados contradit\'{o}rios'', no que se refere ao estado de superposi\c{c}\~{a}o entre os estados ``em jejum'' e ``alimentado''. No entanto, o correto seria utilizar a senten\c{c}a ``estados contr\'{a}rios'', tendo em vista a defini\c{c}\~{a}o de tais termos no cl\'{a}ssico quadrado de oposi\c{c}\~{o}es, em que uma situa\c{c}\~{a}o de contraditoriedade se estabelece quando duas proposi\c{c}\~{o}es n\~{a}o podem ser simultaneamente verdadeiras nem simultaneamente falsas, e uma situa\c{c}\~{a}o de contrariedade se estabelece quando duas proposi\c{c}\~{o}es n\~{a}o podem ser simultaneamente verdadeiras, mas podem ser simultaneamente falsas. Krause prop\~{o}e que a superposi\c{c}\~{a}o seja entendida como um terceiro estado, um estado ``novo'':

\begin{quote}
\textelp{} em certas ``situa\c{c}\~{o}es qu\^{a}nticas'', nomeadamente nas de superposi\c{c}\~{a}o, n\~{a}o podemos de modo algum dizer ---como parece f\'{a}cil de fazer a partir de uma vis\~{a}o ``cl\'{a}ssica''--- que dois objetos qu\^{a}nticos, como dois el\'{e}trons, quando em superposi\c{c}\~{a}o de dois estados $\psi_1$ e $\psi_2$ (ou seja, quando s\~{a}o descritos por uma fun\c{c}\~{a}o de onda $\psi_{12}=\psi_1+\psi_2$) est\~{a}o em um dos dois estados. Nem no outro, nem em ambos, nem em nenhum ---que seriam as quatro situa\c{c}\~{o}es logicamente poss\'{i}veis (de um ponto de vista ``cl\'{a}ssico''---, mas podemos dizer que est\~{a}o em um ``novo'' estado, o de superposi\c{c}\~{a}o de $\psi_1$ e $\psi_2$. \citep[p.~128]{krause2010indiscernibles}.
\end{quote}

No caso do exemplo do gato\index{Schr\"{o}dinger, Erwin!paradoxo do gato} de \citet[p.~157]{schrodinger1935cat}\index{Schr\"{o}dinger, Erwin}, tem-se três estados: o estado ``em jejum'', o estado ``alimentado'' e o estado ``superposto''. No \'{u}ltimo, as proposi\c{c}\~{o}es ``o gato\index{Schr\"{o}dinger, Erwin!paradoxo do gato} est\'{a} em jejum'' e ``o gato\index{Schr\"{o}dinger, Erwin!paradoxo do gato} est\'{a} alimentado'' s\~{a}o simultaneamente falsas, o que parece configurar uma rela\c{c}\~{a}o de contrariedade e n\~{a}o de contraditoriedade. Essa forma de interpretar o estado de superposi\c{c}\~{a}o se coaduna com o fato de que os vetores matem\'{a}ticos que representam os estados ``em jejum'' e ``alimentado'' s\~{a}o ortogonais, e n\~{a}o a nega\c{c}\~{a}o um do outro.\footnote{~Para uma discuss\~{a}o aprofundada e atualizada sobre o assunto, ver tamb\'{e}m \citet{arekrause2016contradictionQM}.} Na interpreta\c{c}\~{a}o de \citet{vNeum1955mathematical}\index{Neumann, John von}, tal quadro se traduziria na afirma\c{c}\~{a}o de que nenhum evento atual ocorreria at\'{e} que o sistema composto ---isto \'{e}, o sistema qu\^{a}ntico e o aparelho de medi\c{c}\~{a}o--- seja percebido pelo ego abstrato do observador. Esse ponto de vista, como vimos no cap\'{i}tulo \ref{CapEPR}, era inaceit\'{a}vel para Einstein\index{Einstein, Albert}. Em uma carta endere\c{c}ada a Schr\"{o}dinger\index{Schr\"{o}dinger, Erwin} em 1950, Einstein reitera tal insatisfa\c{c}\~{a}o (\textit{viz.}, que o resultado \textit{passa a existir} no momento da medi\c{c}\~{a}o, algo que vimos nos cap\'{i}tulos anteriores como ``\textit{medi\c{c}\~{a}o=cria\c{c}\~{a}o}''):

\begin{quote}
    Você \'{e} o \'{u}nico f\'{i}sico contempor\^{a}neo, al\'{e}m de Laue, que percebe que n\~{a}o se pode contornar a suposi\c{c}\~{a}o da realidade ---desde que sejamos honestos. A maioria deles simplesmente n\~{a}o percebe que tipo de jogo arriscado est\~{a}o fazendo com a realidade ---a realidade \'{e} algo independente do que \'{e} estabelecido experimentalmente. De alguma forma, eles acreditam que a teoria qu\^{a}ntica fornece uma descri\c{c}\~{a}o da realidade, e at\'{e} mesmo uma descri\c{c}\~{a}o \textit{completa}; no entanto, essa interpreta\c{c}\~{a}o \'{e} refutada de maneira mais elegante pelo seu sistema de $\text{\'{a}tomo radioativo} + \text{contador Geiger} + \text{amplificador} + \text{[alimentador el\'{e}trico]} + \text{gato em uma caixa}$\index{Schr\"{o}dinger, Erwin!paradoxo do gato}, no qual o estado do sistema cont\'{e}m o gato tanto [em jejum] quanto [alimentado]. O estado do gato deve ser criado apenas quando um f\'{i}sico investiga a situa\c{c}\~{a}o em algum momento definido? Ningu\'{e}m realmente duvida de que a presen\c{c}a ou ausência do gato seja algo independente do ato de observa\c{c}\~{a}o. Mas ent\~{a}o a descri\c{c}\~{a}o por meio do estado qu\^{a}ntico \'{e} certamente incompleta, e deve haver uma descri\c{c}\~{a}o mais completa. \citep[p.~39, ênfase original]{einstein190letter-schrod}.
\end{quote}

\index{metalinguagem e metametalinguagem}Pelo que foi considerado at\'{e} aqui, existem ao menos duas leituras poss\'{i}veis da teoria da medi\c{c}\~{a}o de \citet{vNeum1955mathematical}\index{Neumann, John von}, sendo uma ontol\'{o}gica e outra puramente l\'{o}gica. Considerando a an\'{a}lise l\'{o}gica, fa\c{c}o referência ao estudo de \citet[p.~80--81]{breuer2001vneumannincomplet}, que faz uma aproxima\c{c}\~{a}o entre a hierarquia infinita dos tipos l\'{o}gicos, da linguagem-objeto e das infinitas metalinguagens subjacentes (isto \'{e}, a metametalinguagem, a metametametalinguagem, etc.) de \citet[p.~241--265]{tarski1956concept} e a cadeia infinita de observa\c{c}\~{o}es de \citet{vNeum1955mathematical}\index{Neumann, John von}. \index{G\"{o}del, Kurt!incompletude de G\"{o}del|seealso{metalinguagem e metametalinguagem}}Para \citet[p.~80]{breuer2001vneumannincomplet}, tais hierarquias infinitas est\~{a}o intimamente ligadas com o racioc\'{i}nio da incompletude de \citet[p.~610, nota~48]{godel1967undecidable}\index{G\"{o}del, Kurt}, o qual admite textualmente que ``\textelp{} a verdadeira raz\~{a}o para a incompletude \'{e} que a forma\c{c}\~{a}o de tipos cada vez mais elevados pode ser continuado transfinitamente''.

Na teoria da verdade de \citet{tarski1956concept}, uma predica\c{c}\~{a}o da no\c{c}\~{a}o de verdade aplic\'{a}vel a todas as senten\c{c}as da linguagem-objeto n\~{a}o \'{e} parte da linguagem-objeto, mas de um tipo l\'{o}gico de hierarquia mais alta, isto \'{e}, uma metalinguagem. Se o termo ``verdade'' for intercambiado por ``demonstrabilidade'', o racioc\'{i}nio da incompletude de \citet[p.~592--616]{godel1967undecidable}\index{G\"{o}del, Kurt} poderia ser parafraseado, segundo \citet[p.~80]{breuer2001vneumannincomplet}, da seguinte maneira: ``um conceito de demonstrabilidade que \'{e} formulado dentro de um sistema formal n\~{a}o pode ser aplicado a todas as senten\c{c}as desse mesmo sistema''.

Voltando ao racioc\'{i}nio da hierarquia infinita na teoria da medi\c{c}\~{a}o de \citet{vNeum1955mathematical}\index{Neumann, John von}, uma medi\c{c}\~{a}o n\~{a}o est\'{a} completa no sistema $\mathcal{X} \wedge \mathcal{M}$, $\mathcal{X} \wedge \mathcal{M} \wedge \mathcal{M}^{\prime})$, ou $(\mathcal{X} \wedge \mathcal{M} \wedge \mathcal{M}^{\prime} \wedge \mathcal{M}^{\prime\prime})$ etc. at\'{e} que o colapso ocorra, o que somente aconteceria pela a\c{c}\~{a}o de um agente fora do sistema, ou seja, externo. Nesse preciso sentido, a fun\c{c}\~{a}o de tal observador $O$ externo pode ser aproximada a um funcionamento metate\'{o}rico, isto \'{e}, a um n\'{i}vel l\'{o}gico mais alto (um \textit{meta}-n\'{i}vel). Para \citet[p.~81]{breuer2001vneumannincomplet}, a aproxima\c{c}\~{a}o feita entre a concep\c{c}\~{a}o de ``obter uma prova de uma afirma\c{c}\~{a}o'' e concep\c{c}\~{a}o de ``obter o resultado de uma medi\c{c}\~{a}o'' seria v\'{a}lida na medida em que ```medi\c{c}\~{a}o' e `prova' s\~{a}o ambos conceitos sem\^{a}nticos que estabelecem uma rela\c{c}\~{a}o entre um formalismo f\'{i}sico ou matem\'{a}tico, e aquilo ao qual o formalismo refere''.

Pela senten\c{c}a com um valor de verdade tal como ``completar uma medi\c{c}\~{a}o'', refiro-me a um evento, cuja probabilidade  ``$P$'' de resultado ``$R$'' seja, exclusivamente, ao menos um dos dois resultados poss\'{i}veis, ``$s$'' e ``$s'$'', em que a probabilidade dos dois resultados poss\'{i}veis seja equivalente, de modo que $R(s) = R(s')$. O colapso indica que o estado de $R$ \'{e} (por exemplo) $s'$ (e, consequentemente, n\~{a}o-$s$). Nesse preciso sentido, o observador deve estar fora dos limites da f\'{i}sica. Dito de outro modo, da mesma forma que para Bohr, para von Neumann\index{Neumann, John von} o agente causal da medi\c{c}\~{a}o, isto \'{e}, aquilo que completa uma medi\c{c}\~{a}o est\'{a} para al\'{e}m dos limites da f\'{i}sica qu\^{a}ntica:

\begin{quote}
\textelp{} \'{e} inerentemente inteiramente correto que a medi\c{c}\~{a}o ou o processo relacionado \`{a} percep\c{c}\~{a}o subjetiva seja uma nova entidade em rela\c{c}\~{a}o ao ambiente f\'{i}sico e n\~{a}o pode ser reduzido a esse \'{u}ltimo. De fato, a percep\c{c}\~{a}o subjetiva nos leva para a vida interior intelectual do indiv\'{i}duo que \'{e} extra observacional, por sua pr\'{o}pria natureza. \citep[p.~421]{vNeum1955mathematical}\index{Neumann, John von}.
\end{quote}

Esse \'{e} o motivo pelo qual \citet[p.~79--80]{breuer2001vneumannincomplet} delineia o problema da medi\c{c}\~{a}o em f\'{i}sica qu\^{a}ntica como o problema da compatibilidade entre o que est\'{a} fora da f\'{i}sica (tal como o colapso) e o que est\'{a} dentro da f\'{i}sica (tal como a evolu\c{c}\~{a}o linear). Dessa forma, por mais que a teoria da medi\c{c}\~{a}o de \citet{vNeum1955mathematical}\index{Neumann, John von} incorra na mesma dificuldade de Bohr, no que tange \`{a} arbitrariedade da diferencia\c{c}\~{a}o entre o que \'{e} e o que n\~{a}o \'{e} dom\'{i}nio da mec\^{a}nica qu\^{a}ntica, seu ganho \'{e} de especificar a discuss\~{a}o para os campos l\'{o}gicos e ontol\'{o}gicos e n\~{a}o t\~{a}o somente explicitar uma cis\~{a}o arbitr\'{a}ria entre o que \'{e} um objeto qu\^{a}ntico e o que n\~{a}o \'{e}. Ainda assim, de acordo com \citet[p.~121]{becker2004collapse}, existe uma concep\c{c}\~{a}o recebida acerca da teoria da medi\c{c}\~{a}o de von Neumann\index{Neumann, John von} segundo a qual o colapso \'{e} um processo f\'{i}sico que ``modifica de modo indeterminista o estado do sistema que est\'{a} sendo medido''. Para \citet[p.~123]{becker2004collapse}, o aspecto central dessa concep\c{c}\~{a}o recebida \'{e} considerar o colapso como um \textit{processo f\'{i}sico} ``que ocorre durante o processo de uma medi\c{c}\~{a}o, embora n\~{a}o seja especificado em qual instante''.\footnote{~Sobre a referida ``concep\c{c}\~{a}o recebida'' do colapso, ver \citet{everett1957relative}\index{Everett, Hugh}, \citet{stapp1982mind}, \citet{albert1992quantum}, e \citet{barrett1999quantum}.}

Dadas as caracter\'{i}sticas l\'{o}gicas da teoria da medi\c{c}\~{a}o de \citet{vNeum1955mathematical}\index{Neumann, John von}, passo \`{a} discuss\~{a}o em torno de seus aspectos ontol\'{o}gicos. Foi poss\'{i}vel constatar que a posi\c{c}\~{a}o de \citet{vNeum1955mathematical}\index{Neumann, John von} em rela\c{c}\~{a}o ao problema da medi\c{c}\~{a}o est\'{a} comprometida ontologicamente com um novo objeto que comp\~{o}e o mobili\'{a}rio do mundo,\index{ontologia!comprometimento ontol\'{o}gico} isto \'{e}, com uma nova entidade com poder causal para completar uma medi\c{c}\~{a}o: o ``ego abstrato'', que tem certas caracter\'{i}sticas ontol\'{o}gicas, por exemplo, ser um dom\'{i}nio da existência \textit{distinto} do dom\'{i}nio f\'{i}sico. Tradicionalmente, a entidade do tipo ``ego abstrato'' fora entendida como consciência. No entanto, como observado por \citet{otavio2019consciousnessQM}, essa generaliza\c{c}\~{a}o pode ser apressada, e at\'{e} mesmo equivocada. Essa n\~{a}o foi a \'{u}nica confus\~{a}o conceitual encontrada na literatura.

Conforme aponta \citet[p.~482]{Jammer1974}, a teoria da medi\c{c}\~{a}o formulada por \citet{vNeum1955mathematical}\index{Neumann, John von}, que culmina na tese de que a consciência \'{e} o agente causal respons\'{a}vel pelo ato da medi\c{c}\~{a}o, n\~{a}o seria acess\'{i}vel a grande parte dos f\'{i}sicos experimentais da \'{e}poca na medida em que, sendo demasiadamente formal, requereria dos interlocutores um alto conhecimento de matem\'{a}tica. No entanto, tal teoria foi reelaborada por \citet{LonBau1939theory} em um estudo que \citet[p.~482]{Jammer1974} considera uma apresenta\c{c}\~{a}o ``\textelp{} concisa e simplificada'' da teoria da medi\c{c}\~{a}o de \citet{vNeum1955mathematical}\index{Neumann, John von}.

O interesse de London por filosofia, especificamente pelo problema mente-corpo \'{e} documentado em uma pequena biografia escrita por sua esposa, Edith \citet[X-XIV]{london1961supefluids}. Dentre suas influências filos\'{o}ficas, \citet[p.~482--483]{Jammer1974} destaca Pf\"{a}nder, objeto de an\'{a}lise na tese de doutorado em filosofia de London e, principalmente, seu professor de filosofia em Munique, Erich Becher. De acordo com \citet[p.~482--483]{Jammer1974}, a tese, apresentada no Instituto Arnold Sommerfeld em Munique, trata sobre \citet{pfander1904psychologie}, que influenciara a teoria psicol\'{o}gica de \citet{lipps1907psychologische} que, ent\~{a}o influenciaria a concep\c{c}\~{a}o de medi\c{c}\~{a}o em mec\^{a}nica qu\^{a}ntica de London. \citet[p.~483]{Jammer1974} tamb\'{e}m ressalta que o estudo de \citet{LonBau1939theory} faz referência a duas obras de \citet{brecher1906phenom1,brecher1906phenom2}, para quem o problema mente-corpo seria a quest\~{a}o central em toda a filosofia.

Em rela\c{c}\~{a}o aos problemas da filosofia da mente, Becher rejeitaria, segundo \citet[p.~484]{Jammer1974}, a doutrina do epifenomenalismo, isto \'{e}, o pensamento segundo o qual os processos mentais emergem ou s\~{a}o causados pelos processos cerebrais, e defende o interacionismo, isto \'{e}, o pensamento segundo o qual os processos f\'{i}sicos ``\textelp{} permeiam o c\'{e}rebro em um curso cont\'{i}nuo e produzem, al\'{e}m de efeitos f\'{i}sicos, efeitos ps\'{i}quicos que, por sua vez, afetam de forma decisiva os eventos f\'{i}sicos''. \'{E} natural que London tenha acatado a cr\'{i}tica de Becher acerca do epifenomenalismo, uma vez que tenha dado continuidade \`{a} ideia de que a consciência age sobre a mat\'{e}ria.

Para \citet[p.~484]{Jammer1974}, London teria encontrado na mec\^{a}nica qu\^{a}ntica, especificamente no problema da medi\c{c}\~{a}o, conforme delineado por \citet{vNeum1955mathematical}\index{Neumann, John von}, um campo para aplicar tais ideias filos\'{o}ficas, na medida em que, na interpreta\c{c}\~{a}o de \citet[p.~251]{LonBau1939theory}, a intera\c{c}\~{a}o entre um objeto microf\'{i}sico e um aparelho macrosc\'{o}pico de medi\c{c}\~{a}o n\~{a}o seriam suficientes para produzir uma medi\c{c}\~{a}o, de modo que uma medi\c{c}\~{a}o ocorre somente quando tal sistema composto $\langle \text{objeto} + \text{aparato}\rangle$ \'{e} ``observado'', ou ``medido''. No caso, seria a consciência que de fato causa o colapso, isto \'{e}, completa uma medi\c{c}\~{a}o.

Tal afirma\c{c}\~{a}o deve, no entanto, ser melhor caracterizada, visto que existe um car\'{a}ter ontol\'{o}gico da proposta \citet{LonBau1939theory} que difere da proposta de \citet{vNeum1955mathematical}\index{Neumann, John von}. A interpreta\c{c}\~{a}o de \citet{LonBau1939theory}, como aponta Abner \citet[p.~759]{Shi1963role}, considera que o observador est\'{a} no mesmo n\'{i}vel ontol\'{o}gico que o sistema composto (sistema microsc\'{o}pico e aparato de medi\c{c}\~{a}o), de modo que ``London e Bauer n\~{a}o parecem atribuir uma posi\c{c}\~{a}o transcendente ao observador''. Isto \'{e}, ao passo que \citet{vNeum1955mathematical}\index{Neumann, John von} enfatiza o car\'{a}ter n\~{a}o-f\'{i}sico do observador, \citet[p.~251]{LonBau1939theory} consideram que o observador est\'{a} no mesmo sistema composto que o sistema microsc\'{o}pico e o aparato de medi\c{c}\~{a}o, que pode ser representado como $\langle \text{objeto} + \text{aparato} + \text{observador}\rangle$.

O observador teria, ainda assim, um papel distinto dentro do sistema composto. A tese subjetivista, atribu\'{i}da a von Neumann\index{Neumann, John von} devido \`{a} passagem em que considera o ``ego abstrato'' do observador o agente causal da medi\c{c}\~{a}o, parece se tornar expl\'{i}cita na teoria de London e Bauer quando, em uma passagem decisiva, afirmam que a ``faculdade de introspec\c{c}\~{a}o'' \'{e} central no processo de medi\c{c}\~{a}o:

\begin{quote}
O observador tem uma impress\~{a}o completamente diferente. Para ele, \'{e} apenas o objeto x e o aparelho y que pertencem ao mundo externo, para o que ele chama de ``objetividade''. Por outro lado, ele tem consigo mesmo rela\c{c}\~{o}es de uma maneira muito diferente. Ele possui uma faculdade caracter\'{i}stica e bastante familiar que podemos chamar de ``faculdade de introspec\c{c}\~{a}o''. Ele pode acompanhar cada momento de seu pr\'{o}prio estado. Em virtude desse ``conhecimento imanente'' ele atribui a si o direito de criar a sua pr\'{o}pria objetividade ---ou seja, cortar a cadeia de correla\c{c}\~{o}es estat\'{i}sticas \textelp{}. \'{E} apenas a consciência de um ``eu'' que pode separ\'{a}-lo da fun\c{c}\~{a}o anterior \textelp{} e, em virtude de sua observa\c{c}\~{a}o, configurar uma nova objetividade ao atribuir para o objeto uma nova fun\c{c}\~{a}o dali pra frente \textelp{}. \citep[p.~252]{LonBau1939theory}.
\end{quote}

A consciência individual do observador, sua faculdade interna, de introspec\c{c}\~{a}o, \'{e} considerada por \citet[p.~252]{LonBau1939theory} um sistema distinto do sistema composto material ---que se define pela intera\c{c}\~{a}o entre o objeto microf\'{i}sico e o aparelho medidor macrosc\'{o}pico--- de modo que esse sistema, n\~{a}o sujeito \`{a}s leis da mec\^{a}nica qu\^{a}ntica, \'{e} causal no sistema material. Como aponta \citet[p.~759]{Shi1963role}, o observador ``\textelp{} por possuir a faculdade de introspec\c{c}\~{a}o, pode conceder a si mesmo a abstra\c{c}\~{a}o dos sistemas f\'{i}sicos com os quais interage''. Em outras palavras, a interpreta\c{c}\~{a}o subjetivista parece sugerir um estatuto ontol\'{o}gico privilegiado para a consciência individual do observador humano no universo. Dito ainda de outro modo, essa interpreta\c{c}\~{a}o se compromete ontologicamente\index{ontologia!comprometimento ontol\'{o}gico} com uma entidade mental que causa sobre uma entidade material, ponto em que \citet[p.~484]{Jammer1974} tra\c{c}a a influência de Becher no pensamento de London. Nesse ponto, as teses de \citet{vNeum1955mathematical}\index{Neumann, John von} \textit{parecem} \citet{LonBau1939theory} se alinhar.

\'{E} justamente nesse ponto que muitos comentadores se equivocaram. Como mostraram os estudos de \citet{French2002467,french2020,french2023}, a teoria da medi\c{c}\~{a}o de \citet{LonBau1939theory} n\~{a}o exige que a faculdade de introspec\c{c}\~{a}o do observador \textit{cause} o colapso, mas que \textit{reconhe\c{c}a} o colapso. Esse \'{e} motivo pelo qual a chave filos\'{o}fica de leitura para a teoria da medi\c{c}\~{a}o de \citet[p~252]{LonBau1939theory} esteja na fenomenologia Husserliana, como troca de doa\c{c}\~{a}o de sentido, e n\~{a}o causa ---muito menos subjetivista.\footnote{~Para mais uma breve discuss\~{a}o sobre esse ponto, ver \citet{RaoLau2018principia}.} Essas s\~{a}o, portanto, as duas principais confus\~{o}es conceituais encontradas na literatura: 1) a identifica\c{c}\~{a}o de \citet{vNeum1955mathematical}\index{Neumann, John von} com a tese de que a consciência causa o colapso; e 2) a identifica\c{c}\~{a}o de \citet{LonBau1939theory} com 1). De modo mais preciso, pode-se afirmar que o predecessor da interpreta\c{c}\~{a}o da consciência causal\index{interpreta\c{c}\~{a}o da consciência!causal}, que considera que \'{e} \textit{de fato} a consciência do observador que causa o colapso seria \citet{wigner1961mindbody}\index{Wigner, Eugene}, na situa\c{c}\~{a}o conhecida como o ``amigo de Wigner''\index{Wigner, Eugene!amigo de Wigner, paradoxo do}. Suponha que todas as intera\c{c}\~{o}es poss\'{i}veis entre um indiv\'{i}duo humano com um dado sistema f\'{i}sico se resumam a olhar para certo ponto em certa dire\c{c}\~{a}o nos instantes de tempo $t_0, t_1, t_2, \dots, t_n$, e que as sensa\c{c}\~{o}es poss\'{i}veis que tal indiv\'{i}duo possa vir a ter se resumam \`{a}s de ver ou n\~{a}o ver um flash de luz; suponha, ainda, que a formula\c{c}\~{a}o matem\'{a}tica representando a possibilidade do indiv\'{i}duo ver o flash seja uma fun\c{c}\~{a}o de onda $|\psi_1\rangle$ e que uma fun\c{c}\~{a}o de onda $|\psi_2\rangle$ represente a possibilidade do indiv\'{i}duo n\~{a}o ver o flash.

Assim, a comunicabilidade da fun\c{c}\~{a}o de onda, qualquer que seja o resultado, dependeria daquilo que o indiv\'{i}duo observou. Em outras palavras, ele poderia nos dizer qual das fun\c{c}\~{o}es de onda seria o caso, isto \'{e}, se o indiv\'{i}duo viu ou n\~{a}o viu o flash de luz. Espera-se que o resultado seja objetivo no preciso sentido em que seja comunic\'{a}vel, isto \'{e}, no caso de perguntarmos para um indiv\'{i}duo $\mathcal{X}$ o resultado da intera\c{c}\~{a}o num instante $t$, um outro indiv\'{i}duo, $\mathcal{Y}$, que interagisse com o sistema num instante $t + 1$ poderia se utilizar do resultado obtido em $t$ como se fosse $\mathcal{Y}$, e n\~{a}o $\mathcal{X}$, que tivesse interagido com o sistema no instante $t$.

O racioc\'{i}nio do experimento mental consiste em questionar o estado do indiv\'{i}duo $\mathcal{X}$, que observa o sistema no instante $t$ antes de comunicar o resultado para o indiv\'{i}duo $\mathcal{Y}$. Dito de outro modo, o experimento mental prop\~{o}e uma situa\c{c}\~{a}o em que algu\'{e}m realiza uma observa\c{c}\~{a}o em um sistema. No caso, supondo que $\mathcal{Y}$ seja o pr\'{o}prio Wigner\index{Wigner, Eugene} e que $\mathcal{X}$ seja o \index{Wigner, Eugene!amigo de Wigner, paradoxo do}amigo de Wigner, qual seria o estado do sistema no instante de tempo entre a intera\c{c}\~{a}o de $\mathcal{X}$ em $t$ e a comunica\c{c}\~{a}o do resultado da intera\c{c}\~{a}o para $\mathcal{Y}$ no instante $t + 1$?

Isto \'{e}, se for assumido que o estado inicial seja uma combina\c{c}\~{a}o linear dos dois estados poss\'{i}veis relacionados com a probabilidade de que cada um dos estados seja o caso, o estado do sistema composto na intera\c{c}\~{a}o $\langle \text{objeto} + \text{observador}\rangle$ (em que o termo ``observador'' corresponde ao amigo) poderia ser descrito pela mec\^{a}nica qu\^{a}ntica atrav\'{e}s uma equa\c{c}\~{a}o linear. No entanto, de acordo com a mec\^{a}nica qu\^{a}ntica, tal descri\c{c}\~{a}o pretendida n\~{a}o \'{e} alcan\c{c}ada. A descri\c{c}\~{a}o de objetos f\'{i}sicos por meio da fun\c{c}\~{a}o de onda descreve diversas situa\c{c}\~{o}es poss\'{i}veis, com dado peso estat\'{i}stico de ocorrerem, mas n\~{a}o descreve resultados \'{u}nicos independentes de uma medi\c{c}\~{a}o. Como consequência disso, antes que o amigo diga o resultado (isto \'{e}, se viu ou n\~{a}o viu o flash), a descri\c{c}\~{a}o qu\^{a}ntica \'{e} uma fun\c{c}\~{a}o de onda atribu\'{i}da ao sistema composto $\langle \text{objeto} + \text{amigo}\rangle$.

Assim, Wigner\index{Wigner, Eugene} ($\mathcal{Y}$) pode interagir com o sistema composto $\langle \text{objeto} + \text{amigo}\rangle$ perguntando ao amigo ($\mathcal{X}$) se ele viu algum flash. A t\'{i}tulo de precis\~{a}o, o termo utilizado no texto de Wigner\index{Wigner, Eugene} \'{e} ``mistura''.  Ele se refere, contudo, ao termo t\'{e}cnico chamado ``mistura estat\'{i}stica'', denotado pelo operador $\rho$, utilizado no formalismo da mec\^{a}nica qu\^{a}ntica para designar situa\c{c}\~{o}es de ignor\^{a}ncia. No entanto, como apontou \citet[p.~483, nota 27]{French2002467}, o termo ``mistura'' designava, na \'{e}poca, aquilo que hoje chama-se de ``superposi\c{c}\~{a}o''. Qualquer que seja o caso, a fun\c{c}\~{a}o de onda do sistema composto se modifica para um caso em que o objeto passa a ser descrito por um estado \'{u}nico. Tal mudan\c{c}a ocorre somente em contato com $\mathcal{Y}$:

\begin{quote}
\textelp{A} mudan\c{c}a t\'{i}pica na fun\c{c}\~{a}o de onda ocorre somente quando alguma informa\c{c}\~{a}o (o ``sim'' ou ``n\~{a}o'' do meu amigo) entra na minha consciência. Disso se segue que a descri\c{c}\~{a}o qu\^{a}ntica dos objetos \'{e} influenciada por impress\~{o}es que entram na minha consciência. \citep[p.~173]{wigner1961mindbody}\index{Wigner, Eugene}.
\end{quote}

Wigner considera que a consciência do observador modifica ativamente o conhecimento\footnote{~\citet[p.~169, nota 3]{wigner1961mindbody}\index{Wigner, Eugene} se utiliza dos textos posteriores de \citet{heisen1958physphil}\index{Heisenberg, Werner}, em que o autor se refere ao termo consciência como ``conhecimento''.} do sistema e, com isso, as condi\c{c}\~{o}es de previsibilidade do sistema dos flashes, isto \'{e}, modifica sua representa\c{c}\~{a}o matem\'{a}tica atrav\'{e}s da fun\c{c}\~{a}o de onda:

\begin{quote}
\textelp{} a impress\~{a}o que se obt\'{e}m em uma intera\c{c}\~{a}o, chamada tamb\'{e}m de o resultado de uma observa\c{c}\~{a}o, modifica a fun\c{c}\~{a}o de onda do sistema. A fun\c{c}\~{a}o de onda modificada \'{e}, al\'{e}m disso, em geral imprevis\'{i}vel antes que a impress\~{a}o adquirida na intera\c{c}\~{a}o entrasse em nossa consciência: \'{e} a entrada de uma impress\~{a}o em nossa consciência, que altera a fun\c{c}\~{a}o de onda porque modifica a avalia\c{c}\~{a}o das probabilidades para diferentes impress\~{o}es que esperamos receber no futuro.'' \citep[p.~172--173]{wigner1961mindbody}\index{Wigner, Eugene}.
\end{quote}

A situa\c{c}\~{a}o proposta \'{e} an\'{a}loga \`{a} cadeia infinita de observa\c{c}\~{o}es de \citet{vNeum1955mathematical}\index{Neumann, John von}: enquanto a intera\c{c}\~{a}o do sistema composto $\langle \text{objeto} + \text{amigo}\rangle$ estiver no mesmo n\'{i}vel, n\~{a}o h\'{a}, de fato, uma medi\c{c}\~{a}o. H\'{a} que se perguntar ``quem observa o observador?'', pois at\'{e} que um observador final interaja com o sistema composto, uma medi\c{c}\~{a}o n\~{a}o estar\'{a} completa. Para \citet[p.~176]{wigner1961mindbody}\index{Wigner, Eugene}, quem teria tal posi\c{c}\~{a}o privilegiada seria ele mesmo, isto \'{e}, o amigo, ocupando uma posi\c{c}\~{a}o intermedi\'{a}ria, n\~{a}o poderia ter o resultado da observa\c{c}\~{a}o registrado em sua consciência a despeito do observador final: ``\textelp{} a teoria da medi\c{c}\~{a}o, direta ou indireta, \'{e} logicamente consistente desde que eu mantenha minha posi\c{c}\~{a}o privilegiada de observador final''. Ainda assim, se depois de completada a situa\c{c}\~{a}o proposta acima, \citet[p.~176]{wigner1961mindbody}\index{Wigner, Eugene} perguntar ao amigo sobre o estado do objeto $\mathcal{S}$ antes da intera\c{c}\~{a}o entre $\mathcal{X}$ e $\mathcal{Y}$ proposta no racioc\'{i}nio acima, o amigo responderia (a depender do que tenha sido o caso de $\mathcal{S}$) que ``eu j\'{a} lhe disse, eu vi [n\~{a}o vi] um flash''.

Para ilustrar a problem\'{a}tica que est\'{a} em jogo, \citet[p.~177]{wigner1961mindbody}\index{Wigner, Eugene} prop\~{o}e que o papel do observador intermedi\'{a}rio seja trocado: ao inv\'{e}s do amigo, que se utilize um simples aparelho f\'{i}sico de medi\c{c}\~{a}o, que amplificaria o sinal de um \'{a}tomo que poderia (ou n\~{a}o) ser excitado pela luz do flash no sistema $\mathcal{S}$. Nesse caso, como aponta \citet[p.~499]{Jammer1974}, n\~{a}o haveria d\'{u}vida de que uma representa\c{c}\~{a}o matem\'{a}tica, atrav\'{e}s de uma equa\c{c}\~{a}o linear, poderia descrever o sistema composto $\langle \text{objeto} + \text{aparato}\rangle$ ---contrariamente \`{a} assun\c{c}\~{a}o de que tal intera\c{c}\~{a}o poderia indicar o estado atual de $\mathcal{S}$. Com isso em mente, se modificarmos novamente o observador intermedi\'{a}rio, voltando a consider\'{a}-lo como o amigo, a representa\c{c}\~{a}o matem\'{a}tica, de acordo com \citet[p.~177]{wigner1961mindbody}\index{Wigner, Eugene} ``\textelp{} parece absurda, pois implica que meu amigo estaria em um estado de anima\c{c}\~{a}o suspensa antes de responder \`{a} minha pergunta'', isto \'{e}, parece absurda, por implicar n\~{a}o s\'{o} que o objeto $\mathcal{S}$ n\~{a}o teria seu estado atual desenvolvido (ou seja, o flash n\~{a}o teria nem n\~{a}o teria sido disparado) mas, principalmente, que o amigo n\~{a}o teria sua pr\'{o}pria existência atualizada at\'{e} que houvesse a a\c{c}\~{a}o interativa de $\mathcal{Y}$ sobre o sistema composto $\langle \text{objeto} + \text{amigo}\rangle$.

A fim de esclarecer tal dificuldade, Wigner\index{Wigner, Eugene} conclui que:

\begin{quote}
Segue-se que o ser com uma consciência deve ter um papel diferente na mec\^{a}nica qu\^{a}ntica que o dispositivo de medi\c{c}\~{a}o inanimado: o \'{a}tomo considerado acima \textelp{}. Esse argumento implica que ``meu amigo'' tem os mesmos tipos de impress\~{o}es e sensa\c{c}\~{o}es como eu ---em particular, que, depois de interagir com o objeto, ele n\~{a}o est\'{a} nesse estado de anima\c{c}\~{a}o suspensa \textelp{}. \citep[p.~177--178]{wigner1961mindbody}\index{Wigner, Eugene}.
\end{quote}

Quando \citet[p.~177]{wigner1961mindbody}\index{Wigner, Eugene} descreve que o amigo est\'{a} em um estado de suspens\~{a}o, parece sugerir que no racioc\'{i}nio todo s\'{o} h\'{a} \textit{um} colapso, isto \'{e}, somente um momento em que uma medi\c{c}\~{a}o \'{e} efetivamente realizada: quando Wigner\index{Wigner, Eugene} (e n\~{a}o o amigo) tem consciência de todo o processo atrav\'{e}s da intera\c{c}\~{a}o com o amigo. Um racioc\'{i}nio semelhante foi proposto por \citet[p.~290--293]{penrose1989emperors}, que revisita a situa\c{c}\~{a}o do gato\index{Schr\"{o}dinger, Erwin!paradoxo do gato} de Schr\"{o}dinger, adicionando no racioc\'{i}nio um observador humano ---propriamente vestido com um traje que o proteja do veneno--- dentro da caixa onde se encontra o gato\index{Schr\"{o}dinger, Erwin!paradoxo do gato} e todo o restante do aparato que envolve o experimento mental de \citet{schrodinger1935cat}\index{Schr\"{o}dinger, Erwin}. No experimento revisitado por \citet[p.~293]{penrose1989emperors}, o observador de dentro, que visualiza diretamente o que ocorre com o gato\index{Schr\"{o}dinger, Erwin!paradoxo do gato}, e o observador de fora, que \'{e} limitado pelo c\'{a}lculo das probabilidades sobre o que ocorre com o gato\index{Schr\"{o}dinger, Erwin!paradoxo do gato}, teriam, for\c{c}osamente, impress\~{o}es discrepantes sobre o que acontece com o gato\index{Schr\"{o}dinger, Erwin!paradoxo do gato}. Isso ocorreria at\'{e} que a caixa fosse aberta, quando as impress\~{o}es tornariam-se precisamente as mesmas.

Tal situa\c{c}\~{a}o \'{e} oportuna para visualizarmos a dificuldade apresentada por \citet{wigner1961mindbody}\index{Wigner, Eugene}. Acatando-se a tese de que a consciência humana (individual/subjetiva) \'{e} de alguma maneira causa do que acontece com o gato\index{Schr\"{o}dinger, Erwin!paradoxo do gato}, ter-se-ia a mesma situa\c{c}\~{a}o que ocorre com o racioc\'{i}nio do \index{Wigner, Eugene!amigo de Wigner, paradoxo do}amigo de Wigner: a consciência de quem atuou como agente causal no caso proposto por Penrose? A do observador de dentro ou do observador de fora? \'{E} relevante constatar que \citet[p.~445]{vNeum1955mathematical}\index{Neumann, John von} j\'{a} havia considerado que haveriam dificuldades no caso de mais de um observador concomitante, mas a situa\c{c}\~{a}o fora levada ao limite somente por \citet{wigner1961mindbody}\index{Wigner, Eugene}.

A situa\c{c}\~{a}o, no entanto, \'{e} desconcertante do ponto de vista conceitual. Para enfatizar, vou unir os paradoxos que vimos at\'{e} aqui em uma \'{u}nica situa\c{c}\~{a}o. Vamos dar um nome para a tal amizade de Wigner que deu origem ao paradoxo\index{Wigner, Eugene!amigo de Wigner, paradoxo do}; convenientemente, junto com  \citet{albert1992quantum}, nomearei a amiga de Wigner como ``Martha''. Suponha que Martha est\'{a} interagindo com um gato de Schr\"{o}dinger\index{Schr\"{o}dinger, Erwin!paradoxo do gato}, e que ela possa acompanhar a situa\c{c}\~{a}o do felino por um dispositivo de medi\c{c}\~{a}o. Suponha que o dispositivo aponta para o ponteiro ``$S_0$'' caso o gato permane\c{c}a em jejum (em decorrência do n\~{a}o-decaimento do objeto qu\^{a}ntico) e para o ponteiro ``$S_1$'' caso o gato seja alimentado (em decorrência do decaimento do objeto qu\^{a}ntico). A situa\c{c}\~{a}o em que Wigner se encontra antes de encontr\'{a}-la \'{e} uma superposi\c{c}\~{a}o ``$S_0+S_1$''. E \'{e} precisamente essa superposi\c{c}\~{a}o que \'{e} conceitualmente desconcertante:

\begin{quote}
    \'{E} uma superposi\c{c}\~{a}o de um estado em que Martha pensa que o ponteiro est\'{a} apontando para ``[$S_0$]'' e outro estado em que Martha pensa que o ponteiro est\'{a} apontando para ``[$S_1$]''; \'{e} um estado em que n\~{a}o h\'{a} nenhuma quest\~{a}o de fato sobre se Martha pensa ou n\~{a}o que o ponteiro est\'{a} apontando em uma dire\c{c}\~{a}o espec\'{i}fica. \textelp{} Isso n\~{a}o \'{e} nada parecido com um estado em que Martha est\'{a}, digamos, confusa sobre para onde o ponteiro est\'{a} apontando. Isso (merece ser repetido) \'{e} algo realmente estranho. Este \'{e} um estado em que \textelp{} n\~{a}o \'{e} certo dizer que Martha acredita que o ponteiro est\'{a} apontando para ``[$S_0$]'', e n\~{a}o \'{e} certo dizer que Martha acredita que o ponteiro est\'{a} apontando para ``[$S_1$]'', e n\~{a}o \'{e} certo dizer que ela tem essas duas cren\c{c}as (o que quer que isso signifique), e n\~{a}o \'{e} certo dizer que ela n\~{a}o tem nenhuma dessas cren\c{c}as. \citep[p.~79]{albert1992quantum}.
\end{quote}

\subsection{O problema ontol\'{o}gico}

Atribuir um papel causal \`{a} consciência individual de uma pessoa pode levar a uma dificuldade filos\'{o}fica bastante s\'{e}ria, que \'{e} o solipsismo, isto \'{e}, a implica\c{c}\~{a}o de que exista uma \'{u}nica subjetividade real e que todas as outras subjetividades sejam irreais ou ilus\'{o}rias. \citet[p.~258]{LonBau1939theory} j\'{a} haviam reconhecido essa dificuldade ao reiterar que, em mec\^{a}nica qu\^{a}ntica, a existência de um objeto f\'{i}sico depende do ato da medi\c{c}\~{a}o que, por sua vez, ``\textelp{} est\'{a} intimamente ligado \`{a} consciência da pessoa que realiza [a medi\c{c}\~{a}o], como se a mec\^{a}nica qu\^{a}ntica nos levasse a um completo solipsismo''. Para enfrentar a problem\'{a}tica do solipsismo, os autores argumentam em favor de um consenso intersubjetivo dos fenômenos externos, visto que, na pr\'{a}tica cotidiana, os fenômenos objetivos ocorrem como se fossem de fato objetivos no sentido de serem p\'{u}blicos e comuns a mais de uma subjetividade. Isso se apoiaria no fato de que existe tal coisa como uma comunidade cient\'{i}fica, o que s\'{o} seria poss\'{i}vel mediante tal consenso intersubjetivo. 

\citet[p.~485]{Jammer1974} considera que tal tentativa de superar o solipsismo atrav\'{e}s do consenso intersubjetivo acaba por entrar em contradi\c{c}\~{a}o com a hip\'{o}tese inicial de que os dois componentes do sistema composto $\langle \text{objeto} + \text{aparato}\rangle$ estejam no mesmo n\'{i}vel ontol\'{o}gico. De fato, existe uma dificuldade, pois como poderia um sistema composto, causado por uma consciência individual $\mathcal{C}i_1$, ser objetivo, isto \'{e}, publicamente acess\'{i}vel a outras consciências individuais $\mathcal{C}i_2$ \dots $\mathcal{C}i_n$ numa situa\c{c}\~{a}o em que $\mathcal{C}i_1$ n\~{a}o estivesse ciente do sistema composto? Isto \'{e}, a contradi\c{c}\~{a}o est\'{a} em assumir a existência de um objeto que, num racioc\'{i}nio posterior, n\~{a}o existe por si, mas t\~{a}o somente diante de uma consciência individual. 

Da mesma forma, a situa\c{c}\~{a}o proposta por \citet[p.~173]{wigner1961mindbody}\index{Wigner, Eugene} parece sugerir uma interpreta\c{c}\~{a}o solipsista, como vê-se no trecho: ``[o] solipsismo pode ser logicamente consistente com a mec\^{a}nica qu\^{a}ntica presente; j\'{a} o \index{monismo}monismo, no sentido materialista, n\~{a}o \'{e}''. Claramente, \citet[p.~178]{wigner1961mindbody}\index{Wigner, Eugene} n\~{a}o fica contente com essa implica\c{c}\~{a}o ontol\'{o}gica: ``\textelp{} negar a existência da consciência de um amigo a esse ponto \'{e} certamente uma atitude antinatural que se aproxima do solipsismo, e poucas pessoas, em seus cora\c{c}\~{o}es, ir\~{a}o segui-la''.

No entanto, ao final do racioc\'{i}nio, fica claro que a assun\c{c}\~{a}o do solipsismo parece ter um significado estritamente \textit{metodol\'{o}gico} para \citet[p.~173]{wigner1961mindbody}\index{Wigner, Eugene}. Em outras palavras, \'{e} precisamente a ideia de uma interpreta\c{c}\~{a}o subjetivista para o conceito de consciência na mec\^{a}nica qu\^{a}ntica que \'{e} colocada em xeque com a situa\c{c}\~{a}o paradoxal proposta em tal racioc\'{i}nio, isto \'{e}, a ideia de que a consciência subjetiva, individualizada, seria agente causal na medi\c{c}\~{a}o qu\^{a}ntica.

\subsection{O problema metaf\'{i}sico}

Esse novo objeto ---a consciência--- com poder causal \'{e} introduzido na ontologia da mec\^{a}nica qu\^{a}ntica sem pormenorizadas informa\c{c}\~{o}es acerca de sua natureza. Assim, ao passo que o problema ontol\'{o}gico da consciência na mec\^{a}nica qu\^{a}ntica seja a pr\'{o}pria introdu\c{c}\~{a}o da entidade, o problema metaf\'{i}sico \'{e} justamente a falta de uma metaf\'{i}sica que explique a natureza dessa entidade.\index{ontologia!e metaf\'{i}sica} A necessidade (ou n\~{a}o) de que a lacuna entre ontologia e metaf\'{i}sica seja preenchida tem sido extensamente debatida na literatura recente.\footnote{~Ver \citet{FRENCH2019defending, chakravartty2019challenge, bueno2019viking, jonas2019filomena, RaoJonas2019dualismQM, arenhartarroyo2021meta,arenhartarroyo2021veri}.} No entanto, ao passo que o debate geralmente gire em torno da metodologia da metaf\'{i}sica e do \index{realismo cient\'{i}fico}realismo cient\'{i}fico, trarei um \^{a}ngulo pouco explorado. Farei uma esp\'{e}cie de ``mostru\'{a}rio'' dos perfis metaf\'{i}sicos pouco investigados para a interpreta\c{c}\~{a}o da consciência causal\index{interpreta\c{c}\~{a}o da consciência!causal}.

A introdu\c{c}\~{a}o da no\c{c}\~{a}o de consciência como um ``objeto'' n\~{a}o f\'{i}sico no sentido de n\~{a}o material, na ontologia subjacente a essa interpreta\c{c}\~{a}o da medi\c{c}\~{a}o qu\^{a}ntica vem acompanhada de uma s\'{e}rie de problemas. Dentre eles, destaco a problem\'{a}tica em rela\c{c}\~{a}o \`{a} defini\c{c}\~{a}o do termo consciência, isto \'{e}, como a consciência deve ser entendida em termos metaf\'{i}sicos.\index{ontologia!e metaf\'{i}sica} Qual o lugar de tal consciência no mundo? Ou seja, o problema ontol\'{o}gico da consciência na mec\^{a}nica qu\^{a}ntica pode ser brevemente enunciado com a seguinte quest\~{a}o: ``a consciência existe?''. Buscarei elencar como tal quest\~{a}o \'{e} abordada pela literatura, bem como a problem\'{a}tica suscitada por essa discuss\~{a}o.

Como observa \citet[p.~82]{albert1992quantum}, a tese defendida por Wigner\index{Wigner, Eugene} dependeria de uma diferencia\c{c}\~{a}o entre sistemas inteiramente materiais e sistemas conscientes, isto \'{e}, a diferencia\c{c}\~{a}o entre sistemas n\~{a}o-conscientes e sistemas conscientes, de modo que a evolu\c{c}\~{a}o do estado f\'{i}sico de um dado objeto qu\^{a}ntico seria diferente caso o objeto fosse ou n\~{a}o consciente. Consequentemente, o entendimento do comportamento dos objetos qu\^{a}nticos dependeria da defini\c{c}\~{a}o ou do significado do termo consciência.

No entanto, nenhum dos autores referidos ofereceu uma defini\c{c}\~{a}o do termo consciência \citet[ver][p.~83]{albert1992quantum}, de modo que n\~{a}o fica claro o significado de uma senten\c{c}a tal como a afirma\c{c}\~{a}o de que ``a consciência \'{e} o agente causal na medi\c{c}\~{a}o qu\^{a}ntica''. Assim, a problem\'{a}tica suscitada pela interpreta\c{c}\~{a}o da consciência causal\index{interpreta\c{c}\~{a}o da consciência!causal}, isto \'{e}, de que a medi\c{c}\~{a}o seria completa somente com a introdu\c{c}\~{a}o de um agente causal n\~{a}o-f\'{i}sico, permanece em aberto ---e, como aponta \citet{quentin2003cognitiveQM}, os resultados de tal debate (se a consciência n\~{a}o f\'{i}sica \'{e} realmente um agente causal ou n\~{a}o) seriam definitivos para as discuss\~{o}es contempor\^{a}neas, especialmente nas \'{a}reas da filosofia da mente e nas ciências cognitivas.

Deve ficar claro, nesse aspecto, que a no\c{c}\~{a}o de consciência, conforme apresentada, desempenha um papel fundamentalmente distinto da ordem material, onde se situam os sistemas f\'{i}sicos. Nesse preciso sentido, essa interpreta\c{c}\~{a}o da consciência causal\index{interpreta\c{c}\~{a}o da consciência!causal} \'{e} incompat\'{i}vel com uma metaf\'{i}sica \index{monismo}monista materialista, como sugerido por \citet[p.~173]{wigner1961mindbody}\index{Wigner, Eugene} e demonstrado por \citet{RaoJonas2019dualismQM}.

Conforme a analogia proposta, do estatuto ontol\'{o}gico como o mobili\'{a}rio do mundo, destaco, em espec\'{i}fico, que esta interpreta\c{c}\~{a}o, que caracterizarei como interpreta\c{c}\~{a}o da consciência causal\index{interpreta\c{c}\~{a}o da consciência!causal}, carece de uma formula\c{c}\~{a}o ontol\'{o}gica (do tipo \index{ontologia!tradicional}$\mathscr{O}_T$) que abarque esse novo objeto: a consciência. Como aponta \citet[p.~114]{kohler2001vneumanncarnap}, ``von Neumann\index{Neumann, John von} consistentemente evitava discuss\~{o}es `filos\'{o}ficas' de quest\~{o}es epistemol\'{o}gicas''. Pelo contr\'{a}rio, a \'{u}nica categoriza\c{c}\~{a}o que \'{e} feita em rela\c{c}\~{a}o ao termo consciência \'{e} que se trata de um objeto ontologicamente distinto dos objetos materiais, o que sugere que essa consciência se trata de uma subst\^{a}ncia distinta da subst\^{a}ncia material. Tal proposta, como observam \citet[p.~167]{stapp2007whiteheadQM} e \citet[p.~58--59]{stoltzner2001opportunistic}, se alinha com o \index{dualismo}dualismo do tipo cartesiano, conhecido como ``dualismo de subst\^{a}ncia''\index{dualismo}, que possui diversas dificuldades filos\'{o}ficas ---uma das grandes quest\~{o}es seria o problema mente-corpo. Ainda assim, em termos metaf\'{i}sicos, n\~{a}o \'{e} poss\'{i}vel \textit{determinar} nem mesmo \textit{extrair} a metaf\'{i}sica associada \`{a} ontologia da interpreta\c{c}\~{a}o de von Neumann\index{Neumann, John von}. Para maiores detalhes, ver \citet{RaoJonas2019dualismQM}.

\'{E} poss\'{i}vel delinear a quest\~{a}o da seguinte maneira: da forma como descrevem \citet{vNeum1955mathematical}\index{Neumann, John von} e \citet{wigner1961mindbody}\index{Wigner, Eugene}, a no\c{c}\~{a}o de consciência com poder causal na medi\c{c}\~{a}o qu\^{a}ntica deveria cumprir as seguintes caracteriza\c{c}\~{o}es: a consciência deve ser imaterial, no sentido de que n\~{a}o pertence ao mesmo n\'{i}vel ontol\'{o}gico que os sistemas qu\^{a}nticos, isto \'{e}, deve ser considerada em um n\'{i}vel diferente em rela\c{c}\~{a}o \`{a} aplica\c{c}\~{a}o da mec\^{a}nica qu\^{a}ntica; n\~{a}o deve ser subjetiva, isto \'{e}, individualizada.

\subsection{O problema m\'{i}stico}\label{sec:mistico}

Nos par\'{a}grafos seguintes, elencarei algumas alternativas que preenchem a lacuna metaf\'{i}sica da consciência na mec\^{a}nica qu\^{a}ntica. Isto \'{e}, tratam-se de alternativas que respondem quest\~{o}es de \textit{natureza} para essa entidade, a consciência, obtida como parte da ontologia da interpreta\c{c}\~{a}o da mec\^{a}nica qu\^{a}ntica analisada neste cap\'{i}tulo. No entanto, n\~{a}o o fazem pelo uso de um referencial da metaf\'{i}sica, mas do misticismo ---e aqui o termo n\~{a}o \'{e} utilizado como pejorativo, mas descritivo dessa fam\'{i}lia de propostas que apresento a seguir, \textit{viz.}, a uni\~{a}o entre todos os entes como no Vedanta. Opto pelas propostas de \citet{bass1971mind} e \citet{goswami1989idealistic}, por tratarem diretamente das quest\~{o}es apresentadas e serem alternativas pouco discutidas na literatura espec\'{i}fica, mas que (de algum modo) s\~{a}o bastante difundidas no ide\'{a}rio popular.

Em particular, a proposta de \citet{bass1971mind} se trata de uma generaliza\c{c}\~{a}o do pensamento tardio de \citet{schro1964myview} para solucionar a situa\c{c}\~{a}o paradoxal presente no racioc\'{i}nio do \index{Wigner, Eugene!amigo de Wigner, paradoxo do}amigo de Wigner.\footnote{~Um estudo detalhado sobre a concep\c{c}\~{a}o filos\'{o}fica tardia de Schr\"{o}dinger pode ser encontrado em \citet{murr2014phd}.} Para Schr\"{o}dinger, os debates em rela\c{c}\~{a}o ao conceito de consciência ou mente enfrentariam uma situa\c{c}\~{a}o problem\'{a}tica, devido ao frequente comprometimento ontol\'{o}gico\index{ontologia!comprometimento ontol\'{o}gico} com a existência de m\'{u}ltiplas mentes ---tal como a situa\c{c}\~{a}o do \index{Wigner, Eugene!amigo de Wigner, paradoxo do}amigo de Wigner\index{Wigner, Eugene} parece pressupor:

\begin{quote}
Para a filosofia \textelp{} a dificuldade real est\'{a} na multiplicidade espacial e temporal de observadores e indiv\'{i}duos cognoscentes. Se todos os eventos ocorressem em uma consciência, a situa\c{c}\~{a}o seria extremamente simples. \citep[p.~18]{schro1964myview}.
\end{quote}

Pode-se perceber na passagem anteriormente citada, assim como em diversas outras, como observa \citet{cohen1992schromystic}, o comprometimento ontol\'{o}gico\index{ontologia!comprometimento ontol\'{o}gico} com a existência de uma \'{u}nica mente ---ver tamb\'{e}m \citet[p.~171]{bitbol2004schrocarnap}--- que, conforme observa \citet{bertotti1994schro} \'{e} de not\'{a}vel influência do pensamento indiano, especificamente do Vedanta:

\begin{quote}
O enigma das consciências individuais e sua comunidade levaram ele [Schr\"{o}dinger] a uma posi\c{c}\~{a}o, caracter\'{i}stica da filosofia indiana, que \'{e} o fundamento filos\'{o}fico do cl\'{a}ssico Vedanta: todas as mentes individuais \textelp{} s\~{a}o manifesta\c{c}\~{o}es de uma \'{u}nica Mente que abrange tudo. \citep[p.~91]{bertotti1994schro}.
\end{quote}

Sobre o termo ``Vedanta'', destaco um trecho de uma exposi\c{c}\~{a}o de Conger, que explicita precisamente o aspecto espiritualista do Vedanta que \'{e} abordado na discuss\~{a}o acima:

\begin{quote}
\textelp{} a filosofia central dos Upanixades e do \textit{Ved\={a}nta}, muitas vezes considerada pante\'{i}sta, seria descrita com maior precis\~{a}o como um \index{monismo}monismo espiritualista. Exemplo melhor de pante\'{i}smo \'{e} apresentado pelo Deus de Espinosa com um n\'{u}mero infinito de atributos. No \textit{Advaita Ved\={a}nta}, \textit{Brahman} \'{e} caracterizada por \textit{sat} (ser), \textit{cit} (inteligência) e \textit{\={a}nanda} (bem-aventuran\c{c}a), ao inv\'{e}s de uma gama de atributos pessoais; \textelp{} Brahman \'{e} alcan\c{c}ada pelo indiv\'{i}duo que chega a compreender sua pr\'{o}pria identidade com a Realidade Una. \citep[p.~239]{conger1944eastwest}.
\end{quote}

Schr\"{o}dinger faz uso da no\c{c}\~{a}o de \textsc{maya} ---tamb\'{e}m encontrado como ``Maja'', ou ``\textit{m\=ay\=a}'', em s\^{a}nscrito---, correspondente \`{a} distin\c{c}\~{a}o ---bastante antiga tamb\'{e}m na filosofia grega--- entre o que \'{e} real e o que seria aparente para responder \`{a} quest\~{a}o da multiplicidade das mentes:

\begin{quote}
A \'{u}nica alternativa poss\'{i}vel consiste apenas em reter da experiência imediata que a consciência \'{e} um singular cujo plural \'{e} desconhecido; que existe apenas uma coisa e o que parece ser uma pluralidade \'{e} apenas uma s\'{e}rie de aspectos diferentes dessa mesma coisa, produzidos por um engano (o termo indiano \textsc{maya}). \citep[p.~93]{schro1967whatislife}.
\end{quote}

Devo apontar, conforme \citet[p.~237]{gough1891upanishads}, que ``a doutrina de \textit{m\=ay\=a}, ou a irrealidade do dualismo sujeito/objeto, bem como a irrealidade da pluralidade de almas e seu ambiente, \'{e} a vida da filosofia indiana primitiva''. Assim, \textit{m\=ay\=a} n\~{a}o se remete exclusivamente ao Vedanta. No entanto, conforme afirma \citet{bertotti1994schro}, a influência do pensamento tardio de Schr\"{o}dinger seria primordialmente o Vedanta e, por isso, destaco apenas seu uso dentro do sistema vedantino. Se fosse poss\'{i}vel extrair uma metaf\'{i}sica do Vedanta, ela estaria associada com a identifica\c{c}\~{a}o entre \textit{Atman}, um termo que designa as ``mentes individuais'' e Brahman, que seria algo como uma ``consciência c\'{o}smica''. De acordo com \citet{radhakrishnan1914vedanta}, o termo \textit{m\=ay\=a} se insere no sistema vedantino da seguinte forma:

\begin{quote}
\textelp{} apenas o Absoluto, chamado Brahman, \'{e} real e as manifesta\c{c}\~{o}es finitas s\~{a}o ilus\'{o}rias. H\'{a} apenas uma realidade absoluta e indiferenciada, cuja natureza \'{e} constitu\'{i}da pelo conhecimento. O mundo emp\'{i}rico \'{e} inteiramente ilus\'{o}rio, com suas distin\c{c}\~{o}es de mentes finitas e objetos e os objetos de seu pensamento. Sujeitos e objetos s\~{a}o como imagens fugazes que englobam a alma que sonha, e que se reduzem a nada no momento em que acorda. O termo ``\textit{m\=ay\=a}'' significa o car\'{a}ter ilus\'{o}rio do mundo finito. \textelp{} Os aspectos centrais da filosofia Vedantina, como \'{e} concebida atualmente, s\~{a}o resumidamente explicitados nas seguintes frases: Brahman \'{e} o real e o universo \'{e} falso. O Atman \'{e} Brahman. Nada mais. \citep[p.~431]{radhakrishnan1914vedanta}.
\end{quote}

Dessa forma, a multiplicidade das mentes seria uma aparência ao passo que a unicidade da mente seria real ou, nas palavras de \citet[p.~97--98]{cohen1992schromystic}, ``n\~{a}o existe `realmente' uma multiplicidade de eus. \textelp{} existe uma unidade de todas as consciências''.

No entanto, \citet[p.~18]{schro1964myview} reconhece, conforme explicita na seguinte passagem, que esse n\~{a}o \'{e} um assunto estritamente racional: ``\textelp{} eu n\~{a}o penso que essa dificuldade possa ser resolvida logicamente, atrav\'{e}s de um pensamento consistente, em nossos intelectos'', ao dizer que ``\textelp{} a pluralidade que percebemos \'{e} apenas aparente, n\~{a}o \'{e} real''. De forma mais enf\'{a}tica, Schr\"{o}dinger explicita que esta ideia, pr\'{o}pria do pensamento do Vedanta, \'{e} um pensamento \textit{m\'{i}stico}:

\begin{quote}
Em si, a ideia n\~{a}o \'{e} nova. Os registros mais antigos datam, at\'{e} onde sei, de 2.500 anos atr\'{a}s. Desde os primitivos grandes Upanixades, no pensamento indiano, a identifica\c{c}\~{a}o de ATHMAN = BRAHMAN (o eu pessoal iguala-se ao eu eterno, e onipresente e onisciente), longe de constituir uma blasfêmia, representava a quintessência da mais profunda intui\c{c}\~{a}o quanto aos acontecimentos do mundo. O maior empenho de todos os estudiosos da escola Vedanta era, ap\'{o}s o aprendizado dos movimentos dos l\'{a}bios para a pron\'{u}ncia correta, realmente assimilar em suas mentes este pensamento, o mais grandioso de todos. \citep[p.~92]{schro1967whatislife}.
\end{quote}

\begin{quote}
    Resumidamente, \'{e} a vis\~{a}o de que todos n\'{o}s, seres vivos, somos unidos na medida em que somos, na verdade, lados ou aspectos de um \'{u}nico ser, que talvez na terminologia ocidental possa ser chamado de ``Deus'' enquanto nos Upanixades seu nome \'{e} ``Brahman''. \textelp{} N\'{o}s reconhecemos que estamos lidando aqui n\~{a}o com algo logicamente dedut\'{i}vel, mas com misticismo. \citep[p.~95]{schro1964myview}.
\end{quote}

No entanto, a liga\c{c}\~{a}o desse aspecto de seu pensamento, caracterizado por \citet{bertotti1994schro} como ``misticismo racional'' \'{e} obscura. \citet[p.~98]{cohen1992schromystic} sugere que a ausência de uma liga\c{c}\~{a}o se d\'{a} pela posi\c{c}\~{a}o de Schr\"{o}dinger de que a ciência deve ser fundamentalmente objetiva, isto \'{e}, deve excluir de forma preliminar o sujeito que conhece daquilo que \'{e} conhecido. Ainda assim, Schr\"{o}dinger jamais defendeu uma ideia de ciência subjetiva, tampouco objetiva \`{a} maneira do empirismo moderno, mas impessoal.

Para \citet[p.~212]{murr2014phd}, a vis\~{a}o de mundo de Schr\"{o}dinger, justamente por ter uma estreita rela\c{c}\~{a}o com seu trabalho cient\'{i}fico, n\~{a}o deve ser entendida como um aspecto religioso, mas essencialmente filos\'{o}fico. \citet[p.~161]{poser1992schrodingerconsciousness} aponta, ainda, que sua proposta filos\'{o}fica \'{e} mais do que uma continua\c{c}\~{a}o de seu trabalho cient\'{i}fico; pelo contr\'{a}rio, afirma que seu trabalho na ciência seja \textit{fruto} de suas reflex\~{o}es filos\'{o}ficas.

O posicionamento filos\'{o}fico tardio de Schr\"{o}dinger \'{e} classificado por \citet[p.~163]{poser1992schrodingerconsciousness} como um ``monismo idealista din\^{a}mico'', cuja express\~{a}o m\'{a}xima se encontra na express\~{a}o s\^{a}nscrita ``\textit{tat tvam asi}'', que \citet[p.~8]{huxley1947perene} traduz para o inglês como ``\textit{that art thou}'', que traduzido livremente para o português significaria algo como ``tu \'{e}s isto'', e que \citet[p.~22]{schro1964myview} interpreta como: ``Eu estou no leste e no oeste, eu estou abaixo e acima, eu sou o universo todo''.

\citet[p.~166]{poser1992schrodingerconsciousness} destaca ainda que Schr\"{o}dinger utiliza o pensamento vedantino como referência te\'{o}rica para seu projeto cient\'{i}fico e filos\'{o}fico, e n\~{a}o como autoridade religiosa; ou seja, utiliza a discuss\~{a}o presente no Vedanta para argumentar em favor de sua proposta, de modo que constr\'{o}i um modelo aberto a cr\'{i}ticas e n\~{a}o um dogma incontest\'{a}vel. Dessa forma, \citet[p.~83]{bertotti1994schro} utiliza o termo ``misticismo racional'' para classificar esse tipo de atitude, identificada tamb\'{e}m na posi\c{c}\~{a}o filos\'{o}fica de Einstein.

Como observa \citet[p.~212]{murr2014phd}, o referido sentimento de ``unidade'' pode ser alcan\c{c}ado por diversas vias, sendo a t\'{e}cnica da medita\c{c}\~{a}o uma delas. Wilber vai al\'{e}m e considera que tal unidade \'{e} emp\'{i}rica:

\begin{quote}
A psicologia vedantina funda-se na introvis\~{a}o experimentalmente verific\'{a}vel de que Brahman-Atman \'{e} a \'{u}nica Realidade, e sua preocupa\c{c}\~{a}o prim\'{a}ria consiste em proporcionar uma explica\c{c}\~{a}o pragm\'{a}tica do ``por que'' os seres humanos n\~{a}o compreendem sua b\'{a}sica e suprema identidade com Brahman. Em geral, a cega aceita\c{c}\~{a}o, pelos humanos, de dualismos e distin\c{c}\~{o}es \'{e} a ignor\^{a}ncia (\textit{avidy\=a}) que os fazem pousar diretamente num mundo de ilus\~{o}es (\textit{m\=ay\=a}). \citep[p.~152]{wilber1997spectrum}.
\end{quote}

Tal referencial, que \citet[p.~211--214]{murr2014phd} chama de ``p\'{o}s-objetivado'', \'{e} utilizado por \citet{bass1971mind} em um artigo intitulado ``\textit{The Mind of Wigner's Friend}'' (que traduzido livremente para o português significa ``A Mente do Amigo de Wigner''), na tentativa de solucionar o paradoxo do amigo de \citet{wigner1961mindbody}\index{Wigner, Eugene} com a introdu\c{c}\~{a}o da hip\'{o}tese, inspirada na obra tardia de \citet{schro1964myview}, chamada de ``vis\~{a}o Vedantina'', que remete \`{a} tese da unicidade da consciência.

Para tal racioc\'{i}nio, Bass prop\~{o}e as seguintes premissas:

\begin{quote}
A. Meu corpo, com seu sistema nervoso central (explorado em qualquer grau de completude fisiol\'{o}gica) funciona puramente como um mecanismo, de acordo com as leis da natureza. Al\'{e}m disso, a mec\^{a}nica qu\^{a}ntica \'{e} a base final desse mecanismo.

B. Estou ciente, por evidência direta incontest\'{a}vel, do conhecimento (informa\c{c}\~{a}o) entrando em minha consciência. \citep[p.~56]{bass1971mind}.
\end{quote}

Se aceitarmos que exclusivamente a premissa ``A'' se aplica ao ``observador intermedi\'{a}rio'', ent\~{a}o este observador seria, para os efeitos de medi\c{c}\~{a}o, tal como um aparelho medidor, isto \'{e}, seria incapaz de completar uma medi\c{c}\~{a}o conforme o sentido do termo medi\c{c}\~{a}o proposto por \citet{vNeum1955mathematical}\index{Neumann, John von}; da mesma forma, se aceitarmos que exclusivamente a premissa ``B'' se aplica ao ``observador intermedi\'{a}rio'', ent\~{a}o este observador seria, para os efeitos de medi\c{c}\~{a}o, um observador final na medida em que seria capaz de completar uma medi\c{c}\~{a}o.

As duas premissas, quando aplicadas juntamente ao observador intermedi\'{a}rio, trariam uma situa\c{c}\~{a}o paradoxal, visto que levam a situa\c{c}\~{o}es mutuamente exclusivas. Essa seria a leitura de \citet[p.~57]{bass1971mind} do paradoxo do amigo de \citet{wigner1961mindbody}\index{Wigner, Eugene}. No entanto, o racioc\'{i}nio acima parece levar em considera\c{c}\~{a}o dois observadores, nomeadamente o observador intermedi\'{a}rio e o observador final. Assim, \citet[p.~59]{bass1971mind} \'{e} capaz de enunciar uma terceira premissa subentendida no racioc\'{i}nio que leva \`{a} situa\c{c}\~{a}o paradoxal: ``C. Existem, independentemente, ao menos duas mentes conscientes''.

No entanto, \citet[p.~58--61]{bass1971mind} procura demonstrar que a situa\c{c}\~{a}o paradoxal proposta por \citet{wigner1961mindbody}\index{Wigner, Eugene} s\'{o} ocorre quando as premissas A, B e C s\~{a}o aceitas, de modo que, se somente a premissa ``C'' for negada, as premissas ``A'' e ``B'' podem ser ambas verdadeiras ao mesmo tempo. Para tanto, uma hierarquia das três premissas, do ponto de vista emp\'{i}rico, \'{e} estabelecida por \citet[p.~59]{bass1971mind}: ``mantenho, como Descartes, que a premissa ``B'' \'{e} a mais forte dentre as três: n\~{a}o tenho conhecimento mais direto e menos incerto que esse''.

A premissa ``A'' estaria em segundo lugar na ``hierarquia emp\'{i}rica'' de \citet[p.~59]{bass1971mind}, e \'{e} analisada criticamente: a primeira parte da premissa ``\textelp{} extrapola os avan\c{c}os maravilhosos e cont\'{i}nuos da fisiologia do sistema nervoso'', mas que, ainda assim, permanece v\'{a}lida na medida em que a neurofisiologia n\~{a}o nega que o c\'{e}rebro \'{e} ``uma rede de unidades de opera\c{c}\~{a}o eletroqu\'{i}micas finamente interligadas (c\'{e}lulas, axônios, sinapses)''.

Na an\'{a}lise da premissa ``C'', \citet[p.~59]{bass1971mind} afirma que n\~{a}o \'{e} apoiada por qualquer evidência emp\'{i}rica direta'', utilizando-se do racioc\'{i}nio de \citet[p.~88]{schro1967whatislife}, para quem ``[a] consciência nunca \'{e}
experimentada no plural, apenas no singular'' ---o que \citet[p.~60]{bass1971mind} considera suficiente para afirmar que a premissa ``C'' seria a premissa mais fraca dentre as três, do ponto de vista emp\'{i}rico.

Por outro lado, do ponto de vista l\'{o}gico, Bass aponta que a atualiza\c{c}\~{a}o de uma potencialidade, no caso de uma medi\c{c}\~{a}o efetuada pela consciência, deveria representar ``um efeito espec\'{i}fico da consciência sobre o mundo f\'{i}sico'', de modo que seja precisamente 

\begin{quote}
\textelp{} esse efeito espec\'{i}fico da consciência sobre o mundo f\'{i}sico que pode ser tomado para acoplar a introspec\c{c}\~{a}o [premissa B] na f\'{i}sica [premissa A], de modo a gerar o paradoxo. \citep[p.~60]{bass1971mind}.
\end{quote}

Tal ``efeito espec\'{i}fico'' seria a a\c{c}\~{a}o da premissa ``C'', isto \'{e}, a a\c{c}\~{a}o de uma (dentre uma vasta pluralidade) consciência individualizada sobre o mundo f\'{i}sico.

Assim, Bass resume seu argumento da nega\c{c}\~{a}o da premissa ``C'' da seguinte forma. A faculdade de introspec\c{c}\~{a}o, contida na premissa B:

\begin{quote}
\textelp{} pode envolver apenas uma consciência. O mundo externo (na premissa A) \'{e} introduzido e confrontado com a introspec\c{c}\~{a}o de tal modo que a hip\'{o}tese sobre a pluralidade das mentes conscientes (na premissa C) resulta em uma nega\c{c}\~{a}o. \citep[p.~60]{bass1971mind}.
\end{quote}

Dessa forma, \citet[p.~63]{bass1971mind} assume a ``\textelp{} vis\~{a}o vedantina, que nega a pluralidade das mentes conscientes''. A existência da pluralidade da consciência, contudo, n\~{a}o \'{e} negada em absoluto: ela existiria enquanto aparência, se referindo \`{a} doutrina indiana de \textit{m\=ay\=a}, isto \'{e}, da aparência da pluralidade das consciências, na medida em que realmente s\'{o} existiria uma consciência \citet[p.~61--62]{bass1971mind}. No entanto, Bass reconhece que a emergência de uma dualidade sujeito/objeto, tal como parece ocorrer na percep\c{c}\~{a}o humana, \'{e} um aspecto problem\'{a}tico de sua proposta:

\begin{quote}
Assumindo a pluralidade, deduzi uma contradi\c{c}\~{a}o. Seria desej\'{a}vel complementar tal resultado ao assumir a unidade e deduzir uma consequência espec\'{i}fica que possa ser, ao menos em princ\'{i}pio, observ\'{a}vel. Isso asseguraria que a distin\c{c}\~{a}o entre pluralidade e unidade \'{e} significativa at\'{e} mesmo no \^{a}mbito das ciências naturais. Mas a no\c{c}\~{a}o ordin\'{a}ria de um ato de observa\c{c}\~{a}o envolve um sujeito e um objeto, o que n\~{a}o se coaduna com a hip\'{o}tese da unidade, quando ambos sujeito e objeto envolvem consciência. \citep[p.~65]{bass1971mind}.
\end{quote}

A dualidade sujeito/objeto no ato de observa\c{c}\~{a}o, referida acima, \'{e} mais sutil do que a referida por \citet{bohr1928quantumpostulate}\index{Bohr, Niels}: h\'{a} impl\'{i}cita aqui uma distin\c{c}\~{a}o entre aquilo que conhece e aquilo que \'{e} conhecido. Mantendo o vocabul\'{a}rio \index{monismo}monista da consciência proposta por \citet{bass1971mind}, h\'{a} a distin\c{c}\~{a}o entre o que est\'{a} dentro da consciência e o que est\'{a} fora da consciência. O tema da dualidade, isto \'{e}, a multiplicidade de consciências subsidi\'{a}ria ao \index{monismo}monismo, \`{a} unicidade da consciência, seria, \`{a} luz do Vedanta, abordado pela doutrina da ilus\~{a}o.

Portanto, longe de solucionar os problemas metaf\'{i}sicos da consciência na mec\^{a}nica qu\^{a}ntica, essa hip\'{o}tese daria lugar a outro espectro de problemas conceituais, pr\'{o}prios do pensamento vedantino. Ademais, metodologicamente, essa proposta parece querer impor uma ontologia do tipo \index{ontologia!tradicional}$\mathscr{O}_T$ para a ciência, sem qualquer justificativa aparente para tal.

Ainda assim, essa atitude frente ao problema da medi\c{c}\~{a}o qu\^{a}ntica \'{e} levada adiante por Goswami. Apresentarei resumidamente sua interpreta\c{c}\~{a}o da medi\c{c}\~{a}o qu\^{a}ntica nos par\'{a}grafos seguintes ---j\'{a} adiantando de antem\~{a}o, contudo, que se trata de uma proposta que incorre na mesma dificuldade que a proposta de Bass, conforme apontada no par\'{a}grafo anterior.

A partir de uma generaliza\c{c}\~{a}o da ontologia de \citet{heisen1958physphil}\index{Heisenberg, Werner} acerca da distin\c{c}\~{a}o entre potencialidade e atualidade e da medi\c{c}\~{a}o=cria\c{c}\~{a}o, \citet[p.~534]{goswami2003QMtextbook} afirma que a evolu\c{c}\~{a}o determinista e temporal, descrita atrav\'{e}s da evolu\c{c}\~{a}o linear, ocorre em um dom\'{i}nio transcendente, que define ---utilizando a terminologia de \citet{heisen1958physphil}\index{Heisenberg, Werner}--- como ``\textit{potentia}''.\footnote{~Conforme apresento no cap\'{i}tulo \ref{CapWhitehead}, essa n\~{a}o \'{e} uma interpreta\c{c}\~{a}o apropriada para os escritos tardios de Heisenberg\index{Heisenberg, Werner}.}

A defini\c{c}\~{a}o de \citet[p.~534]{goswami2003QMtextbook} para o dom\'{i}nio ``potentia'', transcendente, seria tamb\'{e}m reminiscente da ontologia processual de \citet[p.~202, nota~2]{whitehead1925enquiry}, que considera que ``espa\c{c}o e tempo precisam resultar de algo em processo que transcenda os objetos''. Assim, \citet[p.~534]{goswami2003QMtextbook} utiliza o termo ``n\~{a}o localidade'' como ``fora do espa\c{c}o-tempo'', de modo que o dom\'{i}nio ``potentia'' seja n\~{a}o local. Aplicando tal aspecto, que \citet[p.~535]{goswami2003QMtextbook} chama de ``ontologia b\'{a}sica de Heisenberg'', \`{a} teoria da medi\c{c}\~{a}o de \citet{vNeum1955mathematical}\index{Neumann, John von}, tem-se que o colapso atualiza, isto \'{e}, traz para a realidade manifesta apenas uma possibilidade dentre diversas outras possibilidades contidas neste dom\'{i}nio transcendente, de modo que a realidade transfenomenal, isto \'{e}, a realidade entre tais atualiza\c{c}\~{o}es, estaria contida no dom\'{i}nio ``\textit{potentia}''.

Um dos aspectos caracter\'{i}sticos da interpreta\c{c}\~{a}o de \citet[p.~385]{goswami1989idealistic} seria a proposta metaf\'{i}sica do ``idealismo \index{monismo}monista'', na qual todos os elementos est\~{a}o dentro da mesma e \'{u}nica consciência: tanto os elementos transcendentes, potenciais, quanto os imanentes s\~{a}o atualizados. Isto \'{e}, tanto o colapso quanto a evolu\c{c}\~{a}o linear acontecem dentro da consciência. No entanto, diferentemente da interpreta\c{c}\~{a}o de Heisenberg, que entende a \textit{potentia} em termos aristot\'{e}licos, Goswami o faz utilizando elementos do platonismo. Como apresentarei adiante, o termo corresponde \`{a}quilo que \citet[p.~239]{conger1944eastwest} chamou de ``monismo espiritualista'': dentre os autores ocidentais que advogam essa corrente de pensamento, Conger destaca os nomes de Plat\~{a}o, Plotino e Espinoza, principalmente. Nas palavras de Goswami:

\begin{quote}
\textelp{} os objetos j\'{a} est\~{a}o na consciência primordialmente, como formas poss\'{i}veis em potentia. O colapso n\~{a}o est\'{a} fazendo algo aos objetos via observa\c{c}\~{a}o, mas consiste em escolher entre as possibilidades alternativas que a fun\c{c}\~{a}o de onda fornece, e em reconhecer o resultado da escolha. \citep[p.~536]{goswami2003QMtextbook}.
\end{quote}

O que est\'{a} em jogo n\~{a}o \'{e} a a\c{c}\~{a}o da consciência sobre a mat\'{e}ria, \textit{e.g.}, o poder de mover algum corpo material com a for\c{c}a do pensamento (algo como a psicocinese ou a telecinesia). Isso, afinal, pressup\~{o}e uma distin\c{c}\~{a}o entre as no\c{c}\~{o}es de consciência e mat\'{e}ria. O que parece estar em jogo aqui \'{e} o postulado de que todos os objetos s\~{a}o objetos dentro da mesma e \'{u}nica consciência. Essa seria uma forma de tratar a no\c{c}\~{a}o de consciência a partir de uma ontologia outra que n\~{a}o a do \index{monismo}monismo materialista ---em que a consciência \'{e} um fenômeno advindo da complexidade do arranjo material (neuronal) e, portanto, sem poder causal--- ou a do \index{dualismo}dualismo ---segundo o qual as no\c{c}\~{o}es de consciência e mat\'{e}ria correspondem a subst\^{a}ncias separadas.

Da mesma forma, Goswami procura demonstrar de que forma a no\c{c}\~{a}o de consciência, quando tratada a partir do idealismo \index{monismo}monista, evita dificuldades filos\'{o}ficas conforme apontadas em situa\c{c}\~{o}es tais como a do \index{Wigner, Eugene!amigo de Wigner, paradoxo do}amigo de Wigner:

\begin{quote}
O problema de Wigner\index{Wigner, Eugene} surge do seu racioc\'{i}nio dualista acerca da sua pr\'{o}pria consciência separada da consciência de seu amigo. O paradoxo desaparece se existir somente um sujeito ---n\~{a}o sujeitos separados como estamos acostumados a pensar. \textelp{} Se a consciência do \index{Wigner, Eugene!amigo de Wigner, paradoxo do}amigo de Wigner\index{Wigner, Eugene} n\~{a}o difere em essência da consciência de Wigner, se for sempre uma consciência causando o colapso da fun\c{c}\~{a}o de onda, n\~{a}o h\'{a} paradoxo. \citep[p.~536]{goswami2003QMtextbook}.
\end{quote}

Essa proposta de solu\c{c}\~{a}o para a situa\c{c}\~{a}o elaborada por \citet{wigner1961mindbody}\index{Wigner, Eugene}, atrav\'{e}s do ``paradoxo do amigo'', \'{e} muito pr\'{o}xima da solu\c{c}\~{a}o proposta por \citet{bass1971mind}, como vimos anteriormente. Revisitando a situa\c{c}\~{a}o do gato\index{Schr\"{o}dinger, Erwin!paradoxo do gato} de \citet{schrodinger1935cat}\index{Schr\"{o}dinger, Erwin}, expandida por \citet{penrose1989emperors}, \citet[p.~390]{goswami1989idealistic} afirma que quest\~{o}es acerca da consciência do gato\index{Schr\"{o}dinger, Erwin!paradoxo do gato} ou a discrep\^{a}ncia entre os humanos de dentro e fora da caixa s\~{a}o dificuldades que acompanham a concep\c{c}\~{a}o dualista da no\c{c}\~{a}o de consciência.

No entanto, \citet[p.~537]{goswami2003QMtextbook} aponta uma dificuldade para essa solu\c{c}\~{a}o do problema da medi\c{c}\~{a}o: se admitirmos que a consciência, unitiva e transcendente, traz \`{a} atualidade manifesta alguns aspectos da sua pr\'{o}pria potencialidade transcendente, ela seria onipresente. No entanto, se aceitarmos tal uso do termo consciência, ela estaria sempre observando, de modo que caberia a pergunta: a que ponto uma medi\c{c}\~{a}o est\'{a} completa? Isto \'{e}, como poderia haver mais do que uma medi\c{c}\~{a}o se a consciência onipresente estaria continuamente medindo? Dessa forma, a simples introdu\c{c}\~{a}o da hip\'{o}tese de uma consciência onipresente como agente causal na medi\c{c}\~{a}o qu\^{a}ntica n\~{a}o resolveria o problema da medi\c{c}\~{a}o.

Na tentativa de resolver tal dificuldade, \citet[p.~537]{goswami2003QMtextbook} afirma que ``a medi\c{c}\~{a}o n\~{a}o est\'{a} completa sem a inclus\~{a}o da percep\c{c}\~{a}o autorreferencial mente-c\'{e}rebro'', o que implicaria numa circularidade causal na medida em que ``a percep\c{c}\~{a}o \'{e} necess\'{a}ria para completar a medi\c{c}\~{a}o, mas sem que uma medi\c{c}\~{a}o esteja completa, n\~{a}o h\'{a} percep\c{c}\~{a}o''. \citet[p.~99]{goswami1993selfaware} afirma que \'{e} dessa autorreferência que surge a percep\c{c}\~{a}o subjetiva, como um epifenômeno da experiência.

Tais ideias acerca do funcionamento autorreferencial entre mente-corpo teriam sido inspiradas na obra de Douglas \citet{hofstadter1979godelescherbach}. Resumidamente, \citet[p.~684--714]{hofstadter1979godelescherbach} considera que uma das caracter\'{i}sticas da autorreferência ---tal como apontada pela no\c{c}\~{a}o de incompletude de \citet{godel1967undecidable}\index{G\"{o}del, Kurt}--- seria a emergência de um n\'{i}vel que a transcenda; em sua terminologia, afirma que a autorreferência forma uma ``hierarquia entrela\c{c}ada'', da qual um ``n\'{i}vel inviolado'' emerge. Para \citet[p.~688]{hofstadter1979godelescherbach}, tais n\'{i}veis s\~{a}o hier\'{a}rquicos, de modo que o n\'{i}vel inviolado governa o que acontece no n\'{i}vel entrela\c{c}ado, mas o n\'{i}vel entrela\c{c}ado n\~{a}o pode afetar o n\'{i}vel inviolado.

Na terminologia de \citet[p.192]{goswami1993selfaware}, a consciência seria an\'{a}loga ao ``n\'{i}vel inviolado'', que governa o aparelho mente-corpo autorreferente, ou em ``hierarquia entrela\c{c}ada''. No entanto, pr\'{o}prio do n\'{i}vel inviolado, a defini\c{c}\~{a}o de consciência, para Goswami, fugiria aos crit\'{e}rios discursivos:

\begin{quote}
O que \'{e} a consciência? Podemos come\c{c}ar a discuss\~{a}o com o que n\~{a}o \'{e}. N\~{a}o \'{e} uma parte da dualidade mente-mat\'{e}ria, interno-externo. N\~{a}o \'{e} um objeto, embora objetos apare\c{c}am nela. Tem algo a ver com o subjetivo, o experienciador, o conhecedor de objetos. \textelp{} Porque a consciência \'{e} a base do ser, tudo mais, incluindo palavras, conceitos e met\'{a}foras, s\~{a}o secund\'{a}rios a ela. N\~{a}o podemos definir a consciência completamente com itens que s\~{a}o secund\'{a}rios a ela, acentuando o mist\'{e}rio. \citep[p.~14]{goswami2001viewofnature}.
\end{quote}

Porder-se-ia, talvez, delinear certa influência da filosofia platônica no pensamento de \citet[p.~14]{goswami2001viewofnature} acerca da (in)defini\c{c}\~{a}o do termo consciência na medida em que, para a ontologia de \citet[VI, \S 509d--511e]{plataorepublica}, a raz\~{a}o discursiva (do grego ``\textit{dian\'{o}ia}'') n\~{a}o seria suficiente para apreender os n\'{i}veis ontol\'{o}gicos mais elevados, tal como a suprema Ideia de Bem ou Sumo Bem. \citet{pereira1990introplatao} comenta esse aspecto da metaf\'{i}sica platônica da seguinte maneira:

\begin{quote}
\textelp{} o mundo vis\'{i}vel (\textit{horata} ou \textit{doxasta}) tem em primeiro lugar uma zona de eikones (``imagens'', ou, como outros preferem, ``ilus\~{a}o''). Num n\'{i}vel mais elevado, temos todos os seres vivos (\textit{zoa}) e objetos do mundo, conhecidos atrav\'{e}s de \textit{pistis} (f\'{e}). O mundo intelig\'{i}vel (\textit{noeta}) tem tamb\'{e}m dois sectores proporcionais a estes, o inferior e o superior, o primeiro apreendido atrav\'{e}s da \textit{dian\'{o}ia} (``entendimento'' ou ``raz\~{a}o discursiva'') e o segundo s\'{o} pela \textit{n\'{o}esis} (``inteligência'' ou ``raz\~{a}o intuitiva''). \citep[\textsc{xxix--xxx}]{pereira1990introplatao}.
\end{quote}

Em seu dicion\'{a}rio etimol\'{o}gico do vocabul\'{a}rio filos\'{o}fico grego, Ivan Gobry reitera essa ideia:

\begin{quote}
Esse termo [``\textit{dian\'{o}ia}''] tem sentido vago; indica habitualmente um modo de pensamento menos elevado que a \textit{n\'{o}esis}. Classicamente, a \textit{di\'{a}noia} \'{e} o conhecimento discursivo, por racioc\'{i}nio. Assim, em Plat\~{a}o, ela \'{e} o grau inferior da ciência, que recorre a conceitos em vez de contemplar diretamente as Essências (v. \textit{dialektik\'{e}}, \textit{psykh\'{e}}). \citep[p.~41]{gobry2007dicionariogrego}.
\end{quote}

Ademais, h\'{a}, na ontologia de \citet[VII, \S 519d-521b]{plataorepublica}, considera\c{c}\~{o}es que pressup\~{o}em a conex\~{a}o entre as no\c{c}\~{o}es de ``unidade'' e a ``Ideia de Bem'', o que atenua a possibilidade de um paralelo com a no\c{c}\~{a}o de consciência em Goswami.

Al\'{e}m da influência na filosofia grega, da mesma forma que Schr\"{o}dinger em seu pensamento tardio, o pensamento de Goswami \'{e} claramente influenciado por diversos aspectos da literatura m\'{i}stica, principalmente no que se refere \`{a} unidade com o n\'{i}vel ontol\'{o}gico mais elevado (a saber, a consciência unitiva):

\begin{quote}
Mas, dizem os s\'{a}bios espirituais, os descobridores da filosofia \index{monismo}monista idealista, embora n\~{a}o possamos defini-la, podemos sê-la, n\'{o}s somos ela. \'{E} nossa ignor\^{a}ncia que nos impede de ver nossa natureza original, nossa interconectividade com a fonte. \citep[p.~14]{goswami2001viewofnature}.
\end{quote}

As propostas de solu\c{c}\~{a}o para o problema do \index{dualismo}dualismo analisadas acima pressup\~{o}em o uso do referencial m\'{i}stico ---abertamente n\~{a}o circunscrito pelo discurso racional. Uma das principais dificuldades de utilizar a sabedoria do vasto oriente para compreender o uso da no\c{c}\~{a}o de consciência na mec\^{a}nica qu\^{a}ntica \'{e} que v\'{a}rias vertentes do pensamento indiano, tal como o Vedanta, pressup\~{o}em a experiência m\'{i}stica, isto \'{e}, parece fugir do escopo de investiga\c{c}\~{a}o circunscrita pelo discurso racional da ciência e pela filosofia ocidental.

Dessa forma, na medida em que fazem uso referencial do Vedanta, as solu\c{c}\~{o}es de \citet{bass1971mind} e \citet{goswami1989idealistic}, bem como o pensamento tardio de \citet{schro1964myview}, a despeito de sua plausibilidade, deveriam ser, no m\'{i}nimo, precedidas por uma discuss\~{a}o acerca da legitimidade do uso da literatura m\'{i}stica como referencial ontol\'{o}gico para as ciências emp\'{i}ricas, como a mec\^{a}nica qu\^{a}ntica ---o que n\~{a}o \'{e} do escopo desta discuss\~{a}o.

Essa seria uma das vantagens de ter filosofia processual de \citet{whitehead1928process} como pano de fundo filos\'{o}fico para as discuss\~{o}es da interpreta\c{c}\~{a}o da consciência causal\index{interpreta\c{c}\~{a}o da consciência!causal}, na medida em que a filosofia de processos tem aberto um frut\'{i}fero campo de investiga\c{c}\~{a}o para os estudos da consciência frente \`{a}s dificuldades da no\c{c}\~{a}o de consciência frente ao \index{dualismo}dualismo e sua rela\c{c}\~{a}o com a mec\^{a}nica qu\^{a}ntica, como apontam os estudos de \citet{eastman2003QMwhitehead}, \citet{epperson2004QMwhitehead}, \citet{stapp2007whiteheadQM}. Tratarei brevemente dessa investiga\c{c}\~{a}o no cap\'{i}tulo \ref{CapWhitehead}.

Para finalizar essa exposi\c{c}\~{a}o, refiro-me \`{a} proposta de \citet{Manousakis2006-MANFQT}, que oferece um modelo em que a teoria qu\^{a}ntica \'{e} fundada sob a base ontol\'{o}gica da consciência sem fazer referência ao pensamento indiano, mas, ainda assim, h\'{a} caracter\'{i}sticas m\'{i}sticas em sua base ontol\'{o}gica. Pode-se constatar diversos pontos em comum com a proposta de \citet{goswami1989idealistic}: para \citet[p.~800]{Manousakis2006-MANFQT}, a consciência teria car\'{a}ter unitivo e seria a base ontol\'{o}gica da realidade; haveria apenas uma \'{u}nica consciência, nomeada de ``fluxo Universal da consciência'', do qual emergiriam ``subfluxos'', como o ``fluxo individual da consciência''.

\section{Interpretando a interpreta\c{c}\~{a}o da consciência}

Aqui \'{e} importante que fique clara a divis\~{a}o da interpreta\c{c}\~{a}o da consciência em duas vertentes. N\~{a}o h\'{a} tal coisa como uma ``interpreta\c{c}\~{a}o da consciência \textit{simpliciter}''; isso deve ser qualificado. N\~{a}o fazê-lo \'{e} incorrer em um erro que, em parte, nos levou at\'{e} os abusos da f\'{i}sica qu\^{a}ntica por parte de uma literatura n\~{a}o-cient\'{i}fica. Ent\~{a}o vamos l\'{a}.

Em contraste \`{a} interpreta\c{c}\~{a}o da consciência causal\index{interpreta\c{c}\~{a}o da consciência!causal}, chamarei as propostas delineadas na se\c{c}\~{a}o \ref{sec:mistico} como ``interpreta\c{c}\~{a}o da consciência m\'{i}stica\index{interpreta\c{c}\~{a}o da consciência!m\'{i}stica}'', ainda que a motiva\c{c}\~{a}o possa ter sida metaf\'{i}sica. At\'{e} o presente, pouco se avan\c{c}ou no debate, e a interpreta\c{c}\~{a}o da consciência m\'{i}stica\index{interpreta\c{c}\~{a}o da consciência!m\'{i}stica} tamb\'{e}m acaba n\~{a}o decolando ---ainda que por motivos distintos daqueles enfrentados pela interpreta\c{c}\~{a}o da consciência causal\index{interpreta\c{c}\~{a}o da consciência!causal}. Talvez a maior dificuldade conceitual das interpreta\c{c}\~{o}es m\'{i}sticas da consciência \'{e} o misteriosismo que envolve a pr\'{o}pria no\c{c}\~{a}o de consciência, central n\~{a}o s\'{o} para o funcionamento da interpreta\c{c}\~{a}o, mas base ontol\'{o}gica de toda uma vis\~{a}o de mundo que depende desse conceito.

Para al\'{e}m dessa dificuldade da interpreta\c{c}\~{a}o da consciência m\'{i}stica\index{interpreta\c{c}\~{a}o da consciência!m\'{i}stica}, pode-se apontar ainda outra, que \'{e} a inadequa\c{c}\~{a}o emp\'{i}rica da mesma. \'{E} importante sempre lembrarmo-nos que as interpreta\c{c}\~{o}es ad mec\^{a}nica qu\^{a}ntica devem ser ---no m\'{i}nimo--- empiricamente adequadas. Isto \'{e}, apesar de que toda interpreta\c{c}\~{a}o v\'{a} al\'{e}m do que o formalismo e os experimentos dizem, nenhuma delas pode \textit{contradizer} os dados experimentais j\'{a} estabelecidos; pelo contr\'{a}rio, esses dados devem ser tomados como ponto de partida. N\~{a}o fosse esse o caso, ter\'{i}amos uma nova teoria f\'{i}sica que deveria poder ser testada experimentalmente ---o que n\~{a}o \'{e} o caso. Assim, se algu\'{e}m disser que observadores humanos podem \textit{alterar o resultado} das medi\c{c}\~{o}es ou dos padr\~{o}es estat\'{i}sticos por meio de sua mera ``vontade consciente'', essa atitude n\~{a}o chega a ser nem mesmo uma interpreta\c{c}\~{a}o da mec\^{a}nica qu\^{a}ntica. 

Infelizmente, essa pr\'{a}tica tem sido comum em atitudes que tentam passar-se por ``interpreta\c{c}\~{o}es'' da mec\^{a}nica qu\^{a}ntica, tais como ``coaches'' qu\^{a}nticos e ``curas'' qu\^{a}nticas.\footnote{~Novemente, ver \citet{osvaldo2011misticismoquantico}, \citet{fred2011coachquantico} e \citet{sandrofred2016misticismo} para um aprofundamento nesse t\'{o}pico.} Como vimos, interpretar a mec\^{a}nica qu\^{a}ntica \'{e} uma atitude filosoficamente e metodologicamente permissiva na medida em que permite-nos ir al\'{e}m do que a f\'{i}sica nos diz ---mas n\~{a}o t\~{a}o permissiva a ponto de contradizê-la. Como um exemplo, digamos que Martha adora gatos, de modo que ela sempre ter\'{a} uma preferência pelo resultado no qual seu gato seja alimentado pelo experimento de Schr\"{o}dinger. Essa preferência n\~{a}o deve poder afetar os resultados. Dito de outro modo, se for constatado que a preferência altera o resultado da medida, ent\~{a}o a mec\^{a}nica qu\^{a}ntica seria falseada. No entanto, \'{e} justamente isso que n\~{a}o acontece.

O fato da interpreta\c{c}\~{a}o da consciência m\'{i}stica\index{interpreta\c{c}\~{a}o da consciência!m\'{i}stica} n\~{a}o ser suficientemente clara sobre essas quest\~{o}es (\textit{viz.}, que nenhuma atitude puramente mental poderia alterar um resultado experimental) evidencia sua inadequa\c{c}\~{a}o emp\'{i}rica. Portanto, podemos consider\'{a}-la descartada do rol de interpreta\c{c}\~{o}es poss\'{i}veis da interpreta\c{c}\~{a}o (\textit{sic}!) da consciência.

As interpreta\c{c}\~{o}es da consciência causal (tamb\'{e}m chamadas de ``interpreta\c{c}\~{o}es subjetivistas''), por outro lado, sofrem com outros tipos de problema. Um deles \'{e} o problema da vagueza, e \citet{becker2018} descreve-o de maneira particularmente eloquente:

\begin{quote}
    Afirmar que a consciência colapsa o estado qu\^{a}ntico resolve o problema da medi\c{c}\~{a}o, mas apenas ao pre\c{c}o da introdu\c{c}\~{a}o de novos problemas. Como a consciência poderia causar o colapso do estado qu\^{a}ntico? Como o colapso do estado qu\^{a}ntico viola a Equa\c{c}\~{a}o de Schr\"{o}dinger\index{Schr\"{o}dinger, Erwin!Equa\c{c}\~{a}o de Schr\"{o}dinger}, isso significa que a consciência tem a capacidade de suspender ou alterar temporariamente as leis da natureza? Como isso poderia ser verdade? E, afinal, o que \'{e} consciência? Quem a possui? Um chimpanz\'{e} pode colapsar um estado qu\^{a}ntico? E quanto a um cachorro? Uma pulga? ``Resolver'' o problema da medi\c{c}\~{a}o abrindo a caixa de Pandora dos paradoxos associados \`{a} consciência \'{e} uma jogada desesperada \textelp{}. \citep[p.~75]{becker2018}.
\end{quote}

Outro problema \'{e} a sua associa\c{c}\~{a}o autom\'{a}tica com o dualismo ---o que por si s\'{o} n\~{a}o \'{e} um problema, mas herda um ônus pesado \textit{viz.}, o de solucionar o problema mente-corpo. Assim, ainda que a interpreta\c{c}\~{a}o da consciência causal\index{interpreta\c{c}\~{a}o da consciência!causal} mantenha-se legitimamente consistente com os resultados experimentais \citep[ver][]{BarOas2016Consc}, sua parte conceitual parece ter chegado a um impasse. Apresentarei uma alternativa para novos estudos dentro da interpreta\c{c}\~{a}o da consciência causal\index{interpreta\c{c}\~{a}o da consciência!causal} no cap\'{i}tulo \ref{CapWhitehead}. Mas, antes, farei um pequeno desvio em uma paisagem panor\^{a}mica com outras interpreta\c{c}\~{o}es. Desta maneira, poderemos apreciar onde a interpreta\c{c}\~{a}o da consciência causal\index{interpreta\c{c}\~{a}o da consciência!causal} se situa diante suas rivais.

\chapter{A paisagem ao redor}\label{CapPaisagem}

Existem in\'{u}meras atitudes frente ao problema da medi\c{c}\~{a}o, e este cap\'{i}tulo re\'{u}ne algumas delas. Fa\c{c}o isso para enfatizar como o problema das interpreta\c{c}\~{o}es da mec\^{a}nica qu\^{a}ntica n\~{a}o \'{e} abordado de forma unilateral pela literatura. A interpreta\c{c}\~{a}o da consciência causal, longe de ser necess\'{a}ria, \'{e} apenas \textit{apenas mais uma} dentre as diversas outras interpreta\c{c}\~{o}es poss\'{i}veis da mec\^{a}nica qu\^{a}ntica.

Esse \'{e} o famoso problema da subdetermina\c{c}\~{a}o, que apresenta s\'{e}rios problemas para a ado\c{c}\~{a}o de uma atitude em dire\c{c}\~{a}o ao \index{realismo cient\'{i}fico}realismo cient\'{i}fico \citep[ver][]{frenchsaatsi2020}. Para entender como isso \'{e} um problema, vejamos uma t\'{i}pica defini\c{c}\~{a}o do que \'{e} o realismo cient\'{i}fico:\footnote{~Ver tamb\'{e}m \citet{chakravartty-vanfraassen2021}.}

\begin{quote}
    Talvez seja apenas um pequeno exagero dizer que o realismo cient\'{i}fico \'{e} caracterizado de maneira diferente por cada uma das pessoas que o discutem, o que apresenta um desafio para quem espera compreender o que seja tal coisa. Felizmente, subjacente \`{a}s muitas qualifica\c{c}\~{o}es variantes da posi\c{c}\~{a}o, h\'{a} um n\'{u}cleo comum de ideias, exemplificado por uma atitude epistêmica positiva em rela\c{c}\~{a}o aos resultados da investiga\c{c}\~{a}o cient\'{i}fica, abrangendo tanto aspectos observ\'{a}veis quanto inobserv\'{a}veis do mundo. \textelp{} O que todas essas abordagens [realistas] têm em comum \'{e} um compromisso com a ideia de que nossas melhores teorias têm um certo ``status'' epistêmico: elas proporcionam conhecimento sobre aspectos do mundo, incluindo aspectos inobserv\'{a}veis. \citep[se\c{c}\~{a}o~1.1]{sep-scientific-realism}.
\end{quote}

Parte do problema para o realismo cient\'{i}fico reside no fato de que n\~{a}o h\'{a} raz\~{o}es definitivas para dizermos que \textit{apenas uma}, entre as v\'{a}rias interpreta\c{c}\~{o}es, com comprometimentos ontol\'{o}gicos diferentes, de fato descreve aspectos \textit{do mundo}. Isso envolve aspectos da no\c{c}\~{a}o de ``verdade''. Em particular, envolve a no\c{c}\~{a}o \textit{correspondentista} da verdade ---segundo a qual, uma interpreta\c{c}\~{a}o \'{e} verdadeira \textit{porque} corresponde aos fatos do mundo. Mas os dados experimentais n\~{a}o determinam qual interpreta\c{c}\~{a}o (se alguma) \'{e} a verdadeira nesse sentido, e isso gera um problema para o realismo, \textit{viz.}, o de escolha de interpreta\c{c}\~{o}es. Assim, ainda que a atitude do realismo cient\'{i}fico seja favor\'{a}vel \`{a} ideia de que nossas melhores teorias cient\'{i}ficas (como a mec\^{a}nica qu\^{a}ntica) devam nos fornecer uma vis\~{a}o verdadeira do mundo, por hora n\'{o}s n\~{a}o sabemos ---n\~{a}o temos como saber!--- \textit{qual} seria essa tal vis\~{a}o de mundo. Isso se deve, em parte por termos \textit{diversas} op\c{c}\~{o}es \`{a} mesa ---o problema da subdetermina\c{c}\~{a}o---, sendo a interpreta\c{c}\~{a}o da consciência causal \textit{uma} delas. A seguir apresento brevemente algumas outras que comp\~{o}em essa paisagem (subdeterminada) de interpreta\c{c}\~{o}es qu\^{a}nticas.

Como toda sele\c{c}\~{a}o, tive que deixar muita coisa de fora. Ent\~{a}o, antes de prosseguirmos, deixe-me esclarecer os crit\'{e}rios que utilizei. A sele\c{c}\~{a}o foi feita com base na repercuss\~{a}o que tiveram na literatura, e as trago apenas a t\'{i}tulo de amostragem. Deve ficar claro que tal n\~{a}o \'{e} meu prop\'{o}sito aprofundar a discuss\~{a}o acerca de todas as interpreta\c{c}\~{o}es selecionadas adiante. Cada uma delas mereceria um estudo \`{a} parte para que se pudesse apresentar sua riqueza e complexidade; limito-me a apresent\'{a}-las muito brevemente, a t\'{i}tulo de amostragem, como interpreta\c{c}\~{o}es poss\'{i}veis dentre as mais influentes e/ou populares. Dessa forma, me limito a uma abordagem bastante resumida e superficial, indicando bibliografias que possam aprofundar a discuss\~{a}o. 

N\~{a}o posso enfatizar o bastante o fato de que mapear \textit{todas} as interpreta\c{c}\~{o}es existentes em um s\'{o} estudo seria uma tarefa herc\'{u}lea para al\'{e}m dos prop\'{o}sitos deste livro. Portanto, devo alertar que algumas omiss\~{o}es gritantes invariavelmente ocorrer\~{a}o. Para mencionar algumas delas: a abordagem ``logos'' formulada por \citet{derondemassri2019synthese} \citep[para uma breve discuss\~{a}o e referências, ver][]{arroyoarenhart2023foda}, as diversas interpreta\c{c}\~{o}es modais \citep[ver][para apresenta\c{c}\~{a}o e referências]{sep-qm-modal}, e o ``QBism'' ---que outrora j\'{a} foi chamado de ``Bayesianismo qu\^{a}ntico'' (do inglês ``Quantum Bayesianism'', abreviado ``QBism'') devido \`{a}s probabilidades bayesianas, mas que hoje considera-se um nome pr\'{o}prio dessa interpreta\c{c}\~{a}o \citep[ver][para apresenta\c{c}\~{a}o e referências]{sep-quantum-bayesian}. Ainda assim, h\'{a} outras que nem sequer foram mencionadas.\footnote{~Ver, por exemplo, \citet{pessoa1992MPatualizado}.}

As leituras selecionadas s\~{a}o, cronologicamente: a interpreta\c{c}\~{a}o estat\'{i}stica que, assim como a interpreta\c{c}\~{a}o de Copenhague, tamb\'{e}m \'{e} amplamente aceita pela comunidade cient\'{i}fica e frequentemente utilizada em diversos livros did\'{a}ticos de mec\^{a}nica qu\^{a}ntica \citep[p.~25, nota 3]{pessoa2003conceitos}; a interpreta\c{c}\~{a}o de David Bohm\index{Bohm, David}, por se tratar de uma abordagem heterodoxa bastante completa do ponto de vista conceitual \citep{freirejr2015}; a interpreta\c{c}\~{a}o dos estados relativos de \citet{everett1957relative}\index{Everett, Hugh}, por ser uma das abordagens heterodoxas mais populares; a interpreta\c{c}\~{a}o dos estados latentes, abordagem cr\'{i}tica de \citet{margenau1963measurement} frente ao conceito de ``colapso'' na medi\c{c}\~{a}o qu\^{a}ntica, bem como sua atitude cr\'{i}tica frente \`{a}s interpreta\c{c}\~{o}es subjetivas, que \citet{Jammer1974} destaca como influente; a abordagem do colapso espont\^{a}neo de \citet*[, abreviada como ``\textsc{grw}'']{grw}, por tamb\'{e}m ser uma das atitudes com colapso mais bem aceitas na comunidade cient\'{i}fica contempor\^{a}nea \citep{albert1992quantum}.

Com exce\c{c}\~{a}o das formula\c{c}\~{o}es \textsc{grw} e ``estat\'{i}stica'', todas as outras atitudes destacadas adiante negam a validade do colapso, se enquadrando nas chamadas ``teorias sem colapso''.

\section{A interpreta\c{c}\~{a}o estat\'{i}stica}

Iniciarei a discuss\~{a}o a partir da interpreta\c{c}\~{a}o estat\'{i}stica, tamb\'{e}m conhecida como ``interpreta\c{c}\~{a}o dos coletivos estat\'{i}sticos'' ou ``interpreta\c{c}\~{a}o dos \textit{ensemble}''. \citet[p.~360]{Ballantine1970ensemble} distingue as interpreta\c{c}\~{o}es da teoria qu\^{a}ntica em dois grupos maiores: as interpreta\c{c}\~{o}es nas quais a mec\^{a}nica qu\^{a}ntica provê uma descri\c{c}\~{a}o completa e exaustiva sobre sistemas individuais e as interpreta\c{c}\~{o}es nas quais a mec\^{a}nica qu\^{a}ntica provê uma descri\c{c}\~{a}o completa e exaustiva sobre sistemas coletivos. A mesma oposi\c{c}\~{a}o \'{e} feita por \citet[p.~440]{Jammer1974}. As interpreta\c{c}\~{o}es do primeiro tipo s\~{a}o consideradas interpreta\c{c}\~{o}es ortodoxas e, as do segundo tipo, s\~{a}o consideradas interpreta\c{c}\~{o}es estat\'{i}sticas. Isso corresponderia \`{a} nega\c{c}\~{a}o premissa $\alpha$ do problema da medi\c{c}\~{a}o (conforme exposto na p\'{a}gina \pageref{trilema:A}). A no\c{c}\~{a}o de ``coletivos estat\'{i}sticos'' ou ``\textit{ensemble}'' remete a um grupo imagin\'{a}rio de diversos sistemas com a mesma estrutura macrosc\'{o}pica e o mesmo sistema microsc\'{o}pico a ser medido.\footnote{~A no\c{c}\~{a}o de ``o mesmo'' experimento \'{e} questionada por \citet*[p.~1]{debarros-holik-krause2017}; para esses autores, ``\textelp{} quando repetimos um experimento, na realidade estamos realizando um experimento que mede uma propriedade que \'{e} indistingu\'{i}vel da primeira, mas n\~{a}o a mesma.''}

Primeiramente, \'{e} relevante destacar a maneira como \citet{Ballantine1970ensemble} define a no\c{c}\~{a}o de interpreta\c{c}\~{a}o ortodoxa da teoria qu\^{a}ntica com um significado distinto e mais abrangente do que aquele que utilizo ao longo deste livro. At\'{e} aqui, a no\c{c}\~{a}o de ``ortodoxia'' tem correspondência exclusiva com a formula\c{c}\~{a}o de Copenhague e suas liga\c{c}\~{o}es com o empirismo l\'{o}gico. Segundo a formula\c{c}\~{a}o de Ballantine, no entanto, at\'{e} mesmo a interpreta\c{c}\~{a}o de \citet{vNeum1955mathematical}\index{Neumann, John von} seria entendida como uma atitude ortodoxa. De fato, \citet[p.~360]{Ballantine1970ensemble} considera que tanto a ``interpreta\c{c}\~{a}o de Princeton'' ---da qual von Neumann\index{Neumann, John von} seria o fundador--- quanto a interpreta\c{c}\~{a}o de Copenhague da mec\^{a}nica qu\^{a}ntica ``\textelp{} reivindicam ortodoxia''.

No entanto, como vimos anteriormente, essas duas interpreta\c{c}\~{o}es ditas ortodoxas têm suas dificuldades no \^{a}mbito filos\'{o}fico. Seja a necessidade de uma ontologia para abarcar a no\c{c}\~{a}o de um observador para causar a medi\c{c}\~{a}o na interpreta\c{c}\~{a}o de Princeton, ou a prioridade ontol\'{o}gica dos objetos cl\'{a}ssicos na medi\c{c}\~{a}o da interpreta\c{c}\~{a}o de Copenhague.

O fato de evitar os paradoxos e os problemas filos\'{o}ficos da teoria qu\^{a}ntica seria uma das três motiva\c{c}\~{o}es principais que \citet[p.~262--264]{Home1992-HOMEIO} destacam para a ado\c{c}\~{a}o das interpreta\c{c}\~{o}es estat\'{i}sticas. Proposta por Einstein em 1927, na ocasi\~{a}o da vig\'{e}sima terceira Conferência de Solvay, tal interpreta\c{c}\~{a}o fora formulada justamente para evitar quase todas as dificuldades filos\'{o}ficas discutidas neste livro--- qui\c{c}\'{a} todas as dificuldades filos\'{o}ficas da mec\^{a}nica qu\^{a}ntica. Isso porque as dificuldades surgem quando os sistemas qu\^{a}nticos s\~{a}o tratados como sistemas individuais, e n\~{a}o como apanhados estat\'{i}sticos. Outra motiva\c{c}\~{a}o destacada por \citet[p.~262]{Home1992-HOMEIO} seria a de erigir a f\'{i}sica sobre uma ontologia realista-objetivista, isto \'{e}, manter na mec\^{a}nica qu\^{a}ntica nossas percep\c{c}\~{o}es intuitivas acerca da realidade que nos cerca. 

Como vimos no cap\'{i}tulo \ref{CapEPR}, essa motiva\c{c}\~{a}o seria compartilhada por Einstein. Talvez o fato da interpreta\c{c}\~{a}o de Copenhague oferecer uma vis\~{a}o contraintuitiva do mundo \`{a} nossa volta seria um dos motivos para que Einstein tivesse tantas obje\c{c}\~{o}es a essa interpreta\c{c}\~{a}o. Para ilustrar esse ponto, \citet{putnam2005lookqm} relata um di\'{a}logo que teve com Einstein\index{Einstein, Albert}; em par\'{a}frase, ele teria afirmado algo como ``olha, eu n\~{a}o acredito que quando n\~{a}o estou no meu quarto minha cama se espalha por todo o cômodo, e sempre que eu abro a porta e entro ela salta novamente para o canto'' (Einstein \textit{apud} \cite[p.~624]{putnam2005lookqm}). Essa quest\~{a}o. como visto no cap\'{i}tulo \ref{CapEPR}, resulta da tens\~{a}o entre uma \index{ontologia!tradicional}$\mathscr{O}_T$ assumida previamente e uma \index{ontologia!naturalizada}$\mathscr{O}_N$ acompanhada pela mec\^{a}nica qu\^{a}ntica.

No entanto, \citet[p.~968]{fine1990einsteinensemble} declara que ``at\'{e} onde eu pude descobrir \textelp{} Einstein n\~{a}o oferece em lugar algum uma descri\c{c}\~{a}o detalhada da \textelp{} interpreta\c{c}\~{a}o estat\'{i}stica''. Ainda assim, a despeito da falta de uma formula\c{c}\~{a}o textual detalhada, diversos f\'{i}sicos teriam utilizado as ideias de Einstein sobre \textit{ensembles} para criar propostas estat\'{i}sticas para a mec\^{a}nica qu\^{a}ntica.

H\'{a} uma grande variedade de abordagens estat\'{i}sticas para a interpreta\c{c}\~{a}o da mec\^{a}nica qu\^{a}ntica, com diferentes nomes e especificidades, e n\~{a}o existe consenso sobre exatamente qual interpreta\c{c}\~{a}o Einstein teria endossado. Contudo, como procurei enfatizar no cap\'{i}tulo \ref{CapEPR}, o comprometimento ontol\'{o}gico de Einstein com uma realidade independente acaba por sugerir que ele endossaria um tipo de interpreta\c{c}\~{a}o na qual todas as vari\'{a}veis, em todos os instantes, possuem valores pass\'{i}veis de serem revelados por meio de medi\c{c}\~{o}es, de modo que todo indeterminismo se dê pelo desconhecimento de algumas vari\'{a}veis envolvidas no processo de medi\c{c}\~{a}o. Tais vari\'{a}veis seriam as vari\'{a}veis ocultas,\footnote{~Para um estudo detalhado das teorias de vari\'{a}veis ocultas, ver \citet{belinfante1973hidden}.} isto \'{e}, s\~{a}o criptodeterministas no sentido de um indeterminismo epistemol\'{o}gico subjacente a um determinismo ontol\'{o}gico.

Para \citet{Ballantine1970ensemble}, essa seria a forma mais natural de pensar a posi\c{c}\~{a}o einsteiniana sobre \textit{ensembles}. Essa posi\c{c}\~{a}o se coaduna com evidência textual, que procurei destacar, do comprometimento ontol\'{o}gico com uma realidade independente e pr\'{e}-existente na obra de Einstein. \citet[p.~263]{Home1992-HOMEIO}, \citet[p.~7]{bunge1967QMreality} e \citet[p.~43]{fine1986shaky} v\~{a}o al\'{e}m e apontam para o fato de que, para muitos, essa interpreta\c{c}\~{a}o seria \textit{a} interpreta\c{c}\~{a}o estat\'{i}stica.

Destaco, no entanto, uma defini\c{c}\~{a}o m\'{i}nima para a atitude estat\'{i}stica, presente em todas as interpreta\c{c}\~{o}es estat\'{i}sticas, formulada por \citet[p.~76]{gibbins1987particles}. De acordo com tal defini\c{c}\~{a}o, uma interpreta\c{c}\~{a}o estat\'{i}stica considera que uma fun\c{c}\~{a}o de onda representa um \textit{ensemble}, isto \'{e}, que a mec\^{a}nica qu\^{a}ntica trataria exclusivamente das estat\'{i}sticas dos resultados obtidos por uma numerosa sequência de medi\c{c}\~{o}es simult\^{a}neas de sistemas coletivos (chamados de ``\textit{ensemble}''), e n\~{a}o sobre quaisquer propriedades dos objetos f\'{i}sicos. Dessa forma, a atitude estat\'{i}stica contrasta com a atitude ortodoxa, para a qual a fun\c{c}\~{a}o de onda forneceria uma descri\c{c}\~{a}o completa de um sistema individual.

De acordo com \citet{park1973contradictionQM}, o conceito de colapso tamb\'{e}m \'{e} rejeitado por essa interpreta\c{c}\~{a}o. Ent\~{a}o deve ficar claro que, para a interpreta\c{c}\~{a}o estat\'{i}stica, o problema da medi\c{c}\~{a}o \textit{n\~{a}o existe}. Para exemplificar a atitude m\'{i}nima da interpreta\c{c}\~{a}o estat\'{i}stica frente \`{a} situa\c{c}\~{a}o do gato\index{Schr\"{o}dinger, Erwin!paradoxo do gato} de \citet{schrodinger1935cat}\index{Schr\"{o}dinger, Erwin}, \citet[p.~22]{ross-boney1974dicegod} escreve que ``Em qualquer experimento, aproximadamente metade dos gatos\index{Schr\"{o}dinger, Erwin!paradoxo do gato} est\~{a}o [alimentados] e metade est\~{a}o [em jejum]''.

Isto \'{e}, todo debate filos\'{o}fico em torno do conceito de medi\c{c}\~{a}o \'{e} evitado. Se trata de uma interpreta\c{c}\~{a}o puramente funcional da teoria qu\^{a}ntica, evitando grande parte dos seus problemas filos\'{o}ficos. Por esse motivo, recebe grande aten\c{c}\~{a}o por parte da comunidade cient\'{i}fica. Da forma como \citet[p.~119]{Jammer1974} descreve, tal interpreta\c{c}\~{a}o seria ``mais palat\'{a}vel para a maioria dos f\'{i}sicos''. Isto \'{e}, tal interpreta\c{c}\~{a}o evita diversos problemas filos\'{o}ficos ao pre\c{c}o de considerar a ciência como um instrumento computacional, e n\~{a}o uma descri\c{c}\~{a}o da realidade objetiva.

Essa concep\c{c}\~{a}o instrumentalista, de acordo com o que vimos anteriormente, parece conflitar diretamente com a concep\c{c}\~{a}o de ciência do pr\'{o}prio \citet[p.~667]{einstein1949remarks}, segundo o qual, reitero, uma teoria f\'{i}sica deveria fornecer ``\textelp{} a descri\c{c}\~{a}o completa de qualquer situa\c{c}\~{a}o real (e individual, que supostamente existe independentemente de qualquer ato de observa\c{c}\~{a}o ou comprova\c{c}\~{a}o)''. Desse modo, parece mais seguro afirmar que as interpreta\c{c}\~{o}es estat\'{i}sticas n\~{a}o solucionam os problemas filos\'{o}ficos nos fundamentos da interpreta\c{c}\~{a}o da teoria qu\^{a}ntica, mas somente evitam-nos para fins heur\'{i}sticos.

\section{A interpreta\c{c}\~{a}o das vari\'{a}veis ocultas}

A interpreta\c{c}\~{a}o de Bohm\index{Bohm, David} (teoria da \textit{onda piloto}, \textit{mec\^{a}nica Bohmiana}, ou teoria de \textit{de Brolie--Bohm})\footnote{~N\~{a}o \'{e} claro que todos esses termos refiram \`{a} mesma coisa
\citep[para uma discord\^{a}ncia a respeito disso, ver, por exemplo,][]{sole2012,matarese2023book,matarese2023foop}; no entanto, por quest\~{o}es did\'{a}ticas, vou trat\'{a}-las como intercambi\'{a}veis.} \'{e} uma solu\c{c}\~{a}o ao problema da medi\c{c}\~{a}o, que, ao inv\'{e}s de negar a assun\c{c}\~{a}o da validade universal das equa\c{c}\~{o}es lineares da mec\^{a}nica qu\^{a}ntica (\textit{e.g.}, postulando o colapso), nega que o estado qu\^{a}ntico fornece uma descri\c{c}\~{a}o f\'{i}sica \textit{completa} sobre as propriedades que objetos f\'{i}sicos possuem. Isso tamb\'{e}m corresponderia \`{a} nega\c{c}\~{a}o premissa $\alpha$ do problema da medi\c{c}\~{a}o (conforme exposto na p\'{a}gina \pageref{trilema:A}). Vejamos brevemente como essa solu\c{c}\~{a}o opera, tanto no n\'{i}vel din\^{a}mico quanto ontol\'{o}gico. 

A teoria da onda piloto \'{e} uma teoria sobre part\'{i}culas bem-localizadas se movendo no espa\c{c}o. Essas part\'{i}culas s\~{a}o guiadas por uma onda de abrangência universal. A onda, por assim dizer, ``pilota'' a part\'{i}cula. Isso nos d\'{a} uma boa ideia sobre a ontologia, e \'{e} por isso que \citet{albert1996} e \citet{norsen2017}, respectivamente, dizem que ``[n]a teoria de Bohm\index{Bohm, David} \textelp{} o mundo consistir\'{a} em exatamente dois objetos f\'{i}sicos. Um deles \'{e} o estado qu\^{a}ntico universal e o outro \'{e} a part\'{i}cula universal'' \citep[p.~278]{albert1996}; ``\textelp{} um \'{u}nico el\'{e}tron, de acordo com a teoria da onda piloto, n\~{a}o \'{e} \textit{uma} \'{u}nica coisa, mas \textit{duas}---uma onda e uma part\'{i}cula (literal, pontual) cujo movimento \'{e} controlado pela onda.'' \citep[p.~178]{norsen2017}. De acordo com \citet*[p.~124]{freirepatybarros2000recbohm}, a mec\^{a}nica bohmiana apresenta ``os mesmos resultados j\'{a} obtidos pela teoria qu\^{a}ntica n\~{a}o relativista, mas em uma interpreta\c{c}\~{a}o distinta daquela usual, a da complementaridade'', distin\c{c}\~{a}o essa que residiria ``na recupera\c{c}\~{a}o de certas premissas epistemol\'{o}gicas pr\'{o}prias da f\'{i}sica cl\'{a}ssica, como o determinismo''. Ainda assim, n\~{a}o se trata de uma recupera\c{c}\~{a}o do quadro cl\'{a}ssico, na medida em que Bohm\index{Bohm, David} propunha a ideia de um chamado ``potencial qu\^{a}ntico'', que seria respons\'{a}vel por efeitos essencialmente qu\^{a}nticos, como a n\~{a}o localidade.

Aqui vale a pena mencionar uma breve digress\~{a}o hist\'{o}rica e conceitual acerca dessa interpreta\c{c}\~{a}o.\footnote{~Por raz\~{o}es de espa\c{c}o, este cap\'{i}tulo n\~{a}o trata da hist\'{o}ria que envolve a vida e obra de David Bohm\index{Bohm, David}. No entanto, recomendo fortemente que ela seja conhecida por quem quer que se interesse pelos fundamentos da mec\^{a}nica qu\^{a}ntica. Bons lugares para come\c{c}ar a fazê-lo s\~{a}o as obras de hist\'{o}ria dos fundamentos da mec\^{a}nica qu\^{a}ntica de \citet{freirejr1999,freirejr2005,freirejr2015,freirejr2019}, mas tamb\'{e}m \citet*{cushing1994,freirejr-etal-1994,becker2018}.} Assim como a interpreta\c{c}\~{a}o estat\'{i}stica, a interpreta\c{c}\~{a}o de Bohm\index{Bohm, David} postula vari\'{a}veis ocultas, \textit{viz.}, as vari\'{a}veis de \textit{posi\c{c}\~{a}o} ---conforme veremos a seguir. Historicamente, ainda que Bohm\index{Bohm, David} tenha sido influenciado diretamente por Einstein na formula\c{c}\~{a}o de sua interpreta\c{c}\~{a}o \citep[ver][\S~5]{becker2018}, o fato de que as vari\'{a}veis ocultas da interpreta\c{c}\~{a}o bohmiana s\~{a}o n\~{a}o locais jamais agradaria Einstein\index{Einstein, Albert}. Como vimos no cap\'{i}tulo \ref{CapEPR}, a n\~{a}o localidade e a n\~{a}o separabilidade eram caracter\'{i}sticas f\'{i}sicas e filos\'{o}ficas as quais Einstein sempre se opôs. Ainda assim, Einstein jamais se posicionou contrariamente \`{a} interpreta\c{c}\~{a}o de Bohm\index{Bohm, David}.

A interpreta\c{c}\~{a}o de Bohm\index{Bohm, David} \'{e} essencialmente determinista, introduzindo vari\'{a}veis ocultas n\~{a}o locais; assim, como observa \citet[p.~7]{freirejr2005exilebohm}, ``os el\'{e}trons de Bohm\index{Bohm, David} tem posi\c{c}\~{o}es e momentos bem definidos; assim, eles têm trajet\'{o}rias cont\'{i}nuas e bem definidas''. De acordo com \citet[p.~5]{cushing1996causalQM}, n\~{a}o h\'{a} um ``problema da medi\c{c}\~{a}o'', na medida em que o colapso n\~{a}o \'{e} admitido; assim, ``uma part\'{i}cula sempre tem uma posi\c{c}\~{a}o definida entre medi\c{c}\~{o}es. N\~{a}o h\'{a} superposi\c{c}\~{a}o de propriedades e `medi\c{c}\~{a}o' \textelp{} \'{e} uma tentativa de descobrir sua posi\c{c}\~{a}o atual''. Vamos aprofundar um pouco esse ponto.

Do ponto de vista din\^{a}mico, como as part\'{i}culas sempre têm posi\c{c}\~{o}es definidas, o estado qu\^{a}ntico n\~{a}o expressa tudo o que h\'{a} para saber sobre el\'{e}trons. O estado qu\^{a}ntico descreve o estado da onda que guia a part\'{i}cula em v\'{a}rios estados, \textit{i.e.}, em v\'{a}rias posi\c{c}\~{o}es. A equa\c{c}\~{a}o usual que rege o processo 2 (\textit{i.e.}, a Equa\c{c}\~{a}o de Schr\"{o}dinger\index{Schr\"{o}dinger, Erwin!Equa\c{c}\~{a}o de Schr\"{o}dinger}) descreve o movimento dessa onda em termos da evolu\c{c}\~{a}o do estado qu\^{a}ntico. J\'{a} as part\'{i}culas possuem uma equa\c{c}\~{a}o pr\'{o}pria, chamada ``f\'{o}rmula-guia''\footnote{~Do inglês ``\textit{guiding equation}'', tamb\'{e}m referida sob a nomenclatura de ``equa\c{c}\~{a}o onda-guia''.}, que evolui em termos do \textit{potencial qu\^{a}ntico}. Esse potencial refere \`{a} posi\c{c}\~{a}o de uma part\'{i}cula arbitr\'{a}ria.

Ambas as equa\c{c}\~{o}es din\^{a}micas s\~{a}o deterministas, ent\~{a}o uma primeira quest\~{a}o que surge de imediato \'{e} a seguinte: como capturar as previs\~{o}es estat\'{i}sticas da mec\^{a}nica qu\^{a}ntica? Essas quest\~{o}es s\~{a}o sens\'{i}veis, pois as previs\~{o}es estat\'{i}sticas da mec\^{a}nica qu\^{a}ntica s\~{a}o grande parte do sucesso experimental da teoria. A resposta da teoria da onda piloto est\'{a} em nossa incapacidade de saber tudo sobre as posi\c{c}\~{o}es das part\'{i}culas: as condi\c{c}\~{o}es iniciais s\~{a}o inacess\'{i}veis, ou---como frequentemente coloca-se---s\~{a}o \textit{vari\'{a}veis ocultas}. As probabilidades s\~{a}o meramente epistêmicas. Isto \'{e}, expressam as limita\c{c}\~{o}es de nosso conhecimento sobre onde est\'{a} o el\'{e}tron: ``[n]\~{a}o sabemos onde o el\'{e}tron vai acabar porque n\~{a}o sabemos onde ele come\c{c}ou'' \citep[p.~197]{barrett2019}.

Vamos recuperar o experimento da fenda dupla para ilustrar o que foi dito (figura \ref{fig:DF}, p\'{a}gina \pageref{fig:DF}). Quando uma part\'{i}cula \'{e} disparada da fonte em dire\c{c}\~{a}o \`{a} tela, a onda piloto \'{e} distribu\'{i}da por \textit{ambos as fendas, $F_1$ e $F_2$}. Cada part\'{i}cula, no entanto, passa por \textit{exatamente apenas uma fenda}, $F_1$ ou $F_2$. \`{A} medida que a part\'{i}cula passa por uma das fendas, a onda piloto \'{e} afetada e modifica seu comportamento, influenciando sua trajet\'{o}ria. A interferência (padr\~{a}o estat\'{i}stico 2) ocorre justamente porque o movimento do el\'{e}tron-enquanto-part\'{i}cula \'{e} \textit{influenciado} por uma onda associada, ou pelo el\'{e}tron-enquanto-onda. 

Essa mesma interferência desaparece (padr\~{a}o estat\'{i}stico 1) porque a medi\c{c}\~{a}o influencia o comportamento da onda. Na detec\c{c}\~{a}o da passagem da part\'{i}cula por um dos caminhos, a onda piloto \'{e} perturbada de tal forma que o termo de interferência \'{e} destru\'{i}do. Isso ocorre porque a informa\c{c}\~{a}o sobre o caminho atrav\'{e}s da qual a part\'{i}cula passou influencia a onda piloto de uma maneira que a interfere de forma destrutiva com a onda que passou pelo outro caminho.

\`{A} primeira vista, tudo isso n\~{a}o parece t\~{a}o distante da solu\c{c}\~{a}o do colapso. No entanto, a quest\~{a}o ontol\'{o}gica \'{e} de suma import\^{a}ncia para a compreens\~{a}o da solu\c{c}\~{a}o oferecida pela teoria da onda piloto. Nas palavras de \citet[p.~128, ênfase original]{bell2004-1981}, ``\textit{ningu\'{e}m pode entender essa teoria at\'{e} que haja a disposi\c{c}\~{a}o de pensar no estado qu\^{a}ntico como um campo objetivo real, em vez de apenas uma {`amplitude de probabilidade'}}''.

Trata-se de uma interpreta\c{c}\~{a}o que se compromete com algum tipo de realismo metaf\'{i}sico, na medida em que a medi\c{c}\~{a}o \'{e} considerada um ato de revela\c{c}\~{a}o de propriedades dos objetos qu\^{a}nticos ---isto \'{e}, opera sob o princ\'{i}pio medi\c{c}\~{a}o=revela\c{c}\~{a}o. Esse \'{e} um dos motivos pelos quais \citet[p.~94]{despagnat1983reality} classifica a mec\^{a}nica bohmiana\index{Bohm, David} como um ``realismo n\~{a}o-f\'{i}sico'', isto \'{e}, porque a realidade transfenomenal dos objetos qu\^{a}nticos (\textit{viz.}, entre observa\c{c}\~{o}es) n\~{a}o corresponde \`{a} ordem f\'{i}sica.

De acordo com \citet[p.~59]{freirejr2015}, Bohm\index{Bohm, David} abandona a interpreta\c{c}\~{a}o causal j\'{a} na d\'{e}cada de 1950; na d\'{e}cada de 1980 desenvolve, com a colabora\c{c}\~{a}o do matem\'{a}tico Hiley, uma interpreta\c{c}\~{a}o ontol\'{o}gica ---tamb\'{e}m chamada de ``teoria da ordem implicada'' \citep{bohm1980implicate}.\footnote{~Ver \citet{bohm2006undivided}\index{Bohm, David}.} Apesar de tal mudan\c{c}a na concep\c{c}\~{a}o da interpreta\c{c}\~{a}o da teoria qu\^{a}ntica, \citet[p.~60]{freirejr2015} aponta que ``houve um comprometimento permanente com um tipo de \index{realismo cient\'{i}fico}realismo cient\'{i}fico. \textelp{} O determinismo, que seria a motiva\c{c}\~{a}o da interpreta\c{c}\~{a}o causal, foi abandonado''.

Em sua interpreta\c{c}\~{a}o ontol\'{o}gica, \citet[p.~218--271]{bohm1980implicate}\index{Bohm, David} postula ordens ontol\'{o}gicas sutis, de modo que a ordem f\'{i}sica, observ\'{a}vel, seria chamada de ``ordem explicada'', que seria determinada por uma ordem sutil mais alta, chamada de ``ordem implicada'' ---em que estariam, por exemplo, fenômenos n\~{a}o locais como a consciência.\footnote{~Ver tamb\'{e}m \citet[p.~381--388]{bohm2006undivided}.} No entanto, conforme expressa em uma entrevista com \citet[p.~140]{reneeweber2003bohm}, quando questionado sobre a existência de uma ``ordem super super-implicada'', Bohm\index{Bohm, David} respondera que ``pode haver uma ordem implicada at\'{e} mesmo maior do que essa [super super-implicada]'' ---o que poderia ser considerado uma dificuldade filos\'{o}fica na medida em que as ``ordens'' ontol\'{o}gicas cada vez mais altas poderiam ser postuladas infinitamente. Tal dificuldade parece se assimilar ao argumento de \citet[I, \S 990b17]{aristotlemeta} do ``terceiro homem'' que deriva de uma redu\c{c}\~{a}o ao infinito da teoria das formas platônicas, que poderiam, de acordo com a interpreta\c{c}\~{a}o aristot\'{e}lica, ser postuladas em graus ontol\'{o}gicos infinitamente mais altos.

\citet[p.~6]{cushing1996causalQM} e \citet[p.~63--64]{freirejr2015} destacam que a interpreta\c{c}\~{a}o de Bohm\index{Bohm, David} n\~{a}o fora aceita nas primeiras d\'{e}cadas desde sua formula\c{c}\~{a}o, por motivos sociol\'{o}gicos, embora \citet[p.~64]{freirejr2015} aponte que tal teoria tem conquistado prest\'{i}gio e popularidade na comunidade dos fundamentos da f\'{i}sica, principalmente a partir dos anos 2001.

\section{A interpreta\c{c}\~{a}o dos estados latentes}

Para \citet{margenau1958philproblemsQM}, a evolu\c{c}\~{a}o linear \'{e} suficiente para descrever os sistemas qu\^{a}nticos, de modo que o colapso introduziria, desnecessariamente, uma assimetria na teoria. As interpreta\c{c}\~{o}es subjetivistas da consciência causando o colapso tamb\'{e}m s\~{a}o rejeitadas por \citet[p.~482]{margenau1963measurement}, sob a acusa\c{c}\~{a}o de tornar a mec\^{a}nica qu\^{a}ntica uma teoria psicol\'{o}gica.\footnote{~Ver \citet[p.~478]{Jammer1974}.} Proponente da ``teoria de latência'', Margenau considera que uma medi\c{c}\~{a}o revela um estado latente de um objeto. \'{E} desafiador classificar essa interpreta\c{c}\~{a}o conforme o trilema exposto nas p\'{a}ginas \pageref{trilema:A}--\pageref{trilema:C}, o que evidencia as limita\c{c}\~{o}es dessa classifica\c{c}\~{a}o. \citet[p.~505]{Jammer1974} chama a aten\c{c}\~{a}o para o fato de que Margenau, mesmo utilizando um referencial epistemol\'{o}gico e metodol\'{o}gico diverso daquele oferecido pela interpreta\c{c}\~{a}o de Copenhague, chega a conclus\~{o}es muito similares.

Um dos aspectos not\'{a}veis seria a interpreta\c{c}\~{a}o sobre os estados latentes, que se tornariam manifestos com o ato da medi\c{c}\~{a}o, que \'{e} muito pr\'{o}xima da posi\c{c}\~{a}o tardia de \citet{heisen1958physphil}\index{Heisenberg, Werner} de que os estados observ\'{a}veis s\~{a}o potencialidades (\`{a} maneira aristot\'{e}lica) pass\'{i}veis de serem atualizadas com o ato da medi\c{c}\~{a}o.\footnote{~Retomarei essa interpreta\c{c}\~{a}o de Heisenberg no cap\'{i}tulo \ref{CapWhitehead}.} Ainda assim, os dois autores diferem em um aspecto ontol\'{o}gico, na medida em que Margenau considera a medi\c{c}\~{a}o um ato de \textit{revela\c{c}\~{a}o}, enquanto \citet[p.~73]{heisenberg1927uncert}, como vimos no cap\'{i}tulo \ref{CapCopenhague} a considera um ato de \textit{cria\c{c}\~{a}o}. Para mais detalhes sobre esse ponto, ver \citet[p.~483]{Jammer1974}. 

Outro aspecto not\'{a}vel seria que Margenau considera a medi\c{c}\~{a}o um fenômeno macrosc\'{o}pico, o que se aproxima da posi\c{c}\~{a}o de Copenhague frente \`{a} interpreta\c{c}\~{a}o da medi\c{c}\~{a}o qu\^{a}ntica. Ao mesmo tempo, a posi\c{c}\~{a}o de Margenau acaba por engendrar a mesma problem\'{a}tica que, do ponto de vista filos\'{o}fico, representa uma dificuldade para a interpreta\c{c}\~{a}o de Bohr\index{Bohr, Niels}: o referido aspecto duplo da ontologia com a qual a interpreta\c{c}\~{a}o se compromete, isto \'{e}, a cis\~{a}o arbitr\'{a}ria entre os dom\'{i}nios cl\'{a}ssico/qu\^{a}ntico, acompanhada por uma ontologia pr\'{o}pria de cada dom\'{i}nio ---especificamente com o comprometimento ontol\'{o}gico com entidades diferentes. Dessa maneira, por mais que evite os problemas ontol\'{o}gicos da consciência, a proposta de Margenau acabaria por herdar problemas fundamentalmente similares aos enfrentados pela interpreta\c{c}\~{a}o de Copenhague, como vimos no cap\'{i}tulo \ref{CapCopenhague}.

\section{A interpreta\c{c}\~{a}o dos estados relativos}
A interpreta\c{c}\~{a}o de \citet{everett1957relative}\index{Everett, Hugh} da mec\^{a}nica qu\^{a}ntica, conhecida como a ``interpreta\c{c}\~{a}o dos estados relativos'' \'{e} uma das interpreta\c{c}\~{o}es heterodoxas da mec\^{a}nica qu\^{a}ntica mais populares. \citet[\S 2]{barrett1999quantum} identifica tal interpreta\c{c}\~{a}o como uma rea\c{c}\~{a}o direta ao problema da medi\c{c}\~{a}o, conforme enunciada por \citet{vNeum1955mathematical}\index{Neumann, John von}.

Essa solu\c{c}\~{a}o leva a s\'{e}rio a superposi\c{c}\~{a}o, aceitando que as equa\c{c}\~{o}es lineares do processo 2 s\~{a}o \textit{corretas e completas};\footnote{~Isso tamb\'{e}m \'{e} chamado de ``hip\'{o}tese de fiss\~{a}o'' (do inglês ``\textit{fission hypothesis}'').} na verdade, s\~{a}o tudo o que precisamos: todos os termos de uma superposi\c{c}\~{a}o de fato ocorrem, mas n\~{a}o no mesmo mundo. Essa \'{e} a solu\c{c}\~{a}o da mec\^{a}nica qu\^{a}ntica everettiana, nomeada ap\'{o}s \citet{everett1957relative}.\index{Bohm, David} \citet[p.~316]{everett1957relative}\index{Everett, Hugh} apresenta tal interpreta\c{c}\~{a}o a partir de dois postulados iniciais: a) a teoria qu\^{a}ntica \'{e} completa sem o colapso, isto \'{e}, funciona inteiramente com as leis din\^{a}micas contidas na evolu\c{c}\~{a}o linear; b) ``todo sistema sujeito a uma observa\c{c}\~{a}o externa pode ser considerado como parte de um sistema isolado maior''. Tal ``sistema maior'', \'{e} chamado por \citet[p.~317]{everett1957relative}\index{Everett, Hugh} de ``estado absoluto'', do qual partem os m\'{u}ltiplos ``estados relativos''. Na formula\c{c}\~{a}o de Everett\index{Everett, Hugh}, no processo de medi\c{c}\~{a}o, o estado absoluto se desdobra em estados relativos paralelos, de modo que cada possibilidade de superposi\c{c}\~{a}o de fato aconte\c{c}a em cada estado relativo:

\begin{quote}
Ao longo de toda sequência do processo de observa\c{c}\~{a}o, existe apenas um sistema f\'{i}sico representando o observador, ainda que n\~{a}o exista um \'{u}nico estado do observador (que se segue das representa\c{c}\~{o}es dos sistemas que interagem). Apesar disso, existe uma representa\c{c}\~{a}o em termos de uma superposi\c{c}\~{a}o, em que cada elemento cont\'{e}m um estado definido do observador e um estado do sistema correspondente. Assim, em cada observa\c{c}\~{a}o (ou intera\c{c}\~{a}o) sucessiva, o estado do observador se ``ramifica''\index{Everett, Hugh!muitos mundos} em um n\'{u}mero de estados diferentes. Cada ramifica\c{c}\~{a}o\index{Everett, Hugh!muitos mundos} representa um resultado diferente da medi\c{c}\~{a}o e do estado correspondendo ao estado do objeto. Todas as ramifica\c{c}\~{o}es\index{Everett, Hugh!muitos mundos} existem simultaneamente na superposi\c{c}\~{a}o ap\'{o}s qualquer sequência de observa\c{c}\~{o}es. A ``trajet\'{o}ria'' da configura\c{c}\~{a}o da mem\'{o}ria de um observador realizando uma sequência de medi\c{c}\~{o}es n\~{a}o \'{e}, portanto, uma sequência linear de configura\c{c}\~{o}es na mem\'{o}ria, mas uma \'{a}rvore que se ramifica,\index{Everett, Hugh!muitos mundos} com todos os resultados poss\'{i}veis existindo simultaneamente em uma superposi\c{c}\~{a}o final com v\'{a}rios coeficientes no modelo matem\'{a}tico. \citep[p.~320--321]{everett1957relative}\index{Everett, Hugh}.
\end{quote}

A solu\c{c}\~{a}o da mec\^{a}nica qu\^{a}ntica everettiana \'{e} tamb\'{e}m conhecida como ``interpreta\c{c}\~{a}o dos muitos mundos''\index{Everett, Hugh!muitos mundos}. A equivalência entre esses nomes \'{e} disputada, mas seguiremos a tradi\c{c}\~{a}o de mantê-la. No entanto, farei uma pequena digress\~{a}o terminol\'{o}gica sobre isso. Est\'{a} para al\'{e}m do escopo deste livro estabelecer se Everett endossava/endossaria a ``interpreta\c{c}\~{a}o dos muitos mundos''\index{Everett, Hugh!muitos mundos}, ou se essa \'{e} a melhor nomenclatura para descrever a solu\c{c}\~{a}o proposta por Everett\index{Everett, Hugh} ao problema da medi\c{c}\~{a}o. Muitas pessoas dizem que n\~{a}o \citep{barrett1999,barrett2011,conroy2012}, e h\'{a} um termo mais neutro para quem concorda com isso: ao inv\'{e}s de chamar de ``muitos mundos'', essas pessoas preferem usar ``estados relativos'' \citep{barrett2022} para descrever a posi\c{c}\~{a}o de Everett\index{Everett, Hugh} ---afinal, foi sob o t\'{i}tulo de ``A formula\c{c}\~{a}o dos `estados relativos' da mec\^{a}nica qu\^{a}ntica'' que \citet{everett1957relative}\index{Everett, Hugh} publicou os resultados da sua tese de doutorado na qual apresentava sua interpreta\c{c}\~{a}o da mec\^{a}nica qu\^{a}ntica. Inegavelmente, quem cunhou o termo ``muitos mundos'' foi \citet{dewitt1970quantum,dewitt1971}--- e n\~{a}o Everett\index{Everett, Hugh}. No entanto, a evidência hist\'{o}rica de correspondências, dentre diversos outros documentos, d\'{a} mais for\c{c}a \`{a} tese de que Everett\index{Everett, Hugh} desde sempre endossou a interpreta\c{c}\~{a}o de muitos mundos \citep[ver][]{bevers2011,barrett-byrne2012}. A quest\~{a}o \'{e} que a literatura padr\~{a}o acerca da mec\^{a}nica qu\^{a}ntica everettiana n\~{a}o est\'{a} interessada, por assim dizer, em fazer uma exegese da obra de Everett.\index{Everett, Hugh} \citet[p.~2]{wallace2012} \'{e} expl\'{i}cito sobre isso ao afirmar que ``eu uso [mec\^{a}nica qu\^{a}ntica] `everettiana' e `a interpreta\c{c}\~{a}o de Everett' livremente, mas para os prop\'{o}sitos atuais, eu n\~{a}o sei nem me importo se estou descrevendo a vis\~{a}o hist\'{o}rica do pr\'{o}prio Everett\index{Everett, Hugh}''. Aqui seguiremos essa mesma tradi\c{c}\~{a}o, e apresentaremos a mec\^{a}nica qu\^{a}ntica everettiana \textit{enquanto} (uma vers\~{a}o da) interpreta\c{c}\~{a}o dos muitos mundos. Ent\~{a}o passemos \`{a} breve apresenta\c{c}\~{a}o de como ela funciona.

A parte din\^{a}mica da mec\^{a}nica qu\^{a}ntica everettiana\index{Everett, Hugh} \'{e} bastante simples: tudo o que ela precisa \'{e} da boa e velha Equa\c{c}\~{a}o de Schr\"{o}dinger\index{Schr\"{o}dinger, Erwin!Equa\c{c}\~{a}o de Schr\"{o}dinger} ---linear, sem modifica\c{c}\~{o}es ou suplementa\c{c}\~{o}es. Para algumas pessoas, essa \'{e} a mec\^{a}nica qu\^{a}ntica \textit{lida literalmente} \citep{wallace2012,dewitt1970quantum}.

Mas lembre-se que esse \'{e} justamente o problema com o qual nos deparamos pra come\c{c}ar: se tomarmos como valor de face o que as equa\c{c}\~{o}es lineares da mec\^{a}nica qu\^{a}ntica est\~{a}o nos dizendo sobre o mundo, iremos inevitavelmente descrever os estados de coisas dos objetos f\'{i}sicos enquanto superposi\c{c}\~{o}es. E o problema disso \'{e} que superposi\c{c}\~{o}es desafiam nossa experiência imediata. Isso porque, relembremo-nos, ela implica que existem situa\c{c}\~{o}es nas quais os estados de coisas de objetos s\~{a}o indeterminados (por exemplo, gatos sem estado determinado de estar em jejum ou alimentados) enquanto nossa experiência diz que isso n\~{a}o \'{e} o caso.\footnote{~Como vimos no cap\'{i}tulo \ref{CapEPR}, essa foi a principal cr\'{i}tica de Einstein em rela\c{c}\~{a}o \`{a} mec\^{a}nica qu\^{a}ntica.} Pior ainda, no cen\'{a}rio do amigo de Wigner, nossa pr\'{o}pria experiência seria indeterminada (no caso de estarmos na superposi\c{c}\~{a}o de termos experienciado ou n\~{a}o um flash, ou no caso da Martha de estar na superposi\c{c}\~{a}o de ter percebido o gato em jejum ou alimentado).

Ciente dessa situa\c{c}\~{a}o, a mec\^{a}nica qu\^{a}ntica everettiana resolve o problema com a parte ontol\'{o}gica: cada termo de uma superposi\c{c}\~{a}o de fato ocorre, mas em \textit{mundos} distintos. Eis uma maneira bastante simplificada de entender a situa\c{c}\~{a}o:

\begin{quote}
    Quando o ponteiro de um dispositivo de medi\c{c}\~{a}o est\'{a} em uma superposi\c{c}\~{a}o de v\'{a}rias dire\c{c}\~{o}es diferentes, por exemplo, devemos entender isso como v\'{a}rios ponteiros, cada um em um mundo diferente, cada um apontando para uma dire\c{c}\~{a}o determinada diferente. \citep[p.~35]{albertbarrett1995}.
\end{quote}

E n\~{a}o apenas ponteiros de aparelhos medidores, mas toda a realidade f\'{i}sica ---incluindo \textit{n\'{o}s}--- tamb\'{e}m. \citet[p.~30]{dewitt1970quantum} cunhou o termo ``mundos'' para a no\c{c}\~{a}o de ``estados relativos'', quando afirmou que, revisitando o paradoxo do gato\index{Schr\"{o}dinger, Erwin!paradoxo do gato}, a interpreta\c{c}\~{a}o dos estados relativos ``\textelp{} considera que os gatos\index{Schr\"{o}dinger, Erwin!paradoxo do gato} habitam dois mundos simult\^{a}neos, que n\~{a}o interagem, mas que s\~{a}o igualmente reais'', o que popularizou a interpreta\c{c}\~{a}o de Everett\index{Everett, Hugh} como a ``interpreta\c{c}\~{a}o dos muitos mundos\index{Everett, Hugh!muitos mundos}''. Jammer ressalta que, nessa interpreta\c{c}\~{a}o dos estados relativos, as superposi\c{c}\~{o}es nunca colapsam. Dessa forma:

\begin{quote}
Para conciliar essa suposi\c{c}\~{a}o com a experiência ordin\'{a}ria, que atribui ao sistema do objeto (ou o sistema de aparelhos correlacionados) ap\'{o}s a medi\c{c}\~{a}o apenas um valor definitivo do observ\'{a}vel, a formula\c{c}\~{a}o dos estados relativos faz a sugest\~{a}o ousada de que o ``mundo''\index{Everett, Hugh!muitos mundos} \textelp{} foi dividido, como consequência da intera\c{c}\~{a}o, para uma multiplicidade de ``mundos'' igualmente reais, cada um dos quais correspondendo a um componente definido pela superposi\c{c}\~{a}o \textelp{}. Assim, em cada ``mundo''\index{Everett, Hugh!muitos mundos} separado uma medi\c{c}\~{a}o tem apenas um resultado, apesar do resultado diferir, em geral, de ``mundo'' para ``mundo''\index{Everett, Hugh!muitos mundos}. \citep[p.~512]{Jammer1974}.
\end{quote}

A simplicidade na din\^{a}mica \'{e} compensada pela complexidade ---ou ``extravag\^{a}ncia'', nos termos de \citet[p.~133]{bell2004-1981}--- na ontologia, e isso gera quest\~{o}es importantes de dois tipos: a primeira diz respeito \`{a} no\c{c}\~{a}o de \textit{probabilidade} e a segunda diz respeito \`{a} no\c{c}\~{a}o de ``\textit{mundo}''.\index{Everett, Hugh!muitos mundos} Comecemos com a \'{u}ltima.

Um ``mundo''\index{Everett, Hugh!muitos mundos} \'{e} definido na mec\^{a}nica qu\^{a}ntica everettiana de acordo com nossa \textit{experiência}. Um mundo \'{e} o lugar onde acontecem todas as coisas que você vê ao seu redor; \'{e} nesse lugar tamb\'{e}m que ocorrem suas experiências, pensamentos, percep\c{c}\~{o}es, etc. Assim, um ``mundo''\index{Everett, Hugh!muitos mundos} s\'{o} se ramifica\index{Everett, Hugh!muitos mundos} em outros mundos em situa\c{c}\~{o}es onde essa experiência seria comprometida com superposi\c{c}\~{o}es de estados macroscopicamente distingu\'{i}veis. Por exemplo, a quantidade de mundos aumenta somente quando estamos sendo descritos por superposi\c{c}\~{o}es de estados macrosc\'{o}picos (por exemplo, estados de gatos em jejum ou alimentados). Em outras palavras, ``mundos''\index{Everett, Hugh!muitos mundos} s\~{a}o mundos \textit{decoerentes}.

No n\'{i}vel fundamental, o estado f\'{i}sico do universo \'{e} descrito por um estado qu\^{a}ntico universal, que evolui segundo a Equa\c{c}\~{a}o de Schr\"{o}dinger\index{Schr\"{o}dinger, Erwin!Equa\c{c}\~{a}o de Schr\"{o}dinger} puramente linear. Esse estado se divide em mundos\index{Everett, Hugh!muitos mundos} distintos em situa\c{c}\~{o}es de decoerência/amplifica\c{c}\~{a}o macrosc\'{o}pica. E, nas palavras de \citet[p.~118]{durrlazarovici2020}, ``[c]ada ramifica\c{c}\~{a}o\index{Everett, Hugh!muitos mundos} do estado qu\^{a}ntico universal descreve uma hist\'{o}ria macroscopicamente bem-definida no espa\c{c}o tridimensional, que chamamos de {`mundo'}''\index{Everett, Hugh!muitos mundos}. Dessa maneira, a no\c{c}\~{a}o de ``mundo''\index{Everett, Hugh!muitos mundos} \'{e} emergente, e n\~{a}o fundamental. O que existe no n\'{i}vel fundamental \'{e} o estado qu\^{a}ntico; \'{e} a fun\c{c}\~{a}o de onda em um espa\c{c}o de alta dimens\~{a}o. O espa\c{c}o tridimensional ao qual chamamos de ``mundo''\index{Everett, Hugh!muitos mundos}, com part\'{i}culas, mesas e cadeiras, s\~{a}o fenômenos \textit{emergentes}.

Outra caracter\'{i}stica dessa interpreta\c{c}\~{a}o \'{e} a impossibilidade de especificar o momento exato em que a ramifica\c{c}\~{a}o\index{Everett, Hugh!muitos mundos} ocorre ---trata-se de um conceito \textit{vago}, assim como tantos outros na filosofia, como o conceito de \textit{observ\'{a}vel} \citep{vanfrass1991quantum}. Ao inv\'{e}s disso a ramifica\c{c}\~{a}o\index{Everett, Hugh!muitos mundos} \'{e} um processo\footnote{~\textit{N.B.}: n\~{a}o trata-se aqui do termo de arte, ``processo'', conforme \'{e} tratado na filosofia de processos de Whitehead. Sobre isso, consultar o cap\'{i}tulo \ref{CapWhitehead}.} gradual:

\begin{quote}
Quando dizemos que um mundo\index{Everett, Hugh!muitos mundos} ``se divide'' ou ``se ramifica''\index{Everett, Hugh!muitos mundos} (por exemplo, no decorrer de um experimento de medi\c{c}\~{a}o), na verdade estamos falando de um processo gradual. \textelp{} N\~{a}o tente pensar em um momento exato em que ``plim'' e o mundo\index{Everett, Hugh!muitos mundos} de repente se multiplica. O conceito de ``mundo''\index{Everett, Hugh!muitos mundos} tem uma certa imprecis\~{a}o--- geralmente n\~{a}o \'{e} poss\'{i}vel dizer exatamente quantos mundos\index{Everett, Hugh!muitos mundos} existem ou em que momento ocorreu uma nova divis\~{a}o. \citep[p.~118]{durrlazarovici2020}.
\end{quote}

Com a no\c{c}\~{a}o de ``mundo''\index{Everett, Hugh!muitos mundos} colocada no lugar, passemos \`{a} discuss\~{a}o acerca da probabilidade. Aqui temos uma situa\c{c}\~{a}o parecida com a que encontramos no caso da teoria da onda piloto: a Equa\c{c}\~{a}o de Schr\"{o}dinger\index{Schr\"{o}dinger, Erwin!Equa\c{c}\~{a}o de Schr\"{o}dinger} \'{e} determinista, mas \'{e} importante recuperar o aspecto probabilista da mec\^{a}nica qu\^{a}ntica haja vista que ele representa boa parte do seu sucesso experimental. Mas \'{e} longe de ser \'{o}bvio como isso pode ser o caso, pois eis como se d\'{a} uma t\'{i}pica descri\c{c}\~{a}o da mec\^{a}nica qu\^{a}ntica everettiana sobre a situa\c{c}\~{a}o em que Martha se encontra.

Aconte\c{c}a o que for, sabemos que o mundo\index{Everett, Hugh!muitos mundos} em que Martha interage com um gato de Sch\"{o}odinger ir\'{a} se multiplicar em outros dois, como ilustrado na Figura \ref{ArrAre:fig:everett-branching}: o \textsc{mundo 1}, no qual o estado de coisas \textsc{jejum} obt\'{e}m, e o \textsc{mundo 2}\index{Everett, Hugh!muitos mundos} no qual o que ocorre \'{e} \textsc{alimentado}. Isso ocorre $100\%$ das vezes.

\begin{figure}[ht]
\centering
\begin{forest} for tree={grow=270}
[Martha
[$\text{Martha}_{\text{\sc mundo 1}}$
]
[$\text{Martha}_{\text{\sc mundo 2}}$
]
]
\end{forest}
\caption{A estrutura da ramifica\c{c}\~{a}o de Martha.}
\label{ArrAre:fig:everett-branching}
\end{figure}

\'{E} importante salientar que na interpreta\c{c}\~{a}o de \citet[p.~320]{everett1957relative}\index{Everett, Hugh} n\~{a}o existe a dicotomia entre estados potenciais e estados atuais, tampouco a transi\c{c}\~{a}o de potência para ato: ``todos os elementos de uma superposi\c{c}\~{a}o (todos as `ramifica\c{c}\~{o}es')\index{Everett, Hugh!muitos mundos} s\~{a}o `atuais'; nenhum \'{e} mais `real' do que os demais'', de modo que todos os elementos de uma superposi\c{c}\~{a}o obede\c{c}am, igual e separadamente, \`{a} evolu\c{c}\~{a}o linear ---o que implicaria, para \citet[p.~320]{everett1957relative}\index{Everett, Hugh}, uma ``total falta de efeito de uma ramifica\c{c}\~{a}o sobre outra'', o que tamb\'{e}m implica que ``nenhum observador jamais estar\'{a} ciente de qualquer processo de `divis\~{a}o'''. A quest\~{a}o da impossibilidade da observa\c{c}\~{a}o de tal ramifica\c{c}\~{a}o dos estados \'{e} salientada por \citet[p.~514]{Jammer1974}, quem afirma que ``nenhum experimento em dada ramifica\c{c}\~{a}o poderia revelar o resultado de uma medi\c{c}\~{a}o obtida em outra ramifica\c{c}\~{a}o do universo''.\index{Everett, Hugh!muitos mundos} Assim, lembrando da taxonomia de \citet{Maudlin1995measurementproblem}, essa interpreta\c{c}\~{a}o nega a assun\c{c}\~{a}o $\gamma$ (p\'{a}gina \pageref{trilema:C}), isto \'{e}, que existam resultados \'{u}nicos de medi\c{c}\~{a}o. Nessa interpreta\c{c}\~{a}o, mantendo a analogia do gato\index{Schr\"{o}dinger, Erwin!paradoxo do gato} de Schr\"{o}dinger, gatos\index{Schr\"{o}dinger, Erwin!paradoxo do gato} em jejum e gatos\index{Schr\"{o}dinger, Erwin!paradoxo do gato} alimentados existem, simultaneamente, em ramifica\c{c}\~{o}es diferentes.

E j\'{a} que as coisas acontecem dessa maneira, como recuperar a ideia de que o gato tem $50\%$ de chances de ser encontrado em jejum, j\'{a} que a teoria diz que com $100\%$ de chances você (e o gato, o aparelho, o el\'{e}tron, e tudo mais) ir\'{a} se bifurcar em outros mundos\index{Everett, Hugh!muitos mundos} para encontrar o o gato seja como for? Todas as possibilidades qu\^{a}nticas se realizam, e isso torna a atribui\c{c}\~{a}o de probabilidades uma quest\~{a}o desafiadora.

Diversas solu\c{c}\~{o}es foram propostas, as quais destaco duas: a interpreta\c{c}\~{a}o da teoria da decis\~{a}o e a interpreta\c{c}\~{a}o da incerteza de autolocaliza\c{c}\~{a}o. De acordo com a primeira, devemos abandonar a ideia de probabilidades pois elas simplesmente n\~{a}o fazem sentido na mec\^{a}nica qu\^{a}ntica; ao inv\'{e}s disso, devemos apenas atentarmo-nos sobre como agentes racionais devem devem fazer escolhas em um cen\'{a}rio de muitos mundos \citep{greaves2007,deutsch1999,wallace2012}. Proponentes dessa solu\c{c}\~{a}o aplicam a \textit{teoria da decis\~{a}o} para argumentar que as escolhas racionais (ou as \textit{apostas}) que agentes devem fazer n\~{a}o s\~{a}o diferentes em cen\'{a}rios de um \'{u}nico mundo e em cen\'{a}rios de muitos mundos. Assim, as probabilidades teriam as mesmas aplica\c{c}\~{o}es pr\'{a}ticas do que outras interpreta\c{c}\~{o}es, de modo que isso n\~{a}o seria mais um problema exclusivo para a interpreta\c{c}\~{a}o de muitos mundos.

De acordo com a segunda, a probabilidade na mec\^{a}nica qu\^{a}ntica everettiana deve ser entendida como uma probabilidade subjetiva, isto \'{e}, que reflete nossa incerteza (em contraste \`{a} probabilidade objetiva de, por exemplo, $\tfrac{1}{6}$ que um dado de $6$ lados tem de cair no n\'{u}mero $6$ em uma rolagem) acerca de qual mundo estamos localizados \citep{saunders1998,sebenscarroll2018}. Vejamos brevemente como isso funciona. Considere novamente o caso de Martha. Ap\'{o}s a medi\c{c}\~{a}o, Martha se dividir\'{a} em duas pessoas: uma pessoa que viu o estado de coisas do \textsc{mundo 1} e outra que viu estado de coisas do \textsc{mundo 2} (ver Figura \ref{ArrAre:fig:everett-branching}). No entanto, antes da medi\c{c}\~{a}o, Martha n\~{a}o viu nenhum desses mundos. Chamemos essa pessoa de $\text{Martha}_{\text{\sc mundo 0}}$, e a quest\~{a}o que se coloca a ela \'{e} a seguinte: qual pessoa ela espera se tornar no futuro? Aqui est\~{a}o as respostas poss\'{i}veis:
\begin{enumerate}
    \item Ambas as pessoas;
    \item Nenhuma delas;
    \item Apenas uma delas.
\end{enumerate}

A primeira alternativa parece absurda, pois n\~{a}o h\'{a} como esperar que uma pessoa veja, por exemplo, um aparelho medidor apontando para duas regi\~{o}es ao mesmo tempo. A segunda alternativa tamb\'{e}m \'{e} descartada pois \textit{algo} ir\'{a} ocorrer com o sistema ap\'{o}s a medi\c{c}\~{a}o ---\textit{i.e.}, ele n\~{a}o vai ficar em superposi\c{c}\~{a}o para sempre. Se as coisas forem assim, resta apenas a terceira op\c{c}\~{a}o. Mas ela n\~{a}o sabe dizer de antem\~{a}o qual das pessoas, $\text{Martha}_{\text{\sc mundo 1}}$ ou $\text{Martha}_{\text{\sc mundo 2}}$, ela ir\'{a} se tornar. Dessa forma, ela \'{e} \textit{incerta} sobre qual \'{e} a localiza\c{c}\~{a}o do mundo em que ela habita dentro no multiverso de muitos mundos--- as probabilidades expressam exatamente isso: a probabilidade dela se encontrar em cada um desses mundos.\footnote{~Uma maneira de recuperar a probabilidade objetiva na mec\^{a}nica qu\^{a}ntica everettiana foi proposta por \citet{wilson2013,wilson2020}; para tanto, uma proposta heterodoxa de entender os muitos mundos enquanto \textit{paralelos} ao inv\'{e}s de \textit{ramificados}\index{Everett, Hugh!muitos mundos} foi colocada. A exposi\c{c}\~{a}o seria, contudo, demasiadamente t\'{e}cnica para ser apresentada em um livro introdut\'{o}rio do assunto.}

Por fim, devemos tratar da quest\~{a}o que nos trouxe at\'{e} aqui, a saber: superposi\c{c}\~{o}es desafiam nossa experiência imediata pois n\~{a}o experienciamos a falta de determina\c{c}\~{a}o de propriedades, por exemplo, de localiza\c{c}\~{a}o. Do ponto de vista da nossa experiência, as coisas est\~{a}o \textit{aqui ou ali} o tempo todo. E nesse sentido, negar que medi\c{c}\~{o}es possuem resultados \'{u}nicos tamb\'{e}m vai de encontro \`{a}quilo que podemos testemunhar com nossa experiência direta. Postular uma multiplicidade de mundos n\~{a}o parece ajudar. Isso porque, do ponto de vista da \textit{nossa} experiência, n\'{o}s n\~{a}o nos sentimos como se vivêssemos em um multiverso. N\~{a}o sentimos que ramificamos.\index{Everett, Hugh!muitos mundos}

Mas como sentir\'{i}amos caso pass\'{a}ssemos o universo se dividisse e se multiplicasse a todo momento? Essa quest\~{a}o \'{e} sens\'{i}vel, pois traz \`{a} tona a import\^{a}ncia das nossas intui\c{c}\~{o}es para as vis\~{o}es de mundo sugeridas pela f\'{i}sica. Vejamos como exemplo o trecho a seguir, da reconstru\c{c}\~{a}o de um di\'{a}logo que Elisabeth \citet[, ``E. A.'']{anscombe1959} supostamente teve com Ludwig Wittgenstein (``L. W.''):

\begin{quote}
    L. W.: Por que as pessoas dizem que era natural pensar que o Sol girava em torno da Terra, em vez do contr\'{a}rio?\\E. A.: Porque \textit{parece} que o Sol gira em torno da Terra.\\L. W.: Como \textit{pareceria} se a Terra girasse em torno do Sol? \citep[p.~151, ênfase original]{anscombe1959}.
\end{quote}

E sabemos por uma quest\~{a}o de fato que a Terra \textit{gira} em torno do Sol. Mas n\'{o}s sentimos o movimento da Terra de modo a poder atestar isso sob nossa experiência direta? Lembre-se que \citet{feyerabend1977:trad} traz essa quest\~{a}o ao analisar o caso de Galileu ao propor a tese do heliocentrismo. Dentre as diversas dificuldades, Galileu teve que se opor ao geocentrismo ao mesmo tempo em que se opunha ao apelo intuitivo do mesmo. N\~{a}o sentimos o movimento, e isso era explicado pela teoria cient\'{i}fica da \'{e}poca (lembre-se tamb\'{e}m que o geocentrismo era uma tese cient\'{i}fica na \'{e}poca). O heliocentrismo nega o lugar natural das coisas, e para isso era preciso incrementar a narrativa com outras coisas: por exemplo, a relatividade do movimento e a lei da in\'{e}rcia circular. Ent\~{a}o, como \textit{pareceria} se o mundo bifurcasse a cada processo de medi\c{c}\~{a}o? Isso mesmo: exatamente assim.

Uma an\'{a}lise panor\^{a}mica das cr\'{i}ticas que a interpreta\c{c}\~{a}o dos estados relativos recebeu pode ser encontrada em \citet[p.~516--519]{Jammer1974}. Ressalto apenas que o aspecto mais criticado de tal interpreta\c{c}\~{a}o \'{e} o comprometimento ontol\'{o}gico com algum tipo de multiverso; \citet[p.~191--192]{desp2006physics} chega a descartar tal interpreta\c{c}\~{a}o mediante tal cr\'{i}tica, na medida em que a interpreta\c{c}\~{a}o dos estados relativos n\~{a}o \'{e} clara quanto ao momento em que o universo se divide, isto \'{e}, exatamente quando uma ramifica\c{c}\~{a}o\index{Everett, Hugh!muitos mundos} ocorreria. Para \citet[p.~313]{belinfante1973hidden}, a interpreta\c{c}\~{a}o dos estados relativos n\~{a}o responde o problema da medi\c{c}\~{a}o, mas somente evita o axioma do ``colapso'' de um ponto de vista pr\'{a}tico. Ainda que os aspectos ontol\'{o}gicos da interpreta\c{c}\~{a}o dos estados relativos n\~{a}o tenham sido o objetivo central da discuss\~{a}o suscitada por Everett\index{Everett, Hugh}, \'{e} not\'{a}vel que suscite outro espectro de problemas ontol\'{o}gicos ---por mais que nenhum deles se relacione com o subjetivismo.

Tamb\'{e}m \'{e} relevante ressaltar que tal interpreta\c{c}\~{a}o recebera diversas releituras, com diversas formula\c{c}\~{o}es ontol\'{o}gicas, nas quais a dos ``muitos mundos\index{Everett, Hugh!muitos mundos}'' referida acima \'{e} apenas uma. Outra formula\c{c}\~{a}o derivada seria a interpreta\c{c}\~{a}o das ``muitas mentes'', sobre as quais pode-se fazer referência aos trabalhos de \citet{AlbertLoewer1988manyminds} e \citet{Loc1989mind}. Outra interpreta\c{c}\~{a}o not\'{a}vel, que a princ\'{i}pio se relaciona com a discuss\~{a}o da se\c{c}\~{a}o anterior, fora suscitada por Euan \citet{squires1991onemind, squires1993mindworld}, na medida em que postula uma ``consciência universal'', que remete ao ``estado absoluto'' de \citet{everett1957relative}\index{Everett, Hugh}. Em um racioc\'{i}nio similar ao de \citet{wigner1961mindbody}\index{Wigner, Eugene}, \citet[p.~285]{squires1991onemind} prop\~{o}e o postulado da ``universalidade da consciência'', isto \'{e}, a existência de uma consciência universal. O racioc\'{i}nio de Squires se d\'{a} da seguinte forma:

\begin{quote}
Se supusermos que a minha e a sua consciência podem selecionar independentemente suas experiências, ent\~{a}o n\~{a}o existiria algo para prevenir que fiz\'{e}ssemos escolhas diferentes. \textelp{} Isso n\~{a}o significa que ir\'{i}amos discordar do resultado das nossas experiências quando nos encontrarmos (\'{e} um fato simples da teoria qu\^{a}ntica que isso n\~{a}o pode ocorrer); ao inv\'{e}s disso, significa que o `você' que eu encontraria n\~{a}o seria escolhido pela sua consciência, isto \'{e}, você n\~{a}o seria mais um ser consciente! Tal possibilidade bizarra deve, certamente, ser exclu\'{i}da. Isso requer que haja somente uma sele\c{c}\~{a}o. A maneira mais simples de assegurar que isso ocorra \'{e} postular que h\'{a} somente uma mente consciente \textelp{}, isto \'{e}, que h\'{a} uma consciência universal. \citep[p.~117--118]{squires1993mindworld}.
\end{quote}

A proposta de Squires, no entanto, se relaciona com teorias da medi\c{c}\~{a}o que n\~{a}o aceitam a existência do colapso e, por isso, se diferencia das demais propostas discutidas anteriormente.

Ainda assim, como lembra \citet[p.~9, nota~5]{saunders2010manyworlds}, Everett\index{Everett, Hugh} jamais teria mencionado o termo consciência em seus escritos, ainda que tenha se referido ao termo ``experiência'', e que \citet{zeh2000conscious} tenha insistido continuamente na necessidade de um postulado especial para a consciência na interpreta\c{c}\~{a}o dos estados relativos.

\section{A interpreta\c{c}\~{a}o do colapso espont\^{a}neo}
Do ponto de vista dos fundamentos da mec\^{a}nica qu\^{a}ntica, seria desej\'{a}vel haver um mecanismo natural pelo qual o estado qu\^{a}ntico entrasse em colapso espontaneamente. Isto \'{e}, que o sistema em superposi\c{c}\~{a}o entrasse em colapso algum tempo antes do momento em que olhamos para aparelho de medi\c{c}\~{a}o. \'{E} exatamente tal mecanismo que \citet*{grw} oferecem: uma maneira de entender o colapso sem que seja necess\'{a}rio postular a no\c{c}\~{a}o vaga de ``medi\c{c}\~{a}o'' ou ``consciência''. A teoria \'{e} frequentemente chamada de ``\textsc{grw}'', devido \`{a} abrevia\c{c}\~{a}o das iniciais de seus proponentes. Com isso, \textsc{grw} oferece uma nova din\^{a}mica que opera em adi\c{c}\~{a}o \`{a} din\^{a}mica linear: o colapso espont\^{a}neo.

Para fazê-lo, \textsc{grw} postula novas constantes na natureza: $\tau$, que \'{e} a frequência de colapsos, e $a$ que especifica a ``largura'' da fun\c{c}\~{a}o de colapso. Assim, a din\^{a}mica usual de Schr\"{o}dinger \'{e} multiplicada pelo ``fator de colapso'' $j(x)-Ke^{-x^2/2a^2}$.\footnote{~A aplica\c{c}\~{a}o desse fator de colapso na Equa\c{c}\~{a}o de Schr\"{o}dinger\index{Schr\"{o}dinger, Erwin!Equa\c{c}\~{a}o de Schr\"{o}dinger} n\~{a}o \'{e} nada trivial e apresent\'{a}-la foge do escopo deste texto introdut\'{o}rio. Para mais detalhes sobre o formalismo de \textsc{grw}, ver \citet{durrlazarovici2020}.}

Se $\tau$ \'{e} suficientemente pequena, \'{e} muito improv\'{a}vel que um \'{u}nico sistema de part\'{i}culas colapse. Tome, novamente, a situa\c{c}\~{a}o da dupla fenda. As descri\c{c}\~{o}es que levam ao padr\~{a}o estat\'{i}stico 2 s\~{a}o suficientemente simples, isto \'{e}, com um alto valor de $\tau$. Colapsos de sistemas assim, \textit{viz.}, part\'{i}culas individuais s\~{a}o extremamente raros; precisamente $1$ a cada $10^{16}$ segundos. De acordo com \citet[p.~52]{lewis2016quaont} ``\textelp{} pode-se observar uma \'{u}nica part\'{i}cula por cem milh\~{o}es de anos sem nunca ver um colapso''. Isso \'{e} justamente o que esper\'{a}vamos para poder explicar, por exemplo, fenômenos de interferência. \`{A} medida que a complexidade do estado qu\^{a}ntico aumenta (o valor $\tau$ aumenta), a probabilidade de um colapso tamb\'{e}m aumenta.

Considere novamente o estado qu\^{a}ntico do gato de Schr\"{o}dinger. Ele denota um estado qu\^{a}ntico extremamente complexo, possuindo trilh\~{o}es de part\'{i}culas que comp\~{o}em, por exemplo, o aparelho medidor. Segundo \textsc{grw}, o colapso desse sistema \'{e} altamente prov\'{a}vel. E isso \'{e} justamente o que esper\'{a}vamos para lidar com o fato intuitivo de que sempre observamos resultados de medi\c{c}\~{a}o bem-definidos. Para sistemas complexos, a taxa alta de colapso explica o fato de n\~{a}o vermos superposi\c{c}\~{o}es. Ao se correlacionar com o aparelho, todo o sistema rapidamente colapsa para um dos estados. No entanto, como aponta \citet[p.~475]{maudlin2003distilling}, abandona a no\c{c}\~{a}o de que haja um agente causal necess\'{a}rio para que uma medi\c{c}\~{a}o seja efetuada: ``nessa teoria, colapsos acontecem aleatoriamente, com uma probabilidade fixa, e n\~{a}o s\~{a}o particularmente associados com qualquer tipo de intera\c{c}\~{a}o''.

Lembre-se que a din\^{a}mica linear de Schr\"{o}dinger implica em superposi\c{c}\~{o}es macrosc\'{o}picas. Para \textsc{grw}, essa superposi\c{c}\~{a}o \'{e} inst\'{a}vel, de modo que o sistema todo colapsa espontaneamente para um dos estados em uma fra\c{c}\~{a}o de segundos. Como diz \citet[p.~53]{lewis2016quaont}, ``\textelp{} podemos simplesmente dizer `a posi\c{c}\~{a}o da part\'{i}cula est\'{a} correlacionada com a posi\c{c}\~{a}o de um objeto macrosc\'{o}pico s\'{o}lido' e o resultado ser\'{a} o mesmo''. Assim, a parte da din\^{a}mica fica clara:

\begin{quote}
    Enquanto um estado qu\^{a}ntico \textelp{} normalmente evolui de acordo com a Equa\c{c}\~{a}o de Schr\"{o}dinger\index{Schr\"{o}dinger, Erwin!Equa\c{c}\~{a}o de Schr\"{o}dinger}, de tempos em tempos ele d\'{a} um salto. \textelp{} A probabilidade por unidade de tempo para um salto \textsc{grw} [\textit{viz.}, um colapso espont\^{a}neo] \'{e} $N/\tau$ onde $N$ \'{e} o n\'{u}mero de elementos no estado qu\^{a}ntico e $\tau$ \'{e} uma nova constante. O salto \'{e} para um estado qu\^{a}ntico ``reduzido'' ou ``colapsado''. \citep[p.~202--203]{bell2004-1989}.
\end{quote}

De acordo com \citet{pessoa1992MPatualizado}, as formula\c{c}\~{o}es que assumem a no\c{c}\~{a}o de colapso espont\^{a}neo funcionariam apenas para sistemas macrosc\'{o}picos.

\begin{quote}
Para sistemas de poucas part\'{i}culas, tal localiza\c{c}\~{a}o [colapso] ocorreria muito raramente, e praticamente n\~{a}o violaria a Equa\c{c}\~{a}o de Schr\"{o}dinger\index{Schr\"{o}dinger, Erwin!Equa\c{c}\~{a}o de Schr\"{o}dinger}. Para um sistema macrosc\'{o}pico, no entanto, composto de um grande n\'{u}mero de part\'{i}culas emaranhadas, tal colapso espont\^{a}neo ocorreria freq\"{u}entemente. Isso explicaria porque a redu\c{c}\~{a}o [ou colapso] s\'{o} ocorre quando um aparelho macrosc\'{o}pico se acopla ao objeto qu\^{a}ntico. \citep[p.~200]{pessoa1992MPatualizado}.
\end{quote}

Assim, \citet[p.~105]{albert1992quantum} relembra que, da mesma forma como a interpreta\c{c}\~{a}o de Copenhague e a interpreta\c{c}\~{a}o de Margenau, a formula\c{c}\~{a}o \textsc{grw} incorreria no problema filos\'{o}fico do macrorrealismo.

J\'{a} na parte ontol\'{o}gica, existem ao menos três op\c{c}\~{o}es \citep[para maiores detalhes, ver][]{esfeld2018,allori2021,barrett2019}. A princ\'{i}pio, \textsc{grw} parece ter comprometimento ontol\'{o}gico com part\'{i}culas, j\'{a} que fala-se da grande quantidade de \textit{part\'{i}culas} que d\~{a}o origem \`{a} superposi\c{c}\~{o}es inst\'{a}veis (isto \'{e}, altamente prov\'{a}veis de colapsar, como de aparelhos medidores macrosc\'{o}picos). No entanto, a teoria n\~{a}o nos fornece nada al\'{e}m do estado qu\^{a}ntico/fun\c{c}\~{a}o de onda dos sistemas compostos ---por exemplo, $\langle\text{objeto}+\text{aparelho}+\text{Martha}\rangle$--- e ``como ele evolui para dar conta de nossa experiência'' \citep[p.~135]{barrett2019}. Essa interpreta\c{c}\~{a}o de \textsc{grw} \'{e} chamada de ``\textsc{grw}r''; nela, o mundo seria composto por ``exatamente um objeto f\'{i}sico ---a fun\c{c}\~{a}o de onda universal'' \citep[p.~54]{albert2013}. A segunda op\c{c}\~{a}o \'{e} interpretar \textsc{grw} como uma teoria sobre uma ontologia de \textit{densidade de massa}, de acordo com a qual o que existe \'{e} um campo de densidade de massa no espa\c{c}o ordin\'{a}rio, tridimensional. Essa op\c{c}\~{a}o \'{e} chamada de ``\textsc{grw}m''. Uma terceira op\c{c}\~{a}o, chamada de ``\textsc{grw}f'', \'{e} interpretar \textsc{grw} a partir de uma ontologia de \textit{eventos}, denominados ``\textit{flashes}''. Nada de ondas ou part\'{i}culas em nenhuma dessas interpreta\c{c}\~{o}es, ao menos no n\'{i}vel fundamental \citep{emery2017}.

\section{Uma escolha filos\'{o}fica}

Analisei, nos dois \'{u}ltimos cap\'{i}tulos, o problema da medi\c{c}\~{a}o. Introduzido propriamente por \citet{vNeum1955mathematical}\index{Neumann, John von}, esse problema se origina em conflito axiom\'{a}tico entre as equa\c{c}\~{o}es din\^{a}micas e o fato emp\'{i}rico da observa\c{c}\~{a}o. A posi\c{c}\~{a}o de von Neumann\index{Neumann, John von} foi endossada durante os anos seguintes, atingindo seu \'{a}pice na formula\c{c}\~{a}o subjetivista de \citet{LonBau1939theory} e em sua maior dificuldade com a situa\c{c}\~{a}o solipsista proposta atrav\'{e}s do experimento de pensamento do amigo de \citet{wigner1961mindbody}\index{Wigner, Eugene}. \citet{bass1971mind} tentou superar tal dificuldade utilizando a concep\c{c}\~{a}o de consciência oferecida por \citet{schro1964myview} que, por sua vez, seria baseada nos escritos indianos do Vedanta.\footnote{~Ver tamb\'{e}m \citet{schro1967whatislife}.} \citet{goswami1989idealistic} levou a cabo a formula\c{c}\~{a}o de uma interpreta\c{c}\~{a}o para a mec\^{a}nica qu\^{a}ntica com base no pensamento ved\^{a}ntico, baseado numa ontologia na qual a consciência (\`{a} maneira ved\^{a}ntica) \'{e} a base do ser.

Conforme procurei expor, os debates filos\'{o}ficos suscitados pelas dificuldades conceituais acerca da interpreta\c{c}\~{a}o da no\c{c}\~{a}o de medi\c{c}\~{a}o deram origem a diversas interpreta\c{c}\~{o}es da teoria qu\^{a}ntica em que, como observa \citet[p.~4]{pessoa2003conceitos} ``\textelp{} cada uma dessas interpreta\c{c}\~{o}es \'{e} internamente consistente e, de modo geral, consistente com experimentos qu\^{a}nticos''. Todavia, observa-se que, dentre as interpreta\c{c}\~{o}es que abordam o problema, nenhuma \'{e} livre de dificuldades filos\'{o}ficas.

Parece seguro classificar tais dificuldades em dois grupos maiores: 1) o macrorrealismo, pr\'{o}prio das interpreta\c{c}\~{o}es que separam o dom\'{i}nio cl\'{a}ssico do dom\'{i}nio qu\^{a}ntico em dois dom\'{i}nios ontol\'{o}gicos diferentes, em que o primeiro \'{e} agente causal sobre o segundo; 2) a introdu\c{c}\~{a}o de agentes metate\'{o}ricos para a causa\c{c}\~{a}o da medi\c{c}\~{a}o; nos casos estudados, a introdu\c{c}\~{a}o e comprometimento ontol\'{o}gico com consciência por duas vias: 2a) subjetiva/m\'{u}ltipla, numa concep\c{c}\~{a}o dualista, que herda os problemas da teoria cartesiana; 2b) unitiva, \`{a} maneira do pensamento vedantino, que tamb\'{e}m se compromete com a problem\'{a}tica pr\'{o}pria dessa linha.

Poder-se-ia organizar num terceiro grupo as teorias que n\~{a}o admitem a descontinuidade da medi\c{c}\~{a}o, isto \'{e}, o colapso, como as teorias de Bohm\index{Bohm, David} e Everett\index{Everett, Hugh}, que tamb\'{e}m suscitam problemas ontol\'{o}gicos na tentativa de solucionar o problema da medi\c{c}\~{a}o. As interpreta\c{c}\~{o}es estat\'{i}sticas estariam num outro grupo, no qual a quest\~{a}o da medi\c{c}\~{a}o n\~{a}o \'{e} abordada.

Dessa forma, a pluralidade de op\c{c}\~{o}es n\~{a}o torna f\'{a}cil a vida de quem afirma que existe \textit{uma} interpreta\c{c}\~{a}o correta da mec\^{a}nica qu\^{a}ntica ---``\textit{a mais correta} que as outras''. Esse \'{e} o famoso problema da subdetermina\c{c}\~{a}o: h\'{a} diversas alternativas para interpretar os fenômenos descritos pela mec\^{a}nica qu\^{a}ntica, e n\~{a}o h\'{a} raz\~{o}es dispon\'{i}veis, sejam cient\'{i}ficas ou filos\'{o}ficas, para escolher uma em detrimento de outras ---seja \textit{verdadeira} ou \textit{mais adequada} \citep[para um estudo detalhado sobre o problema da subdetermina\c{c}\~{a}o nas interpreta\c{c}\~{o}es da mec\^{a}nica qu\^{a}ntica, ver][]{arroyo2020phd, arroyoarenhart2020realismo}.

O alto grau de humildade epistêmica gerado pela subdetermina\c{c}\~{a}o, caso n\~{a}o seja percebido, pode esconder atitudes dogm\'{a}ticas mascaradas por senten\c{c}as do tipo: ``a mec\^{a}nica qu\^{a}ntica (\textit{simpliciter}) \textit{implica} que \dots''. Conforme visto neste cap\'{i}tulo, ao menos em rela\c{c}\~{a}o ao dom\'{i}nio ontol\'{o}gico, frases assim carecem de justifica\c{c}\~{a}o epistêmica.

\chapter{Novos horizontes}
\label{CapWhitehead}

\noindent O sucesso emp\'{i}rico da mec\^{a}nica qu\^{a}ntica est\'{a} fora de quest\~{a}o. Seu sucesso conceitual, por outro lado, \'{e} uma hist\'{o}ria totalmente diferente. Devido ao problema da medi\c{c}\~{a}o, a mec\^{a}nica qu\^{a}ntica \'{e} conceitualmente incompleta; completar conceitualmente a mec\^{a}nica qu\^{a}ntica \'{e} oferecer as chamadas ``interpreta\c{c}\~{o}es da mec\^{a}nica qu\^{a}ntica''. No entanto, as interpreta\c{c}\~{o}es da mec\^{a}nica qu\^{a}ntica s\~{a}o fortemente marcadas por hip\'{o}teses \textit{ad hoc}, no sentido proposto por \citet[p.~986]{popper1974replies}, ou seja, ``uma conjectura [\'{e}] `\textit{ad hoc}' se for introduzida \textelp{} para explicar uma dificuldade espec\'{i}fica, mas \textelp{} n\~{a}o pode ser testada de forma independente''.

Meu foco aqui tem sido uma solu\c{c}\~{a}o espec\'{i}fica para o problema da medi\c{c}\~{a}o, a saber: a \textit{interpreta\c{c}\~{a}o da consciência causal\index{interpreta\c{c}\~{a}o da consciência!causal}}. O motivo pelo qual fiz esse recorte \'{e} duplo. Por um lado, essa interpreta\c{c}\~{a}o foi, em grande parte, deixada de lado pela literatura especializada;\footnote{~Ainda que isso esteja mudando, ver \citet{debarrosmontemayir2019ed,gao2022}.} em contrapartida, ela foi apropriada por uma literatura n\~{a}o-cient\'{i}fica (o assim-chamado ``misticismo qu\^{a}ntico''). Espero ter fechado a porta para a \'{u}ltima no cap\'{i}tulo \ref{CapvNeumann}, e espero abrir um caminho para a primeira neste cap\'{i}tulo.

Como vimos, essa interpreta\c{c}\~{a}o defende a agência causal da consciência humana como um recurso fundamental nos processos de medi\c{c}\~{a}o. Seguindo as metaontologias tradicionais na filosofia anal\'{i}tica,\footnote{~Isto \'{e}, as metodologias carnapiana e quineana \citep[ver][]{berto2015ontology,arenhartarroyo2021manu,arroyodasilva2022,arenhart2023}.} poder\'{i}amos dizer que essa interpreta\c{c}\~{a}o \'{e} ontologicamente comprometida com a existência de uma entidade espec\'{i}fica no mobili\'{a}rio do mundo: a ``consciência''\index{ontologia!comprometimento ontol\'{o}gico}. Ao longo deste livro, chamei essa abordagem de \textit{ontologia naturalizada}\index{ontologia!naturalizada}, ou $\mathscr{O}_N$. Frequentemente presume-se, no entanto, que essa entidade deva ser compreendida metafisicamente sob a \'{e}gide da metaf\'{i}sica da subst\^{a}ncia ---e, por consequência, v\'{a}rias vertentes do dualismo\index{dualismo} \citep[ver][]{RaoJonas2019dualismQM}.\index{ontologia!e metaf\'{i}sica}

O problema \'{e} que o dualismo\index{dualismo} n\~{a}o vem de gra\c{c}a: ele herda o problema mente-corpo, que tem assombrado a filosofia (ao menos) desde Descartes. Portanto, \'{e} seguro dizer que a interpreta\c{c}\~{a}o da consciência causal\index{interpreta\c{c}\~{a}o da consciência!causal} \textit{tamb\'{e}m} herda tal problema. Assim, \'{e} compreens\'{i}vel o porquê dela ser uma perspectiva t\~{a}o impopular entre a comunidade de fundamentos qu\^{a}nticos. Ao mesmo tempo, torna-se intrigante compreender os motivos para que algu\'{e}m adotar essa interpreta\c{c}\~{a}o. Em particular, porque ---como se o problema mente-corpo n\~{a}o fosse o bastante--- as tentativas de evitar o problema mente-corpo dentro da interpreta\c{c}\~{a}o da consciência causal\index{interpreta\c{c}\~{a}o da consciência!causal} acabaram por introduzir um \textit{outro} problema, que \'{e} o da inadequa\c{c}\~{a}o emp\'{i}rica da interpreta\c{c}\~{a}o da consciência m\'{i}stica\index{interpreta\c{c}\~{a}o da consciência!m\'{i}stica}. Vimos tudo isso no cap\'{i}tulo \ref{CapvNeumann}.

Como consequência dessas duas frentes (\textit{viz.}, o problema mente-corpo na interpreta\c{c}\~{a}o da consciência causal\index{interpreta\c{c}\~{a}o da consciência!causal} e o problema da inadequa\c{c}\~{a}o emp\'{i}rica na interpreta\c{c}\~{a}o da consciência m\'{i}stica\index{interpreta\c{c}\~{a}o da consciência!m\'{i}stica}), temos a impopularidade dessa fam\'{i}lia de interpreta\c{c}\~{o}es ---erroneamente, na minha opini\~{a}o, identificadas como ``a interpreta\c{c}\~{a}o da consciência'' \textit{simpliciter}.

Para ilustrar essa impopularidade, aponto o estudo de \citet*{poll2013schlo}. Eles apresentaram uma pesquisa aos participantes de uma conferência em fundamentos qu\^{a}nticos, em 2013, contendo perguntas de m\'{u}ltipla escolha sobre diversos t\'{o}picos em aberto. Houve uma pergunta sobre o papel do observador na f\'{i}sica, e apenas seis (de trinta e cinco) afirmaram acreditar que a consciência desempenha um papel crucial no processo de medi\c{c}\~{a}o. Os resultados obtidos pela pesquisa, embora n\~{a}o muito expressivos dado o n\'{u}mero de participantes, s\~{a}o bastante emblem\'{a}ticos em termos da atitude em rela\c{c}\~{a}o ao conceito de consciência nos fundamentos da mec\^{a}nica qu\^{a}ntica.

O plano deste cap\'{i}tulo \'{e} questionar a tradicional ---por vezes, at\'{e} mesmo \textit{autom\'{a}tica!}--- liga\c{c}\~{a}o direta entre a interpreta\c{c}\~{a}o da consciência causal\index{interpreta\c{c}\~{a}o da consciência!causal} e a metaf\'{i}sica do dualismo\index{dualismo} de subst\^{a}ncia. Ao propor uma abordagem baseada em processos para a interpreta\c{c}\~{a}o da consciência causal\index{interpreta\c{c}\~{a}o da consciência!causal}, busco eliminar os enigmas filos\'{o}ficos cruciais associados a ela, \textit{vide} o problema mente-corpo. Notavelmente, o trabalho de \citet{whitehead1928process} parece ser um bom ponto de partida, uma vez que j\'{a} existem v\'{a}rias tentativas de compreender a mec\^{a}nica qu\^{a}ntica sob a metaf\'{i}sica de Whitehead, baseada em processos \citep{malin2001naturelovestohide,shimony1964QMwhitehead}. No entanto, nenhuma dessas tentativas focou explicitamente na interpreta\c{c}\~{a}o da consciência causal\index{interpreta\c{c}\~{a}o da consciência!causal}. E, como a metaf\'{i}sica de Whitehead oferece uma solu\c{c}\~{a}o n\~{a}o eliminativista para o problema mente-corpo \citep{weekes2012mindbody}, pensei que poderia ser frut\'{i}fero tentar conectar ambas as coisas.

\section{Antiga abordagem: consciência como subst\^{a}ncia}

Lembre-se de que a interpreta\c{c}\~{a}o da consciência causal\index{interpreta\c{c}\~{a}o da consciência!causal} resolve o problema da medi\c{c}\~{a}o qu\^{a}ntica ao colocar a mente fora do escopo da din\^{a}mica qu\^{a}ntica linear, \textit{viz.}, fora do processo 2. Uma consequência direta disso \'{e} coloc\'{a}-la fora do escopo das superposi\c{c}\~{o}es. O colapso/processo 1 seria, ent\~{a}o, causado pela intera\c{c}\~{a}o com uma mente humana \citep{wigner1961mindbody}. Lembre-se que essa n\~{a}o \'{e} uma proposta popular ---ao menos n\~{a}o nos fundamentos da f\'{i}sica \citep*{poll2013schlo}--- pelos motivos apresentados acima. Minha hip\'{o}tese de trabalho para explicar isso \'{e}: talvez a alegada implausibilidade da interpreta\c{c}\~{a}o da consciência causal\index{interpreta\c{c}\~{a}o da consciência!causal} ---e, portanto, sua impopularidade--- esteja intimamente relacionada \`{a} ausência do seu desenvolvimento metaf\'{i}sico.

Isso mesmo, sugiro que uma maior aten\c{c}\~{a}o na \textit{metaf\'{i}sica} da interpreta\c{c}\~{a}o da consciência causal\index{interpreta\c{c}\~{a}o da consciência!causal} talvez possa torn\'{a}-la mais plaus\'{i}vel ---e, por conseguinte, mais popular. Assim, formular teorias metaf\'{i}sicas sob medida para a mec\^{a}nica qu\^{a}ntica parece ser uma tarefa essencial para a filosofia contempor\^{a}nea. De fato, h\'{a} at\'{e} mesmo uma metodologia para isso: a Abordagem \textit{Toolbox} para a metaf\'{i}sica (ver \cite{french2012toolbox, FrenMcken2015toolboxagain}; \cite{french2014structure, french2018toolbox}, e \textit{cf.} com \cite{arenhartarroyo2021veri}) sugere que a filosofia da ciência deve utilizar os dispositivos te\'{o}ricos produzidos pela metaf\'{i}sica anal\'{i}tica como uma fonte conceitual para obter uma melhor compreens\~{a}o das teorias cient\'{i}ficas.

Isso j\'{a} foi feito em outros contextos. O mais recente que conhe\c{c}o \'{e} o de \citet{wilson2020}. Ele desenvolveu uma vers\~{a}o do realismo modal, inspirado em David Lewis, levando em considera\c{c}\~{a}o (outra) interpreta\c{c}\~{a}o espec\'{i}fica da mec\^{a}nica qu\^{a}ntica ---a interpreta\c{c}\~{a}o dos muitos mundos\index{Everett, Hugh!muitos mundos}. O que se resultou foi uma teoria metaf\'{i}sica, chamada ``realismo modal qu\^{a}ntico'', feita sob medida\footnote{~Para uma discuss\~{a}o sobre esse ponto, ver \cite{arroyo2021}, mas \textit{cf.} com \cite{arroyoarenhart2022foop}.} para a interpreta\c{c}\~{a}o dos muitos mundos (que vimos no cap\'{i}tulo \ref{CapPaisagem}).

Se esse diagn\'{o}stico estiver correto, aqui est\'{a} \textit{outra} hip\'{o}tese de trabalho que vou explorar ao longo deste cap\'{i}tulo: uma metaf\'{i}sica substancialista/dualista foi automaticamente associada \`{a} interpreta\c{c}\~{a}o da consciência causal\index{interpreta\c{c}\~{a}o da consciência!causal}, sem maiores debates ou desenvolvimentos na \'{a}rea da metaf\'{i}sica da ciência; assim, sugiro que a elabora\c{c}\~{a}o de uma metaf\'{i}sica para a no\c{c}\~{a}o de ``consciência'', inspirada na metaf\'{i}sica dos processos apresentada por Alfred North \citeauthor{whitehead1928process} em sua \textit{magnum opus} ``\citetitle{whitehead1928process}'' \citeyearpar{whitehead1928process}, poderia lan\c{c}ar uma nova luz sobre a interpreta\c{c}\~{a}o da consciência causal\index{interpreta\c{c}\~{a}o da consciência!causal} ---e, com isso, abrir novos horizontes investigativos para a mesma. Trata-se, portanto, de uma proposta fundamentada na esperan\c{c}a de que ``mesmo que a mec\^{a}nica qu\^{a}ntica n\~{a}o explique a consciência, talvez uma teoria da consciência possa lan\c{c}ar luz sobre os problemas da mec\^{a}nica qu\^{a}ntica'' \citep[p.~311]{chalmers1995consciousmind}.

Como n\~{a}o sou a primeira pessoa a pensar sobre isso, vamos ver onde outras pararam para que possamos retomar a partir desse ponto. \citet[p.~271]{shimonymalin2006dialogue} ponderam diferentes atitudes em rela\c{c}\~{a}o \`{a} interpreta\c{c}\~{a}o do conceito de medi\c{c}\~{a}o e consideram que a interpreta\c{c}\~{a}o da consciência causal\index{interpreta\c{c}\~{a}o da consciência!causal} \textit{poderia ser} favor\'{a}vel a uma filosofia whiteheadiana. No entanto, isso foi apenas sugerido por esses autores. Eles acabam negando a plausibilidade dessa interpreta\c{c}\~{a}o devido \`{a} sua ---aparente--- implica\c{c}\~{a}o necess\'{a}ria na ideia de \textit{consciência subjetiva}. Assim, \citet[p.~763--767]{Shi1963role} explicitamente rejeitou como vi\'{a}veis as interpreta\c{c}\~{o}es que consideram a consciência subjetiva do observador como agente causal do colapso na medi\c{c}\~{a}o qu\^{a}ntica. Vou retomar a partir de onde \citet{shimonymalin2006dialogue} pararam. Ao contr\'{a}rio deles, vou assumir alguns riscos a mais, e tentar desenvolver mais a fundo as perspectivas de uma interpreta\c{c}\~{a}o whiteheadiana da interpreta\c{c}\~{a}o (\textit{sic}) da consciência causal.

Enquanto a da consciência causal tem sido descartada por diversos motivos,\footnote{~Veja, por exemplo, \citet{albert1992quantum,lewis2016quaont,chalmers1995consciousmind}.} n\~{a}o existem argumentos fatais contra ela \citep[ver][]{RaoJonas2019dualismQM,BarOas2016Consc}. No entanto, ela herda um problema filos\'{o}fico grave, \textit{vide} o problema mente-corpo. Talvez a afirma\c{c}\~{a}o mais expl\'{i}cita da conex\~{a}o tradicional entre a interpreta\c{c}\~{a}o da consciência causal\index{interpreta\c{c}\~{a}o da consciência!causal} e o problema mente-corpo venha de \citet*{shimony-etal1977}:

\begin{quote}
\textelp{} se [o problema da medi\c{c}\~{a}o] de fato representa um problema genu\'{i}no, ent\~{a}o \'{e} um problema muito dif\'{i}cil, e alguns f\'{i}sicos e fil\'{o}sofos passaram a acreditar que nenhuma solu\c{c}\~{a}o f\'{a}cil, n\~{a}o radical, ter\'{a} sucesso. J\'{a} que o problema mente-corpo \'{e} um problema perene n\~{a}o resolvido (que a f\'{i}sica cl\'{a}ssica de alguma forma conseguiu contornar sem resolver), pode-se conjecturar que os dois problemas est\~{a}o entrela\c{c}ados. \citep*[p.~761]{shimony-etal1977}.
\end{quote}

Portanto, eis como as coisas est\~{a}o para a interpreta\c{c}\~{a}o da consciência causal\index{interpreta\c{c}\~{a}o da consciência!causal}: mesmo que n\~{a}o tenhamos fundamentos f\'{i}sicos nem metametaf\'{i}sicos para descart\'{a}-la objetivamente \citep[ver todo o argumento em][]{BarOas2016Consc,RaoJonas2019dualismQM}, e mesmo que se aceite todas as consequências filos\'{o}ficas dessa interpreta\c{c}\~{a}o, uma coisa de fato permanece: \textit{se} a interpreta\c{c}\~{a}o da consciência causal\index{interpreta\c{c}\~{a}o da consciência!causal} est\'{a} vinculada ao dualismo\index{dualismo} de subst\^{a}ncia, \textit{ent\~{a}o} ela fica presa ao problema mente-corpo. Ou seja, uma vez que essa interpreta\c{c}\~{a}o \'{e} adotada, ela herda tradicionalmente o ônus da prova de resolver o problema mente-corpo, \textit{vide}, fornecer uma explica\c{c}\~{a}o precisa de como uma mente n\~{a}o f\'{i}sica pode interagir com um sistema f\'{i}sico, \textit{e.g.}, um aparato de medi\c{c}\~{a}o, ou qualquer outro sistema qu\^{a}ntico. No entanto, ao fazer isso, a interpreta\c{c}\~{a}o da consciência causal\index{interpreta\c{c}\~{a}o da consciência!causal} troca o problema da medi\c{c}\~{a}o na mec\^{a}nica qu\^{a}ntica pelo problema mente-corpo na filosofia. As probabilidades est\~{a}o contra tal interpreta\c{c}\~{a}o, uma vez que o problema mente-corpo permanece sem solu\c{c}\~{a}o por \textit{muito mais tempo} do que o problema da medi\c{c}\~{a}o.

Bem, talvez.

\section{Novas dire\c{c}\~{o}es: consciência como processo}
Aqui est\'{a} outra maneira de formular o problema. A literatura que defende a interpreta\c{c}\~{a}o da consciência causal\index{interpreta\c{c}\~{a}o da consciência!causal} na mec\^{a}nica qu\^{a}ntica pressup\~{o}e, direta ou indiretamente, uma metaf\'{i}sica dualista para o conceito de consciência que \'{e}, ao mesmo tempo, $(i)$ dualista, na medida em que separa consciência e ``mat\'{e}ria'' em subst\^{a}ncias distintas, e $(ii)$ subjetivista, na medida em que a no\c{c}\~{a}o de consciência se baseia no ``Eu'', que pensa e, portanto, existe.

Agora, aqui est\'{a} a poss\'{i}vel sa\'{i}da. Ao contr\'{a}rio da metaf\'{i}sica materialista, a metaf\'{i}sica whiteheadiana \'{e} considerada n\~{a}o reducionista, uma vez que n\~{a}o nega a efic\'{a}cia causal entre os polos material e n\~{a}o material (mental) da existência. Ao contr\'{a}rio do dualismo\index{dualismo}, ela tamb\'{e}m n\~{a}o os considera ontologicamente separados. No modelo metaf\'{i}sico de Whitehead, a consciência cont\'{e}m e \'{e} contida pelo conceito de ``mat\'{e}ria''; de uma perspectiva de processos (e n\~{a}o de objetos), a consciência transcende e \'{e} transcendida pela mat\'{e}ria. Assim, pode-se afirmar que, a partir de uma perspectiva de metaf\'{i}sica de processos, o mundo \'{e} tanto imanente quanto transcendente. Inicialmente, tais categoriza\c{c}\~{o}es eliminam as principais dificuldades enfrentadas pelo conceito de consciência. No entanto, o aspecto do subjetivismo considerado acima $(ii)$ precisa ser levado em conta, uma vez que uma interpreta\c{c}\~{a}o subjetivista \'{e} indesej\'{a}vel em uma teoria cient\'{i}fica, e Whitehead considera que o conceito de consciência possui um aspecto subjetivo ---n\~{a}o \'{e}, no entanto, \textit{reduzido} \`{a} subjetividade, como na metaf\'{i}sica dualista \citep[ver][]{griffin2001process}. Tendo em mente que o modelo de Whitehead oferece uma maneira original ---e pouco mencionada na literatura espec\'{i}fica, como apontam \citet{weberweekes2009neglected}--- de lidar com o problema mencionado anteriormente, argumento que uma metaf\'{i}sica semelhante \`{a} de Whitehead pode ser frut\'{i}fera para a no\c{c}\~{a}o de consciência aplicada \`{a} interpreta\c{c}\~{a}o da mec\^{a}nica qu\^{a}ntica. Esta se\c{c}\~{a}o introduz um quadro para tal desenvolvimento na metaf\'{i}sica da ciência.

Embora o uso da metaf\'{i}sica de processos de Whitehead para interpretar a rela\c{c}\~{a}o entre consciência e mec\^{a}nica qu\^{a}ntica seja inovador,\footnote{~Compare, por exemplo, com os ensaios presentes nos volumes de \citet{gao2022,debarrosmontemayir2019ed}.} a tentativa mais ampla de interpretar a mec\^{a}nica qu\^{a}ntica a partir de certos aspectos da filosofia de Whitehead n\~{a}o \'{e} nova. Na verdade, os resultados da f\'{i}sica foram um dos principais pontos de partida para a teoria de \citet[p.~121--122]{whitehead1928process}, que visava fornecer uma base conceitual para o que ele se refere como ``teoria qu\^{a}ntica''. No entanto, como observou \citet[p.~240]{shimony1964QMwhitehead}, a mencionada ``teoria qu\^{a}ntica'' em Whitehead \'{e} a teoria qu\^{a}ntica inicial, \textit{vide} a teoria desenvolvida pela primeira vez no in\'{i}cio de 1900. O per\'{i}odo em que a filosofia whiteheadiana estava sendo desenvolvida antecedeu um per\'{i}odo de mudan\c{c}as significativas na mec\^{a}nica qu\^{a}ntica, incluindo debates sobre os fundamentos e a ontologia associada \`{a}s suas interpreta\c{c}\~{o}es ---especialmente na d\'{e}cada de 1930. Portanto, \'{e} muito improv\'{a}vel que Whitehead tenha mencionado em seus escritos os desenvolvimentos mais ``recentes'' na mec\^{a}nica qu\^{a}ntica, relativos \`{a} sua contemporaneidade. Levando isso em considera\c{c}\~{a}o, \'{e} natural que autores como \citet{shimony1964QMwhitehead} e \citet{malin1988whiteheadbell} proponham algumas modifica\c{c}\~{o}es nos conceitos da metaf\'{i}sica de Whitehead para acomodar a interpreta\c{c}\~{a}o da mec\^{a}nica qu\^{a}ntica.

Talvez a primeira proposta documentada de usar a filosofia de Whitehead para elucidar o debate em torno das interpreta\c{c}\~{o}es de uma teoria qu\^{a}ntica relativamente mais estabelecida tenha sido a de \citet{burgers1963qm, burgers1965qm}, seguida principalmente por \citet{Shi1963role, shimony1964QMwhitehead,stapp1979bell, stapp1982mind,malin1988whiteheadbell, malin1993collapse, malin2001naturelovestohide,epperson2004QMwhitehead,ferrari2021}. Vale ressaltar que todos os autores mencionados usam os mesmos conceitos para estabelecer um paralelo entre a mec\^{a}nica qu\^{a}ntica e a metaf\'{i}sica de \citet{whitehead1928process}:

\paragraph{1)} Em rela\c{c}\~{a}o \`{a} mec\^{a}nica qu\^{a}ntica, destaco o conceito de ``potência'' (no original escrito como ``\textit{potentia}'') contido nos escritos tardios de \citet{heisen1958physphil}\index{Heisenberg, Werner}, que interpreta o conceito de ``estado qu\^{a}ntico'' como uma tendência, algo entre a ideia do fenômeno (ou evento) e sua atualidade. Conforme \citet{heisen1958physphil}\index{Heisenberg, Werner} explica, a no\c{c}\~{a}o de \textit{potentia} \'{e}:

\begin{quote}
    \textelp{} uma vers\~{a}o quantitativa do velho conceito de potência da filosofia aristot\'{e}lica, que introduziu algo entre a ideia de evento e o evento real, um tipo estranho de realidade f\'{i}sica a mediar entre possibilidade e realidade. \citep[p.~32]{heisen1958physphil}\index{Heisenberg, Werner}.
\end{quote}

Isto \'{e}, trata-se de uma reinterpreta\c{c}\~{a}o do conceito aristot\'{e}lico de \textit{dynamis}. \citet[p.~263]{shimonymalin2006dialogue}, no entanto, garantem que tal proposta heisenbergiana \'{e} original, j\'{a} que nenhuma outra metaf\'{i}sica at\'{e} ent\~{a}o teria proposto essa modalidade para a realidade. Na concep\c{c}\~{a}o de \citet[p.~134]{heisen1958physphil}\index{Heisenberg, Werner}, at\'{e} mesmo potencialidades contr\'{a}rias poderiam coexistir, como no caso da superposi\c{c}\~{a}o ---e essa modifica\c{c}\~{a}o conceitual traria maior inteligibilidade \`{a} empreitada da mec\^{a}nica qu\^{a}ntica:\footnote{~Esse \'{e}, inclusive, o ponto de partida da abordagem ``logos'', mencionada no cap\'{i}tulo \ref{CapPaisagem} \citep[ver][]{deronde2019probing}.}

\begin{quote}
    \textelp{} se considerarmos a palavra ``estado'' como descrevendo descrevendo mais apropriadamente uma ``potencialidade'' do que uma ``realidade'' (poder\'{i}amos mesmo, sem mais nem menos, substituir ``estado'' por ``potencialidade'') torna-se bastante plaus\'{i}vel o conceito de ``potencialidades coexistentes'', pois uma potencialidade pode superpor-se a outras potencialidades; em outras palavras, potencialidades distintas podem ter algo em comum. \citet[p.~134]{heisen1958physphil}.\index{Heisenberg, Werner}
\end{quote}

Como apontam \citet[p.~264]{shimonymalin2006dialogue}, o pr\'{o}prio conceito de ``superposi\c{c}\~{a}o'' seria ``derivado da inova\c{c}\~{a}o metaf\'{i}sica fundamental da potencialidade''. Em tal interpreta\c{c}\~{a}o, uma medi\c{c}\~{a}o consiste na atualiza\c{c}\~{a}o, por meio do colapso, de uma entre muitas possibilidades sobrepostas.

    \begin{quote} A interpreta\c{c}\~{a}o te\'{o}rica de uma experiência requer, portanto, três est\'{a}gios distintos: (1) traduzir a situa\c{c}\~{a}o experimental inicial em uma fun\c{c}\~{a}o de probabilidade; (2) seguir a evolu\c{c}\~{a}o temporal dessa fun\c{c}\~{a}o; (3) escolher uma nova medida a ser feita no sistema f\'{i}sico considerado, cujo resultado poder\'{a} ent\~{a}o ser calculado da fun\c{c}\~{a}o de probabilidade. \textelp{} E \'{e} somente no terceiro est\'{a}gios que mudamos novamente, passando do ``poss\'{i}vel'' ao ``real''. \citep[p.~36--37]{heisen1958physphil}.\index{Heisenberg, Werner}\end{quote}

Nessa interpreta\c{c}\~{a}o, uma medi\c{c}\~{a}o consiste, atrav\'{e}s do colapso, na atualiza\c{c}\~{a}o de uma (dentre diversas) possibilidades superpostas ---o que torna mais plaus\'{i}vel a afirma\c{c}\~{a}o metaf\'{i}sica de \citet[p.~73]{heisenberg1927uncert}\index{Heisenberg, Werner} de que um evento ``passa a existir somente quando a observamos'', e que chamei no cap\'{i}tulo \ref{CapCopenhague} de ``\textit{medi\c{c}\~{a}o=cria\c{c}\~{a}o}''. No contexto whiteheadiano, considero mais apropriada a nomenclatura ``medi\c{c}\~{a}o=\textit{atualiza\c{c}\~{a}o}''. \citet[p.~76--77]{malin2003col} aponta que as potencialidades n\~{a}o seriam eventos no espa\c{c}o-tempo ---o que seria uma propriedade das atualidades.

\paragraph{2)} Em rela\c{c}\~{a}o \`{a} metaf\'{i}sica whiteheadiana, o conceito de ``entidades atuais'' \'{e} central para a interpreta\c{c}\~{a}o da mec\^{a}nica qu\^{a}ntica. Whitehead enuncia esse conceito pela primeira vez da seguinte forma: \begin{quote} ``Entidades atuais'' ---tamb\'{e}m denominadas ``ocasi\~{o}es atuais''--- s\~{a}o as coisas reais finais das quais o mundo \'{e} composto. N\~{a}o h\'{a} como ir al\'{e}m das entidades atuais para encontrar algo mais real. \citep[p.~18]{whitehead1928process}. \end{quote} De acordo com Malin, o conceito de ``entidades atuais'' seria a base da metaf\'{i}sica proposta por Whitehead. Dadas a abrangência e os objetivos deste livro, \'{e} imposs\'{i}vel resumir toda a constru\c{c}\~{a}o filos\'{o}fica de Whitehead. Sigo o esbo\c{c}o proposto por \citet[p.~77--78]{malin1993collapse};\footnote{~Veja tamb\'{e}m \citet[p.~266--267]{shimonymalin2006dialogue}.} que destaca oito aspectos centrais, relevantes para o debate sobre a interpreta\c{c}\~{a}o da mec\^{a}nica qu\^{a}ntica; dos oito aspectos, seleciono apenas quatro que considero especificamente relevantes para o conceito de medi\c{c}\~{a}o:

\begin{enumerate}
    \item Uma entidade atual \'{e} um processo de ``auto-cria\c{c}\~{a}o'' atemporal e criativa, que leva a uma apari\c{c}\~{a}o moment\^{a}nea das entidades atuais no espa\c{c}o-tempo;
    \item As entidades atuais s\~{a}o instant\^{a}neas; ap\'{o}s o \'{u}nico instante em que emergem no espa\c{c}o-tempo por meio da auto-cria\c{c}\~{a}o, elas se fundem novamente (na terminologia whiteheadiana, elas ``preendem'') em um dom\'{i}nio atemporal e fora do espa\c{c}o com todas as entidades atuais (passadas e futuras), como potencialidades;
    \item Cada entidade atual est\'{a} relacionada e interconectada (na terminologia whiteheadiana, forma um ``\textit{nexus}'') com todas as entidades atuais;
    \item O fim do processo de auto-cria\c{c}\~{a}o de uma entidade atual, ou seja, sua apari\c{c}\~{a}o moment\^{a}nea no espa\c{c}o-tempo, \'{e} a auto-cria\c{c}\~{a}o de uma nova entidade atual ou um ``pulso de experiência'', de modo que o universo whiteheadiano n\~{a}o \'{e} um universo real de ``objetos'', mas um universo de ``experiências''.
\end{enumerate}

Como aponta \citet[p.~92]{stapp2007whiteheadQM}, o paralelo entre a metaf\'{i}sica de \citet[p.~72]{whitehead1928process}, na qual ``as entidades atuais \textelp{} tornam real o que era anteriormente apenas potencial'' e Heisenberg\index{Heisenberg, Werner}, na qual ``\textelp{} a transi\c{c}\~{a}o do `poss\'{i}vel' para o `real' ocorre durante o ato de observa\c{c}\~{a}o'' \'{e} muito sugestivo. Para Shimony, tal paralelo pode ser visualizado da seguinte maneira:

\begin{quote}
Considere, por simplicidade, duas part\'{i}culas emaranhadas. Se s\~{a}o consideradas, juntas, como uma \'{u}nica entidade atual, sua dependência m\'{u}tua \'{e} natural: ambas surgem de um \'{u}nico campo de potencialidade. Quando uma medi\c{c}\~{a}o ocorre em qualquer part\'{i}cula, ela quebra a conex\~{a}o, criando um relacionamento entre duas entidades atuais \textelp{}. \citep[p.~274]{shimonymalin2006dialogue}.
\end{quote}

Para \citet[p.~81]{malin2003col}, o ganho dessa interpreta\c{c}\~{a}o \'{e} oferecer um novo horizonte de respostas para a seguinte pergunta ---ainda n\~{a}o respondida--- no debate sobre a interpreta\c{c}\~{a}o da medi\c{c}\~{a}o qu\^{a}ntica: \textit{qual \'{e} o mecanismo do colapso?}. Na metaf\'{i}sica whiteheadiana, o universo n\~{a}o seria um universo de objetos (ou campos), mas um universo de experiências ou processos, de modo que se o axioma do colapso for interpretado como o processo de auto-cria\c{c}\~{a}o de uma entidade atual, tal processo n\~{a}o poderia ser um mecanismo que exclui a possibilidade de criatividade. Nessa leitura, o conceito de ``mecanismo'' parece n\~{a}o ter lugar. Esse \'{e} o ponto central que desejo enfatizar nesta leitura espec\'{i}fica da consciência e seu papel na medi\c{c}\~{a}o qu\^{a}ntica: se n\~{a}o h\'{a} necessidade de um mecanismo, tamb\'{e}m n\~{a}o h\'{a} necessidade de procurar a causa externa do colapso. A consciência, como uma ocasi\~{a}o de experiência, pode auto-criar atualidades. Em rela\c{c}\~{a}o \`{a} interpreta\c{c}\~{a}o da consciência causal\index{interpreta\c{c}\~{a}o da consciência!causal}, \citet[p.~260--261]{malin2001naturelovestohide} rejeita a interpreta\c{c}\~{a}o de que a consciência desempenha um papel causal no colapso.

\'{E} importante destacar que o estudo da no\c{c}\~{a}o de consciência tem sido permeado por uma literatura na qual a figura mais citada \'{e} Descartes, legando \`{a} discuss\~{a}o contempor\^{a}nea o mesmo escopo de op\c{c}\~{o}es te\'{o}ricas dado s\'{e}culos atr\'{a}s: seja uma forma de monismo\index{monismo} reducionista (das quais as teses do materialismo e do epifenomenalismo eliminativista s\~{a}o apopulares) ou dualismo\index{dualismo}.\footnote{~De maneira muito geral, essas s\~{a}o as duas posi\c{c}\~{o}es que mencionei: ou a mente e o corpo/c\'{e}rebro s\~{a}o duas coisas diferentes (dualismo), ou ambas s\~{a}o a mesma e \'{u}nica coisa (monismo) ---aqui est\~{a}o, em particular, as abordagens segundo as quais a mente n\~{a}o existe, e s\'{o} o que existe \'{e} o c\'{e}rebro.} Para Shimony, uma metaf\'{i}sica inspirada em Whitehead pode oferecer uma abordagem frut\'{i}fera para o tradicional problema mente-corpo:

\begin{quote}
N\~{a}o h\'{a} nada que sabemos melhor do que isso, que temos experiências conscientes. N\~{a}o h\'{a} nada que sabemos muito melhor do que a mat\'{e}ria de que o mundo \'{e} feito \'{e} inanimada. \textelp{} Coloque os juntos; você n\~{a}o tem uma solu\c{c}\~{a}o, você tem um quebra-cabe\c{c}a, um quebra-cabe\c{c}a terr\'{i}vel. \textelp{} Eu sou muito simp\'{a}tico com Whitehead porque ele d\'{a} uma resposta a isso postulando um universo primitivo que n\~{a}o \'{e} totalmente inanimado; ele chama sua filosofia de ``filosofia do organismo''. Isso \'{e} t\~{a}o promissor quanto qualquer coisa que eu conhe\c{c}o para uma solu\c{c}\~{a}o para o problema mente-corpo, mas \'{e} terr\'{i}vel como deixa de fora os detalhes. \citep[p.~451--452]{shimonysmolin2009dialogue}.
\end{quote}

Em resumo, \citet{whitehead1928process} resolve o problema mente-corpo ao propor uma teoria hol\'{i}stica da realidade que reconhece a interdependência entre os aspectos f\'{i}sicos e mentais da experiência. Ao rejeitar a no\c{c}\~{a}o dualista de que o mental e o f\'{i}sico s\~{a}o \textit{subst\^{a}ncias distintas}, argumenta-se que ambos os polos mental--f\'{i}sico fazem parte de um continuum de experiência que abrange todos os aspectos da realidade. De acordo com essa vis\~{a}o, as unidades fundamentais da realidade n\~{a}o s\~{a}o part\'{i}culas materiais substanciais, mas processos ou \textit{ocasi\~{o}es de experiência} que incluem aspectos f\'{i}sicos e mentais. Essas ocasi\~{o}es de experiência interagem constantemente entre si, formando uma teia em constante evolu\c{c}\~{a}o de experiências interconectadas. No geral, a teoria de realidade de Whitehead oferece uma vis\~{a}o integrada da mente e do corpo, evitando o dualismo\index{dualismo} que assombrou a filosofia por s\'{e}culos. No entanto, devo abordar os ``detalhes'' aos quais Shimony se refere na passagem acima ---que tamb\'{e}m s\~{a}o mencionados por Malin na forma de problemas ainda em aberto dentro da metaf\'{i}sica whiteheadiana:

\begin{quote}
A filosofia de processos de Whitehead fornece uma base metaf\'{i}sica para a compreens\~{a}o da realidade. No entanto, quest\~{o}es essenciais s\~{a}o deixadas sem resposta: A realidade consiste em n\'{i}veis, alguns dos quais s\~{a}o ``superiores'' a outros em um sentido profundo? Os seres humanos têm um lugar e um papel a desempenhar no esquema cosmol\'{o}gico? \textelp{} surpreendentemente, o misterioso ``colapso dos estados qu\^{a}nticos'' continua sendo uma rica fonte de sugest\~{o}es. O colapso, o processo de transi\c{c}\~{a}o do potencial para o real, envolve uma sele\c{c}\~{a}o: Existem muitas possibilidades, das quais apenas uma \'{e} atualizada. Como \'{e} feita a sele\c{c}\~{a}o? \citep[p.~189]{malin2001naturelovestohide}.
\end{quote}

A proposta apresentada por \citet[p.~93]{malin2003col} seria seguir a m\'{a}xima, atribu\'{i}da a Paul Dirac, de que ``A Natureza faz a escolha'', ou seja, que ``Natureza'' causa o colapso. Embora a defini\c{c}\~{a}o dessa ``Natureza'' n\~{a}o seja especificada com letra mai\'{u}scula, em sua leitura, isso corresponde \`{a} atualiza\c{c}\~{a}o de potencialidades, ou at\'{e} mesmo, a sua auto-cria\c{c}\~{a}o, com aleatoriedade intr\'{i}nseca ---da\'{i} a indetermina\c{c}\~{a}o qu\^{a}ntica. Dado o car\'{a}ter investigativo dessa proposta, parece prematuro alinhar-me com tal perspectiva antecipadamente. Ent\~{a}o, vejamos outro candidata.

Outra tentativa de interpretar a mec\^{a}nica qu\^{a}ntica, em espec\'{i}fico, o papel causal da consciência na medi\c{c}\~{a}o qu\^{a}ntica, \'{e} feita por Henry Stapp. Sua proposta vai no caminho inverso daquele proposto pela interpreta\c{c}\~{a}o da consciência causal\index{interpreta\c{c}\~{a}o da consciência!causal}, que procurou utilizar a consciência para compreender a mec\^{a}nica qu\^{a}ntica; \citet{stapp2007whiteheadQM} procura utilizar a mec\^{a}nica qu\^{a}ntica para compreender a consciência ---caminho esse que tamb\'{e}m \'{e} tra\c{c}ado por \citet{penrose1994shadows}. No entanto, como observa \citet[p.~172]{landau1998errorpenrose}, ``Penrose aceita que a mente consciente surge como um funcionamento do c\'{e}rebro f\'{i}sico \textelp{}'', tese que n\~{a}o \'{e} endossada por \citet{stapp2006quantumdualism}, que prop\~{o}e uma metaf\'{i}sica que chama de ``dualismo interativo''\index{dualismo}. Como aponta \citet{mohrhoff2002stapperror}:

\begin{quote}
A teoria que ele [Stapp]\footnote{~\'{E} justo dizer que o pr\'{o}prio \citet[p.~264]{stapp2002response} afirma que ``[essa] n\~{a}o \'{e} minha teoria final''. Ainda assim, quando questionado por Malin se a teoria de Stapp considera, como consequência, que a consciência causa o colapso, Stapp responde categoricamente que n\~{a}o endossa tal interpreta\c{c}\~{a}o \citep[ver o di\'{a}logo completo em][p.~110,~ff.]{eastman2003QMwhitehead}.} acaba formulando \'{e} completamente diferente da teoria que ele inicialmente professa formular, pois no come\c{c}o a consciência \'{e} respons\'{a}vel pelas redu\c{c}\~{o}es de vetores de estado [colapso], enquanto no final uma nova lei f\'{i}sica \'{e} respons\'{a}vel ---uma lei que de forma alguma depende da presen\c{c}a da consciência. \citep[p.~250]{mohrhoff2002stapperror}.
\end{quote}

No entanto, tamb\'{e}m h\'{a} ind\'{i}cios de que a ontologia whiteheadiana foi interpretada a partir de uma metaf\'{i}sica \index{dualismo}dualista. Conforme a leitura apontada por \citet[p.~169]{lovejoy1960againstdualism}, Whitehead seria ``um advers\'{a}rio do \index{dualismo}dualismo com o qual estamos preocupados aqui, mas apenas um \index{dualismo}dualista com uma diferen\c{c}a''; como aponta \citet{shimony1964QMwhitehead}, a leitura \index{dualismo}dualista, se leg\'{i}tima, seria fundamentalmente contr\'{a}ria \`{a} pr\'{o}pria proposta whiteheadiana que, como enfatiza \citet{weekes2009caus}, \'{e} essencialmente \index{monismo}monista.

Entendendo a pluralidade de leituras (dualistas\index{dualismo} e \index{monismo}monistas) da metaf\'{i}sica whiteheadiana, procurei utilizar a chave de leitura \index{monismo}monista, oferecida por \citet{weekes2012mindbody}, \citet{griffin2009whiteheadconsc} e \citet{nobo2003whiteheadQM} para compreender o conceito de consciência no que se relaciona com a no\c{c}\~{a}o de ``colapso'' na interpreta\c{c}\~{a}o do conceito de medi\c{c}\~{a}o em mec\^{a}nica qu\^{a}ntica. Como aponta \citet{griffin2009whiteheadconsc}, a concep\c{c}\~{a}o whiteheadiana de consciência difere radicalmente da posi\c{c}\~{a}o cartesiana (dualista) e materialista (reducionista) ---que s\~{a}o as leituras predominantes para o conceito de consciência na filosofia da f\'{i}sica--- ainda que mantenha alguns aspectos dessas concep\c{c}\~{o}es metaf\'{i}sicas:

\begin{quote}
Com os \index{dualismo}dualistas, Whitehead concorda que a consciência pertence a uma entidade ---uma mente ou psique--- que \'{e} distinta do c\'{e}rebro, e que a liberdade genu\'{i}na pode, em parte por essa raz\~{a}o, ser atribu\'{i}da \`{a} experiência consciente. Com os materialistas, Whitehead compartilha uma sensibilidade naturalista, evitando assim qualquer solu\c{c}\~{a}o impl\'{i}cita sobrenaturalista para problemas filos\'{o}ficos, e, em parte por essa raz\~{a}o, rejeita qualquer \index{dualismo}dualismo entre dois tipos de realidades. Como materialistas, em outras palavras, ele afirma um \index{monismo}monismo pluralista. Assim, ele considera a consciência como uma fun\c{c}\~{a}o de algo mais fundamental. \citep[p.~175]{griffin2009whiteheadconsc}.
\end{quote}

\citet[p.~225]{nobo2003whiteheadQM} tamb\'{e}m enfatiza que a no\c{c}\~{a}o de consciência, na metaf\'{i}sica whiteheadiana, n\~{a}o se reduz \`{a} experiência humana ou \`{a} subjetividade ---o que acaba por evitar a dificuldade antropomorfista das leituras utilizadas at\'{e} ent\~{a}o para o conceito na filosofia da f\'{i}sica, e parece oferecer, tamb\'{e}m, uma chave de leitura para evitar a dificuldade do solipsismo que pode emergir de uma leitura subjetivista do conceito de consciência na metaf\'{i}sica whiteheadiana.

Al\'{e}m disso, como observa \citet[p.~206--208]{katzko2009processconsc}, o debate contempor\^{a}neo na filosofia da mente, especificamente para a leitura da no\c{c}\~{a}o de consciência, est\'{a}, em sua parte mais expressiva, circunscrito em metaf\'{i}sicas materialistas ou \index{dualismo}dualistas. A t\'{i}tulo de amostragem: existem os proponentes de uma metaf\'{i}sica fisicalista que, assim como \citet{stapp1982mind}, consideram a causa\c{c}\~{a}o mental sobre o f\'{i}sico mas, ao mesmo tempo, consideram a estrutura cerebral como definitivamente importante para a ocorrência do aspecto mental; \citet{dennet1991consciousnessexpl}, ainda mais radical, defende a tese do ``funcionalismo'' de que a mente \'{e} um produto do arranjo cerebral, n\~{a}o podendo ter a\c{c}\~{a}o causal sobre o c\'{e}rebro, situando-se entre os materialistas ou epifenomenalistas; \citet{chalmers1995consciousmind} considera ambos os polos, material e mental, igualmente importantes, o que o aproxima dos \index{dualismo}dualistas atrav\'{e}s daquilo que chama de ``dualismo interativo''\index{dualismo}; em todos os casos, um dos questionamentos centrais seria de causa\c{c}\~{a}o, isto \'{e}: \textit{como o aspecto f\'{i}sico da realidade poderia dar origem ao aspecto mental?}

Como afirma \citet{weekes2012mindbody}, a metaf\'{i}sica whiteheadiana sugere uma metaf\'{i}sica \index{monismo}monista, o que tamb\'{e}m acaba por desfazer a dificuldade do \index{dualismo}dualismo no caso de utiliz\'{a}-la para interpretar a no\c{c}\~{a}o de consciência na mec\^{a}nica qu\^{a}ntica.

Neste cap\'{i}tulo, argumentei que a interpreta\c{c}\~{a}o da consciência causal\index{interpreta\c{c}\~{a}o da consciência!causal} pode se beneficiar de uma metaf\'{i}sica baseada em processos inspirada em Whitehead. A metodologia para tal tarefa \'{e} a Abordagem \textit{Toolbox} para a metaf\'{i}sica, ou seja, utilizar os dispositivos desenvolvidos pela metaf\'{i}sica anal\'{i}tica a fim de interpretar a empreitada cient\'{i}fica.

Tradicionalmente, tal interpreta\c{c}\~{a}o est\'{a} vinculada ao dualismo de subst\^{a}ncias e ao problema mente-corpo. A metaf\'{i}sica de mente baseada em processos n\~{a}o sofre do mesmo destino, o que poderia ser uma vantagem para a interpreta\c{c}\~{a}o da consciência causal\index{interpreta\c{c}\~{a}o da consciência!causal}, se compreendida dentro do proposto quadro metaf\'{i}sico baseado em processos.

Desenvolver ainda mais esse arcabou\c{c}o metaf\'{i}sico \'{e} uma tarefa deixada para futuras pesquisas no campo da metaf\'{i}sica da ciência. Ainda assim, acredito que essa proposta seja um importante primeiro passo em dire\c{c}\~{a}o ao objetivo de compreender melhor nossas op\c{c}\~{o}es metaf\'{i}sicas na interpreta\c{c}\~{a}o da mec\^{a}nica qu\^{a}ntica. Encerro este cap\'{i}tulo com a esperan\c{c}a de que tal proposta possa vir a incentivar novas pesquisas na filosofia da mec\^{a}nica qu\^{a}ntica. At\'{e} mesmo ---e aqui est\'{a} uma ideia para o seu projeto de doutorado!--- para testar se tal proposta acaba mostrando-se infrut\'{i}fera.

\chapter{Quest\~{o}es de formalismo}\label{Cap:formalismo}

\noindent Os objetos qu\^{a}nticos n\~{a}o podem ser visualizados diretamente, da mesma maneira como este livro diante de nossos olhos. S\~{a}o de tal magnitude que n\~{a}o podem sequer ser visualizados em microsc\'{o}pio. Por isso, o formalismo \'{e} de extrema import\^{a}ncia para as discuss\~{o}es sobre mec\^{a}nica qu\^{a}ntica: \'{e} somente por meio do formalismo que os objetos qu\^{a}nticos s\~{a}o tratados. O termo `formalismo', adverte \citet[p.~27]{KrauseAlgebraLinear}, conforme empregado na literatura da f\'{i}sica, designa a formula\c{c}\~{a}o matem\'{a}tica da mec\^{a}nica qu\^{a}ntica ``\textelp{} e n\~{a}o se relaciona, a princ\'{i}pio, com sistemas formais que s\~{a}o tratados em l\'{o}gica e em fundamentos da matem\'{a}tica''. Como pontuam \citet[p.~2]{susskind2014quantum}, j\'{a} que n\~{a}o somos biologicamente aptos a perceber os objetos da mec\^{a}nica qu\^{a}ntica com nossos \'{o}rg\~{a}os sensoriais, ``o melhor que podemos fazer \'{e} tentar entender os el\'{e}trons e seus movimentos como abstra\c{c}\~{o}es matem\'{a}ticas''.

Nesse preciso sentido, \citet[p.~137]{paty1995materiaroubada} considera que a mec\^{a}nica qu\^{a}ntica, ``\textelp{} uma vez estabelecida, prop\~{o}e-se, antes de qualquer interpreta\c{c}\~{a}o, como um formalismo''. Em tal formalismo, como observa Paty, os estados qu\^{a}nticos:

\begin{quote}
\textelp{} s\~{a}o representados numa formula\c{c}\~{a}o te\'{o}rica, em termos de operadores que se aplicam a vetores de estado e, para realiz\'{a}-lo, recorremos a entidades matem\'{a}ticas apropriadas. As propriedades dos objetos ou conceitos f\'{i}sicos assim designados s\~{a}o, consequentemente, determinados, de um lado, pela coerência l\'{o}gico-matem\'{a}tica do esquema e da formula\c{c}\~{a}o \textelp{}; e, de outro, pela transcri\c{c}\~{a}o das observa\c{c}\~{o}es matem\'{a}ticas em quest\~{a}o. \citep[p.~237]{paty1995materiaroubada}.
\end{quote}

De modo geral, o formalismo da mec\^{a}nica qu\^{a}ntica descreve os estados de um sistema f\'{i}sico, considerando os aspectos que podem ser medidos, chamados de observ\'{a}veis (posi\c{c}\~{a}o, momento, spin, etc.). Aqui, o termo `estado' \'{e} um conceito primitivo, meta-axiom\'{a}tico, cuja defini\c{c}\~{a}o (chamada `defini\c{c}\~{a}o operacional') \'{e} dada pelos postulados. No formalismo usual da mec\^{a}nica qu\^{a}ntica, os estados s\~{a}o representados pela no\c{c}\~{a}o de `vetor'. Uma fun\c{c}\~{a}o de onda, frequentemente notada pelo caractere grego $\psi$, onde $\psi(a,b,c\dots)$ s\~{a}o os coeficientes que se movimentam ---se expandem--- em um espa\c{c}o vetorial complexo $n$-dimensional, nomeado por \citet{vNeum1955mathematical}\index{Neumann, John von} de ``Espa\c{c}o de Hilbert'', notado pelo caractere $\mathscr{H}$, onde $\mathscr{H}\in\mathbb{C}^n$, por sua vez, \'{e} caracterizado por um conjunto de vetores chamado ``base'' do espa\c{c}o. Para \citet[p.~2]{Jammer1974}, ``a ideia de von Neumann de formular a mec\^{a}nica qu\^{a}ntica como um c\'{a}lculo de operador no espa\c{c}o de Hilbert foi, sem d\'{u}vida, uma das grandes inova\c{c}\~{o}es da f\'{i}sica matem\'{a}tica moderna''.

O formalismo, quando tomado isoladamente, sugere que a mec\^{a}nica qu\^{a}ntica trata exclusivamente do resultado de medi\c{c}\~{o}es, mantendo-se silencioso em rela\c{c}\~{a}o a no\c{c}\~{o}es tais como `realidade f\'{i}sica' e, como tal, n\~{a}o favorece nem rejeita uma ou outra interpreta\c{c}\~{a}o particular. Ainda assim, para que se possa tratar do formalismo, parece necess\'{a}rio assumir uma ``interpreta\c{c}\~{a}o m\'{i}nima'', que considera o car\'{a}ter probabil\'{i}stico da teoria qu\^{a}ntica. Hughes considera que tal atitude \'{e} uma premissa necess\'{a}ria para que a teoria qu\^{a}ntica possa ser uma teoria f\'{i}sica:

\begin{quote}
Ao desenvolver nossa representa\c{c}\~{a}o geral de uma teoria f\'{i}sica, partimos de uma suposi\c{c}\~{a}o, de que o mundo \'{e} tal que, em certas circunst\^{a}ncias especific\'{a}veis, v\'{a}rios eventos podem receber probabilidades definidas, eu considero essa suposi\c{c}\~{a}o m\'{i}nima, se quisermos ter alguma teoria f\'{i}sica: assumimos que existem liga\c{c}\~{o}es, embora apenas probabil\'{i}sticas, entre um conjunto de ocorrências (as circunst\^{a}ncias iniciais) e outro (os eventos resultantes). \citep[p.~85]{hughes1989structure}.
\end{quote}

\citet*{buschmittelstaedt1996quantum} v\~{a}o al\'{e}m, e caracterizam a ``interpreta\c{c}\~{a}o m\'{i}nima'' no sentido probabil\'{i}stico:

\begin{quote}
Na interpreta\c{c}\~{a}o m\'{i}nima, a mec\^{a}nica qu\^{a}ntica \'{e} considerada uma teoria f\'{i}sica probabil\'{i}stica, consistindo de uma linguagem (proposi\c{c}\~{o}es sobre resultados de medi\c{c}\~{o}es), uma estrutura de probabilidade (um conjunto convexo de medidas de probabilidade representando as poss\'{i}veis distribui\c{c}\~{o}es de resultados de medi\c{c}\~{a}o) e leis probabil\'{i}sticas. Al\'{e}m disso, as probabilidades s\~{a}o interpretadas como limites das frequências relativas dos resultados das medi\c{c}\~{o}es, ou seja, no sentido de uma interpreta\c{c}\~{a}o estat\'{i}stica epistêmica. \citep*[p.~4]{buschmittelstaedt1996quantum}.
\end{quote}

Ademais, \citet*[p.~8]{buschmittelstaedt1996quantum} constatam que ``essa interpreta\c{c}\~{a}o m\'{i}nima est\'{a} contida em qualquer interpreta\c{c}\~{a}o mais detalhada da mec\^{a}nica qu\^{a}ntica''. \citet[p.~44]{redhead1987inc} nomeia essa atitude de ``interpreta\c{c}\~{a}o instrumentalista m\'{i}nima'':

\begin{quote}
\textelp{} como o formalismo est\'{a} relacionado aos poss\'{i}veis resultados da medi\c{c}\~{a}o e \`{a}s frequências estat\'{i}sticas com as quais esses resultados aparecem quando uma medi\c{c}\~{a}o \'{e} repetida v\'{a}rias vezes (em princ\'{i}pio, um n\'{u}mero infinito de vezes) em sistemas preparados em estados qu\^{a}nticos idênticos. \citep[p.~44]{redhead1987inc}.
\end{quote}

Aquilo que esses autores chamam de `interpreta\c{c}\~{a}o m\'{i}nima' se relaciona com a chamada `interpreta\c{c}\~{a}o estat\'{i}stica' de Max Born\footnote{~Que n\~{a}o deve ser confundida com a `interpreta\c{c}\~{a}o dos \textit{ensembles} estat\'{i}sticos' idealizada por Einstein ---ver \citet{Home1992-HOMEIO}. A interpreta\c{c}\~{a}o estat\'{i}stica (tamb\'{e}m conhecida como interpreta\c{c}\~{a}o dos \textit{ensembles}) \'{e} tratada no cap\'{i}tulo \ref{CapPaisagem}.} que, de acordo com \citet{griffiths1995introduction}, a teoria qu\^{a}ntica fornece, dado um determinado estado, o valor de um observ\'{a}vel no intervalo $x$ e $x+dx$, em um tempo $t$. De acordo com Griffiths, essa particularidade da descri\c{c}\~{a}o qu\^{a}ntica introduz a no\c{c}\~{a}o de ``indeterminismo'' na mec\^{a}nica qu\^{a}ntica, pois:

\begin{quote}
\textelp{} se você sabe tudo o que a teoria tem a lhe dizer sobre a part\'{i}cula (a saber: sua fun\c{c}\~{a}o de onda), você n\~{a}o pode prever com certeza o resultado de um experimento simples para medir sua posi\c{c}\~{a}o ---tudo que a mec\^{a}nica qu\^{a}ntica tem a oferecer \'{e} uma informa\c{c}\~{a}o estat\'{i}stica sobre resultados poss\'{i}veis. \citep[p.~2--3]{griffiths1995introduction}.
\end{quote}

As quest\~{o}es relativas \`{a} realidade transfenomenal dos objetos qu\^{a}nticos s\~{a}o quest\~{o}es que dependem estritamente da interpreta\c{c}\~{a}o adotada, motivo pelo qual postergo tal discuss\~{a}o para as pr\'{o}ximas se\c{c}\~{o}es. Ainda que, como afirma \citet[p.~45]{redhead1987inc} teorias sem interpreta\c{c}\~{a}o ``\textelp{} simplesmente n\~{a}o contribuem para a nossa compreens\~{a}o do mundo natural'', e \citet[p.~343]{Jammer1974} ``\textelp{} um formalismo, ainda que completo e logicamente consistente, ainda n\~{a}o \'{e} uma teoria f\'{i}sica'', reitero: ater-me-ei, nesta se\c{c}\~{a}o, somente \`{a}quilo que denomino ``interpreta\c{c}\~{a}o m\'{i}nima''.

\section{A interpreta\c{c}\~{a}o m\'{i}nima}

Cada base pode ser escolhida em fun\c{c}\~{a}o de um observ\'{a}vel que se quer medir sobre o sistema em um dado estado, a partir do qual \'{e} pos\'{i}vel designar infinitos vetores, de modo que, por exemplo, para um observ\'{a}vel de posi\c{c}\~{a}o, $|\psi\rangle$ denota um coeficiente do vetor de estado na base da posi\c{c}\~{a}o da seguinte maneira:

\begin{equation}
|\psi\rangle=\langle x_{1} | \psi\rangle|x_{1}\rangle+\langle x_{1} | \psi\rangle|x_{1}\rangle+\cdots+\langle x_{n} | \psi\rangle|x_{n}\rangle
\end{equation}

Ou seja, $\langle x_j|\psi\rangle$ denota o $j$-\'{e}simo coeficiente do vetor de estado $|\psi\rangle$ na base da posi\c{c}\~{a}o. Em termos de uma densidade de probabilidade denotada por $\rho_m$, a probabilidade de que uma medi\c{c}\~{a}o efetuada sobre um observ\'{a}vel $A$ no tempo $t$ tenha como resultado o valor $a_m$ \'{e} igual a (utilizarei a nota\c{c}\~{a}o de Paul Dirac dos `bra-kets' para expressar o vetor de estado $\psi$, de modo que `$\langle\psi|$' seja um bra e `$|\psi\rangle$' seja um ket):

\begin{equation}
\rho_{m}(t)=|\langle a_{m} | \psi(t)\rangle|^{2}
\end{equation}

Uma medi\c{c}\~{a}o do observ\'{a}vel $A$ no tempo $t$ representa o valor esperado (que envolve o conceito estat\'{i}stico de `esperan\c{c}a matem\'{a}tica') $\langle A\rangle(t)$, dado pela soma das densidades de probabilidade $\rho_m$ para o resultado $a_m$ no tempo $t$, que por sua vez \'{e} equivalente ao produto interno das fun\c{c}\~{o}es de onda poss\'{i}veis, de modo que:

\begin{equation}
\langle A\rangle(t)=\sum_{m} \rho_{m}(t) a_{m}
\end{equation}

Ou mais especificamente, conforme a regra de Born, a probabilidade de se encontrar o valor da medida de um observ\'{a}vel f\'{i}sico $A$ em um sistema qu\^{a}ntico descrito por $\psi(x,t)$ em um dado intervalo $[a,b]$ de uma reta em $\mathbb{R}$ \'{e}

\begin{equation}
Prob_{[a, b]}^{\psi(x, t)}(A)=\int_{a}^{b}|\psi(x, t)|^{2} d x
\end{equation}

Como pontua \citet[p.~32]{KrauseAlgebraLinear}, isso significa que quando o observ\'{a}vel a ser medido tem dimens\~{a}o unit\'{a}ria, isto \'{e}, normalizada, a probabilidade de encontrar o sistema representado pela fun\c{c}\~{a}o de onda $\psi(x, t)$ no intervalo $[a,b]$ \'{e} dada pela express\~{a}o acima (a regra de Born). O valor $|\psi(x, t)|^{2}$ \'{e} denotado pela densidade de probabilidade $\rho(x,t)$. Reiterando: a mec\^{a}nica qu\^{a}ntica \'{e} uma teoria probabil\'{i}stica no sentido de que fornece apenas probabilidades para os estados dos sistemas qu\^{a}nticos. Como recorda \citet[p.~5--6]{KrauseAlgebraLinear}, somente os estados que obede\c{c}am a uma condi\c{c}\~{a}o de normaliza\c{c}\~{a}o s\~{a}o relevantes para a problem\'{a}tica em quest\~{a}o, visto que, ao representarem probabilidades, os escalares $x_i$ devem ter soma igual \`{a} unidade, tal que:

\begin{equation}
\sum_{i=1}^{n}|x_{i}|^{2}=1
\end{equation}

As probabilidades na teoria qu\^{a}ntica s\~{a}o dadas na forma de express\~{o}es como $|x|^{2}$ e, por isso, \'{e} importante que os coeficientes sejam normalizados, para que as express\~{o}es relativas \`{a}s probabilidades possam assumir valores entre zero e um \citep[p.~28]{hughes1989structure}.

O valor esperado \'{e} tudo o que se pode conhecer sobre um sistema qu\^{a}ntico. Como os estados que interessam \`{a} problem\'{a}tica da medi\c{c}\~{a}o qu\^{a}ntica devem ser normalizados, \'{e} necess\'{a}ria a utiliza\c{c}\~{a}o da no\c{c}\~{a}o de ``norma'', uma aplica\c{c}\~{a}o que associa um escalar a cada vetor, de modo que o vetor \'{e} unit\'{a}rio se $\|\psi\|=1$; em espec\'{i}fico, para tratar do problema da medi\c{c}\~{a}o, interessam as normas advindas do produto interno $\langle\psi | \psi\rangle$, em que:

\begin{equation}
\|\psi\|=\sqrt{\langle\psi | \psi\rangle}
\end{equation}

O quadrado da norma dessa fun\c{c}\~{a}o de onda fornecer\'{a} uma densidade de probabilidade de encontrar um sistema qu\^{a}ntico em certa situa\c{c}\~{a}o (como uma posi\c{c}\~{a}o definida para uma part\'{i}cula, por exemplo). O termo ``part\'{i}cula'' deve ser tomado com cautela, uma vez que n\~{a}o h\'{a} visibilidade ou analogia poss\'{i}vel com qualquer objeto macrosc\'{o}pico. \'{E} relevante ressaltar que, como um instrumento heur\'{i}stico, as part\'{i}culas em mec\^{a}nica qu\^{a}ntica s\~{a}o tomadas como pontos sem extens\~{a}o.

Sintetizando o que foi dito at\'{e} ent\~{a}o, pode-se afirmar que, no formalismo usual da mec\^{a}nica qu\^{a}ntica, s\~{a}o particularmente importantes as equa\c{c}\~{o}es do tipo:

\begin{equation}
T\xi=\lambda \xi
\end{equation}

$T$ representa um observ\'{a}vel de um sistema, cujo estado \'{e} representado por $\xi$, sendo $\lambda$ o valor poss\'{i}vel para a medida desse observ\'{a}vel.

\subsection{Evolu\c{c}\~{a}o temporal dos estados via Equa\c{c}\~{a}o de Schr\"{o}dinger}

Tendo esclarecido tais pontos, passo \`{a} discuss\~{a}o acerca da evolu\c{c}\~{a}o temporal dos estados dos observ\'{a}veis. Muito embora a Equa\c{c}\~{a}o de Schr\"{o}dinger\index{Schr\"{o}dinger, Erwin!Equa\c{c}\~{a}o de Schr\"{o}dinger} n\~{a}o seja a \'{u}nica equa\c{c}\~{a}o de movimento da teoria qu\^{a}ntica (embora seja a mais utilizada), de fato, o formalismo da teoria qu\^{a}ntica \'{e} sempre determinista. \'{E} not\'{a}vel que, embora a teoria qu\^{a}ntica seja essencialmente probabilista, as leis din\^{a}micas que descrevem a evolu\c{c}\~{a}o (ou movimento) temporal dos estados s\~{a}o deterministas.

A Equa\c{c}\~{a}o de Schr\"{o}dinger\index{Schr\"{o}dinger, Erwin!Equa\c{c}\~{a}o de Schr\"{o}dinger}, especificamente, \'{e} determinista no sentido de que sua solu\c{c}\~{a}o no tempo $t=0$ determina a solu\c{c}\~{a}o para todos os outros valores de $t$ (positivos ou negativos, isto \'{e}, \'{e} uma equa\c{c}\~{a}o cujo valor temporal \'{e} revers\'{i}vel). Assim, o valor da medi\c{c}\~{a}o em um observ\'{a}vel $A$ em um tempo $t$, ainda que n\~{a}o forne\c{c}a valores determinados para o estado qu\^{a}ntico $|\psi\rangle$, fornece elementos para a distribui\c{c}\~{a}o estat\'{i}stica de resultados para medi\c{c}\~{o}es futuras.

As leis din\^{a}micas da mec\^{a}nica qu\^{a}ntica s\~{a}o frequentemente expressas sob a Equa\c{c}\~{a}o de Schr\"{o}dinger\index{Schr\"{o}dinger, Erwin!Equa\c{c}\~{a}o de Schr\"{o}dinger}, cuja nota\c{c}\~{a}o \'{e} a seguinte:

\begin{equation}
i \hbar \frac{\partial|\psi\rangle}{\partial t}=H|\psi\rangle
\end{equation}

Trata-se de uma equa\c{c}\~{a}o linear, pois envolve derivadas primeiras somente, isto \'{e}, n\~{a}o envolve deriva\c{c}\~{o}es de en\'{e}sima potência; na medida em que suas vari\'{a}veis s\~{a}o fun\c{c}\~{o}es, \'{e} uma equa\c{c}\~{a}o diferencial. A constante $i \hbar$ trata-se de um coeficiente complexo expl\'{i}cito pelo n\'{u}mero $i$, multiplicada pela constante de Planck $\hbar=h / 2 \pi$, representando a constante do movimento de circunferência em $\mathscr{H}$. A taxa de varia\c{c}\~{a}o, representada pelo `$\partial$', indica uma derivada parcial cuja opera\c{c}\~{a}o $\partial / \partial t$ incide em $|\psi\rangle$ para determinar a evolu\c{c}\~{a}o temporal, fornecendo o estado da fun\c{c}\~{a}o de onda $|\psi\rangle$, isto \'{e}, suas coordenadas no tempo, de modo que tal varia\c{c}\~{a}o \'{e} igual ao c\'{a}lculo do operador de energia $H$, chamado `Hamiltoniano', multiplicado \`{a} fun\c{c}\~{a}o de onda. \'{E} relevante constatar que a Equa\c{c}\~{a}o de Schr\"{o}dinger\index{Schr\"{o}dinger, Erwin!Equa\c{c}\~{a}o de Schr\"{o}dinger}, conforme enunciada acima, \'{e} de f\'{a}cil resolu\c{c}\~{a}o para apenas uma part\'{i}cula (como na simplifica\c{c}\~{a}o do \'{a}tomo de hidrogênio), mas, em realidade, ela funciona para qualquer n\'{u}mero arbitr\'{a}rio de part\'{i}culas.\footnote{~Como advertem \citet[p.~134]{susskind2014quantum}, n\~{a}o tem conex\~{a}o direta com o comportamento ondulat\'{o}rio, sendo apenas um nome atribu\'{i}do por conven\c{c}\~{a}o. No entanto, apesar da nomenclatura corpuscular \textit{i.e.}, ``part\'{i}cula'' ser frequentemente utilizada, o pr\'{o}prio Schr\"{o}dinger advogou ---tardiamente--- uma interpreta\c{c}\~{a}o ondulat\'{o}ria, inclusive para o uso da fun\c{c}\~{a}o de onda.} Em $H$ est\'{a} previsto o potencial, que substitui a influência do n\'{u}cleo como uma ferramenta heur\'{i}stica que possibilita o c\'{a}lculo do movimento do el\'{e}tron desprezando suas rela\c{c}\~{o}es com uma segunda part\'{i}cula, \textit{i.e.}: o n\'{u}cleo.

A solu\c{c}\~{a}o da Equa\c{c}\~{a}o de Schr\"{o}dinger\index{Schr\"{o}dinger, Erwin!Equa\c{c}\~{a}o de Schr\"{o}dinger} pode admitir dois ou mais estados $|\psi\rangle$  poss\'{i}veis, cuja soma \'{e} tamb\'{e}m um estado poss\'{i}vel. Tal \'{e} o `princ\'{i}pio de superposi\c{c}\~{a}o', que de acordo com \citet[p.~23]{pessoa2003conceitos} pode ser enunciado da seguinte maneira: ``dados dois estados admiss\'{i}veis de um sistema qu\^{a}ntico, ent\~{a}o a soma desses dois estados tamb\'{e}m \'{e} um estado admiss\'{i}vel do sistema'', o que pode ser descrito da seguinte maneira:

\begin{equation}
|\psi_{AB}\rangle=|\psi_{A}\rangle+|\psi_{B}\rangle
\end{equation}

Para que os vetores sejam unit\'{a}rios, utiliza-se o fator $1/ \sqrt{2}$, chamado fator de normaliza\c{c}\~{a}o. Quando uma superposi\c{c}\~{a}o envolve certos vetores que podem assumir valores complexos $\mathbb{C}^{\textit{n}}$, introduz-se o n\'{u}mero imagin\'{a}rio $i$, tal que $i \equiv \sqrt{-1}$. Assim,

\begin{equation}
|\psi_{AB}\rangle=\frac{1}{\sqrt{2}}|\psi_{A}\rangle+\frac{i}{\sqrt{2}}|\psi_{B}\rangle
\end{equation}

Os estados acima s\~{a}o ditos estados puros, em que $|\psi\rangle$ descreve toda a informa\c{c}\~{a}o que pode ser obtida sobre o estado de uma \'{u}nica part\'{i}cula. N\~{a}o \'{e} necess\'{a}rio que os vetores dos estados em superposi\c{c}\~{a}o sejam ortogonais, isto \'{e}, vetores $|\psi_{A}\rangle$ e $|\psi_{B}\rangle$ cujo produto interno $\langle\psi_{A} | \psi_{B}\rangle= 0$; ainda assim, a ortogonalidade \'{e} utilizada em racioc\'{i}nios de situa\c{c}\~{o}es limite, sendo uma caracter\'{i}stica importante para a discuss\~{a}o acerca do gato\index{Schr\"{o}dinger, Erwin!paradoxo do gato} de Schr\"{o}dinger. Assim, supomos que os estados tratados aqui sejam ortogonais, expressos como $|\psi_{A}\rangle|\psi_{B}\rangle^\perp$. Quando os estados s\~{a}o ortogonais e normalizados, tais estados s\~{a}o chamados de ortonormais. Uma caracter\'{i}stica importante da ortogonalidade \'{e} a exclusividade de seus estados: dois estados s\~{a}o ortogonais em rela\c{c}\~{a}o um ao outro se n\~{a}o possuem o mesmo valor.

Ambos os estados $1$ e $2$ podem ser descritos separadamente como $|\psi_{A}\rangle$ e $|\psi_{B}\rangle$, ainda que sua soma dê origem a um novo estado $|\psi_{AB}\rangle$ poss\'{i}vel. \'{E} importante salientar que, no princ\'{i}pio de superposi\c{c}\~{a}o, os estados s\~{a}o fator\'{a}veis, isto \'{e}, separ\'{a}veis, sendo apenas o produto tensorial dos componentes da equa\c{c}\~{a}o, de modo que:

\begin{equation}
|\psi_{AB}\rangle=|\psi_{A}\rangle \otimes|\psi_{B}\rangle
\end{equation}

O vetor $|\psi_{AB}\rangle$ pode ser decomposto em um produto de vetores, cada um em um espa\c{c}o (possivelmente infinitos), tal que \begin{equation}\mathscr{H}=\bigotimes^{i}_{n}\mathscr{H}_n\end{equation} de modo que se pode dizer que os vetores agem independentemente. Vale ressaltar que \'{e} bastante comum a seguinte generaliza\c{c}\~{a}o: 
\begin{equation}
|\psi\rangle\otimes|\phi\rangle=|\psi\rangle|\phi\rangle=|\psi \phi\rangle\end{equation}e que os produtos tensoriais n\~{a}o s\~{a}o comutativos, de modo que: $|\psi \phi\rangle \neq|\phi \psi\rangle$. Ressalto tamb\'{e}m que neste livro uso apenas a nota\c{c}\~{a}o $|\psi\rangle\otimes|\phi\rangle$, valendo a diferen\c{c}a $|\psi\rangle\otimes|\phi\rangle\neq|\phi\rangle\otimes|\psi\rangle$.

\subsection{O colapso}

Se uma medi\c{c}\~{a}o for efetuada sobre $|\psi_{AB}\rangle$, apenas um dos estados superpostos $|\psi_{A}\rangle$ ou $|\psi_{B}\rangle$ ser\'{a} obtido. Se o estado do sistema \'{e} $|\psi\rangle=\Sigma_{j} c_{j}|a_{j} \rangle$, e se a medida fornece o valor $a_n$, ap\'{o}s a medida o sistema colapsa para o estado $|a_{n}\rangle$ com a probabilidade $|c_{n}|^{2}=|\langle\psi_{n} | \psi\rangle|^{2}$. Quando isso ocorre, o vetor \'{e} projetado de maneira descont\'{i}nua em um desses valores, chamados `autovalores'. O colapso, contudo, n\~{a}o \'{e} determinado pela evolu\c{c}\~{a}o temporal prevista pela Equa\c{c}\~{a}o de Schr\"{o}dinger\index{Schr\"{o}dinger, Erwin!Equa\c{c}\~{a}o de Schr\"{o}dinger}, sendo que a tentativa de conciliar tais dois aspectos seja uma via de abordar o problema da medi\c{c}\~{a}o, conforme explicitado acima.

Uma caracter\'{i}stica bastante importante para a presente discuss\~{a}o \'{e} que os produtos tensoriais s\~{a}o utilizados no formalismo da mec\^{a}nica qu\^{a}ntica para representar sistemas compostos, ou seja, sistemas envolvendo mais de um sistema f\'{i}sico.

Remontarei um exemplo dado por \citet[p.~52--54]{redhead1987inc} acerca de uma `medi\c{c}\~{a}o ideal' e suas problematiza\c{c}\~{o}es, conforme o esquema oferecido at\'{e} aqui. Suponha que $Q$ \'{e} um observ\'{a}vel com um espectro discreto $\{q_{i}\}$. Suponha que o estado de um sistema qu\^{a}ntico $S$ \'{e} um autoestado $|q_{i}\rangle$ de $Q$, e que $S$ interaja com um aparato de medi\c{c}\~{a}o $A$. Suponha, ainda, que o autoestado de $A$ seja $|r_{0} \rangle$ na quantidade $R$, e que o autoestado de $A$ passe, em decorrência da intera\c{c}\~{a}o, de $|r_{0} \rangle$ para $|r_{i} \rangle$, ao passo que $S$ permane\c{c}a em $|q_{i}\rangle$. Assim, o sistema composto $S+A$ vai de $|q_{i}\rangle|r_{0}\rangle$ para $|q_{i}\rangle|r_{i}\rangle$ ap\'{o}s a intera\c{c}\~{a}o. Como observ\'{a}veis do sistema composto, os operadores para $Q$ e $R$ devem ser designados por $Q\otimes I$ e $I\otimes R$ respectivamente, onde o primeiro produto tensorial corresponde ao sistema $S$ e o segundo a $A$.

O estado inicial da situa\c{c}\~{a}o proposta, denotando que o estado de $S$ \'{e} uma superposi\c{c}\~{a}o de autoestados de $Q$ com amplitude de probabilidade $c_i$, \'{e} dado por:

\begin{equation}
|\psi\rangle=\left(\sum_{i} c_{i}|q_{i}\rangle\right)|r_{0}\rangle
\end{equation}

Dada a linearidade da evolu\c{c}\~{a}o temporal do sistema, em que se sup\~{o}e que todos os $r_i$ s\~{a}o distintos, tem-se que:

\begin{equation}
|\psi^{\prime}\rangle=\sum_{i} c_{i}|q_{i}\rangle|r_{i}\rangle
\end{equation}

Em termos de operadores estat\'{i}sticos, antes da medi\c{c}\~{a}o, o operador para o sistema composto \'{e}:

\begin{equation}
W=P_{|\psi\rangle}=P_{(\sum_{i} c_{1}|q_{1}\rangle|r_{0}\rangle.}
\end{equation}

Ap\'{o}s a medi\c{c}\~{a}o \'{e} o estado puro:

\begin{equation}
W^{\prime}=P_{\sum\limits_{i}^{} c_i|q_i\rangle|r_i\rangle}
\end{equation}

O valor esperado seria:

\begin{equation}
W^{\prime \prime}=\sum_{i}|c_{i}|^{2} P_{|q_{i}\rangle|r_{i}\rangle}
\end{equation}

Nesse caso, $W''$ \'{e} um estado misto que descreve um ensemble de sistemas nos estados $|q_{i}\rangle|r_{i}\rangle$, tal que a probabilidade de achar o estado $|q_{i}\rangle|r_{i}\rangle$ na mistura seja $|c_{i}|^{2}$.

\'{E} importante ressaltar que o formalismo da mec\^{a}nica qu\^{a}ntica \'{e} muito mais rico e complexo do que foi apresentado neste breve apêndice. Essa sucinta apresenta\c{c}\~{a}o serve exclusivamente para uma compreens\~{a}o mais aprofundada das quest\~{o}es filos\'{o}ficas tratadas neste livro.

\printbibliography[title={Refer\^{e}ncias~bibliogr\'{a}ficas},heading=bibintoc]	

\printindex

\end{document}